\documentclass{aa}
 \usepackage{graphicx}
  \date{[Draft compiled for AA]}
  \begin{document}


 \title{A tentative double-jet model for blazar OJ287}
 \author{S.J.~Qian\inst{1}}

 \institute{National Astronomical Observatories,
    Chinese Academy of Sciences, Beijing 100101, China}

  \abstract{We try to propose a relativistic jet model for 
  explaining the entire radio-optical phenomena observed in blazar OJ287, 
  which has been observed quasi-periodically with a cycle of $\sim$12\,yr
   in its optical light curve.}
   {We investigate the currently available theoretical and observational studies
   on the phenomena observed in OJ287
  and try to find clues to its double-jet structure.}{It is found that the 
  kinematic features of its superluminal components observed at 
   43\,GHz and 15\,GHz  could be well interpreted in terms of a precessing 
  double-jet nozzle model with non-ballistic trajectories. And the light 
  curves of a few double-peaked optical outbursts  could  be interpreted in
  terms of relativistic jet models.} 
  {Both jets precess with the same period of 12\,yr, equal to 
   the optical period and the precession of the jets could be originated from
  the orbital motion of the binary. Thus the masses of the binary black holes
  could be estimated.}{We also tentatively suggest a comprehensive framework 
  for understanding the entire phenomena in OJ287 in terms of relativistic jet
   models.}
  \maketitle
  
  \section{Introduction}
   Research on blazars is an important field of extragalactic astrophysics.
   A number of prominent blazars (e.g., 3C279, 3C345, 3C454.3, 1510-089,
   S5 0836+710, OJ287, BL Lacertae etc.)
   have been studied through extensive observations
   of their emission at multi-wavelengths from radio through to
  $\gamma$-rays and very long-baseline interferometry (VLBI)
   observations of their relativistic jets. It is found that studies on
   the connection
   between the inner jet kinematics (formation and evolution of
   superluminal components) and broadband (flux and polarization) variability
   is particularly useful for understanding the
    mechanisms of blazar radiation and determining the location of emission
    regions in the relativistic jets.

    Relativistic jets of blazars may be  produced by the
    supermassive black hole-accretion disk systems existing
   in their nuclei and thus studies on relativistic jets
    of blazars involve black hole physics and general relativity.

  OJ287 (z=0.306) is one of the prominent blazars which have been extensively
   studied. It is a low synchrotron peaked BL Lacertae object (BLO). It 
  radiates across the entire electromagnetic spectrum from
  radio through optical and X-ray to $\gamma$-rays. Its emissions in all 
  wavebands are strongly variable in brightness and polarization 
   with timescales of hours/days to years.

   OJ287 is one of the
  bright Fermi $\gamma$-ray sources (Nolan et al.\cite{No12},
   Ackermann et al. \cite{Ac11}). Multifrequency
  observations and  studies of its spectral energy distribution (SED) have been
  used to investigate the emission mechanisms of its X-rays and $\gamma$-rays.
   Studies of the correlation between its $\gamma$-ray emission and the 
 emergence of superluminal components have shown that $\gamma$-ray outbursts
 can be produced in the core region or in stationary features away from the 
 core (Agudo et al. (\cite{Ag11b}, Hodgson et al. \cite{Hod17}). OJ287 has a
    X-ray jet on Mpc scale discovered by Marscher \& Jorstad (\cite{Ma10}).

   OJ287 has been monitored for long times at multi-frequencies, especially in 
  optical bands (e.g., Agudo et al. \cite{Ag11a}, \cite{Ag11d},
  Villforth et al. \cite{Vil10}, Valtaoja et al. \cite{Val20},
   Sillanp\"a\"a et al. \cite{Si88},
  Marscher \& Jorstad \cite{Ma10},
   Valtonen \& Sillanp\"a\"a \cite{Va11},
   Valtonen et al. \cite{Va09};
  Ciprini et al. \cite{Ci07}, Agudo et al. \cite{Ag11b},
  Agudo et al. \cite{Ag11c}). Its optical variability behavior is remarkable
  and extraordinary, showing that
   optical outbursts in OJ287 occurred quasi-periodically with a period
   of $\sim$12\,yr. Up to now five periodic outbursts with double-peaked 
   flares have been observed
   in 1971--73, 1983--84, 1994--96, 2005--2007 and 2015--2019 (Valtonen et al.
    \cite{Va16}).\footnote{According to Valtonen et al. (\cite{Va16}) the 
    second flaring will appear in $\sim$2019.7 with a time interval of 
   $\sim$3.8\,yr rather than $\sim$1-2\,yr for the previous double-peaked
    outbursts.} The long-lasting quasi-periodicity is believed to be related to
    the orbital motion of the binary black holes in its nucleus. A number of
   models have been proposed to interpret this phenomenon (e.g.,
   Sillanp\"a\"a et al. \cite{Si88}, Lehto \& Valtonen \cite{Le96}, Katz
  \cite{Ka97}, Sundelius et al. \cite{Su97}, Villata et al.
   \cite{Villa98}, Valtaoja et al. \cite{Val20}, Villforth et al.
  \cite{Vil10}, Tanaka \cite{Ta13}, Qian \cite{Qi15}). Until now, 
    the precessing binary black
   hole model proposed by Lehto \& Valtonen (\cite{Le96})
   (and its improved versions) may be the
    most detailed one  to interpret the optical double-peaked
    outbursts observed in OJ287. The authors assumed that the  secondary 
    black hole impacts into the accretion disk of the primary hole twice per
   pericenter passage, causing the 12\,yr periodicity and 1--2\,yr 
   time-intervals.
   This model requires (i) a large inclination of the orbital plane 
   (${>}50^{\circ}-90^{\circ}$); (ii) a total mass
   of ${\geq}{10^{10}}{M_{\odot}}$ with a mass ratio m:M $\sim$0.007:1; (iii)
   an orbital eccentricity e$\sim$0.7. Recently, Tanaka 
   (\cite{Ta13}) proposed  an alternative model to explain the optical double
   peaked outbursts, suggesting that these outbursts are originated from
   the cavity-accretion processes in the binary system. This model has two
   substantial differences from the disk-impact model of Lehto \& Valtonen
   (\cite{Le96}): (i) the total mass of the binary is in the order of
    ${10^9}{M_{\odot}}$ with a mass ratio m:M=0.25:1; (2) the orbital plane
    is coplanar with the 
   circumbinary disk. 

    Based on VLBI-observations, OJ287 has a core-jet structure with
   superluminal knots  emerging from the core steadily.  The structure
    and the kinematics of the jet are very complex: It has both stationary
    and superluminal features. Some superluminal knots  move
   via the stationary ones. Jet position angle swings (with long-term 
   and short-term time-scales)
    have been observed and studied (e.g.,
    Agudo et al. \cite{Ag12}, Tateyama \& Kingham (\cite{Ta04}, Mo\'or et al.
   \cite{Mo11}, Katz \cite{Ka97}, Valtonen \& Pihayoki \cite{Va13},
   Valtonen \& Wiik \cite{Va12}, Cohen \cite{Co17}). 
   But no certain results of the jet precession have been obtained.
    
    The correlation between the optical and radio variability and the 
   connection between the optical outbursts and the emergence of superluminal
   radio components from the core have been extensively studied (e.g, Tateyama 
   et al. \cite{Ta99}, Valtaoja et al. \cite{Val20}, Villata et al. 
   \cite{Villa98}, Vicente et al. \cite{Vic96}). However, it has been proved
   that this is a difficult task. Only recently, Britzen et al.
    (\cite{Br18}) re-analyzed the
    MOJAVE data (Monitoring of Jets in Active galactic nuclei with Very
    Long Baseline Array Experiments, Lister et al. \cite{Li09}) and found
   for the first time that the kinematics of the superluminal components can be
   explained in terms of jet precession plus rotation and the optical and 
   radio light curves can be interpreted in terms of geometric effects under
   a  black hole binary scenario. 

    In this paper we perform model fittings of the kinematics of the radio
    superluminal components and the light curves of optical outbursts 
    in OJ287, and discuss an alternative 
   possibility that OJ287 might have a double-jet structure, providing some
   new viewing aspects on the OJ287 phenomena. Observational data at optical, 
   43\,GHz and 15\,GHz are collected from the literature (Valtonen et al. 
   \cite{Va08}, Agudo et al. \cite{Ag12}, Tateyama et al. \cite{Ta99},
   Britzen et al. \cite{Br18} and others).
     \section{Clues to a double jet and precession}
    We first investigate the theoretical and observational clues to the
   jet precession and double jet structure in OJ287. 
   \subsection{Theoretical clues}
    We first discuss the theoretical clues for double precessing jets
    in blazars.

    (1) According to the $\Lambda$-cold dark matter cosmological paradigm,
    galaxies grow hierarchically through mergers.  (Kormendy \& Richtone
     \cite{Ko95}, Ferrarese \& Ford \cite{Fe05}).
    Hierarchical structure formation
    inevitably leads to the formation of supermassive binary black holes
    with subparsec separation in galactic nuclei (Roedig et al \cite{Ro12},
    Cuadra et al. \cite{Cu09}, Shi et al. \cite{Sh12}, Hayasaki et al.
   \cite{Ha08}, D'Orazio et al. \cite{Do13}). If there is sufficient 
   accretion onto both SMBHs, then two jets would
    be formed and the source would become a blazar if these jets occasionally
    direct toward us.

    (2) The optical light curve of OJ287 shows prominent flarings with roughly
    12\,yr cycles. This quasi-periodicity has been suggested to be
    related to the orbital motion of a binary black hole at its center.
    Thus OJ287 is  one of the most prominent  candidates of black hole binary.
    If both holes produce a relativistic jet forming a double jet 
   structure, there will naturally exist two groups of superluminal knots
    ascribed to the two jets, respectively.
   
    (3) According to the interpretation of the double-peaked optical 
    outbursts in terms of the precessing binary model, Valtonen et al.
    (\cite{Va16}) found that the strong thermal optical outburst in 2015.87
    was followed by a synchrotron flare within $\sim$20 days. Since the
   impacting of the secondary hole onto the primary hole disk occurred near its
    apocenter passage located at about 20,000\,AU (0.1\,pc) away from the 
   primary hole, thus this short time delay seems implying that the 
   synchrotron flare was originated from the relativistic jet produced by the
    secondary black hole. Such an interpretation might be regarded as 
    a useful clue to the double jet structure in OJ287, although the 
   disk-impact model proposed by Lehto \& Valtonen (\cite{Le96}) still 
    needs to be confirmed. 
 
     (4) A further clue comes from the studies of the origin of the
    precursor flares occurred before the major double-peaked outbursts in the
    optical light curve (Kidger et al. \cite{Kid95}, Valtonen et al.
     \cite{Va06}, Pihajoki et al. \cite{Pi13a}, \cite{Pi13b}). These authors
    suggested that the prominent flares prior to the major double-peaked
    outbursts are produced in the jet of the secondary black hole. 
    Pietil\"a (\cite{Pie98}) investigated the essential aspects of the 
   precessing binary model: the kinematics of the binary orbital motion and 
   the disk impactings. The author suggested that the secondary black hole may
    be a source emitting optical synchrotron, because it undergoes enhanced
    accretion when it penetrates the primary disk and passes the pericenter,
    leading to the formation of a temporary accretion disk and jet in the
    secondary hole.

   (5) Villata (\cite{Villa98}) proposed a double jet model to 
    interpret the 
    quasi-periodic optical double-peaked outbursts, assuming that the two 
    relativistic jets are created by the putative binary black holes. This
    model simulated the observed light curve quite well. This is the first 
    model attempting to interpret the phenomena in OJ287 in terms of a double
    jet scenario.
   
   (6) As a supplementary clue,
    Qian et al. (\cite{Qi18}) find some evidence of the possible
    existence of a double-jet structure in blazar 3C279 through the
    model-fitting of the kinematics for its thirty-one superluminal components
    in terms of a precessing double-jet nozzle model. Having obtained 
    some new significant findings for the QSO-type 
   blazar 3C279, we may pose the question: if the BLO-type blazar OJ287 could
    also  have a double-jet structure? We shall try to discuss this possibility
   in this paper.
    \subsection{observational clues}
     We shall summarize the various results from the previous
   studies on  multi-frequency variability  and VLBI observations of the
    jet in OJ287 and show some observational clues for the existence
   of a double jet and the 12\,yr precession.
   \begin{figure}
   \centering
   \includegraphics[width=6cm,angle=-90]{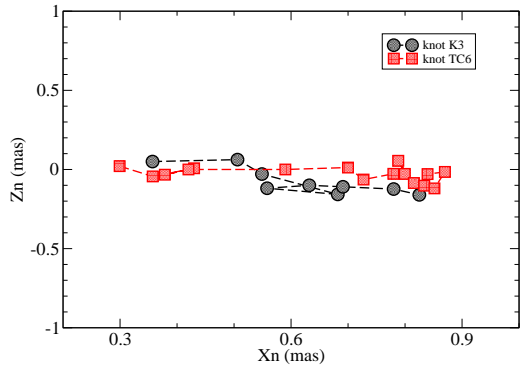}
    \caption{Quasi-periodic ejections of superluminal components: knot-K3
   ejected at 1984.0 (Vicente et al. \cite{Vic96}) and knot-C6 ejected
   at 1995.4 (Tateyama et al. \cite{Ta99}),
    having very similar trajectories with a
   ejection time difference of $\sim$11.4\,yr, closely similar to the optical
    variability period (Sillanp\"a\"a et al. \cite{Si88}).}
   \end{figure}
  
  (1) The ejection of the superluminal components in OJ287 has been monitored
    for a quite long period. Similar trajectories have been registered
    for ejection times different by 11--12\,yrs. An example is shown in 
    Figure 1: Vicente
   et al. (\cite{Vic96}) reported the ejection of knot K3 which was ejected at
   1984.0. Tateyama et al. (\cite{Ta99}) reported a superluminal knot
    (designated as TC6 here)
   ejected at 1995.4. Interestingly, the two knots moved along very
   similar position angles and trajectories with an ejection time difference of
   $\sim$11.4\,yr, almost equal to the optical period of $\sim$11--12\,yr found
   in the optical light curve. 
  
   (2)  Tateyama \& Kingham (\cite{Ta04}) investigated the VLBI structure of
    OJ287 at 8\,GHz during the period [1994,2002], showing that the jet
    position angle swung in the range (from ${\sim}{-90^{\circ}}$ to
    $\sim{-124^{\circ}}$). They interpreted this behavior in terms of
     a ballistic precessing jet model with a precession period of
    $\sim$11.6\,yr, similar to the optical variability period. This precessing
    jet model was confirmed by Mo\'or et al. (\cite{Mo11}) and Piner et al.
    (\cite{Pin07}) for the similar period (before 2002). 

   (3) The phenomenon of "sudden jump of JPA (jet position angle)"
    observed in OJ287 could be another clue to the activity of a double jet.
    Based on the 43\,GHz VLBA (Very
    Long Baseline Array) observations during [1995, 2011], Agudo et al. (2012)
     found a sharp swing of the jet PA (position angle) in 2004
   from PA$\approx{-140^{\circ}}$ to PA$\approx{-20^{\circ}}$. 
   They explained this phenomenon as the innermost jet having a very small
     viewing angle and swinging across the line of sight during a relatively
     short time interval (${<}$1\,yr). However, following the method of Jorstad
   et al. (\cite{Jor05}), they derived the viewing angle
    range ${\theta}_{var}$=[$0.7^{\circ}$, $3.4^{\circ}$], which did not
   verify the crossing of the jet from one side of the line of sight 
   to the other. Similar position PA jumps were observed  by D'Arcangelo et al.
   (\cite{Da09}) and Britzen et al. (\cite{Br18}).
    As for the explanation of the jet position angle jump 
   observed in 3C279, we interpret this behavior in OJ287 as the ejections of 
   superluminal components from two jets along different directions. 

    (4) Tateyama (\cite{Ta13}) investigated the structure of the
   inner jet of OJ287 at 15\,GHz (during time-interval 1995--2012) in
    a super-resolution mode (Tateyama et al. \cite{Ta99}) along with
    the 43\,GHz VLBA maps. The author identified the core position different
   from that identified by Agudo et al. (\cite{Ag12}) and suggested that
    the innermost jet  has a southeast/core/northwest configuration of a
    fork-like shape. Superluminal knots are ejected from the core via 
   stationary components in both southeast and northwest directions, forming 
   a very broad opening angle (or jet cone aperture) at its base. 
   This study implicitly indicated that two jets may possibly be
    produced at the center of OJ287 and is fully consistent with our
    assumptions in our double jet scenario below. 
    The broad fork-like jet morphology is extraordinary and could be formed
   by the superposition of two parabolic jets (e.g.,
   in M87, Hada et al. \cite{Had11}, Asada \& Nakamura \cite{As12}, Doeleman
   et al. \cite{Do12}, Nakamura \& Asada \cite{Na13} for single jets; also
    Polko et al. \cite{Po13}).

  (5) Cohen (\cite{Co17}) investigated the jet structure and analyzed
     the evolution of the jet ridgelines, based on the 15\,GHz VLBA images
     obtained during the time interval [1995, 2015]. The author
    suggested that the jet of OJ287 is rotating with a period of
     $\sim$30 years. Although the jet ridgelines do not reflect the real
    distribution of the jet flows in the 3-dimensional space (distinct from the
    trajectories of the superluminal components), the observed distribution and
    evolution of the jet ridgelines seems  revealing some clues for a double
    jet structure. For example, (i) The observed distribution of the ridgelines
    show two bundles of ridgelines (northern and southern bundles)
    within a core separation of $\sim$1.2\,mas which are roughly
    divided by the "source axis" (designated by Cohen \cite{Co17})
    at a position angle
    $\sim{112^{\circ}}$  with a gap where the density of ridgelines is
     very low (Figure 2 of that paper). This division of ridgeline
    bundles  is also prominent in Figure 4 of Britzen et al. (\cite{Br18});
    (ii) Cohen (\cite{Co17}) showed that
    the jet ridge close to the core appears to split into  two ridges
    at different position  angles and sustained for several years;
    (iii) The 15\,GHz polarization map (Figure 5 of that paper) shows two
    highly-polarized regions respectively at the northern ridge and southern
    ridge. Cohen suggested that
    the southern highly-polarized ridge  may be produced by a separate jet.
   Thus the division of the ridgeline bundles within
   separation $<$1.2\,mas could represent a double jet structure. And the 
   superposition of the two ridgeline bundles leads to an apparent rotation 
   period of $\sim$30\,yr, as suggested by Cohen (\cite{Co17}). 

   (6) Hodgson et al. (\cite{Hod17}) reported that the 43\,GHz
   VLBI-observations during the period of 2007--2013 reveal two strong, 
   compact and highly variable stationary
   components:  southern-most component-C and  northern
   component-S. They noted that there exists two trajectories
   at position angles different by $\sim{100^{\circ}}$. If taking component-C
   as the only core, then the ejection direction of the superluminal components
   would have changed from --$110^{\circ}$ to --$10^{\circ}$ during a
   period of $<$1\,yr. Alternatively, this sudden ``jet PA Jump'' 
   could imply a double jet structure in OJ287: both component-C and
   component-S are jet cores, from which superluminal components are 
   ejected respectively. Thus the problem of differential trajectories
    could  be removed: knot-X1 ejected from core C and knot X2/X3 from core S.
    Observational facts
   given in Hodgson et al. seem to support
   this assumption: (i) Component-S and component-C have very similar
   properties in compactness and brightness temperature, spectrum (15--86\,GHz),
   and flux/spectral variability; (ii) both were regarded as standing shocks 
   with some wanderings; (iii) both have similar  $\gamma$-ray activities. 
   This putative double-core structure could be understood in terms of the
    "Phoenix fire mechanism" proposed by Meier (\cite{Mei13}), if the 
    source has two jets.
     
   (7) Krichbaum et al. (\cite{Kr13}) analyzed
    the  parsec scale jet structure observed in October 2009 at
   15, 43, and 86\,GHz. They found that at 15 and 43\,GHz, OJ287 shows a bent
   core-jet structure curved toward the southwest direction. At 86\,GHz,
   its core region reveals a double structure consisting of component C1
    and component C2. According to this bent jet structure the northern
   component C1 should be identified as the core, because it is the
   unresolved base of a synchrotron self-absorbed jet. However, the southern
    component C2 has a more inverted spectrum, implying that C2 might be the
    core. Using the 43GHz and 86\,GHz data given in Hodgson et al.
    (\cite{Hod17}) we found that during the period
   of 2007.45--2010.75, in 65\% cases component C1 were  stronger and  had 
   smaller angular sizes (thus higher brightness temperatures) than component
    C2, suggesting that component C1 should be more qualified as the core. 
   Thus the dilemma in the identification of the core might be regarded as 
   a clue to the double core structure in OJ287.
   
   (8) As a supplementary clue,
   Qian et al. (\cite{Qi18}) have recently found evidence of
   a supermassive black hole binary with two radio jets in blazar 3C279. They
   showed that the parsec-scale kinematics of 31 superluminal knots could well
   be  interpreted in terms of a  precessing double jet-nozzle model: the jets
   produced by the primary and secondary supermassive black holes are
   precessing with the same precession period of $\sim$25\,yr. The
   model-predicted properties of the relativistic jet produced by the
   secondary black hole are well consistent with those of its kpc-jet
   counterpart and the milliarcsecond jet observed in 1970s and 1980s, 
   strongly supporting the precessing double jet
   scenario.
 
   (9) In this paper we would perform detailed analysis and
   model-fittings of the parsec-scale kinematics for the 14 superluminal
   components observed in OJ287 and show that a double jet scenario can well
  interpret the source kinematics and can provide important information on the
   the physical processes occurring in OJ287. We emphasize that in both 
   3C279 and OJ287 the properties of double-jet structure are consistent with
   the HD/MHD theories and numerical simulations for the cavity-accretion
  in binary black hole systems.
   \section{Working assumptions}
   The arguments given above indicate the possibility that
   OJ287 may harbor a black hole binary and both holes (primary and secondary)
   produce a respective relativistic jet, which ejects superluminal
   components. A schematic plot for the double jet structure is shown in
   Figure 2. In the left panel the two jets have a single (common) apex which
   is designated as k-component in Britzen et al. (\cite{Br18}).
     In the right panel the two jets have
   their respective apex: one is at k-component in BFW18 (for the northern jet)
    and the other is at C-component (for the southern jet) designated in
   Hodgson et al. (\cite{Hod17}).\footnote{We assume that the mean position of
   the S-component (Hodgson et al. \cite{Hod17}; 43\,GHz) relative
   to the C-component is
    ($X_n$=0.091\,mas, $Z_n$=0.173\,mas). We have found that the mean
    position of the S-component relative to C-component is coincided with
    the position of the k-component.} These schemes represent two comparatively
    extreme cases to investigate the double jet structure. We have performed
   the model-fitting of the source kinematics under both schemes and
    obtained very similar results. Thus for brevity and clarity, here we
   will only present the model-fitting results for the first scheme:
   two jets with a single apex. The mode fitting results for the second scheme
   will be presented elsewhere.

    \begin{figure*}
    \centering
    \includegraphics[width=5cm,angle=-90]{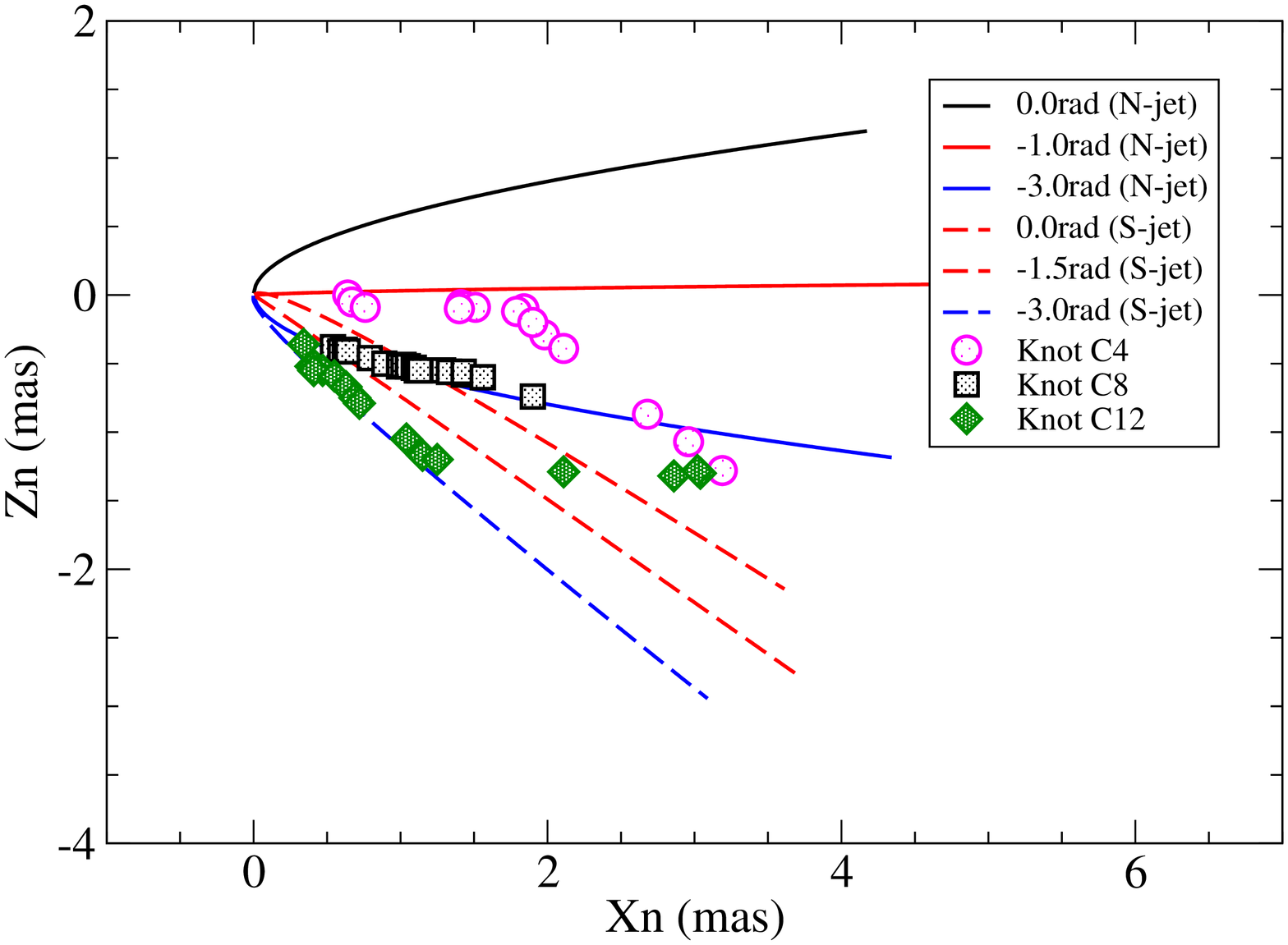}
    \includegraphics[width=5cm,angle=-90]{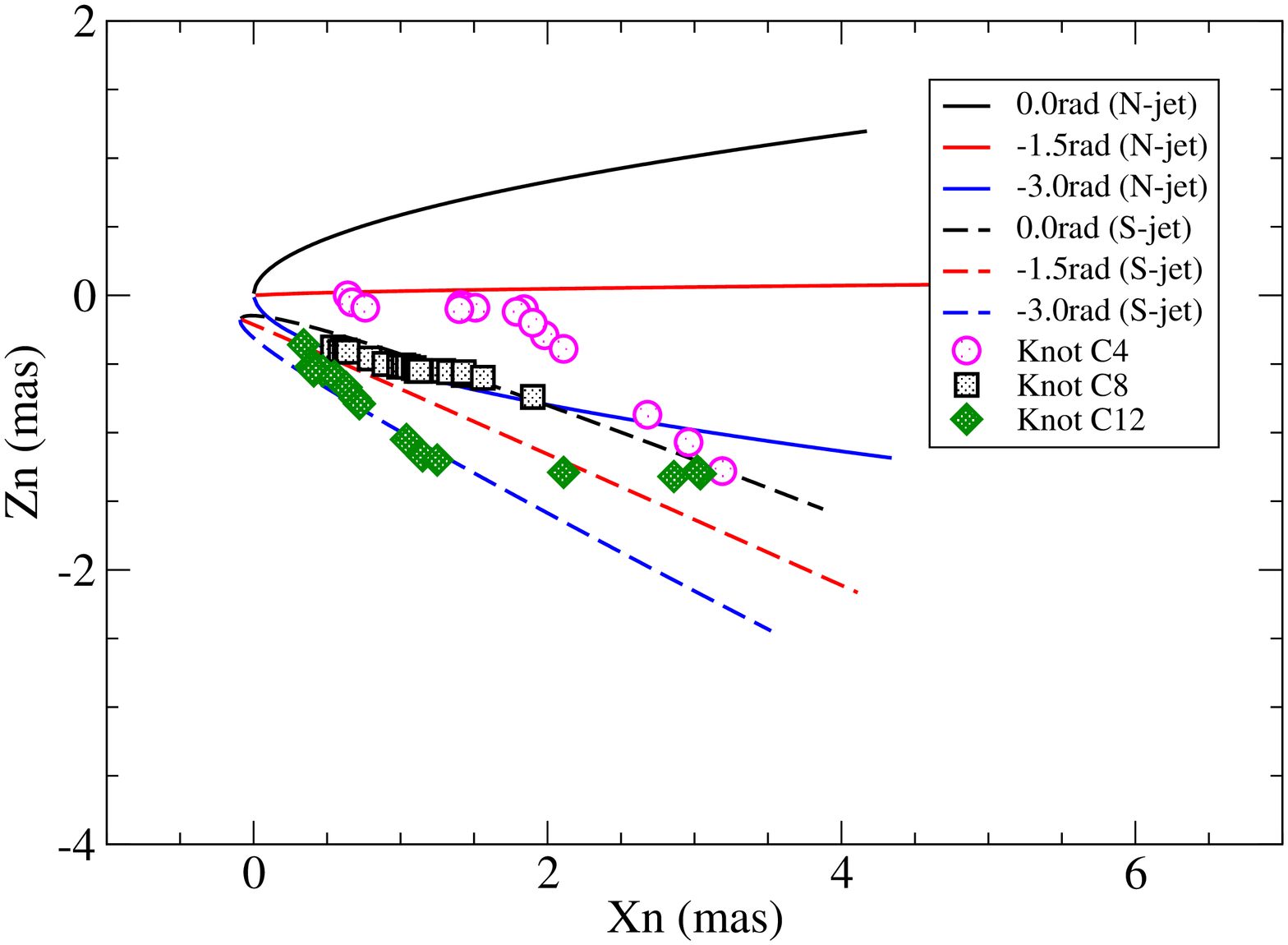}
    \caption{Two schemes for the double jet structure in
    blazar OJ287. Left panel: two jets with a single apex. Right panel:
   two jets with different apexes. Solid lines represent the northern jet
    cone and the dashed lines denote the southern jet cone. Numbers indicate
   the precession phases of the modeled trajectories. The observed 
   trajectories of the knots (C4, C8 and C12) are marked by the symbols.}
    \end{figure*}
  We will divide the fourteen superluminal components into two groups,
  ascribed to two jets: northern jet and southern jet, respectively.
   Model fitting of the kinematics will be performed for each group
  in terms of the  precessing nozzle model, which originally proposed by
  Qian et al. (\cite{Qi91}) and has been applied to study the parsec-scale
  kinematics for a few blazars (e.g., 3C345, Qian et al. \cite{Qi09}; 3C279,
  Qian \cite{Qi13}, \cite{Qi19}; 3C454.3, Qian et al. \cite{Qi14}; NRAO 150,
  Qian \cite{Qi16};  B1308+326, Qian et al. \cite{Qi17}; and PG1302-302,
   Qian et al. \cite{Qi18}).

   The precessing nozzle model
   contains a number of assumptions: (1) superluminal components are ejected
   from the jet nozzle and move along the jet axis which may
    be described as
   having rectilinear, helical or parabolic shapes; (2) the jet axis precesses
   with a certain precession period and the knots ejected at different times
   are moving along a precessing common trajectory; (3)  The change of the
    ejection direction and trajectory of the knots lead to the observed
   evolution of their position angle; (4) The structure and evolution of
   the whole jet (jet body) exhibited on the VLBI maps
    are constructed by the distribution of the isolated knots sequentially
   ejected by this precessing nozzle; (5) In general, this regular kinematic
    pattern can only be applied to the inner jet regions and the outer
   trajectories  may deviate from this regular pattern with curvatures
   occurring at different separations; (6) in addition to the ejection of
   superluminal knots from the precessing nozzle, magnetized plasmas could 
   also be ejected from the precessing nozzle. The assembly of the kinematics 
    and brightness evolution of these superluminal knots and magnetized 
   plasmas will form the entire jet structure revealed on VLBI-maps; 
  (7) Following the MHD
   (magnetohydrodynamic) nozzle
   model proposed by Nakamura \& Asada (\cite{Na13}), we will assume that
   the precessing common trajectory has a parabolic form, that is, we will
   utilize a parabolic precessing nozzle model. The double jet structure is
   specified through the trial model fittings described in the text and shown
   in Figure 2.

     Precessing jet models have been investigated and applied to interpret
  the parsec-scale kinematics in blazars by 
   many authors (e.g., Stirling et al. \cite{St03};
     Roland et al. \cite{Rol08}; Britzen et al. \cite{Br18}; Qian et al.
    \cite{Qi18} and others). While most of these models deal with ballistic 
   motions in the sources, our precessing jet models (including the precessing
   double-jet model for 3C279 and OJ287 (this paper)) deal with non-ballistic 
   (helical or parabolic) motions.

    We point out that the results obtained in this paper are only tentative
   and our aim is to demonstrate an alternative
   scenario to understand the VLBI-kinematics in OJ287.
   Our study may have some advantages, helping to solve some unsolved
   issues about the jet structure, jet precession, the connection of the
  optical outbursts with the emergence of the superluminal components and
  others, leading to a deeper understanding of the entire phenomena in OJ287.
   \section{Formalism of  Model Simulation}
       \begin{figure*}
    \centering
    \includegraphics[width=8cm,angle=0]{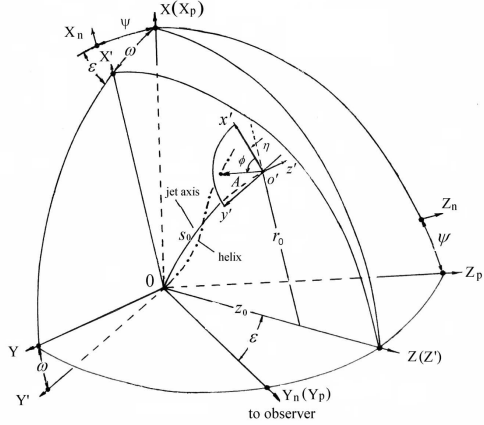}
    \caption{Geometry of the precessing nozzle model. Five
    coordinate systems are introduced. Z-axis denotes the precession
     axis (defined by parameters ($\epsilon$, $\psi$))
    around  which the jet axis (denoted by curve $s_0$ and defined by
     a function $r_0(z_0)$) precesses. In the observer system
    ($X_n$, $Y_n$, $Z_n$) the knot motion
    is defined by parameters ($\epsilon$, $\psi$, $\omega$, $r_0$, $z_0$).}
    \end{figure*}
    In order to investigate the kinematics of the superluminal components in
    blazar OJ287 in terms of a parabolic precessing jet model, we have to 
    introduce a special geometry.  Qian et al.
    (\cite{Qi91}) proposed  a precessing helical jet model, which has been
    applied to study the trajectory distribution/evolution  and jet swing in
    a few blazars: e.g., 3C345 (Qian et al. \cite{Qi09}), 3C454.3 (Qian et al.
    \cite{Qi14}), 3C279 (Qian \cite{Qi12}, \cite{Qi13}, Qian et al.
    \cite{Qi19}), NRAO 150
     (Qian \cite{Qi16}), B1308+326 (Qian et al. \cite{Qi17}), PG 1302-102 (Qian
    et al. \cite{Qi18}).
    Here we  further generalize the model as follows.

     Five coordinate systems are introduced: ($\rm{X_n,Y_n,Z_n}$),
     ($\rm{X_p,Y_p,Z_p}$), ($\rm{X,Y,Z}$),
    ($\rm{X',Y', Z'}$)  and ($\it{x',y',z'}$).
     The geometry of the model  is shown in Figure 3.
   $\rm{Y_n}$($\rm{Y_p}$) axis directs toward the observer.
   The planes ($\rm{X_n,Z_n}$) and ($\rm{X_p,Z_p}$) define
     the plane of the sky with
     $\rm{X_n}$-axis directing the
   negative right ascension and $\rm{Z_n}$-axis the north pole. The angle
   between the $\rm{X_n}$ and $\rm{X(X_p)}$ is $\psi$. We assume that the
    jet-axis locates in the plane ($\rm{X',Z'}$) and is described by
    a function $r_0(z_0)$
    which is assumed to be a parabolic function. Here we choose a parabolic
    shape for the precessing jet axis, following  Asada \& Nakamura's
   observations of the giant radio galaxy M87 and their modeling results
   (Asada \& Nakamura \cite{As12}, Nakamura \& Asada \cite{Na13}, Polko et al.
   \cite{Po13}):
    \begin{equation}
     {r_0}={a}{{z_0}^{x}} ,
    \end{equation}
    $a$ and $x$ are constants.\footnote{Nakamura \& Asada
    (\cite{Na13})
    take $x $=1.0--0.5 for their magnetic nozzle model
    and Polko et al. (\cite{Po13}) adopt $\alpha$=3/4 in their model
    simulations.}
    Angle $\omega$(t) between the plane ($\rm{X,Z}$) and
    plane ($\rm{X',Z'}$) represents the precession of the jet axis around the
    $\rm{Z}$-axis. The angle between $\rm{Y_n(Y_p)}$ and $\rm{Z(Z')}$ is
    $\epsilon$,
    describing the viewing angle of the of precession axis $\rm{Z(Z')}$.
    The precession axis is defined by parameters ($\epsilon$, $\psi$).

    Generally, helical motion of a knot around the precessing jet axis can be
    described by the parameters ($A(s_0)$, $\phi(s_0)$) in the
     coordinate system
    ($\it{x',y',z'}$).
     The $\it{z'}$ axis is along the tangent to the jet axis and
   the ($\it{x',y'}$) plane is perpendicular to the local jet axis.
    $\phi$
   represents the phase of the helical motion of a knot.
    The trajectory of a
   superluminal knot is described in cylinder coordinates
    ($Z,A(s_0)$, $\phi(s_0)$):
   $Z$ -- distance from the origin along the precession axis (Z$\equiv$$z_0$).
    $A(s_0)$ represents the amplitude of the
   knot's path; $\phi(s_0)$ is the azimuthal angle or the phase of the knot.
   $s_0$ denotes the arc length along the jet axis:
    \begin{equation}
       {s_0}={\int_{0}^{z_0}}{\sqrt{[{1+(d{r_0}/d{z_0})^2]}}d{z_0}} ,
    \end{equation}
   $z_0$ and $A(s_0)$ are measured in units of milliarcsecond (mas) and
    $\phi(s_0)$ is
   measured in units of radian. For studying helical motion of a knot
    around the jet axis  the orbital phase $\phi(s_0)$ and amplitude function
   $A(s_0)$ should be given. In this paper we do not discuss knot's
   helical motion and set $A(s_0)$=0.

   In the case knots move along the jet axis, when their coordinates
   $(X,Y,Z,)$=$(X_j,Y_j,Z_j)$ and the coordinates of the jet axis
    $(X_j, Y_j, Z_j)$ are:
    \begin{equation}
     {{X_j}({r_0},{\omega(t)})}={r_0}{\cos}\,{\omega(t)} ,
    \end{equation}
    \begin{equation}
     {{Y_j}({r_0(t)},{\omega(t)})}={r_0}{\sin}\,{\omega(t)} ,
    \end{equation}
    \begin{equation}
     {Z_j}={z_0} ,
    \end{equation}
   When  parameters $\epsilon$, $\psi$, $p$, $\alpha$,
    and $\Gamma$ (bulk Lorentz factor of the knot) are set, the
   kinematics of the knot (projected trajectory, apparent velocity and
   Doppler factor, viewing angle as functions of time ) can then be
   calculated. The formulas are listed as follows.

   The projected trajectory on the plane of the sky is represented by:
     \begin{equation}
      {X_n}({z_0},{\omega})={X_j}{\cos}{\psi}-
             {[{z_0}{\sin}{\epsilon}-{Y_j}{\cos}{\epsilon}]}{\sin}{\psi} ,
      \end{equation}
     \begin{equation}
     {Y_n}({z_0},{\omega})={X_j}{\sin}{\psi}+
            {[{z_0}{\sin}{\epsilon}-{Y_j}{\cos}{\epsilon}]}{\cos}{\psi} ,
     \end{equation}
   Introducing the following functions:\\
     \begin{equation}
   {\Delta}={\arctan}{\left[\left(\frac{dX}{d{z_0}}\right)^2+
                  \left(\frac{dY}{d{z_0}}\right)^2\right]}^{\frac{1}{2}} ,
     \end{equation}
     \begin{equation}
    {{\Delta}_p}={\arctan}{\left[\frac{dY}{d{z_0}}\right]} ,
     \end{equation}
    \begin{equation}
    {{\Delta}_s}={\arccos}
                 {\left[1+\left(\frac{dX}{d{z_0}}\right)^2+
                    \left(\frac{dY}{d{z_0}}\right)^2\right]}^{-\frac{1}{2}} ,
    \end{equation}
    We then can calculate the elapsed time $T_0$ (at which the knot
    reaches an axial distance Z), apparent velocity
    ${\beta}_a$, Doppler factor $\delta$, and viewing angle $\theta$
    of the knot:\\
     \begin{equation}
   {T_0}={\int_{0}^{{z_0}}}{\frac{1+z}{{\Gamma}{\delta}{v}{\cos}{{\Delta}_s}}}
                              {d{z_0}} ,
     \end{equation}
     \begin{equation}
    {\theta}={\arccos}{[\cos{\Delta}(\cos{\epsilon}+\sin{\epsilon}
                     \tan{{\Delta}_p})]} ,
     \end{equation}
   \begin{equation}
    {\delta}={\frac{1}{{\Gamma}(1-{\beta}{\cos}{\theta})}} ,
   \end{equation}
   \begin{equation}
   {{\beta}_a}={\frac{{\beta}{\sin}{\theta}}{1-{\beta}{\sin}{\theta}}} ,
   \end{equation}
   where $\beta$=v/c (v--speed of the knot)
     and $\Gamma$=$(1-{\beta}^2)^{-\frac{1}{2}}$.

   We point out that in the scenario of the precessing nozzle model described
   in Figure 3, the precessing common trajectory is defined in the coordinate
   system (X, Y, Z) and described by three parameters ($a$, $x$, $\omega$).
    But in the observer system ($X_n$, $Y_n$, $Z_n$), the trajectory is
   defined  by   five parameters ($a$, $x$, $\omega$, $\epsilon$, $\psi$).
    Generally, changes
   in any parameter or in their combination will introduce the change of the
   trajectory pattern with respect to the observer's system. In particular,
    for simplicity, in the following model
   simulations of the superluminal knots, changes in single parameter $\psi$
   will be introduced to study the knots' trajectory curvatures in the outer
    jet regions,  while in their inner jet regions parameter $\psi$ will remain
   to be a constant value to demonstrate the jet  precession.
    The change in single parameter parameter $\psi$ implies that the knot's
   trajectory rotates around the viewing axis (Note that parameter $\epsilon$
   remains to be a constant).

   In this paper, we adopt the concordant cosmological model ($\Lambda$CDM
   model) with ${\Omega}_m$=0.27, ${\Omega}_{\lambda}$=0.73 and Hubble constant
   $H_0$=$71{\rm{km{s^{-1}}{Mpc^{-1}}}}$ (Spergel et al. \cite{Sp03}). Thus for
    OJ287, z=0.306, its luminosity distance is ${\rm{D_L}}$=1.58{\rm{Gpc}}
   (Hogg \cite{Ho99}, Pen \cite{Pe99}) and  angular diameter distance
   ${\rm{D_A}}$=0.9257{\rm{Gpc}}. The angular scale 1{\rm{mas}}=4.487\,pc,
   and the proper motion of {\rm{1mas/yr}} is equivalent to an apparent
   velocity of 19.1c (c is the speed of light).
    \section{Selection of model parameters}
    In this paper we try to interpret the source kinematics  of OJ287 in terms
   of the precessing jet model originally proposed by Qian et al.
    (1991, 2009, 2014) in the framework of a double jet scenario.
    In order to perform the model-fitting of the kinematics of
   the knots, we need  to select model parameters for both the
   jets, separately. For each jet two sets of model parameters are required.

    For a single jet the approach of selecting the model parameters has been
    described in detail in Qian et al. (\cite{Qi17}, \cite{Qi18}) where the
     VLBI-kinematics of the superluminal components of QSO B1308+326 and 
   PG 1302-102 were model fitted. In the
   case of OJ287 it involves more and new parameters (e.g.,
    parameters $a$ and $x$ for the common parabolic trajectory pattern)
   and  we have to define them separately for the two jets. Here we briefly
   iterate our procedure of selecting model parameters as follows (see
   Qian et al. \cite{Qi17}, \cite{Qi18}, \cite{Qi19}).
    \begin{itemize}
    \item  Geometric and kinematic parameters: these include
     parameters $\epsilon$ and $\psi$ defining the orientation of the
     precession axis, parameters $a$ and $x$ defining the shape of the common
     precessing trajectory and Lorentz factor for each of the knots. We can 
    derive a preliminary set of the parameters. For example, if the viewing 
    angle of the jet is given, then (i) the parameters $a$ and $x$
      can be approximately determined from the observed trajectories of 
   the knots; (ii) from the observed distribution of the knots' trajectories,
     the position angle of the precession axis  and its orientation in space
    (parameter $\psi$) can be approximately derived; and (iii)
    Lorentz factors of the knots can be
    estimated from their observed apparent velocities. The selection of
    these parameters are not unique, mostly depending on the viewing angle
     (parameter $\epsilon$) of the precession axis. Using the formalism
    described in Sect. 3, appropriate parameters can finally be chosen 
    through trial model fittings of the kinematics of the knots (see Qian
    et al. \cite{Qi19}). Since different viewing
    angles chosen for the precession axis  would lead to different projection
    effects we would firstly select an appropriate value for
    parameter $\epsilon$. In this paper we take $\epsilon$=$3^{\circ}$ (see
    Cohen 2017, Hovatta et al. 2009, Agudo et al. 2012).
    \item Parameters describing the time-dependent kinematic behaviors of the
    knots: ejection times ($t_0$) of the knots and the jet precession period 
    $T_p$ . In the case of OJ287, the 12\,yr periodicity found in  the optical
    light curve might be used as the precession period of
    the jet axis for both the jets, because this period is the only one having
    been determined in the optical variability observed in OJ287. And
    this assumption will be justified by the studies in this paper (see 
    below). Thus for OJ287 the precession period
    is a known parameter which constrains the observed 
    aperture of the jet cone and the distribution of the trajectories of
    the knots. What
    should be done is  to model fitting the observed knots' trajectories and
     their ejection times under the precessing jet nozzle scenario. Although
    the geometric and kinematic parameters are not uniquely chosen,
    the modeled ejection epochs ($t_0$) and the precession period $T_p$ are
    strictly constrained  by the measured ejection epochs ($t_{0,VLBI}$)
    and the observed distribution of the knots' tracks  for both jets.
     The ejection times $t_{0,VLBI}$ measured from
    the VLBI observations can be used  as the initial values
     for the  modeled times $t_0$, and then we can determine their
    final values through trial
    model fitting of the observed distribution of the trajectories of the
    knots. Usually, $t_{0,VLBI}$ of a knot is measured by extrapolating its
     separation from the core to zero through  linear regression. If the motion
     of the knot is  non-radial (having a curved trajectory) or initially
    accelerated (or decelerated), this method could induce  significant
    errors in $t_{0,VLBI}$.\footnote{Opacity effects of a knot
    at its  emergence from the core could also induce
    errors in the measurements of $t_{0,VLBI}$ by this method
     (see Qian et al. 2017).} Therefore, during
    the process of trial model fitting for the kinematics of the knots
    we determine the model ejection times $t_0$ to be close to the values
    $t_{0,VLBI}$ derived from the VLBI measurements as possible and constrain
    their differences  from the $t_{0,VLBI}$ within $\sim$1\,yr. It can be seen
     below (Tables 1 and 2) that for the twelve knots which are
     model fitted, only two
    knots (C6 and C9) ${\mid}{t_0}-{t_{0,VLBI}}{\mid}$$\sim$0.5--1.0\,yr.
    This difference can be regarded as the maximal uncertainties of the
    model fitting results in this paper.
    \end{itemize}
    \begin{figure}
    \centering
    \includegraphics[width=6cm,angle=-90]{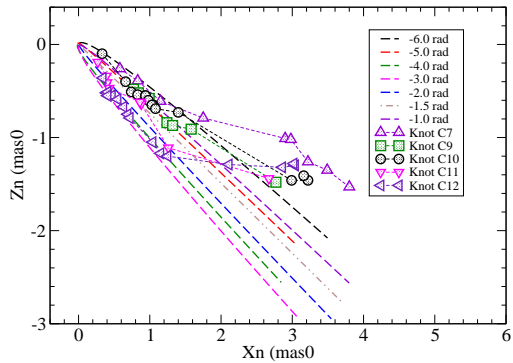}
    \caption{Modeled southern jet cone: The distribution of the precessing
    parabolic trajectories (precession phase $\omega$=--1.0 to --6.0\,rad) and
    the observed trajectories of the superluminal components (C7, C9, C10, C11,
    C12). The precession axis (projected)
    is at $\sim{-130^{\circ}}$.}
    \end{figure}
     \section{Model-fitting results for southern jet}
    Through analyzes and trial model-fittings we have found that the
   southern jet comprises six knots
    (C7, C9, C10, C11, C12 and C13L), forming the southern group of
      superluminal components.
      We select the model parameters for the southern jet as:
    \begin{table}
    \centering
    \caption{Model parameters for the southern jet:
   $T_p$, $\epsilon$, $\psi$, $a$ and $x$.}
    \begin{flushleft}
    \centering
    \begin{tabular}{lr}
    \hline
    $T_p$ & 12\,yr \\
    $\epsilon$ & 0.0524\,rad=$3.0^{\circ}$ \\
    $\psi$   & 0.65\,rad=$37.2^{\circ}$  \\
    $a$    & 0.0536$[\rm{mas}]^{1/2}$ \\
    $x$    &  0.5 \\
    \hline
    \end{tabular}
    \end{flushleft}
    \end{table}
     $T_p$=12\,yr, $\epsilon$=$3^{\circ}$, $\psi$=0.65\,rad,
     $a$=0.0536$[{\rm{mas}}]^{\frac{1}{2}}$, $x$=0.5 which are 
     listed in Table 1.

     The ejection time $t_0$ is related to the precession phase:
     \begin{equation}
      {t_0} = 1998.55-({\omega}+5.70){\rm{T_p}}/2{\pi}.
     \end{equation}
     Here $\omega$=$-$5.70\,rad corresponds to the ejection epoch for knot C7 at
     1998.55. $T_p$=12\,yr. In Table 2 some modeled parameters and relevant
     observation data are listed. Our modeled ejection times ($t_0$)
   are consistent with the ejection times ($t_{0,obs}$) derived from
    VLBI measurements.
   
    The entire structure of the southern jet and the distribution of the
     knots' trajectories are shown in Fig.4.
     \begin{table*}
     \centering
     \caption{Model parameters for the superluminal components of the
      southern jet: modeled ejection time $t_0$, precession phase $\omega$
     (rad), range of Lorentz factor $\Gamma$. VLBI-measured quantities at
     15\,GHz:
    ejection epoch $t_{0,obs}$ and  apparent speed ${\beta}_a$,
     average position angle ${\overline{PA}}$, curvature in trajectory
     defined as $\Delta{\overline{PA}}$= $\overline{PA}$($r_n{>}$2\,mas)
     --$\overline{PA}$($r_n{<}$2\,mas).}
     \begin{flushleft}
     \centering
     \begin{tabular}{lrrrrrrr}
     \hline\noalign{\smallskip}
     Knot & $t_0$ & $\omega$(rad) & $\Gamma$  &  $t_{0,obs}$ & ${\beta}_a$
           & $\overline{PA}$(deg.) &  ${\Delta}{\overline{PA}}$(deg.)\\
  C7 &   1998.55 & --5.70 & 10.3 & 1998.9 & 8.8$\pm$0.2 & -115.3 & +4.8\\
  C9 &   2001.80 & --7.40 & 13.0--9.0 & 2000.8 & 7.5$\pm$0.2 & -121.9 & +3.8\\
     C10 &  2002.34 & --7.68 & 9.2  & 2002.5 & 8.2$\pm$0.2 & -120.2 & +5.3\\
     C11 &  2003.10 & --8.08 &  8.0  & 2002.9 &  6.1$\pm$0.2 & -129.8 & +11.4\\
     C12 &  2006.90 & --10.08 & 8.8  & 2007.3 & 6.5$\pm$0.4 & -137.2 & +23.6 \\
     C13L & 2004--2005 &  --   &   --   &   --  &  --         & -185.7 & --\\
     \noalign{\smallskip}\hline
    \end{tabular}
    \end{flushleft}
     \end{table*}
     \begin{figure*}
     \centering
     \includegraphics[width=5cm,angle=-90]{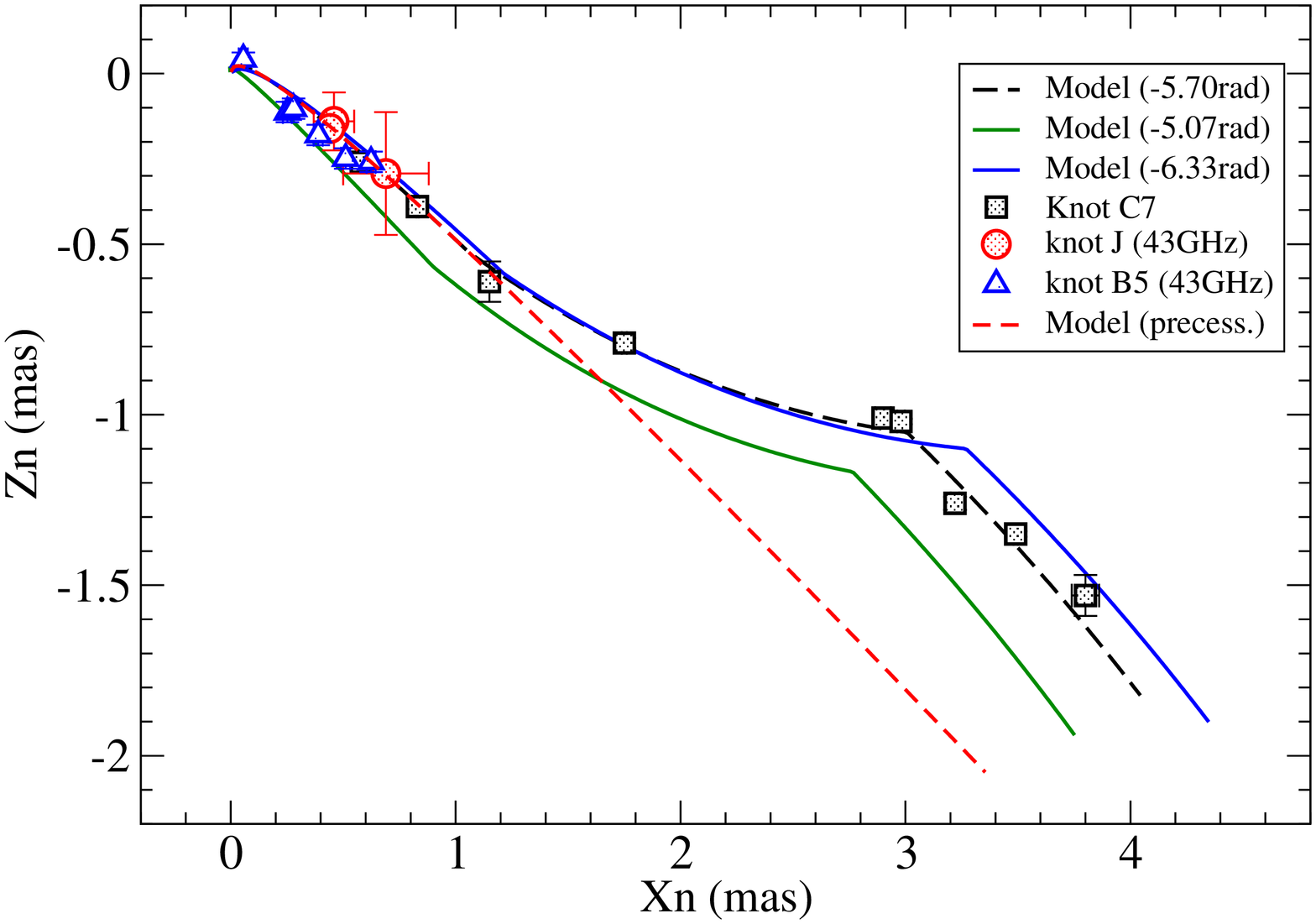}
     \includegraphics[width=5cm,angle=-90]{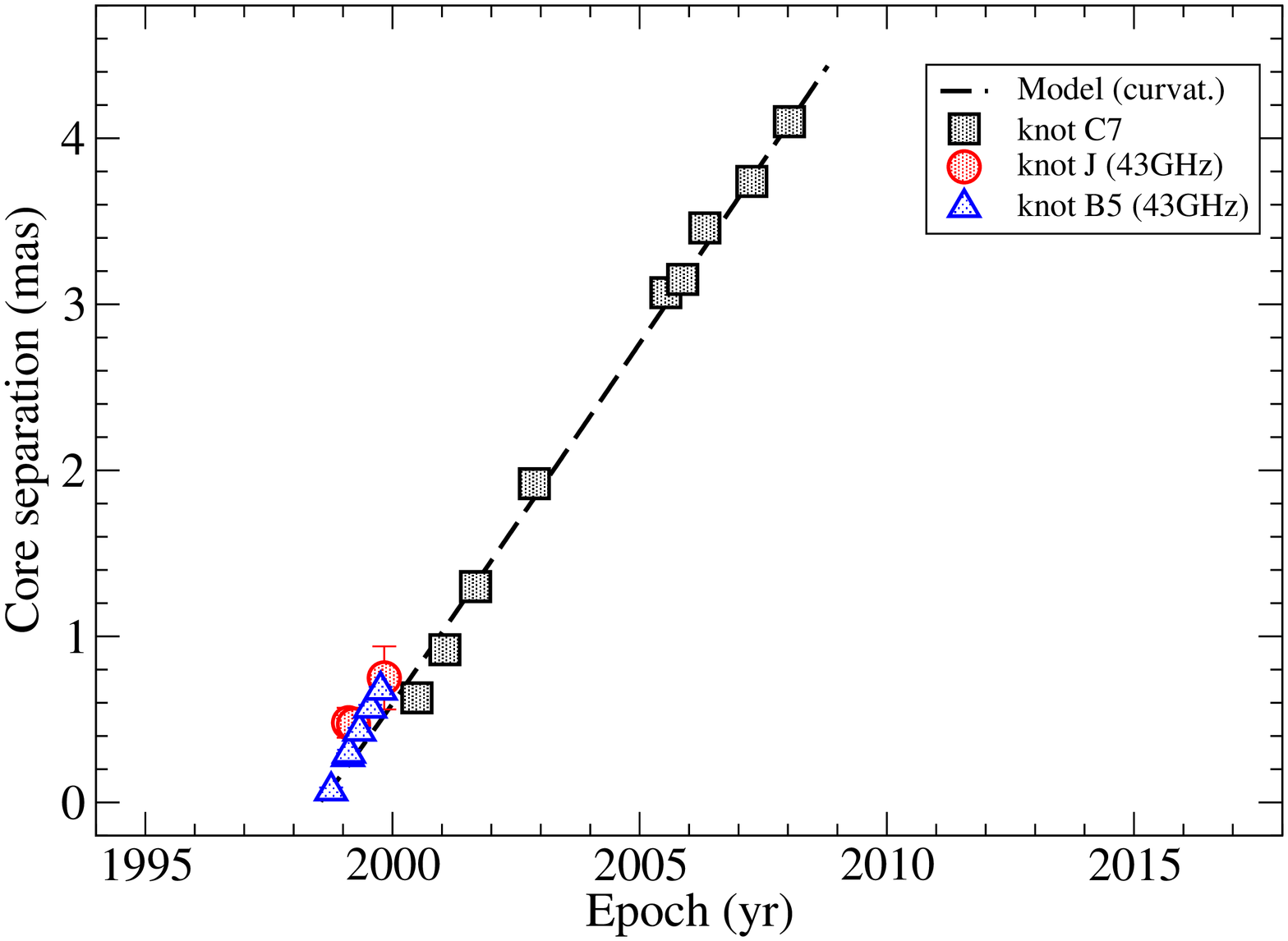}
     \includegraphics[width=5cm,angle=-90]{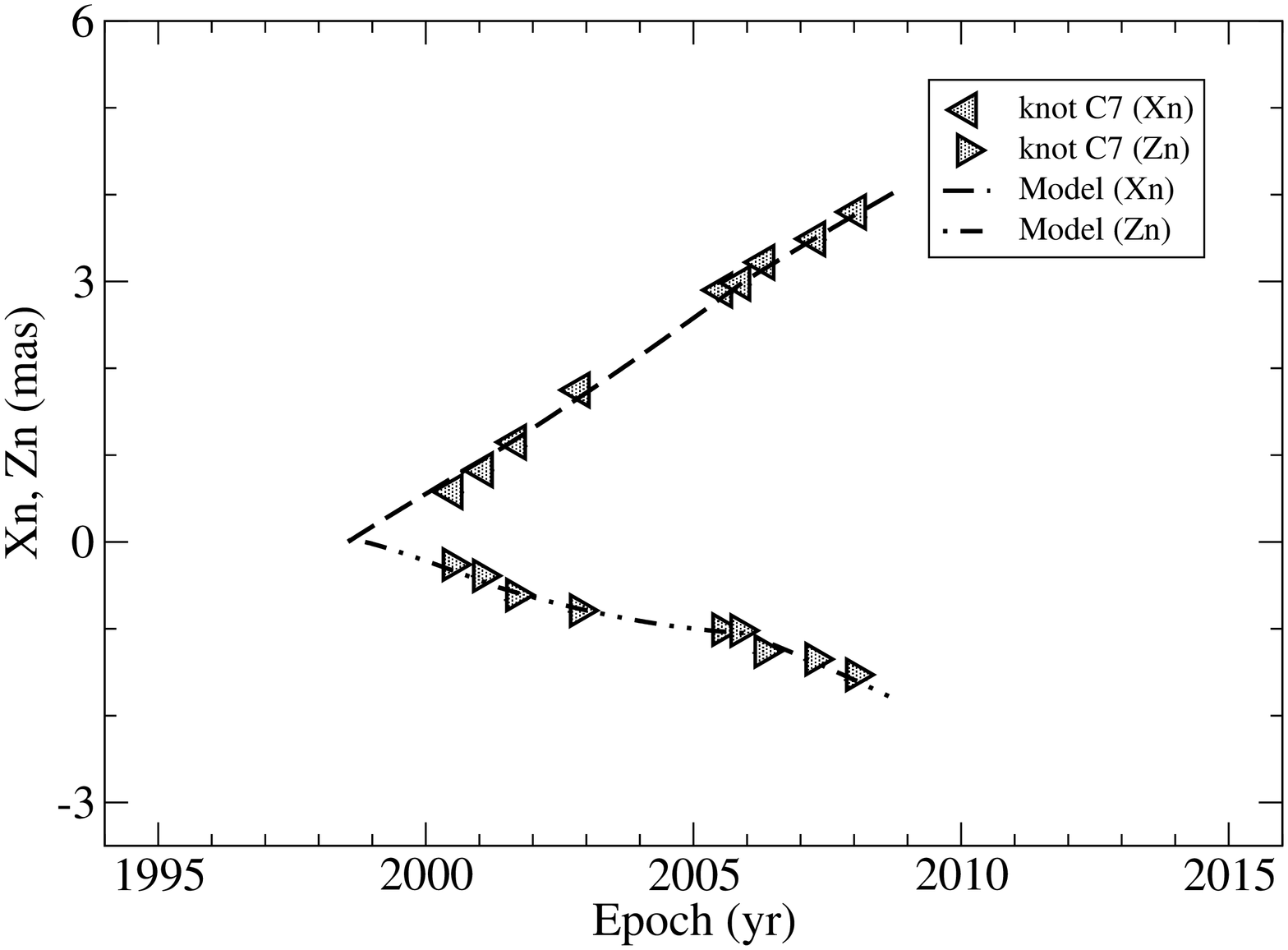}
     \includegraphics[width=5cm,angle=-90]{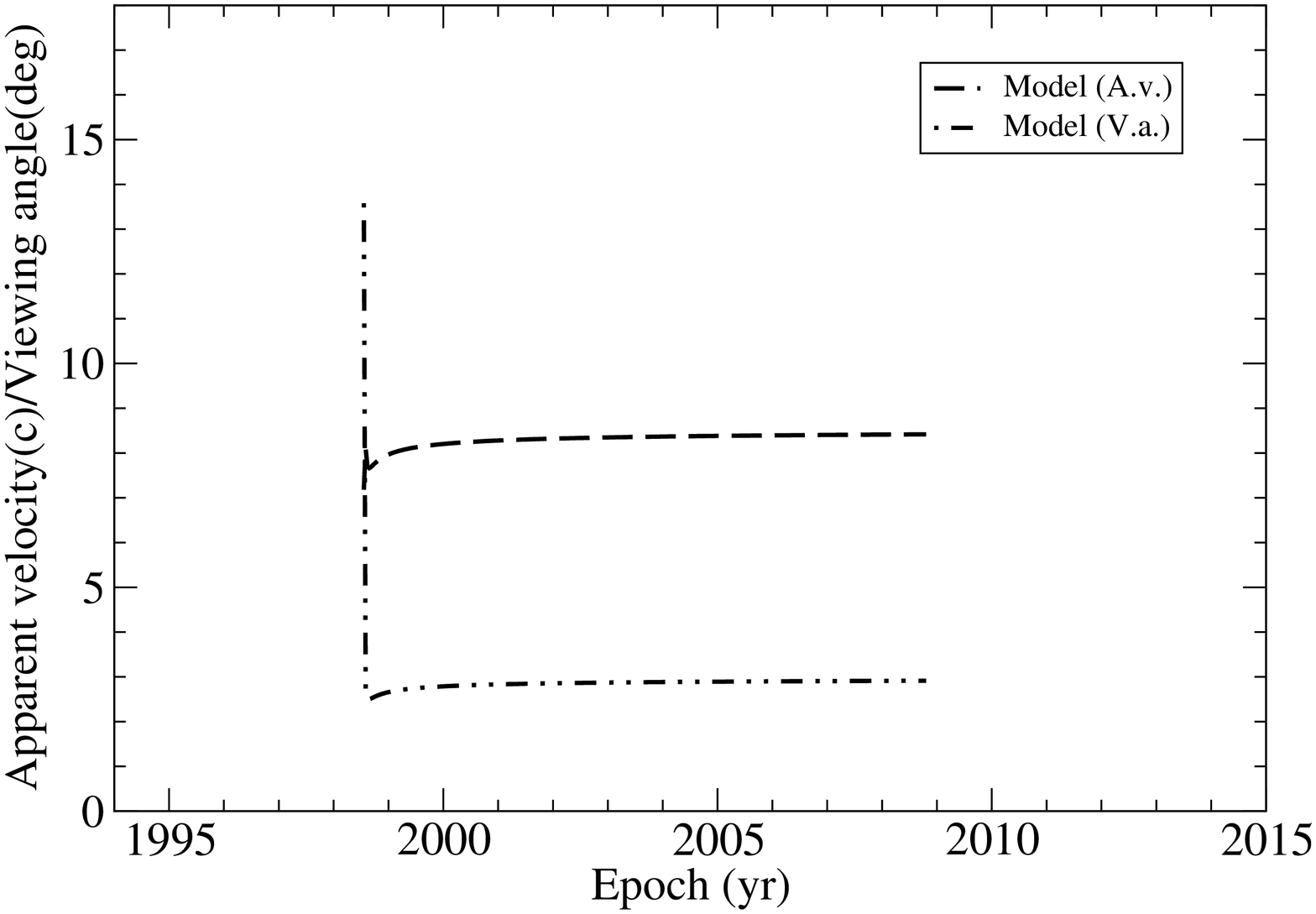}
     \includegraphics[width=5cm,angle=-90]{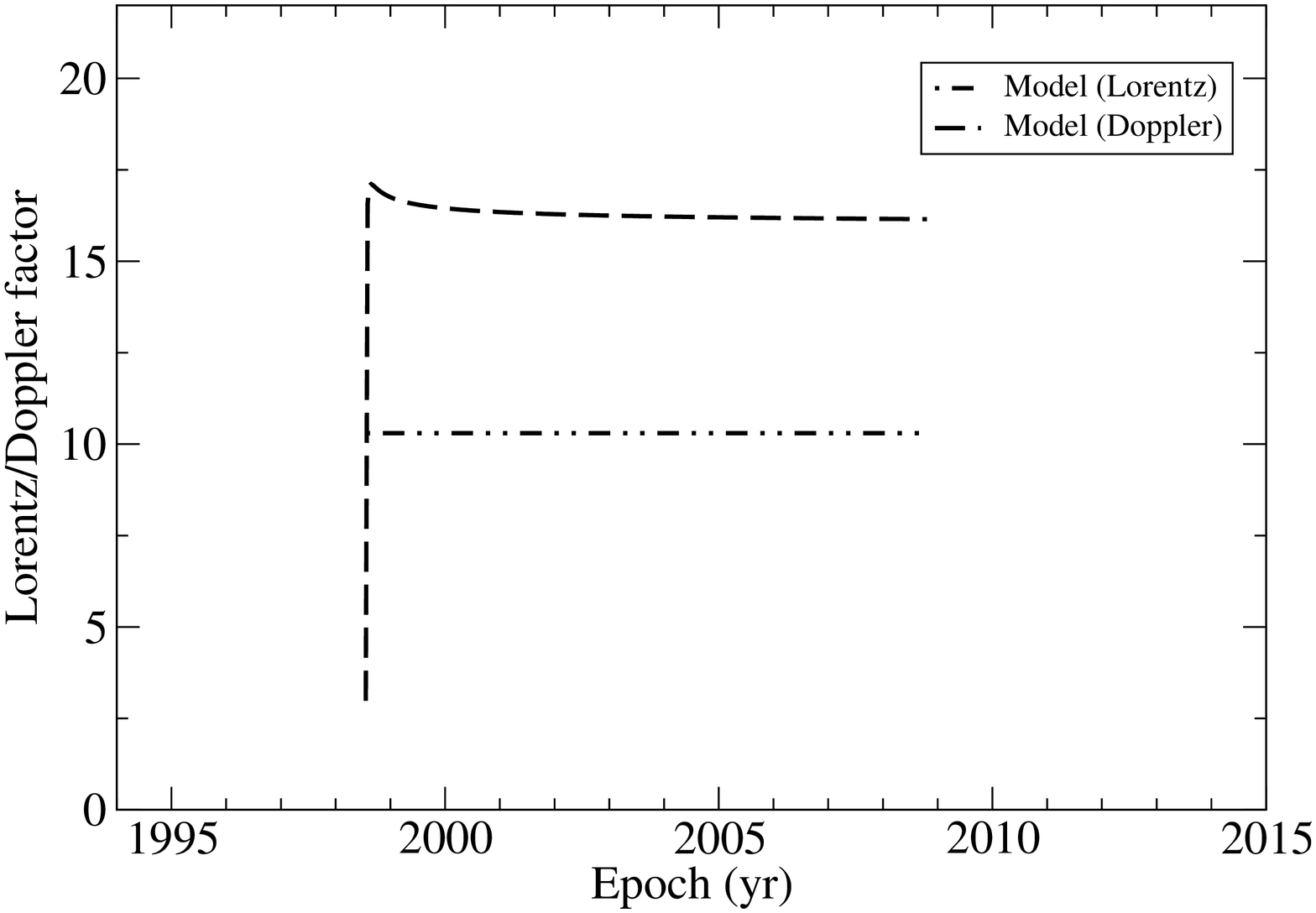}
     \includegraphics[width=5cm,angle=-90]{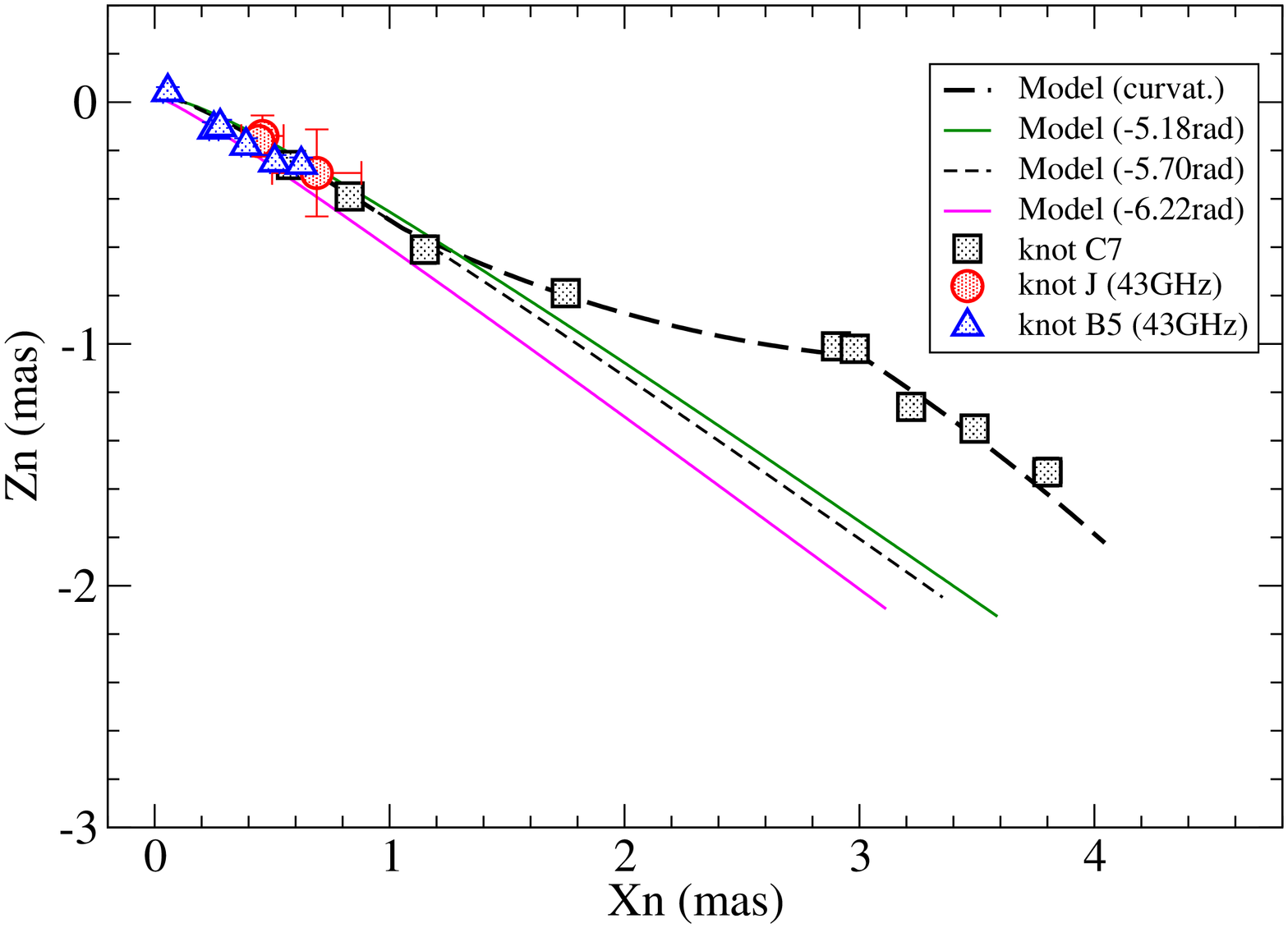}
     \caption{Model-fitting results of the kinematic features for knot C7.
    The entire modeled trajectory is shown by the black dashed line in 
    upper left panel. The green and blue lines in bottom right panel show its
    innermost precessing parabolic trajectory having been observed. The 
    43\,GHz data given in Jorstad et al. (\cite{Jor05}, for knot-B5) and 
    Agudo et al. (\cite{Ag12}, for knot-J) are also well fitted by the model.}
     \end{figure*}
     \begin{figure*}
     \centering
     \includegraphics[width=5cm,angle=-90]{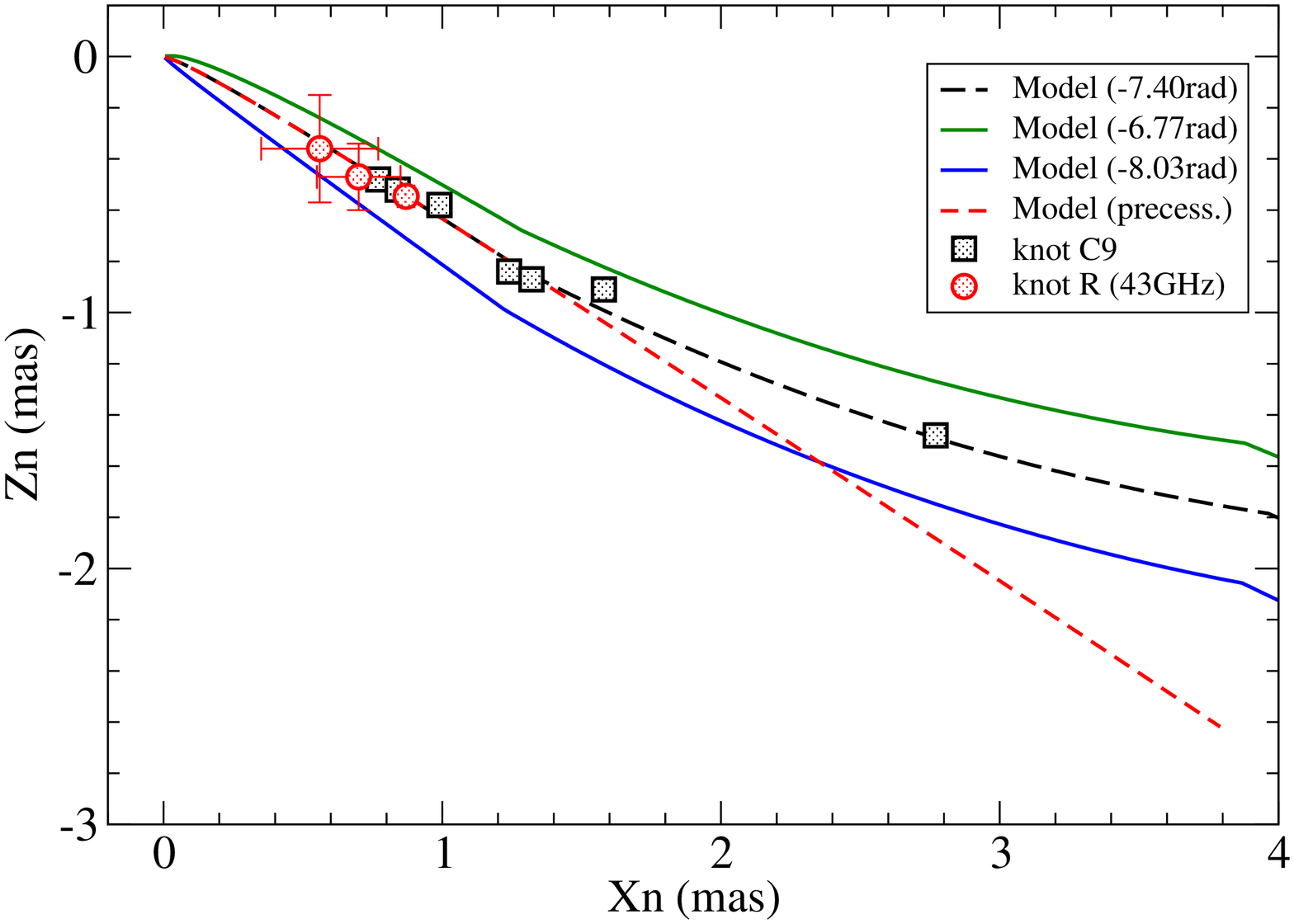}
     \includegraphics[width=5cm,angle=-90]{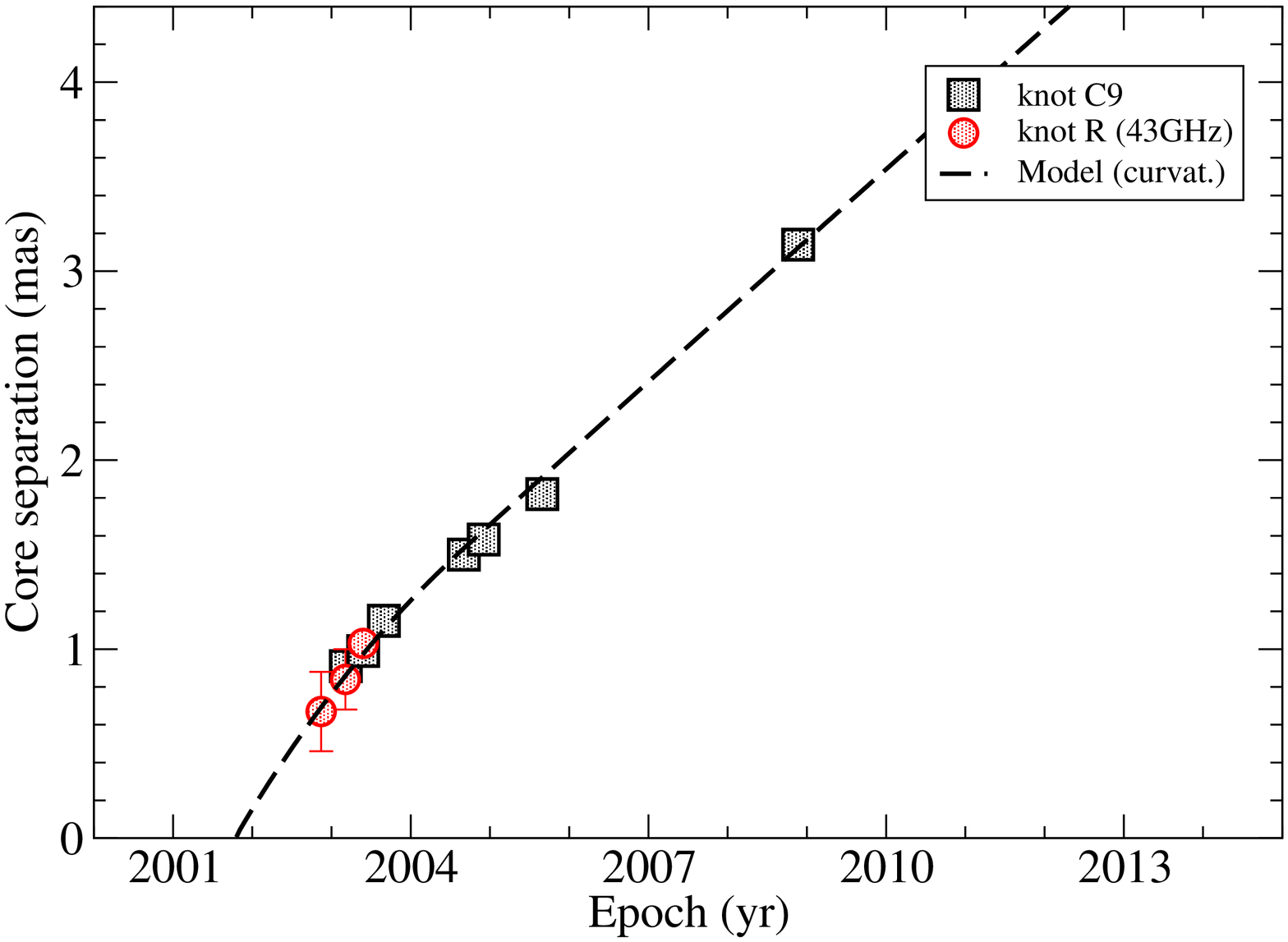}
     \includegraphics[width=5cm,angle=-90]{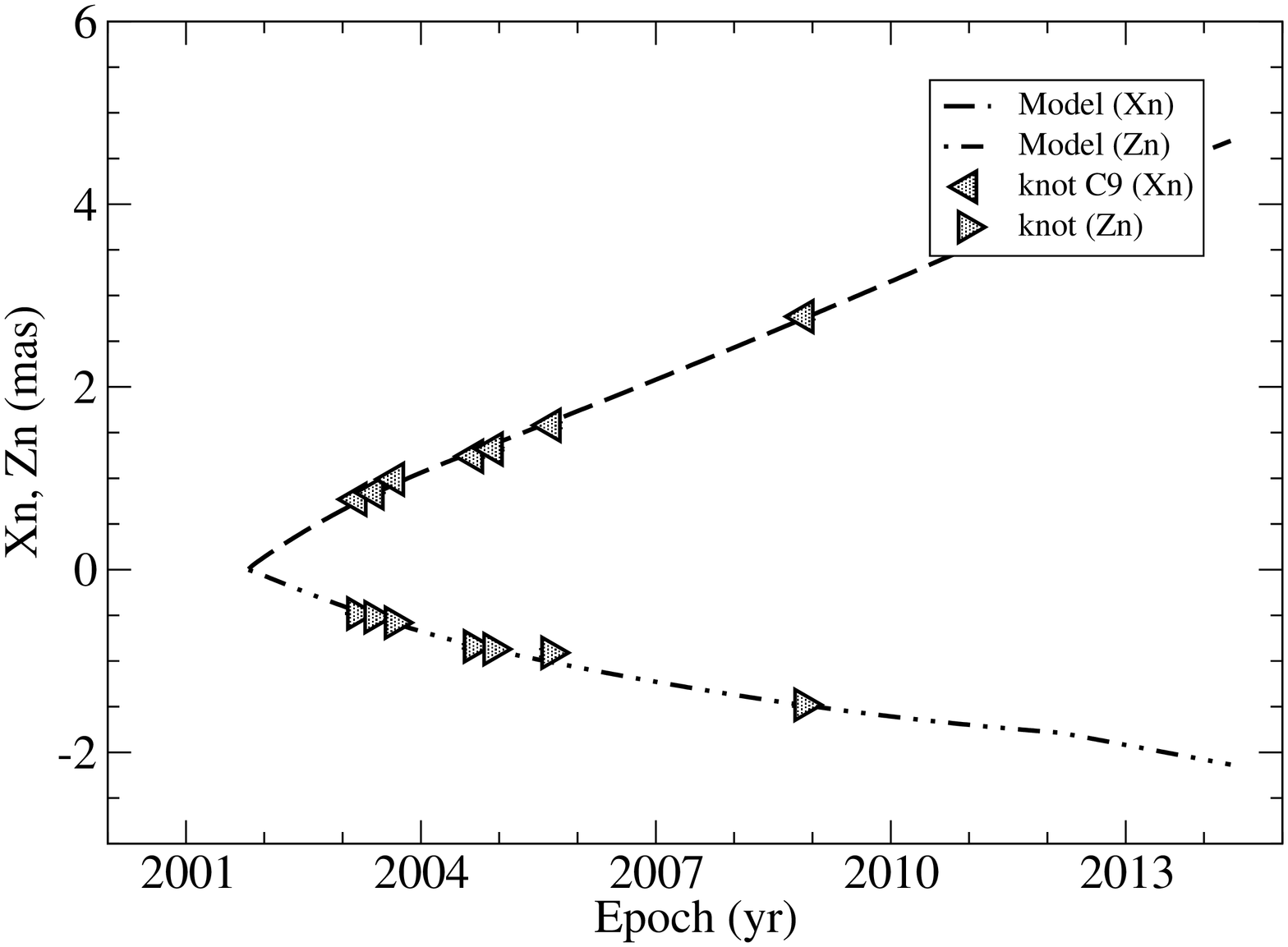}
     \includegraphics[width=5cm,angle=-90]{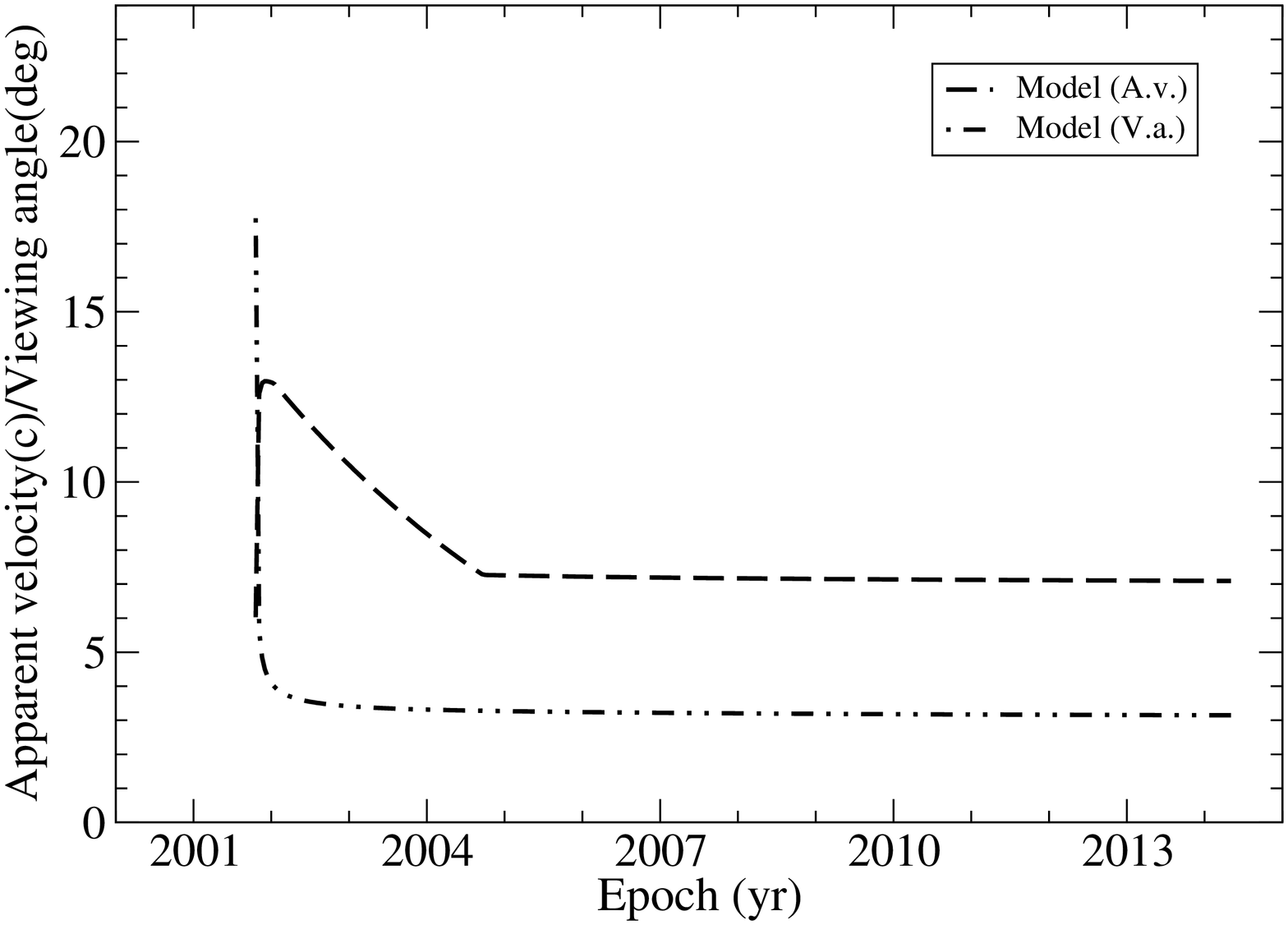}
     \includegraphics[width=5cm,angle=-90]{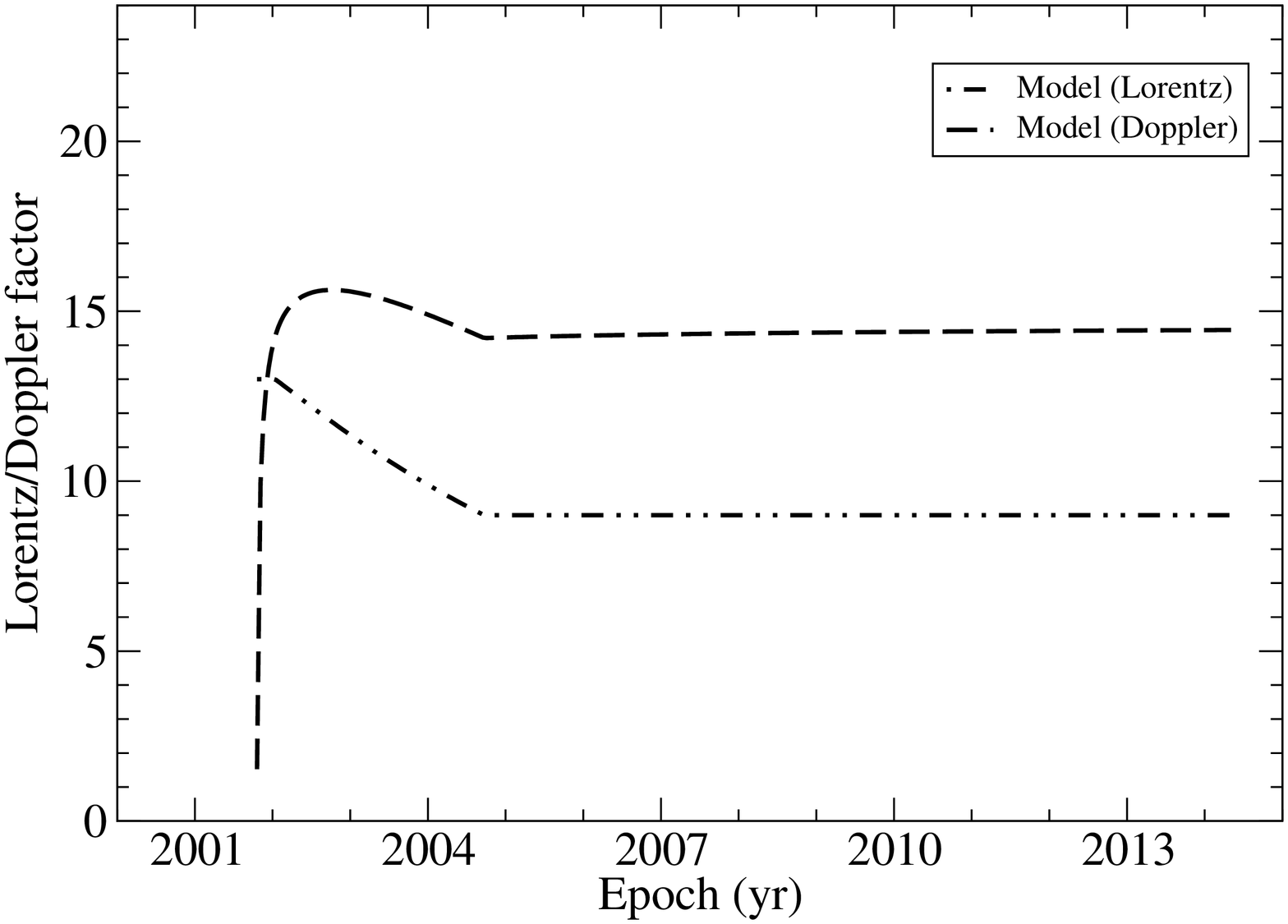}
     \includegraphics[width=5cm,angle=-90]{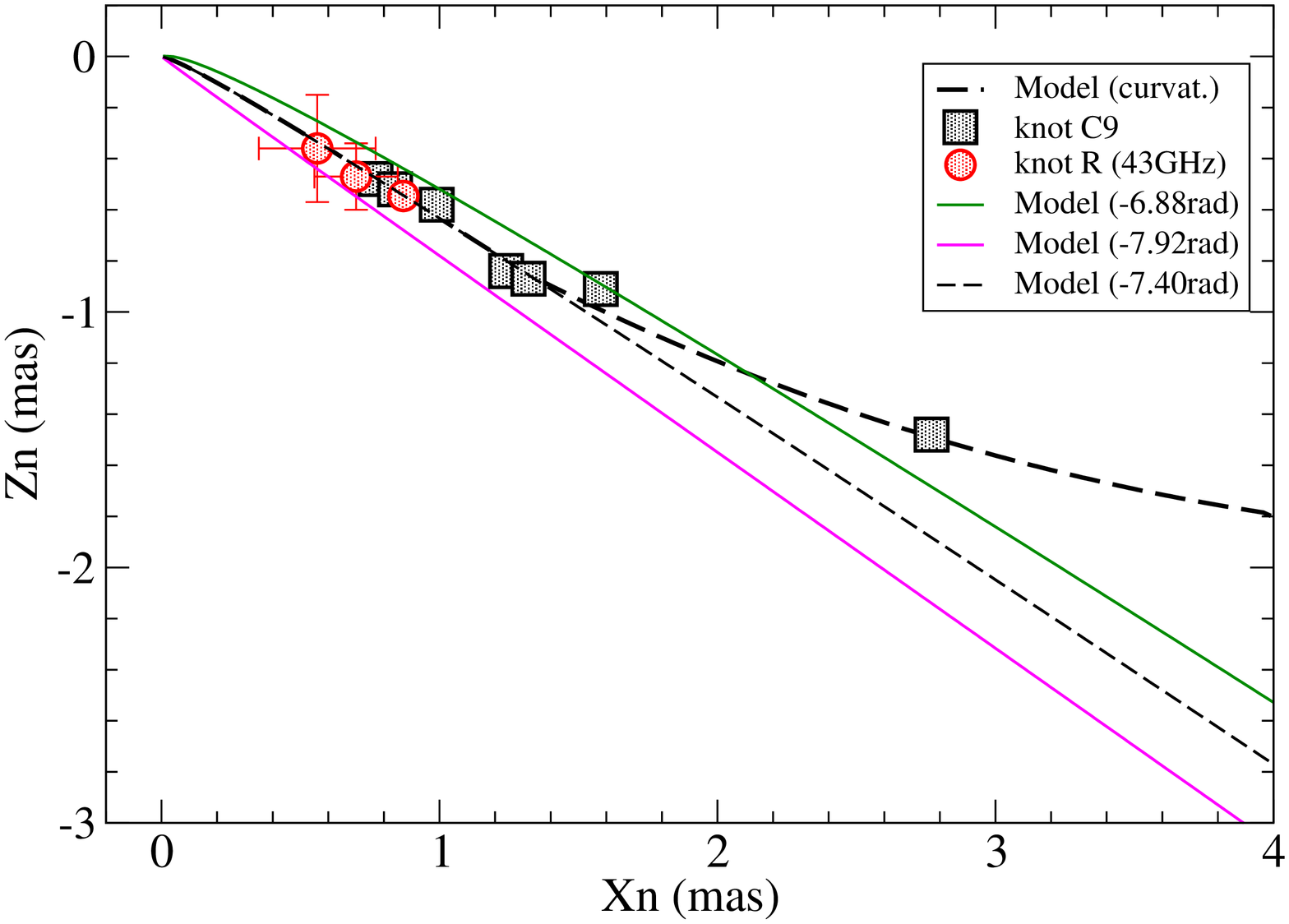}
      \caption{Model-fitting results of the kinematic features for knot C9.
      The entire modeled
      trajectory is shown by the black dashed line in upper left panel.
      The 43\,GHz data given in Agudo et al. (\cite{Ag12}, for knot-R) are also
     well fitted by the model. Its innermost 
     precessing  trajectory has been observed as shown in bottom right
     panel.}
      \end{figure*}
     \subsection{Knot C7}
      The model-fitting results are shown in Figure 5 ,
     including trajectory
    $Z_n(X_n)$, coordinates $X_n(t)$ and $Z_n(t)$, core separation $r_n$(t),
    the modeled apparent velocity and viewing angle, the modeled Lorentz
    factor and Doppler factor.

      Its modeled  ejection time $t_0$=1998.55 and the  corresponding
    precession phase $\omega$=--5.70\,rad. Within core separation
    ${r_n}$=1.18\,mas (or radial distance Z=25\,mas=112\,pc)
     knot C7 is modeled to move along the precessing common parabolic
    trajectory ($\psi$=0.65\,rad) and its kinematics can be well interpreted
   in terms of the precessing nozzle model with a precessing period of 12\,yr.
    The 43\,GHz data points given by  Agudo et al. (\cite{Ag12}, for knot-J)
    and by Jorstad et al. (\cite{Jor05} for knot B5) are also well fitted
    by the precessing model.
     In particular, Agudo et al. and Jorstad et al. measured its ejection
     epoch to be 1998.02$\pm$0.24 and 1998.61$\pm$0.12. The latter is
     extremely well consistent with our fitted value 1998.55.
    As Agudo et al. commented that the 43\,GHz observations made by Jorstad
   et al. had better time  sampling producing better kinematic estimates.

     Beyond separation $r_n$=1.18\,mas its trajectory deviates from the
      precessing nozzle model and changes in parameter $\psi$ (or trajectory
     curvatures) have to be  introduced to fit the outer trajectory:
     For Z=25--65\,mas $\psi$(rad)=0.65--0.20(Z--25)/(65--25);
      For Z=65--90\,mas  $\psi$(rad)=0.45+0.07(Z--65)/(90--65); For
    Z${>}$90\,mas $\psi$=0.52\,rad.

     Its motion is assumed to be uniform: The modeled Lorentz factor
     $\Gamma$=const.=10.3. The modeled apparent speed (shown in middle right 
     panel of Figure 5) is consistent with the VLBI-measured proper speed
     0.46$\pm$0.01 mas/yr (8.8$\pm$0.2c) given in Britzen et al.
    (\cite{Br18}).

    In the top left panel of Figure 5, the green
    and blue lines represent the modeled
    trajectories calculated for precession phases $\omega{\pm}$0.63\,rad,
    showing the data-points being within the position angle range defined by
    the two lines and indicating the precession period having been determined
    within an uncertainty of $\pm$1.2\,yr. In the bottom right panel, 
   the green
    and blue lines represent the precessing common trajectories calculated for
   $\omega{\pm}$0.52\,rad, showing a number of the data-points being within
   the position angle range defined by the two lines and indicating its
   innermost precessing common parabolic trajectory having been observed. Thus
   knot C7 is designated by symbol ``+'' in Table 3. We note that these fitting
   results have been used as criteria in this paper to judge the validity
   of our precessing nozzle scenario to investigate the kinematics of all the
   knots.
     \subsection{Knot C9}
     The model-fitting results are shown in Figure 6,
     including trajectory
     $Z_n(X_n)$, coordinates $X_n(t)$ and $Z_n(t)$, core separation $r_n(t)$,
     the modeled apparent velocity and viewing angle and the modeled Lorentz
     factor and Doppler factor. Its ejection time is modeled as $t_0$=2001.80
     and the corresponding precession phase $\omega$=--7.40\,rad.

     Within core separation $r_n$=1.55\,mas (radial distance Z=25\,mas=112\,pc)
     the observed trajectory can be well
     fitted by the precessing common parabolic trajectory ($\psi$=0.65\,rad;
     red dashed line in top left panel)
     and its kinematics can be well interpreted in terms of the precessing
     nozzle model. Interestingly, the 43\,GHz data points
     measured by Agudo et al. (\cite{Ag12}, for knot-R) are also well fitted.
     These
     43\,GHz data points extend its trajectory to the innermost region.
     Moreover,  Agudo et al. derived the ejection epoch to be 2001.92$\pm$0.12,
     which is extremely well consistent with our modeled ejection time 2001.80,
     verifying the validity of our precessing nozzle model and confirming the
     precession period of 12\,yr. 

     Thus combined with the fitting results for
     knot C7, we have found that for two knots
     (C7 and C9), both the 43\,GHz and 15\,GHz observations yielded  similar
     ejection epochs, verifying the applicability of the precessing model
     with a precession period of 12\,yr. This is really encouraging,
      because we only made use of the 15\,GHz data to construct our precessing
     nozzle model and did not use the 43\,GHz data in the fitting process.
     So the confirmation of the model-fits for knots C7 and C9 by the 43\,GHz
      observation data points  are really posterior confirmation.
    \subsection{Knot C10}
      The model-fitting results of the kinematic features for knot C10
   are shown in Figure 7, including trajectory $Z_n(X_n)$, coordinates $X_n(t)$
   and $Z_n(t)$, core separation $r_n(t)$, modeled apparent velocity and
   viewing angle, and  modeled Lorentz factor and Doppler factor.

     The ejection time is  modeled as $t_0$=2002.34, the corresponding
     precession phase $\omega$=--7.68\,rad.
     Within core separation $r_n$=0.81\,mas (radial distance Z=12\,mas=
    53.9\,pc) knot C10 is modeled to move along the precessing common parabolic
    trajectory (red dashed line in top left panel). The entire modeled
   trajectory is shown by the black dashed line. The green and blue lines 
   represent the modeled trajectories calculated for precession phases
    $\omega{\pm}$0.63\,rad, showing all the data-points being within the 
    position angle range defined by
    the two lines and indicating the precession period having been determined
    within an uncertainty of $\pm$1.2\,yr. In the bottom right panel the green
    and blue lines represent the precessing common trajectories calculated
    for precession phases $\omega{\pm}$0.52\,rad, showing a number of
    data-points being within the position angle range defined by the two lines
    and indicating its innermost precessing common parabolic trajectory having
    been observed. Thus knot C10 is designated by symbol ``+'' in Table 3.
     \begin{figure*}
     \centering
     \includegraphics[width=5cm,angle=-90]{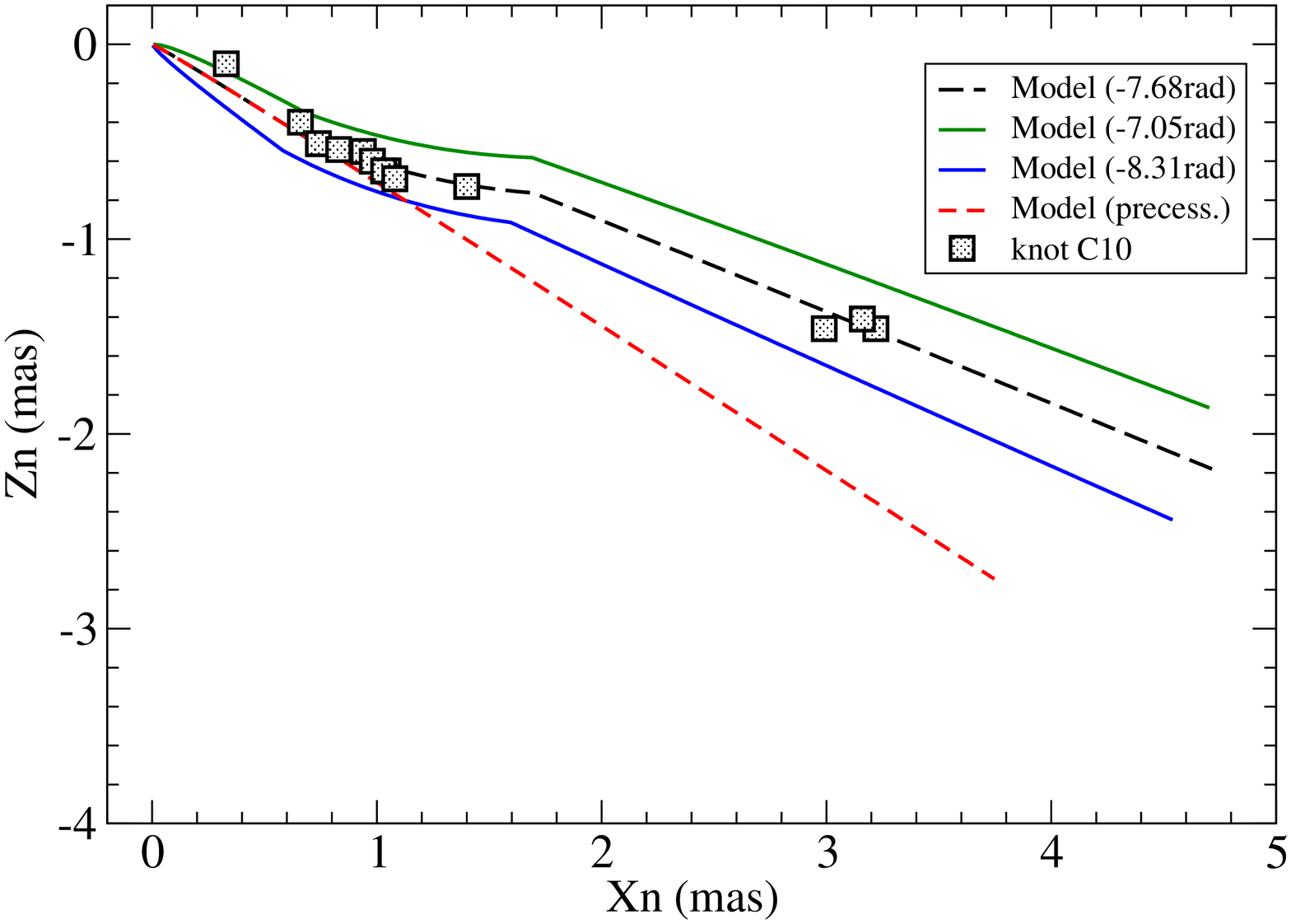}
     \includegraphics[width=5cm,angle=-90]{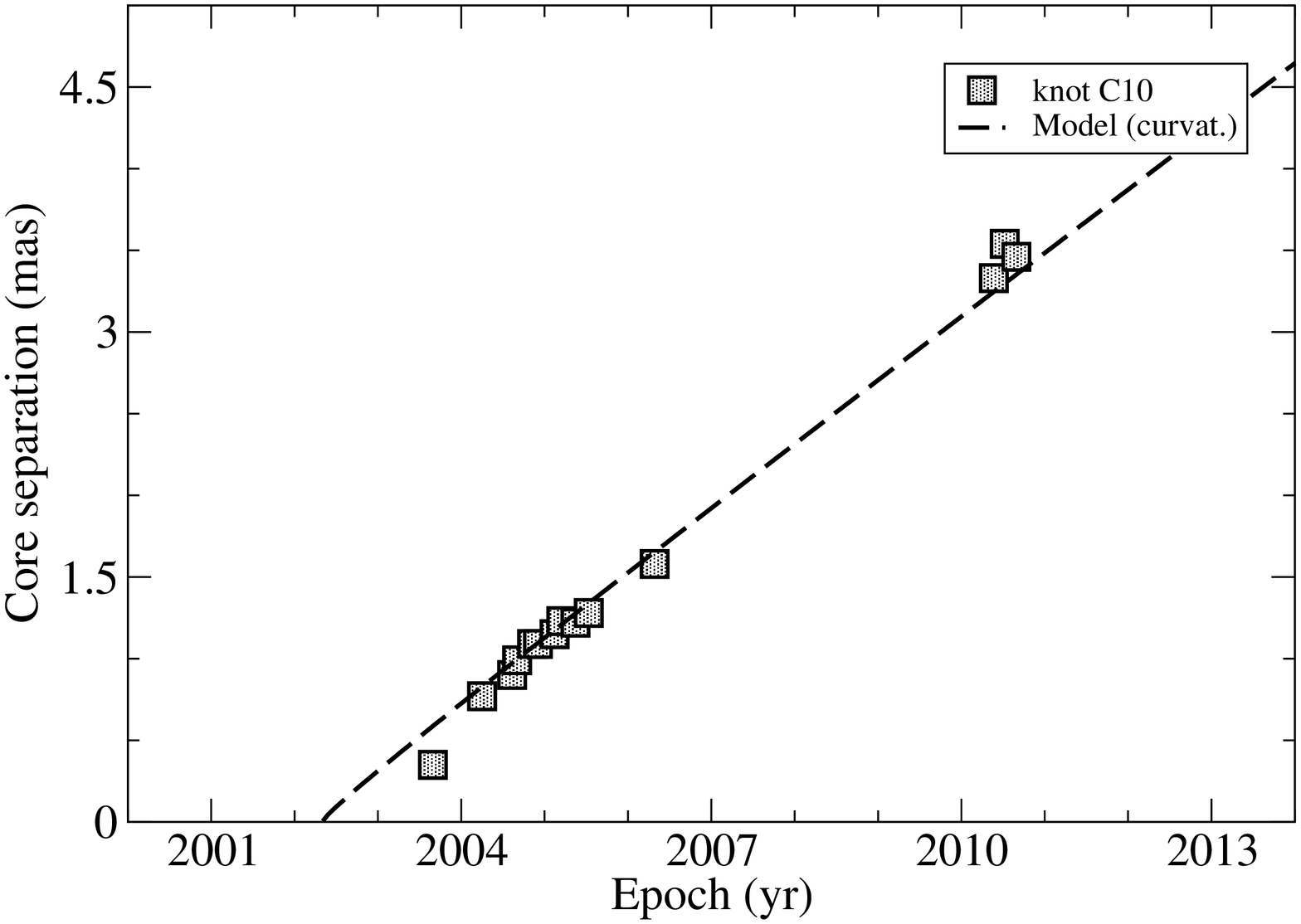}
     \includegraphics[width=5cm,angle=-90]{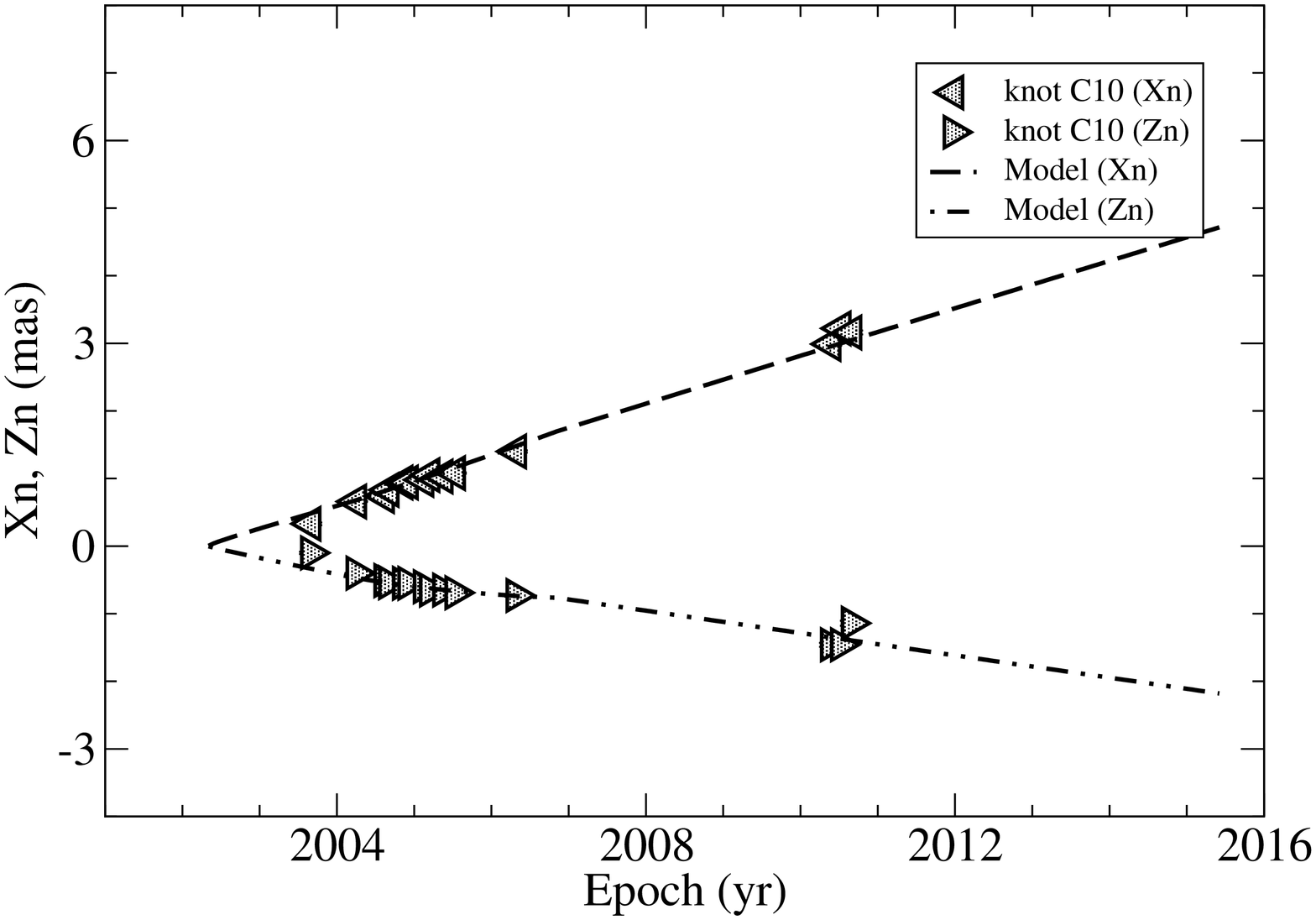}
     \includegraphics[width=5cm,angle=-90]{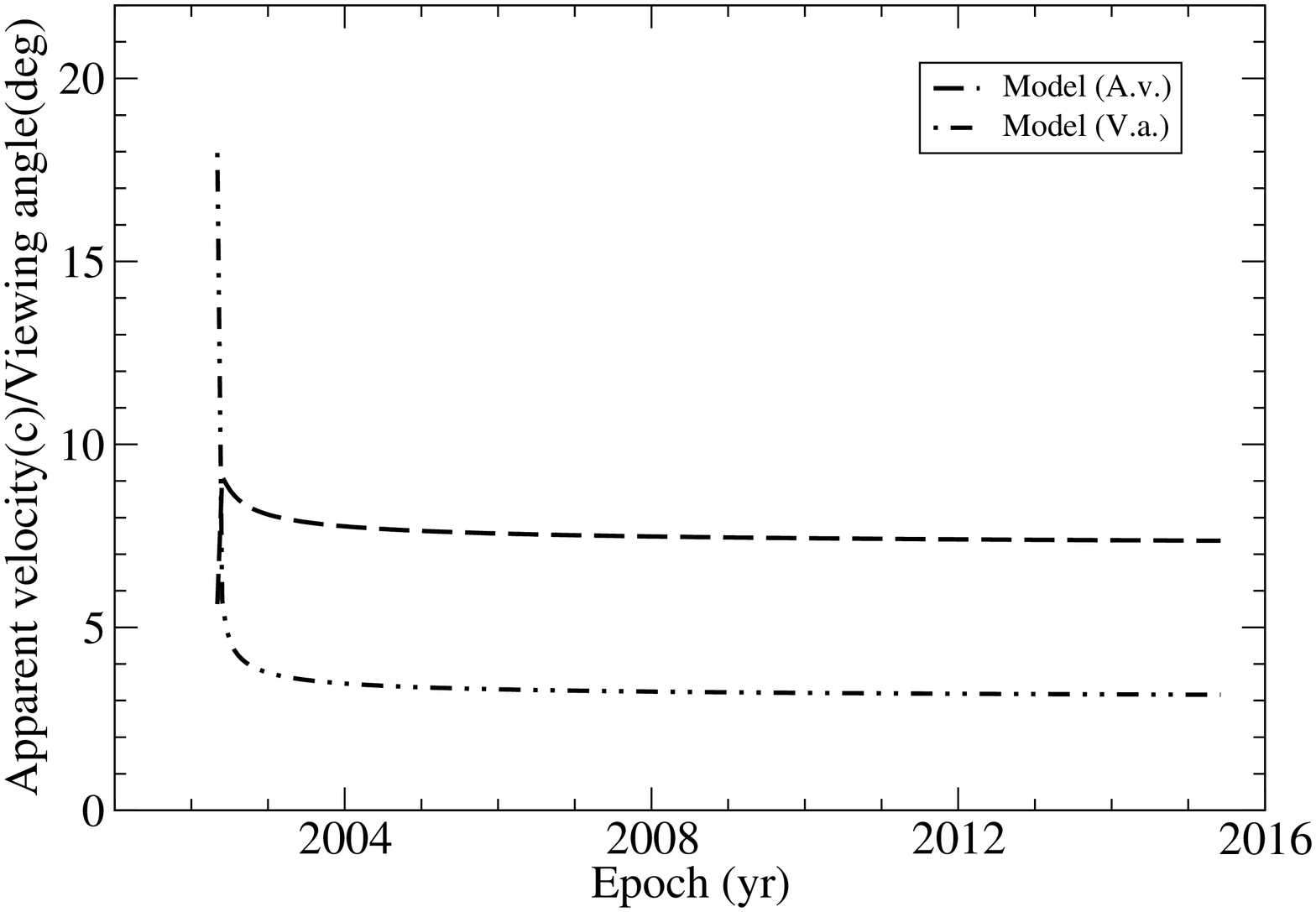}
     \includegraphics[width=5cm,angle=-90]{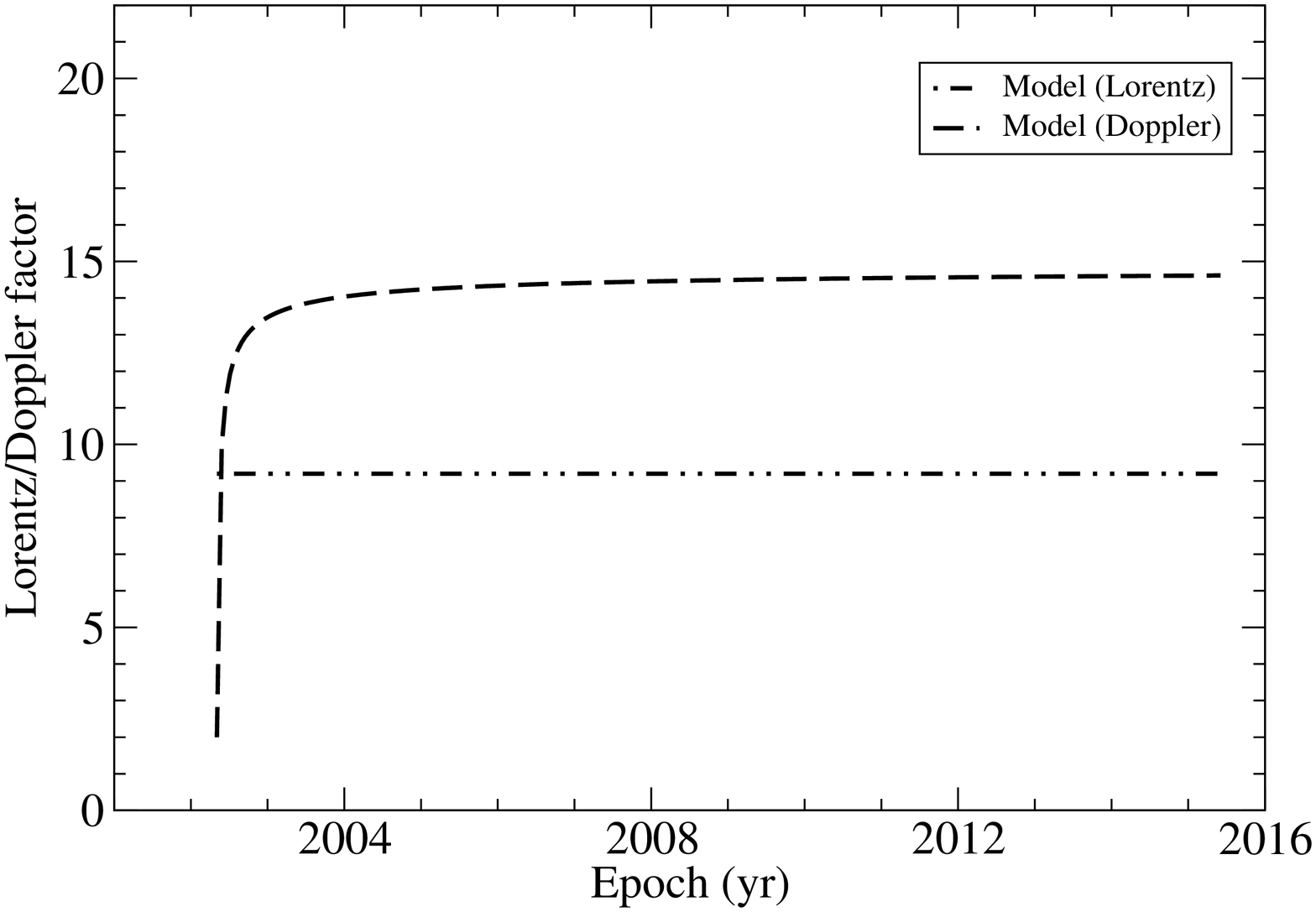}
     \includegraphics[width=5cm,angle=-90]{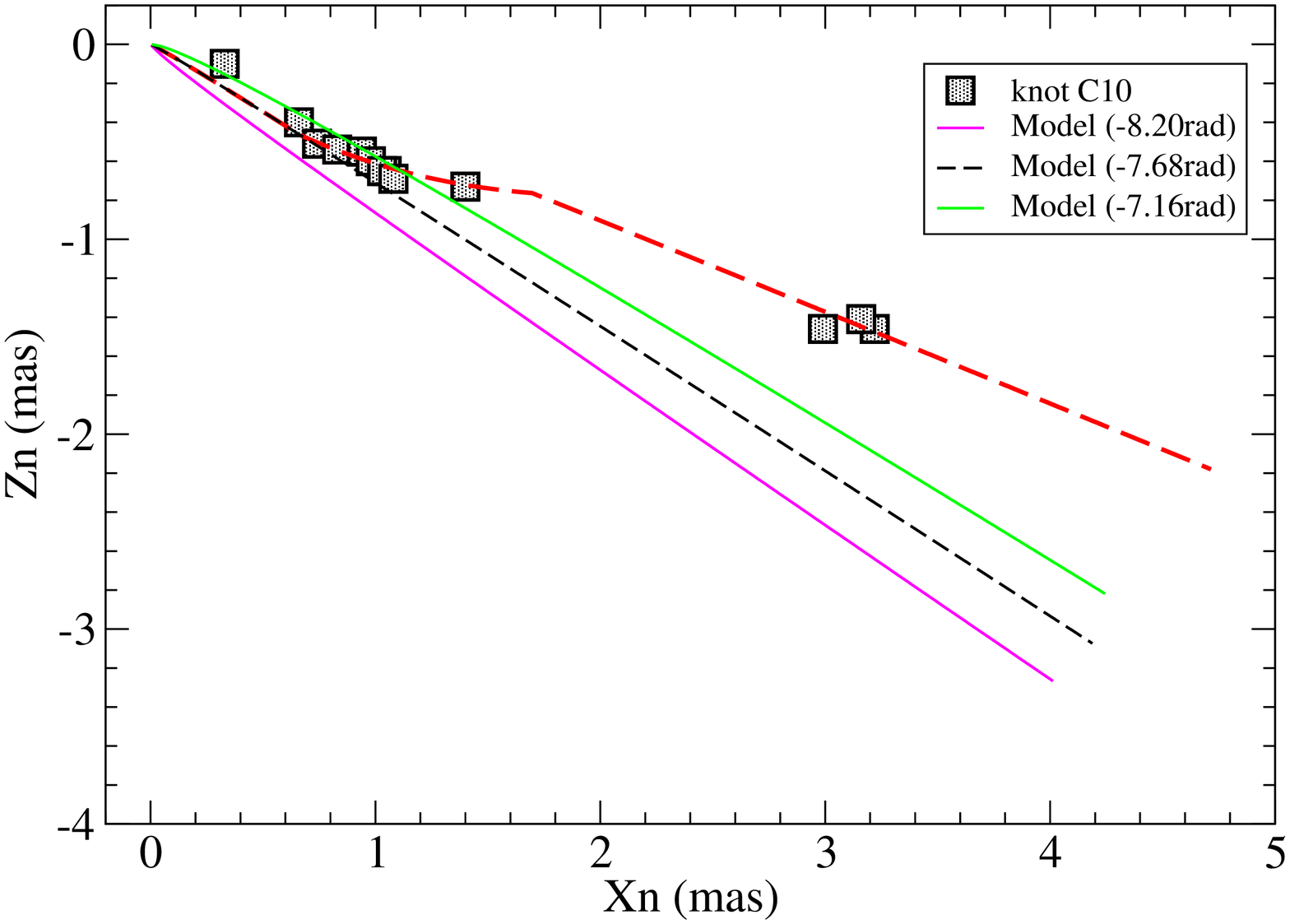}
     \caption{Model-fitting results of the kinematic features for knot C10.
    The entire modeled trajectory is shown by the black dashed line in
      the top left panel. The green and blue lines represent the modeled
    trajectories calculated for precession phases $\omega{\pm}$0.63\,rad,
    showing the precession period having
    been determined within an uncertainty of $\pm$1.2\,yr. In the bottom right
    panel the green and blue lines represent the precessing common trajectories
   calculated for precession phases $\omega{\pm}$0.52\,yr, showing
    its innermost precessing common parabolic trajectory
   having been observed.}
       \end{figure*}
     \begin{figure}
     \centering
     \includegraphics[width=6cm,angle=-90]{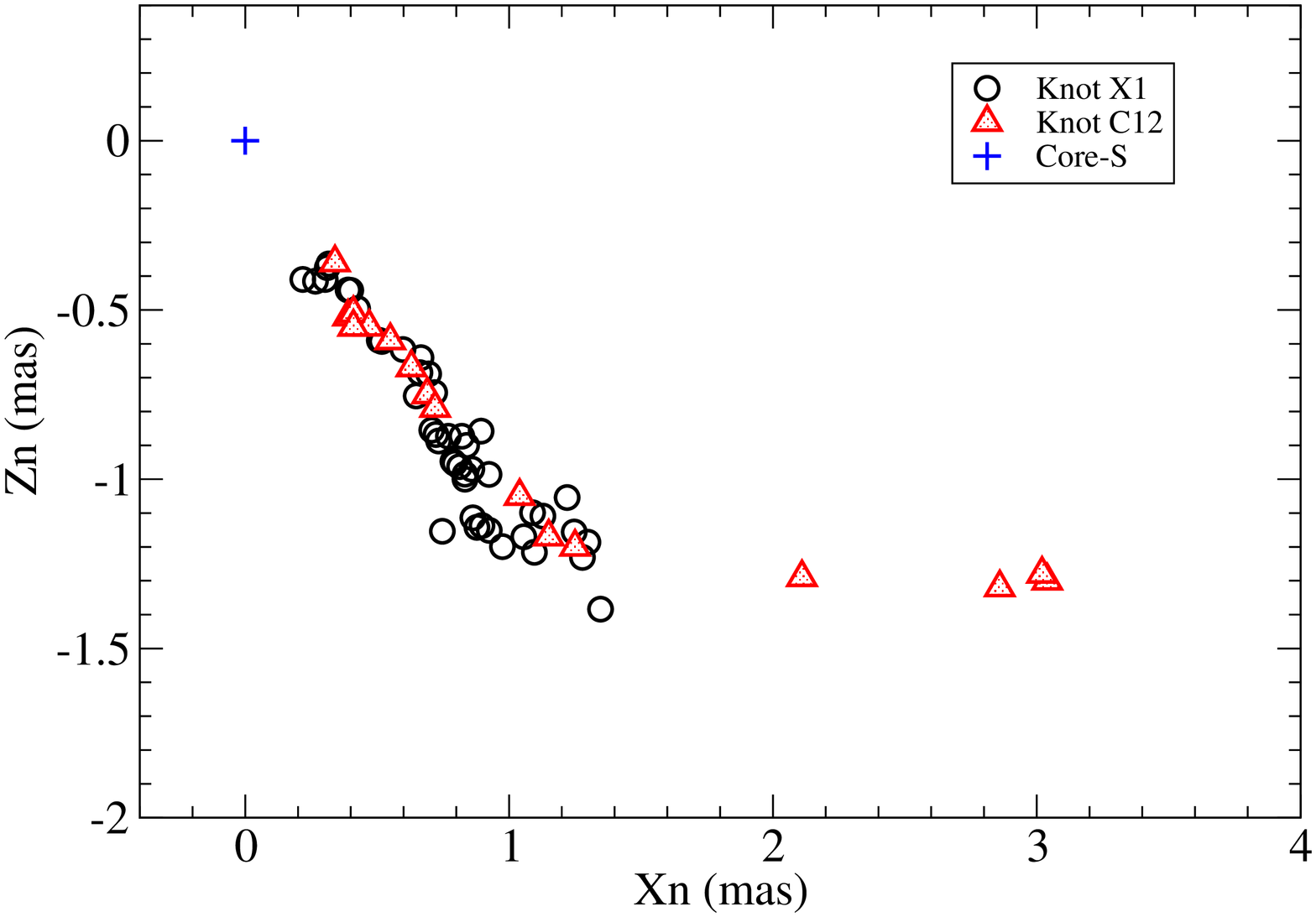}
     \caption{Coincidence of the trajectory of knot C12 and  that of
     knot X1 (43\,GHz; Hodgson et al. \cite{Hod17}): assuming that
     the 15\,GHz core (component-k, Britzen et al. \cite{Br18}) coincides
    with the component-S
     (43\,GHz). The average position of S-component with respect to 43\,GHz 
     core (component-C) is assumed to be (0.091\,mas, 0.173\,mas).}
     \end{figure}
     \subsection{Knots C11 and C12}
     We now come to discuss the model fitting of the kinematic features
     of knots C11 and C12.

     The model-fitting of the kinematics of both knots C11 and C12 is
     a real challenge to our precessing jet scenario for the southern jet.
     The fitting results are particularly important and also a good example
      to reveal the advantage of our precessing jet model.

      According 15\,GHz VLBI-observations, knots C11 and C12 have initial
     position angles of
     ${\sim}-130^{\circ}$  and ${\sim}-135^{\circ}$ (within core separation
      $r_n{\sim}$1.5\,mas), differing only by ${\sim}5^{\circ}$,
     but their ejection time differs by $\sim$4.4 years. This feature
      (the small change in position angle during a large time-interval)
      seems very difficult to be explained in a single jet scenario with a
     rotation period of $\sim$20--30\,yr. However,
       in our double jet scenario knots C11 and C12 belong to
      the southern jet which precesses with a period of 12\,yr and
       the kinematics of both knots C11 and C12 can be consistently
     interpreted.

       The model-fitting results of the kinematic features for knots C11 
     and C12 are shown in Figures A.1 and A.2 of the Appendix, respectively.
    
       For knot C11, the model-fitting results are shown in Fig.~A.1,
      including trajectory
      $Z_n$($X_n$), coordinates $X_n$(t) and
     $Z_n$(t), core separation $r_n(t)$, the modeled apparent velocity and
     viewing angle, the modeled Lorentz factor and Doppler factor.
     Its  ejection time is modeled as $t_0$=2003.10 and the corresponding
     precession phase $\omega$=--8.08\,rad.
   
      Within  core separation $r_n$=1.57\,mas (or radial Z=25\,mas=112\,pc)
      the motion of knot C11 is modeled to follow the precessing common
     parabolic trajectory ($\psi$=0.65\,rad) and its inner kinematics can be
     interpreted in terms of the precessing nozzle model with a precession
     period of 12\,yr. The entire modeled trajectory is shown by the black
     dashed line in the top left panel. Interestingly, 
     the 43\,GHz VLBI-observation data-points
     measured by Agudo et al. (\cite{Ag12}, for knot-T) are also well fitted.
     \footnote{Here we only adopted the three data points for knot-T
     before the appearance of knot-a in Agudo et al., because the appearance of
     knot-a might cause an uncertainty in the core identification,
      see Tateyama \cite{Ta13}.}
     Agudo et al. derived the ejection epoch 2003.22$\pm$0.34 is well consistent
     with our fitted value 2003.10.

     For knot C12, the model-fitting results are shown in Fig.~A.2,
       including trajectory
     $Z_n(X_n)$, coordinates $X_n$(t) and $Z_n$(t), core separation $r_n(t)$,
     the modeled apparent velocity and viewing angle, and the modeled
     Lorentz factor and Doppler factor. Its ejection epoch is modeled as
     $t_0$=2006.90 and the corresponding precession phase
     $\omega$=--10.08\,rad. Within core separation $r_n$=1.56\,mas
    (radial distance Z=33\,mas=148\,pc) the motion of knot C12 is modeled to
    follow the precessing common parabolic trajectory ($\psi$=0.65\,rad; red
   dashed line in the top left panel). Its entire modeled trajectory is
   shown by the black dashed line.
    
     For both knots C11 and C12, their innermost precessing parabolic 
    trajectories have been well observed, as shown in the bottom
     right panel
   of Fig.A.1 and Fig.A.2, respectively. Thus both knots are designated 
   by symbol ``+'' in Table 3.

    In Figure 8 we  show that the trajectory of knot C12 observed at 15\,GHz 
    is coincided with that of the component-X1
    observed at 43\,GHz (Hodgson et al. \cite{Hod17}), if the 15\,GHz core
    component-k (Britzen et al. \cite{Br18}) is assumed to be coincided with 
    the 43\,GHz component-S. The average position of component-S relative to 
    the 43\,GHz core (component-C) is taken to be (0.091\,mas, 0.173\,mas).
     \subsection{Knot C13L}
      In our  double jet scenario
      component C13 designated by Britzen et al. (\cite{Br18})
      has been divided into two
      components C13L and C13U, which are attributed to the southern and
      northern jets, respectively.
   
      We will show that the
      kinematics of knot C13U can be consistently fitted with the other
      components of the northern jet in a precessing nozzle
      model. Component C13L was observed only at three
      epochs at position angles ${\sim}{-180^{\circ}}$ during
      (2009.09--2009.41) and
       seems to be quasi-stationary and disappeared rapidly. Thus it could not
      be taken into the model-fitting together with the other knots of 
      the southern jet. As a possible explanation, knot C13L and the core
      component-k observed at 15\,GHz might form a pair of cores, 
      corresponding to 
      the 43\,GHz pair cores formed by components-C and -S designated 
      in Hodgson et al. (\cite{Hod17}) and the component C13L was not observed
      after 2009.4 due to its opacity at 15\,GHz.
      \begin{table}
      \centering
      \caption{Southern jet (knots C7 to C12): core separation ($r_n$)
       and the corresponding axial distance (Z) within which the knots are
       modeled to move along the precessing
      common parabolic trajectory. Symbol ``+'' denotes that the knots
      have been observed to follow the precessing common parabolic trajectory.}
      \begin{flushleft}
      \centering
      \begin{tabular}{lrrrr}
      \hline
      Knot & $r_n$(mas) & Z(mas) & Z(pc) & status \\
      \hline
      C7 & 1.18 & 25 & 112 & + \\
      C9 & 1.55 & 25 & 112 & + \\
      C10 & 0.81 & 12 & 53.9 & + \\
      C11 & 1.57 & 25 &  112 & + \\
      C12 & 1.56 & 33 & 148 & + \\
      \hline
      \end{tabular}
      \end{flushleft}
      \end{table}
     \subsection{A brief summary for the southern jet}
      The kinematic features of all the superluminal components of the
      southern jet (C7, C9, C10, C11 and C12) are consistently fitted in terms
      of the precessing parabolic jet-nozzle  model with a precession period
    of 12\,yr, thus providing clear evidence for the jet precession
      and supporting the  double jet scenario.
       We summarize briefly the model-fitting results as follows.
     \begin{itemize}
     \item Within core separations $r_n{\leq}$0.8$-$1.5\,mas all the
       superluminal components (C7, C9, C10, C11 and C12) can be well
    modeled to move along the precessing
      common parabolic trajectory, indicating that their inner trajectories
      consistently to follow the precessing nozzle model (referring
     to Qian et al.
     \cite{Qi91}, \cite{Qi09}, \cite{Qi14}, \cite{Qi17}, Qian \cite{Qi11},
      \cite{Qi12}, \cite{Qi13}, \cite{Qi15}, \cite{Qi16}).
    \item Beyond these core separations their paths
      deviate from the model and changes in parameter $\psi$ are introduced to
     explain their outer trajectories. As shown in Table 1, their
     trajectory curvatures are all positive:
     $\Delta{\overline{PA}}$=[$\overline{PA}$($r_n{>}$2\,mas)--
     $\overline{PA}$($r_n{<}$2\,mas)]$>$0, i.e., their outer trajectories
     are curved upward.
     This is just opposite to the curvature direction for the
     superluminal components of the northern jet (see below).
     \item The modeled precession period is 12\,yr,
      similar to the period determined from the optical light curve
       (Sillanp\"a\"a, \cite{Si88}, Valtonen et al. \cite{Va16}), indicating
     that the periodic behavior in optical and radio regimes may originate from
      a common mechanism. As viewed along the line of sight
     the jet precesses  clockwise (Figs 4 and 8).
    \item The (projected) jet cone spans a position angle range from
      $-110^{\circ}$ to $-145^{\circ}$ with an aperture of $\sim{35^{\circ}}$
      and the viewing angle varies in a range from $2.5^{\circ}$
      to $3.5^{\circ}$ (see Figure 14 below).
       The position angle of the precession axis is ${\sim}-130^{\circ}$.
     \item The modeled ejection times for knots (C7, C10, C11 and C12) are
     (1998.55, 2002.33, 2003.10, 2006.90), which are well consistent with those
     (1998.9, 2002.5, 2002.9, 2007.3) obtained from the 15\,GHz
      VLBI- measurements with differences
      ${\mid}{t_0}-{t_{0,obs}}{\mid}$$<$0.2$-$0.4\,yr
      (see Table 1). In particular, the modeled ejection epochs
       for knots C7 and  C9 are confirmed by the 43\,GHz observations
       (Agudo et al. \cite{Ag12}, Jorstad et al. \cite{Jor05}). This strongly
       supports the modeled precession period of 12\,yr for the southern jet.
      \item For knots C7, C9 and C11 the 43\,GHz data given in Agudo et al.
       (\cite{Ag12}) and Jorstad et al. (\cite{Jor05}) were used to extend
       the observations to smaller core separations and they were also well
       fitted by our precessing nozzle model. This is really a posterior test
       supporting the precessing parabolic trajectory model, because we did
       not use the 43\,GHz data to construct the model. In particular,
      our modeled ejection epoch for knot C9 is 2001.80, being one year 
      later than that (2000.80) obtained  from the 15\,GHz-VLBI
       measurements. But the modeled ejection epoch for
       knot C9 is closely consistent with the ejection epoch obtained 
       by the 43\,GHz VLBI measurements (Agudo et al. 2012, knot-R:
       $t_{0,obs}$=2001.92$\pm$0.12). This consistency of ejection epochs
      is significant, verifying our precessing jet nozzle model and the 12\,yr
      precession period.
      \item  Knots C11 and C12 were observed at similar position angles,
      but their ejection epochs differ by four years. The interpretation
     of the kinematic features of both knots C11 and C12 in terms of our
     precessing jet nozzle model is a real success and particularly instructive,
      demonstrating the validity and flexibility of the model.
     \item We find that knot C8 does not belong to the southern jet and reveals
     kinematic features distinct from  knots C7 to C12. Its kinematic features
      will be interpreted consistently with the superluminal components of
     the northern jet in terms of
     another precessing nozzle model (see Sect.7 below).
     The division of knot C8 from the group of superluminal knots
     (C7, C9, C10, C11 and C12) is significant for understanding
     the entire phenomenon observed in OJ287.
    \item In Table 3, we summarize the core separations $r_n$
    and the corresponding axial distance (Z) within which the knots are modeled
    to follow the precessing common parabolic trajectory. It can be seen that
    for all the superluminal components (C7 to C12) their
    inner trajectories (in the axial distance range from $\sim$50\,pc to
    $\sim$150\,pc) have been observed to follow the precessing common parabolic
    trajectory. This indicates that the 12\,yr precession period and the
    parabolic trajectory pattern may be really applicable to the southern jet.
      \end{itemize}
     \begin{figure}
     \centering
     \includegraphics[width=6cm,angle=-90]{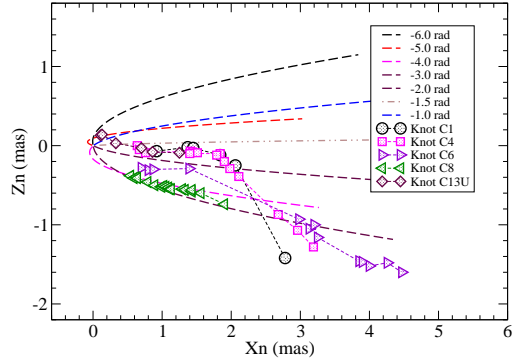}
     \caption{Modeled apparent northern jet cone: Distribution of
    the precessing common parabolic trajectories with precession phases
    $\omega$=--1.0\,rad to --6.0\,rad
    and that of the observed trajectories of the superluminal components (C1,
    C4, C6, C8 and C13U).  The axis of the jet cone is at position angle of
    ${\sim}{-80^{\circ}}$.} 
      \end{figure}
     \section{Model-fitting results for northern jet}
      We now come to discuss the model fitting results for the northern jet
      components.

     The entire structure of the northern jet is shown in Figure 9. It
     consists of nine superluminal components (C1, C2, C3, C4, C5, C6, C8,
     C13U and C14). We note that knots C1, C2, C3, C4, C5 and C6 all have
     downward curvatures in their trajectories beyond a core separation
     $r_n{>}{\sim}$2\,mas
     ($\Delta{\overline{PA}}$=[$\overline{PA}$($r_n{>}$2\,mas)--
      $\overline{PA}$($r_n{<}$2\,mas)]$<$0).
     This is different from that in the southern
     jet, where all the knots have upward curvatures in trajectory. This
     can be regarded as another clue for the double jet structure.

     Through analyzes and trial model-fittings of the kinematics for the
   northern jet components we select the model parameters as: $T_p$=12\,yr,
     $\epsilon$=$3^{\circ}$,
     $\psi$=0.0\,rad, $a$=0.1340$[{\rm{mas}}]^{\frac{1}{2}}$ and $x$=0.5,
     which are listed in Table 4.  The values  for $\psi$ and $a$
     are different from those assumed for the southern jet
     ($\psi$=0.65\,rad and
      $a$=0.0536$[{\rm{mas}}]^{\frac{1}{2}}$), implying the northern jet having
    a different orientation in space and a different precessing common 
    parabolic trajectory pattern.

     The ejection epoch ($t_0$) of the knots is related to the precession
      phase ($\omega$) as:
    \begin{equation}
     {t_0}=1994.10 - ({\omega}+0.755){T_p}/2{\pi} ,
    \end{equation}
     Here $\omega$=--0.755 represents the precession phase of knot C1,
     corresponding to its ejection epoch $t_0$=1994.10. Some modeled parameters
      and relevant observation data
     are listed in Table 5.\\
      \begin{table}
      \centering
    \caption{Model parameters for the northern jet
    (knots C1, C2, C3, C4, C5, C6, C8, C13U and C14):
     $T_p$, $\epsilon$, $\psi$, $a$ and $x$.}
    \begin{flushleft}
    \centering
    \begin{tabular}{lr}
    \hline
    $T_p$ & 12\,yr \\
    $\epsilon$ & 0.0524\,rad=$3.0^{\circ}$ \\
    $\psi$ & 0.0\,rad=$0.0^{\circ}$ \\
    $a$  &   0.1340$[\rm{mas}]^{1/2}$ \\
    $x$   &   0.5 \\
    \hline
    \end{tabular}
    \end{flushleft}
    \end{table}
     \begin{table*}
     \centering
     \caption{Modeled parameters for the superluminal components of the northern
      jet and some relevant observation data (taken from Britzen et al.
      \cite{Br18}): modeled ejection time $t_0$,
      precession phase $\omega$, initial Lorentz factor $\Gamma$, measured
      ejection time $t_{0,obs}$, apparent velocity ${\beta}_a$,
      average position angle $\overline{PA}$, curvature in trajectory
      $\Delta{\overline{PA}}$ (see Table 2).}
     \begin{flushleft}
     \centering
     \begin{tabular}{lrrrrrrr}
     \hline\noalign{\smallskip}
     Knot & $t_0$ & $\omega$ &  $\Gamma$ & $t_{0,obs}$ & ${\beta}_a$ &
                    $\overline{PA}$(deg.) & $\Delta$${\overline{PA}}$(deg.)\\
     C1 & 1994.10 & --0.755 & 13.5 & 1994.1 & 13.4$\pm$1.0 &-92.4 & -14.6\\
     C2 & 1994.80 & --1.12 & 13.0 & 1994.8 & 13.0$\pm$0.4 & -91.2 & -12.6\\
     C3 & 1995.29 & --1.38 & 12.5 & 1995.3 & 12.0$\pm$0.4 & -91.5 & -18.6\\
   C4 & 1995.50 & --1.49 & 11.4 & 1995.5 & 11.1$\pm$0.2 & -93.5 & -12.8 \\
   C5 & 1997.05 & --2.30 & 9.5--11.0  & 1997.1 & 9.0$\pm$0.2 & -107.1 & -4.3 \\
   C6 & 1997.24 & --2.40 & 6.8--10.3  & 1997.9 & 8.8$\pm$0.2 & -108.4 &  -0.9 \\
     C8 & 2000.20 & --3.95 & 12.5 & 2000.1 & 8.6$\pm$0.2 & -117.2 & +5.8\\
     C13U & 2005.45 & --6.70  & 1.1--7.5 & --   & --          & -75.3  & --   \\
     C14 & 2006.03  & --7.00  & 2.0--7.0 & -- & --            & -64.1  & --   \\
     \noalign{\smallskip}\hline
     \end{tabular}
     \end{flushleft}
     \end{table*}
     \begin{figure*}
     \centering
     \includegraphics[width=5cm,angle=-90]{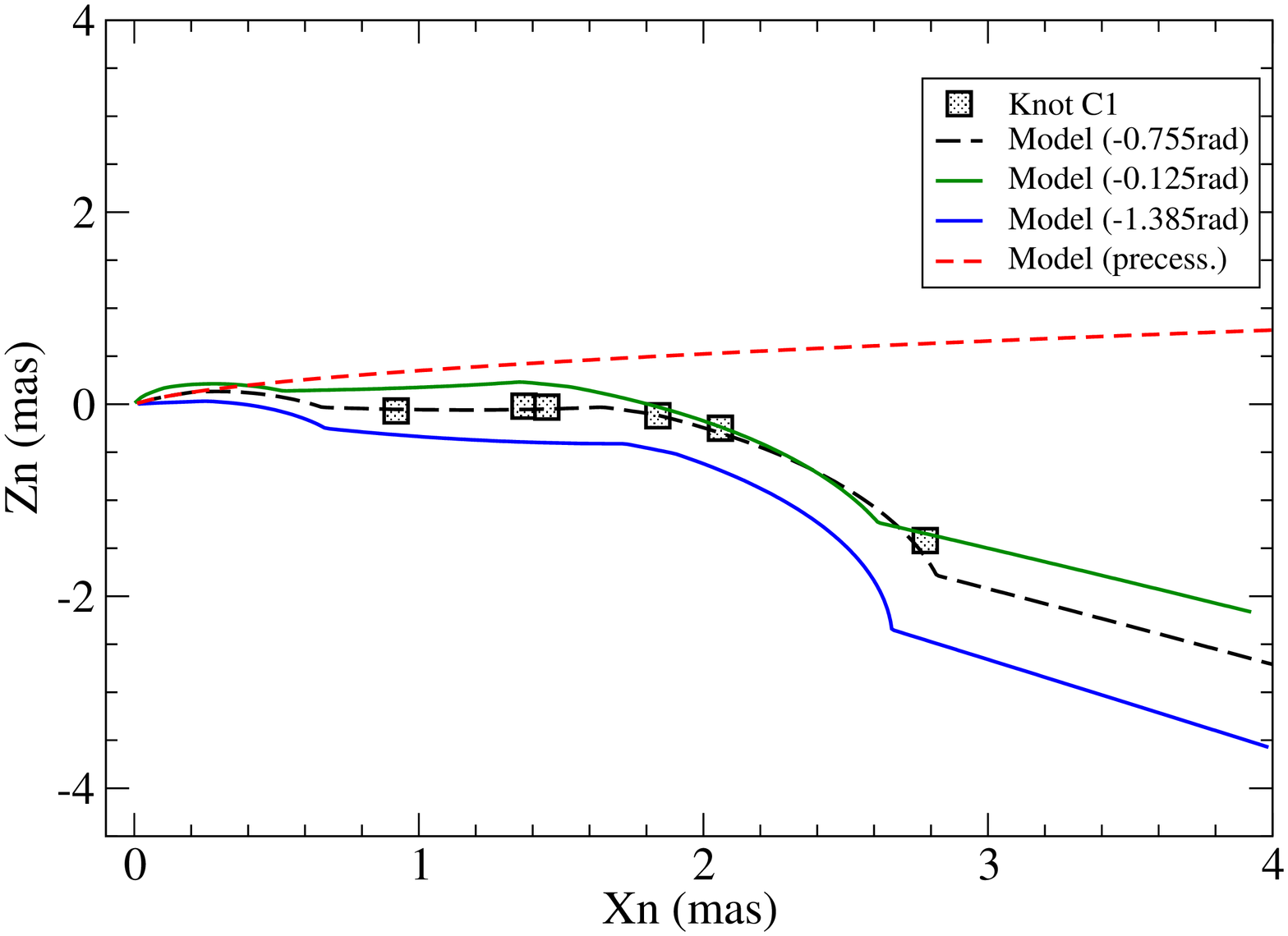}
     \includegraphics[width=5cm,angle=-90]{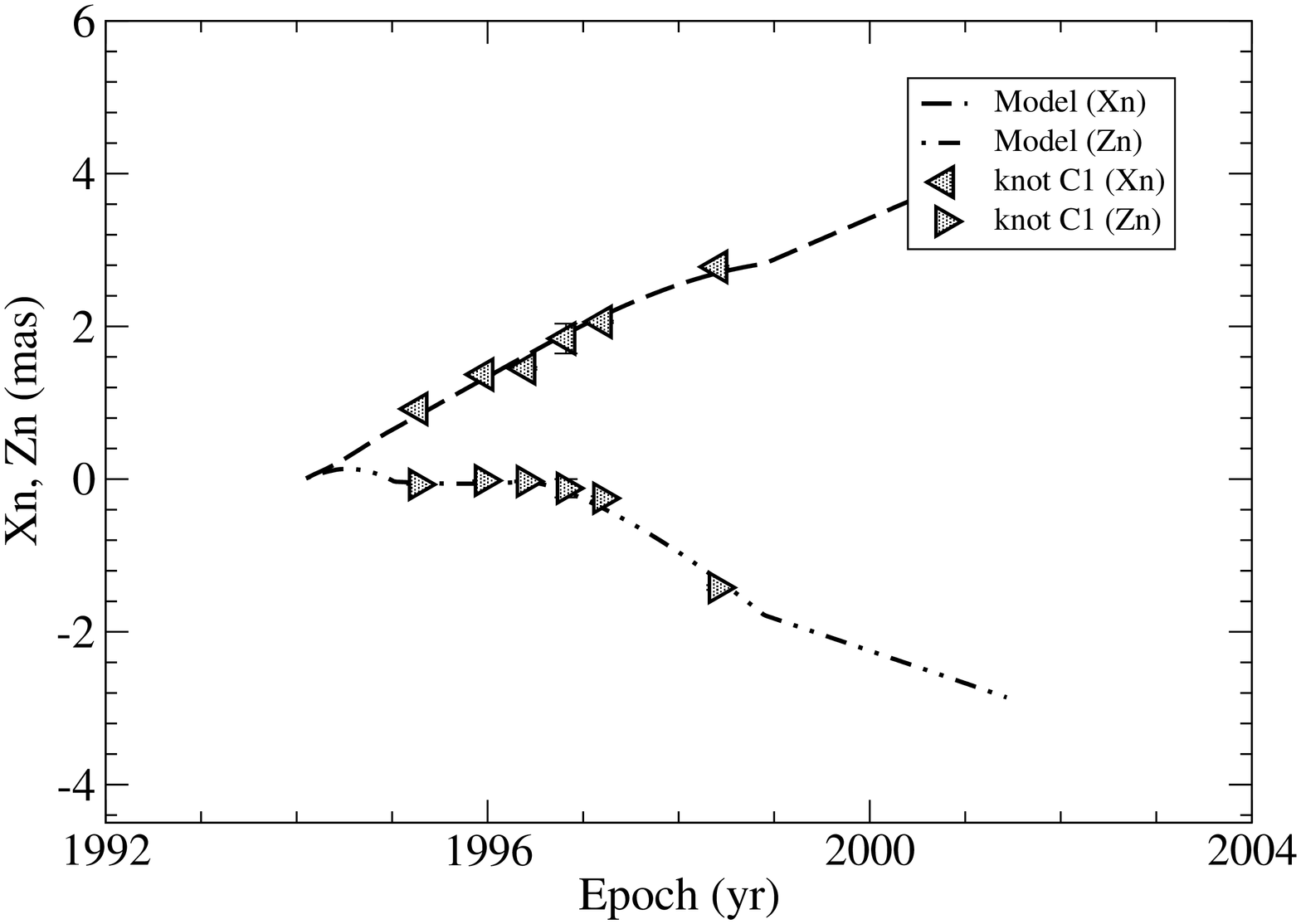}
     \includegraphics[width=5cm,angle=-90]{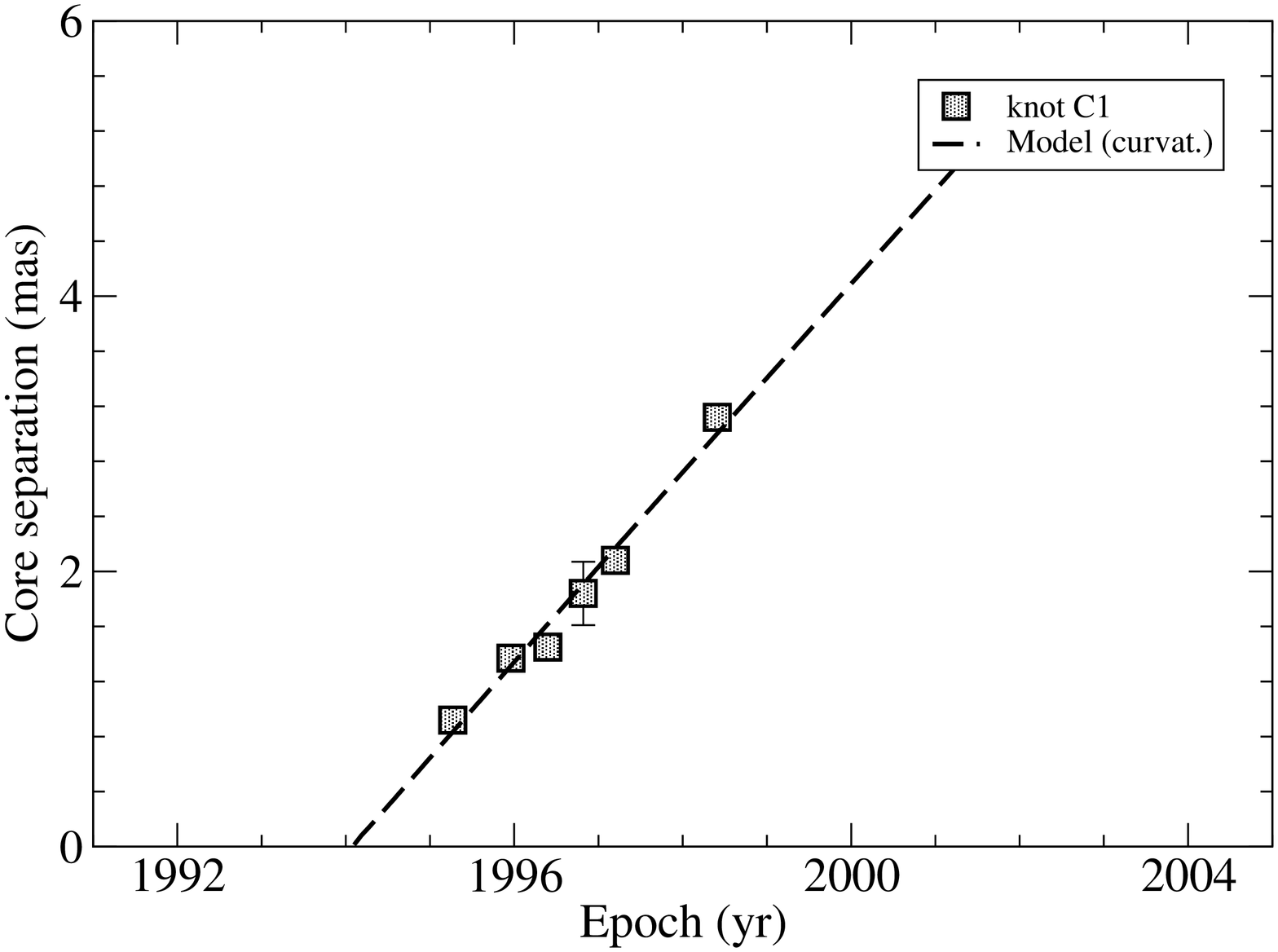}
     \includegraphics[width=5cm,angle=-90]{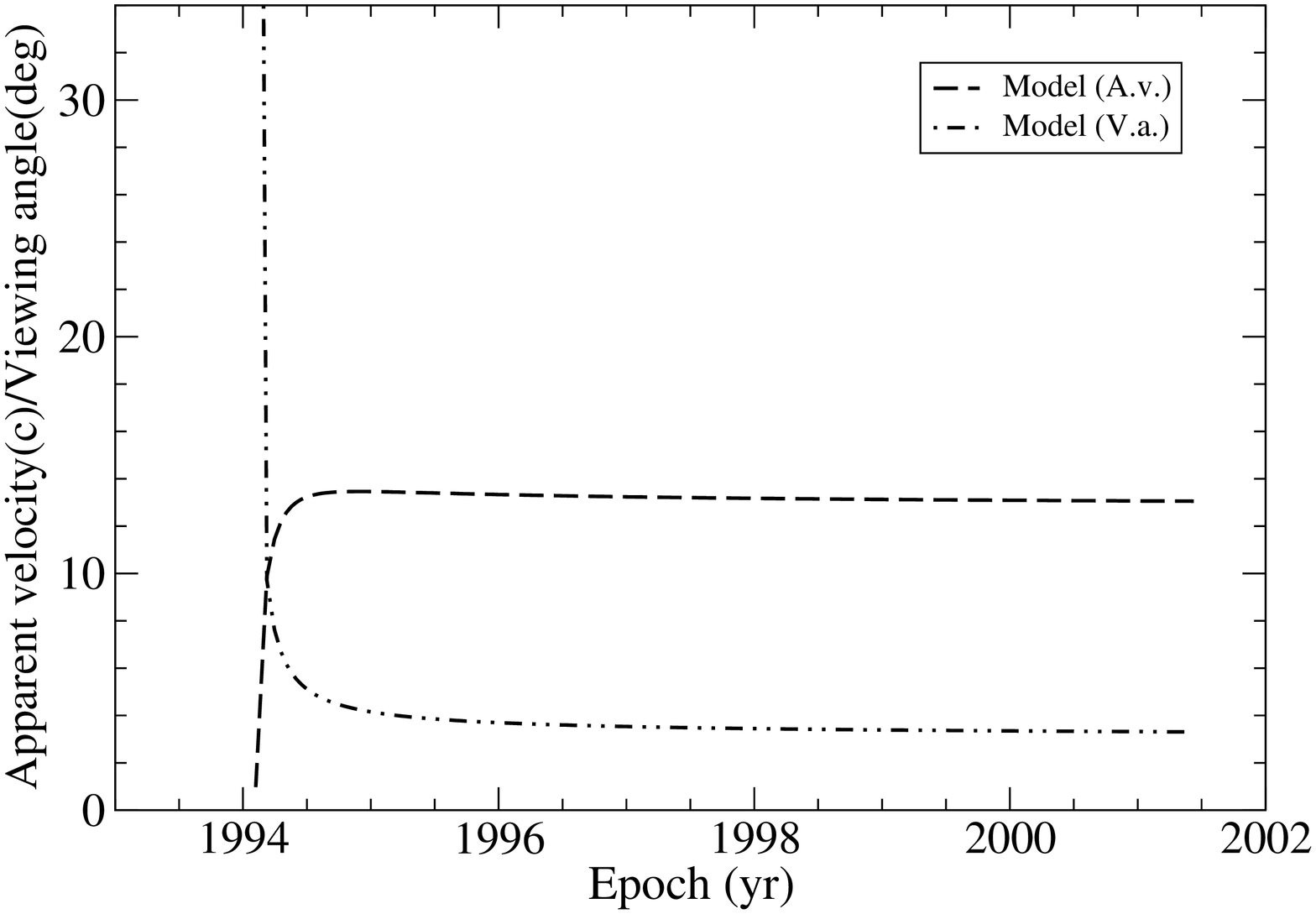}
     \includegraphics[width=5cm,angle=-90]{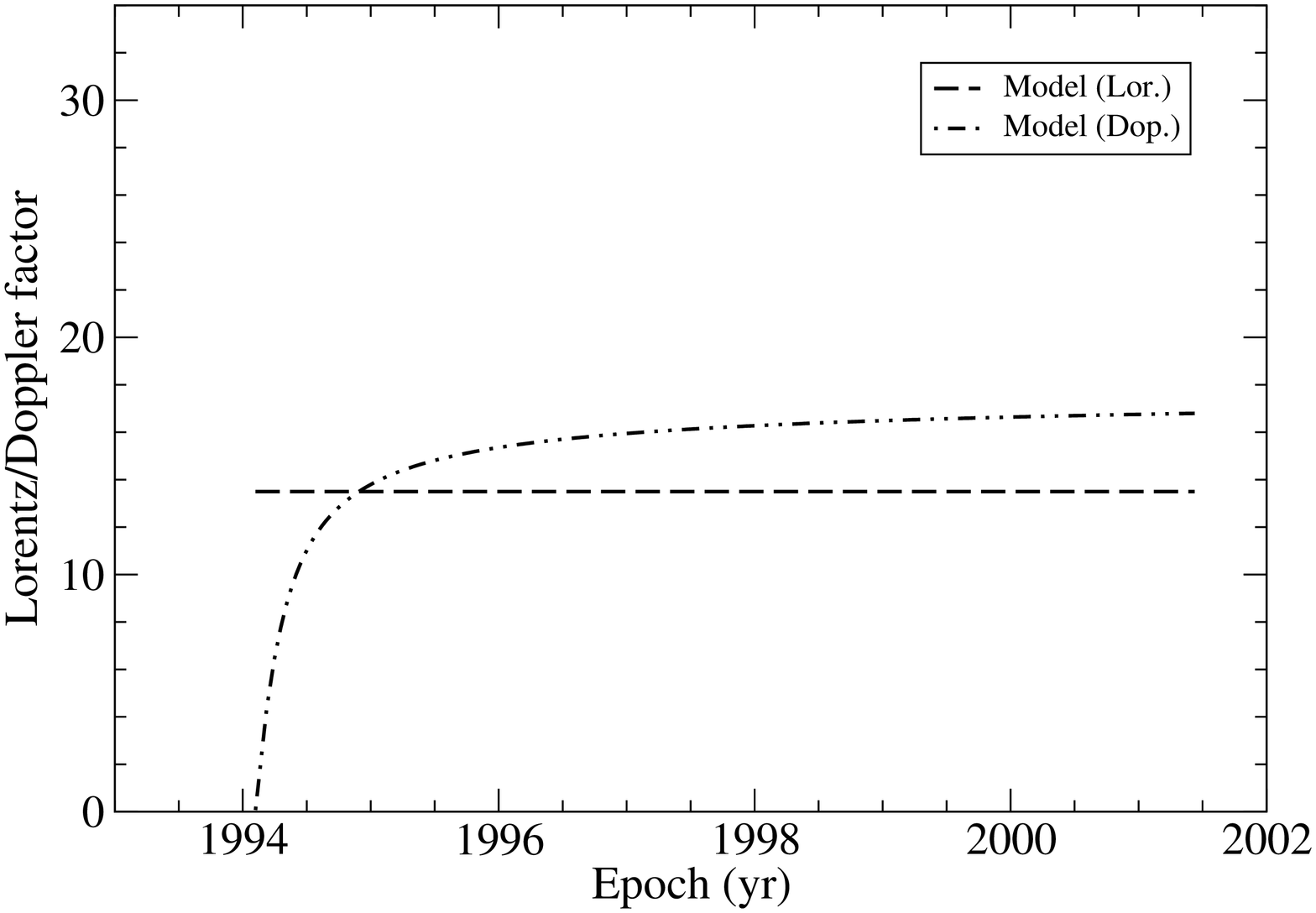}
     \includegraphics[width=5cm,angle=-90]{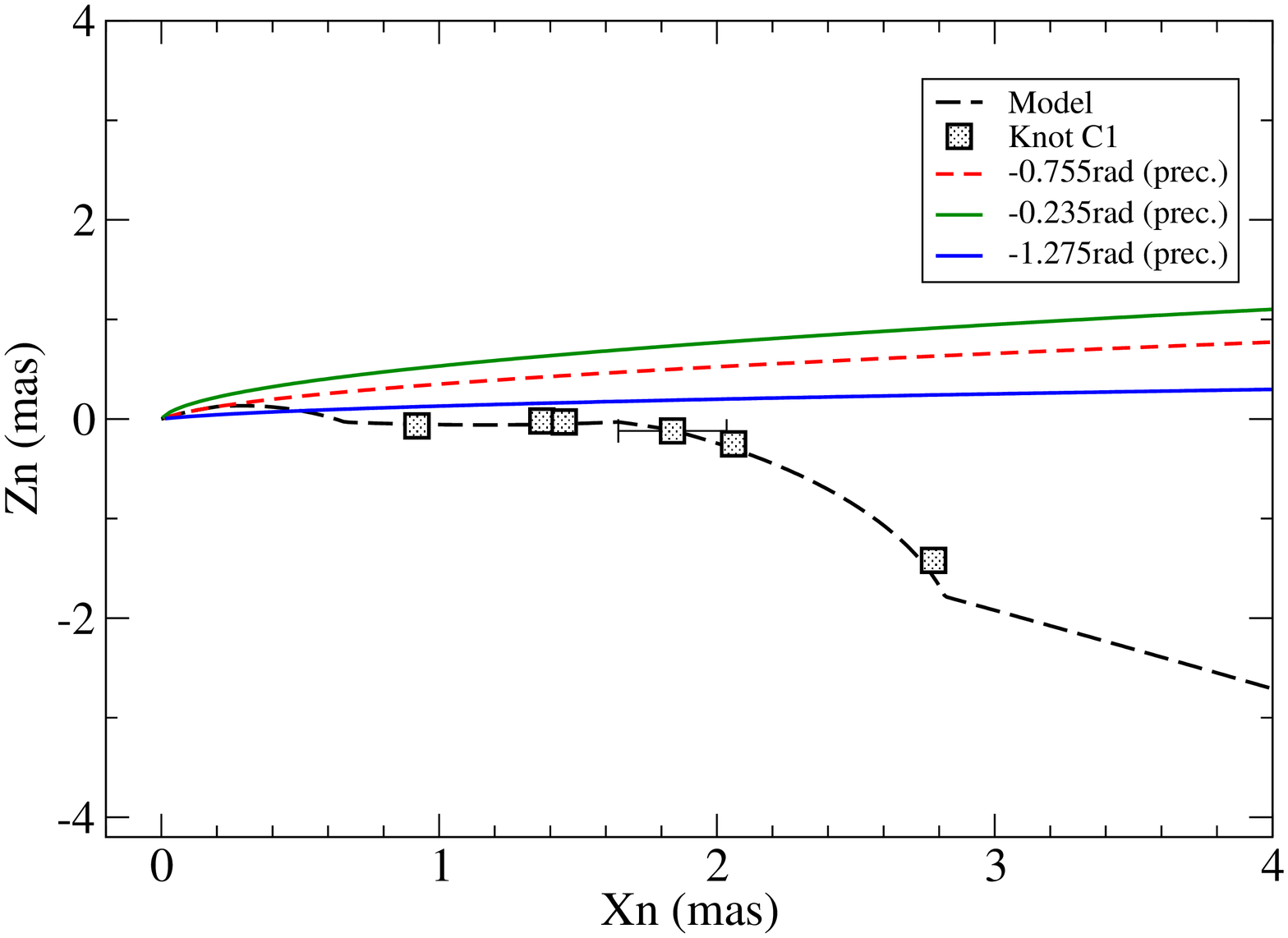}
     \caption{Model fitting results of the kinematic features for knot C1.
     The entire modeled trajectory is indicated
    by the black dashed line in the top left panel. The green and
   blue lines show the  precession period having been determined within 
   an uncertainty of $\sim$1.2\,yr. In the bottom right panel, the green and
    blue lines show  its initial precessing common trajectory having not been
     observed (no observation data available). See text.}
     \end{figure*}
        \subsection{Knot C1}
     It is found that knots C1, C2, C3 and C4 have very similar trajectories
     within core separation $\sim$2\,mas. This might
     imply that in order to explain their ejections at different times,
     their innermost tracks  should deviate from the precessing common
     trajectory at different separations.

     The model-fitting results for knot C1 are shown in Figure 10, including
     trajectory $Z_n(X_n)$, coordinates $X_n(t)$ and $Z_n(t)$, core separation
     $r_n(t)$, the modeled apparent velocity and viewing angle, the modeled
      Lorentz/Doppler factor. Its ejection time is modeled as $t_0$=1994.10
     and the corresponding precession phase $\omega$=-0.755\,rad.

   Within core separation $r_n$=0.23\,mas (radial distance Z=1.6\,mas=7.18\,pc)
    knot C1 is modeled to move along the precessing common parabolic trajectory
     ($\psi$=0.0\,rad; red dashed line in the top left panel of Fig.10).
    Beyond this separation changes in parameter $\psi$
     (i.e. trajectory curvatures) are introduced  to explain its outer path:
     For Z=1.6--7.0\,mas $\psi$(rad)=0.45(Z--1.6)/(7.0--1.6); For Z=7--22\,mas
     $\psi$(rad)=0.45--0.15(Z--7)/(22-7); For Z=22--25\,mas
      $\psi$(rad)=0.30+0.028(Z--22)/(25--22); For Z=25--50\,mas
      $\psi$(rad)=0.328+0.444(Z--25)/(50--25); For Z$>$50\,mas
     $\psi$=0.772\,rad. The entire  modeled trajectory is shown by the black
     dashed line.

     The modeled Lorentz factor $\Gamma$=const.=13.5. The modeled apparent
     velocity is well consistent with the VLBI-measured proper speed
     0.70$\pm$0.05\,mas/yr (13.4$\pm$1.0\,c) given in Britzen et al.
    (\cite{Br18}).

     In the top left panel of Figure 10  two additional lines (green and
      blue) represent the modeled trajectories calculated for precession phases
      $\omega{\pm}$0.63\,rad, showing that most of the data-points are within
   the position angle range defined by the two lines and the precession period
   having been determined within an uncertainty of $\pm$1.2\,yr. In the bottom
   right panel the green and blue lines represent the precessing
     common trajectories calculated for $\omega{\pm}$0.52\,rad, showing that no
     data-points are within the position angle range defined by the two lines
     and its innermost precessing common trajectory having not been observed
     (no observation data available).
     Thus knot C1 is designated by symbol ``--'' in Table 6.
       \subsection{Knots C2/C3 and C5/C6}
      The model-fitting results  of the kinematic features for knots C2/C3 and 
      C5/C6  are shown in Figures A.3/A.4 and A.5/A6 of the Appendix,
        respectively.
      These include trajectory $Z_n(X_n)$, coordinates
      $X_n(t)$ and $Z_n(t)$, core separation $r_n(t)$, the modeled apparent
      velocity and viewing angle, the modeled Lorentz factor and Doppler
      factor.

      The ejection times are modeled as $t_0$=1994.80, 1995.29, 1997.05 
      and 1997.24 and the corresponding precession phases are
       $\omega$=--1.12\,rad, --1.38\,rad, --2.30\,rad and --2.40\,rad.

      As the model-fitting for knot C1, their motion in the innermost regions
      are modeled to follow the precessing
     common parabolic trajectory. But in the outer regions changes in parameter
     $\psi$ (or trajectory  curvatures) are introduced to explain  their outer 
     trajectories. These model-fittings also indicate the precession period 
     has been determined within an uncertainty of $\sim$1.2\,yr and their
     initial parabolic trajectories have been observed. (See captions of 
     Figs.~A.3/A.4 and A.5/A.6).
      \subsection{Knot C4}
    The model-fitting results for knot C4 are shown in Figure 11. Its
     ejection epoch is modeled as $t_0$=1995.50 and
    the corresponding precession phase $\omega$=$-$1.49\,rad. The entire 
    trajectory is model-fitted by the black dashed line in top left
    panel of Fig.11.

    To fit its core separation versus time,  its motion is modeled to be
    uniform: $\Gamma$=const.=11.4. The modeled apparent velocity  is
   consistent with the  VLBI-measured proper speed 0.58$\pm$0.01\,mas/yr
   (${\beta}_a$=11.1$\pm$0.2) given in Britzen et al. (\cite{Br18}).
  
  Interestingly, the data given in Homan et al. (\cite{Hom01}, 22\,GHz, for
   knot-K3) and in Lister et al. (\cite{Li98}, 43\,GHz for knot-C3) are 
   also well fitted by the model. Moreover, Tateyama et al. (\cite{Ta99})
   derived the ejection epoch to be $t_{0,obs}$=1995.4,
   which is quite similar to our modeled ejection epoch $t_0$=1995.50.
  The consistency of these observations provides a strong support to our
   precessing nozzle model with a precession period of 12\,yr.
  
 In top left panel of Figure 11, the green and blue lines represent the
  modeled trajectories calculated for precession phases $\omega{\pm}$0.63\,rad,
    showing most of the data-points within the position angle range defined by
    the two lines and the precession period having been determined within
    an uncertainty of $\pm$1.2\,yr. In bottom right panel of Fig.11,
    the green and blue lines represent the precessing common parabolic
     trajectories calculated for   $\omega{\pm}$0.52\,rad, showing a number
    of data-points within the position
    angle range defined by the two lines and its innermost precessing common
    trajectory having been observed.  Thus knot C4 is designated by
    symbol ``+'' in Table 6.
     \begin{figure*}
     \centering
     \includegraphics[width=5cm,angle=-90]{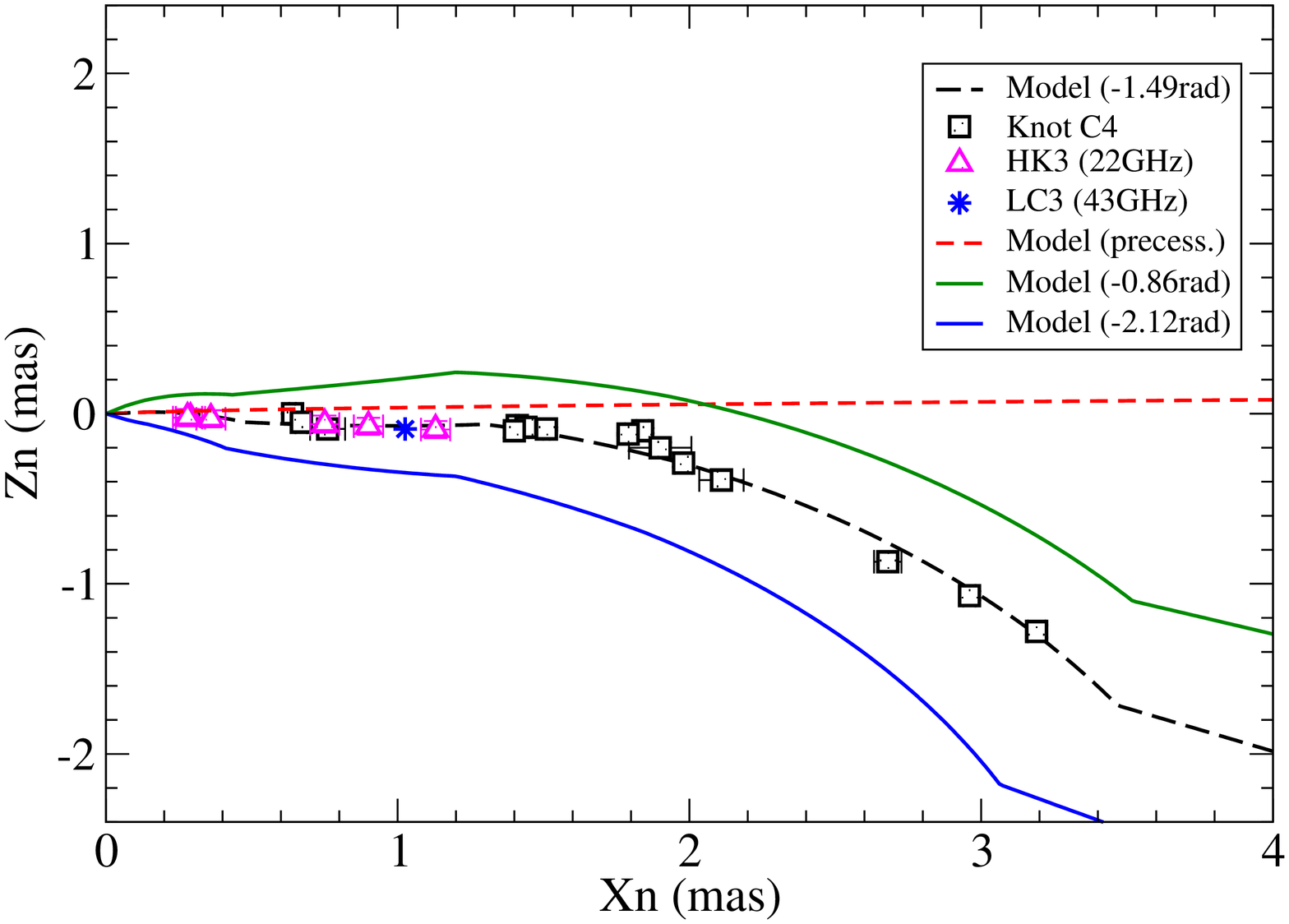}
     \includegraphics[width=5cm,angle=-90]{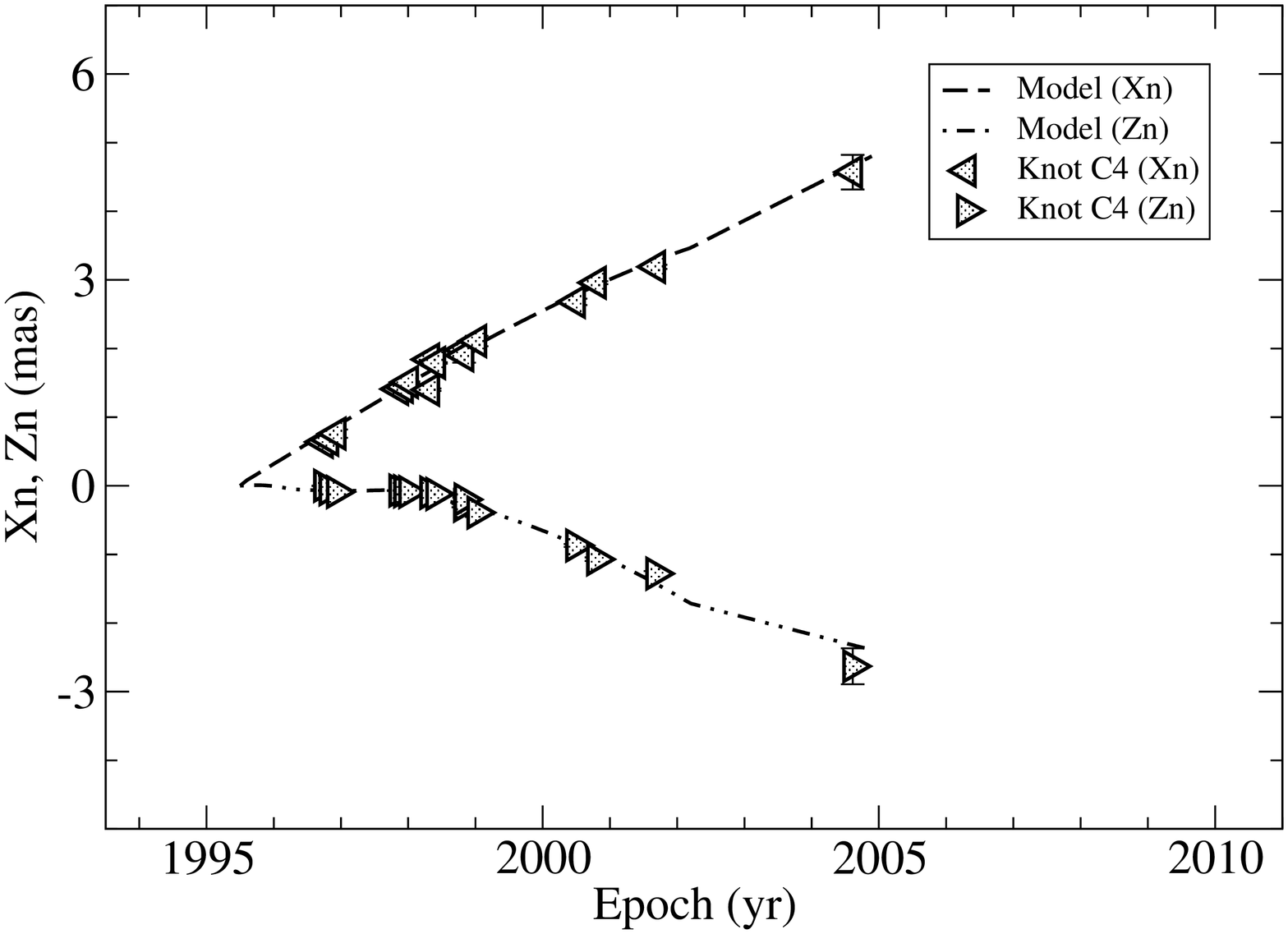}
     \includegraphics[width=5cm,angle=-90]{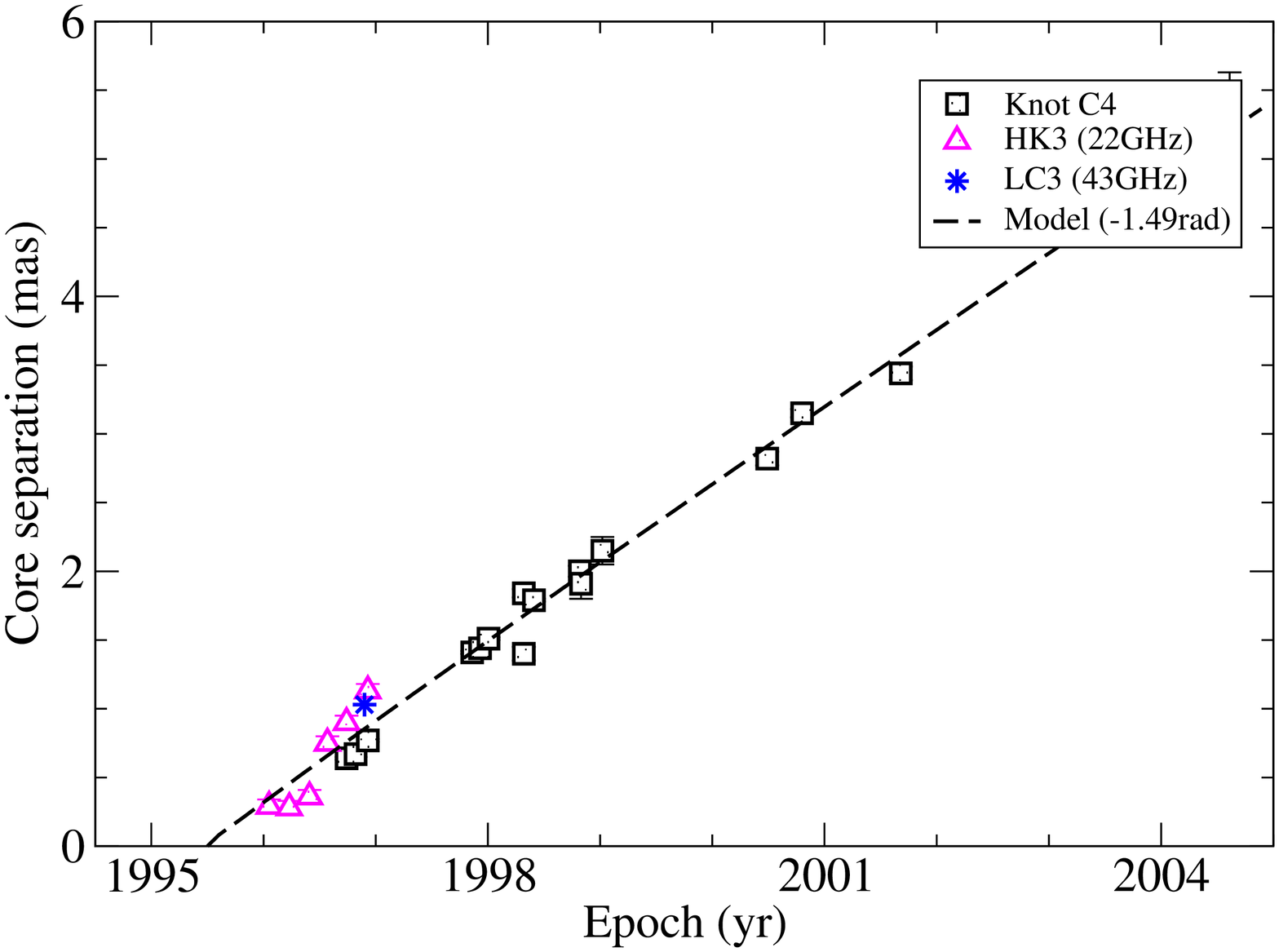}
     \includegraphics[width=5cm,angle=-90]{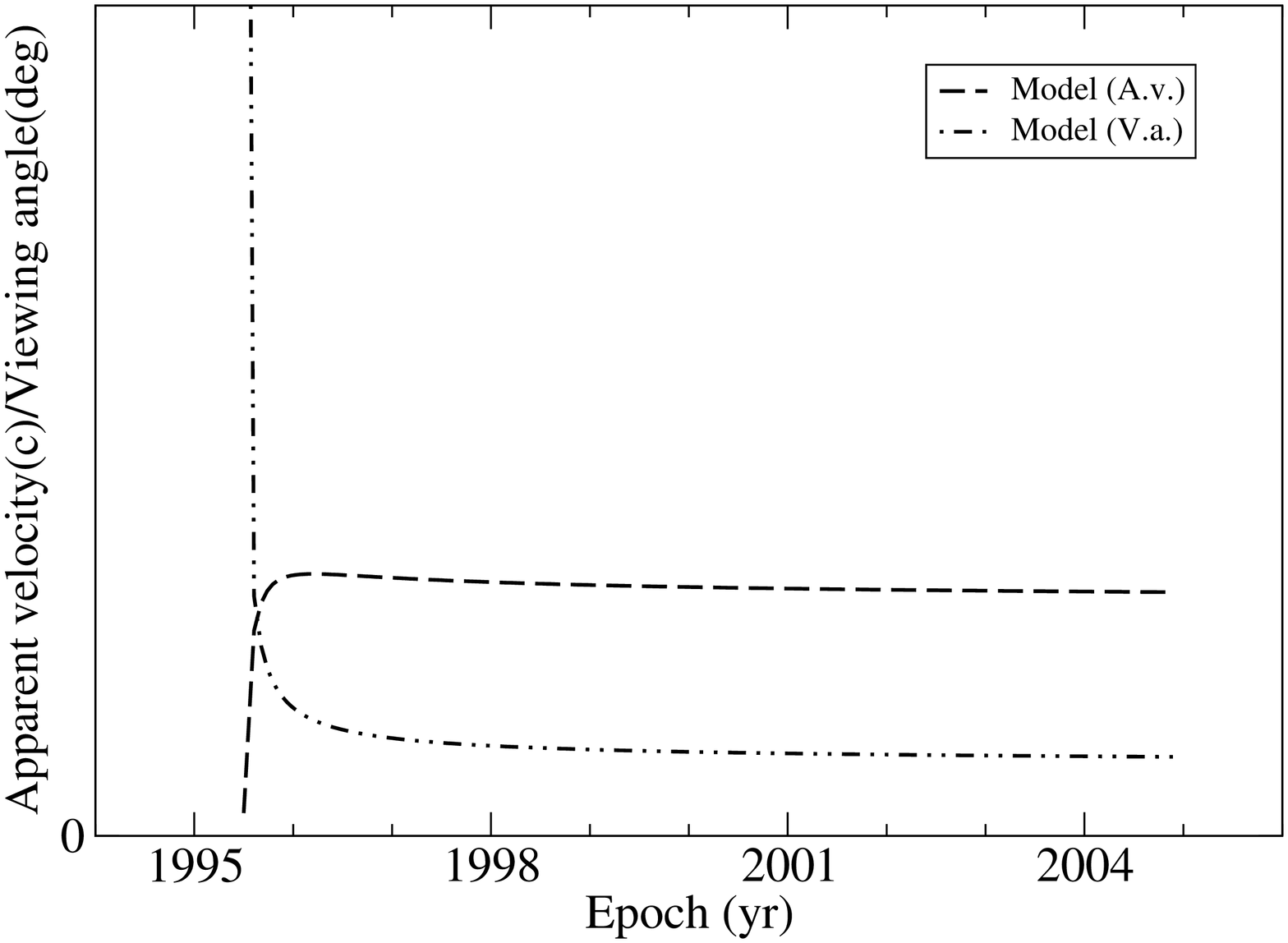}
     \includegraphics[width=5cm,angle=-90]{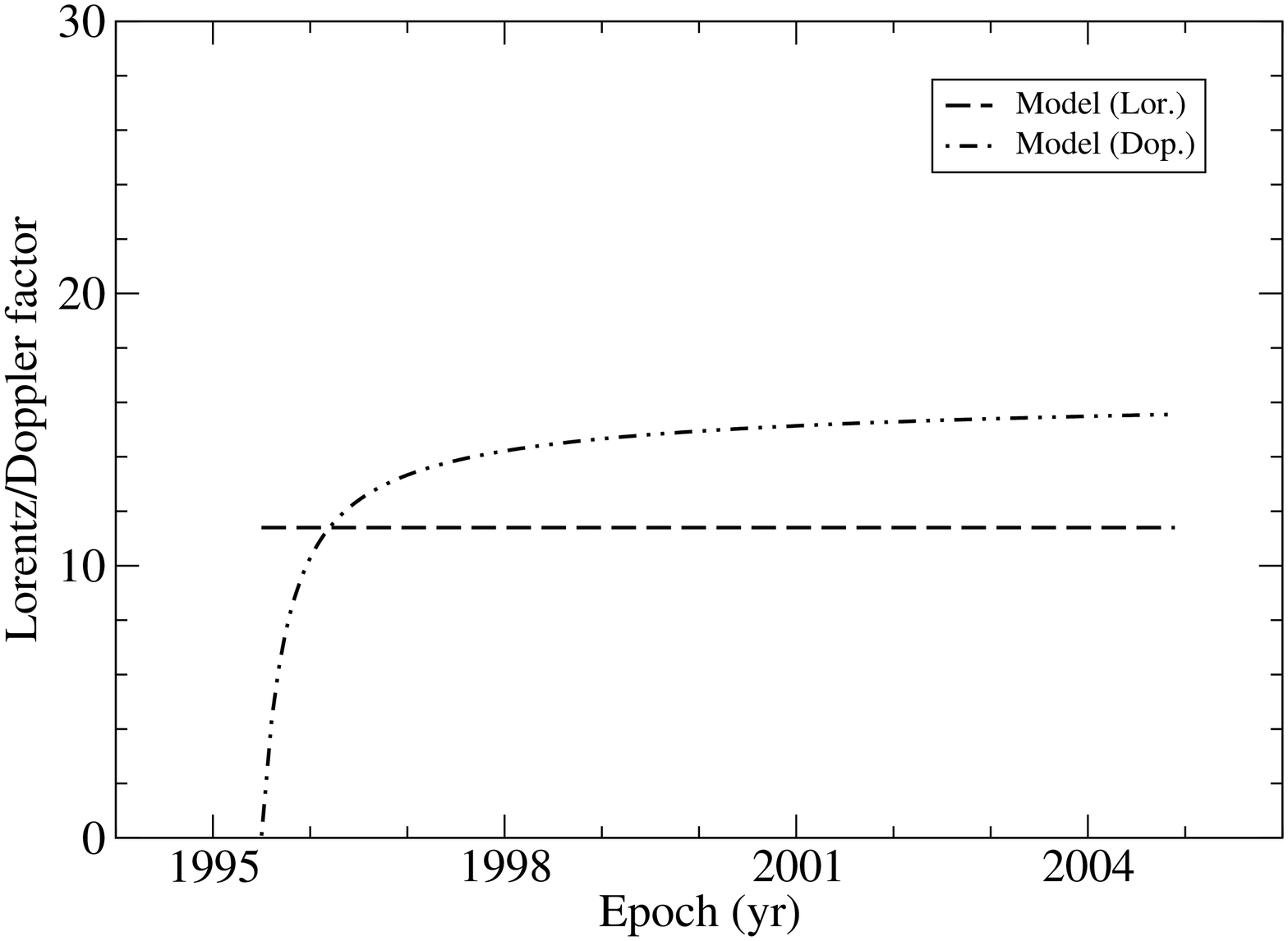}
     \includegraphics[width=5cm,angle=-90]{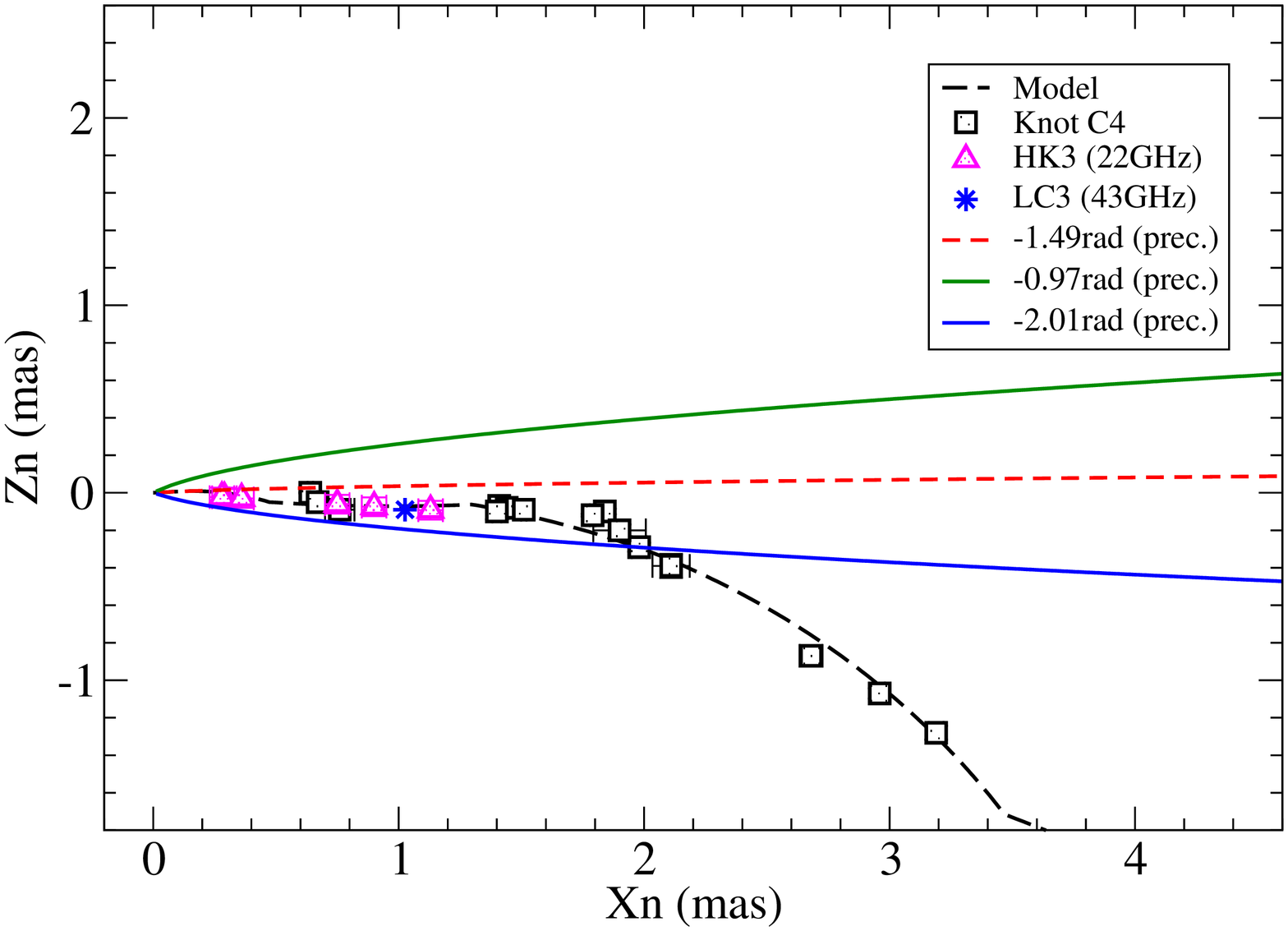}
     \caption{Model-fitting results for knot C4. As shown in top left
   panel, the observation data given 
   in Homan et al. (\cite{Hom01}, 22\,GHz for knot-K3) and in Lister et al.
   (\cite{Li98}, 43\,GHz for knot-C3) are also well fitted. }
     \end{figure*}
     \begin{figure*}
     \centering
     \includegraphics[width=5cm,angle=-90]{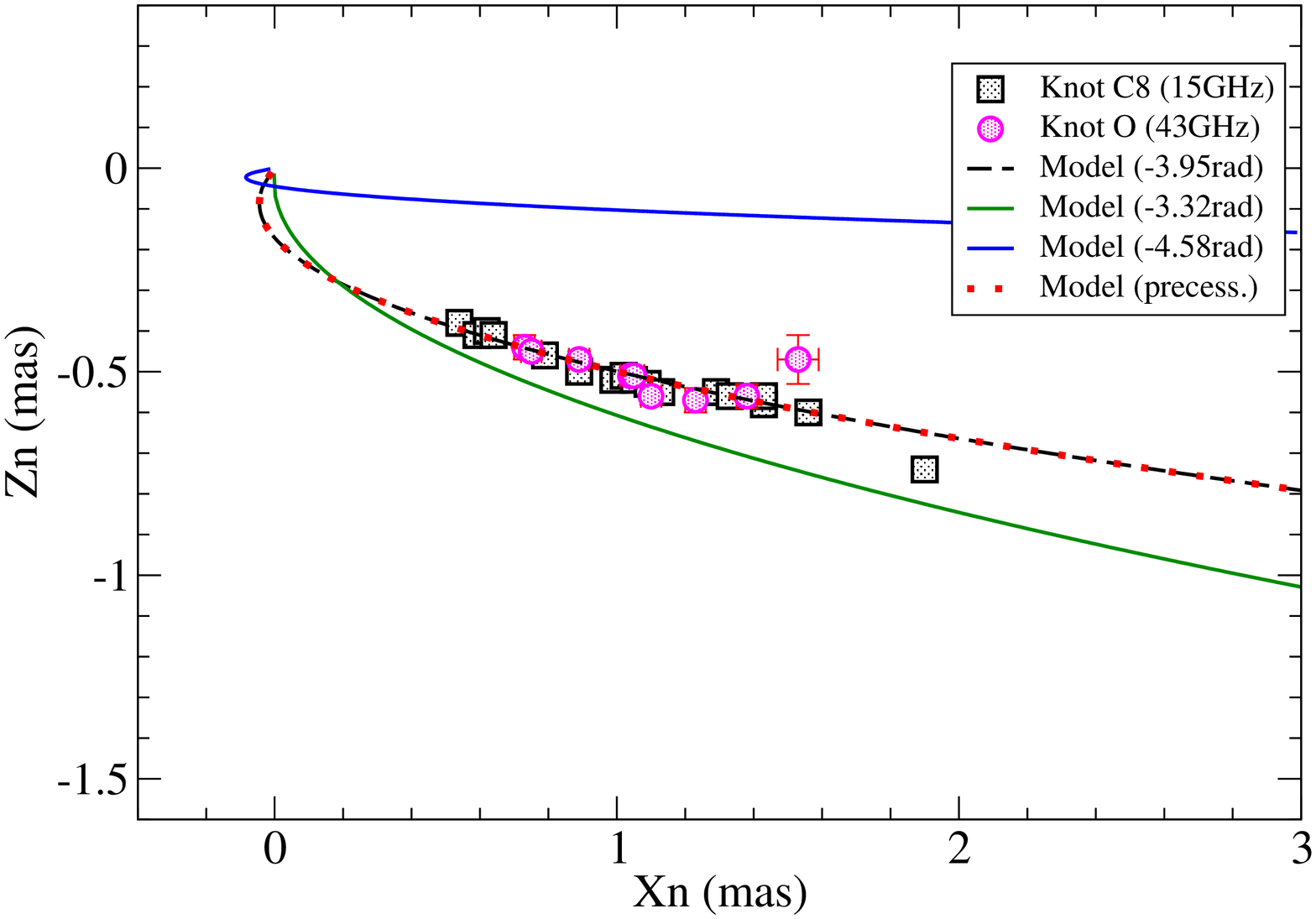}
     \includegraphics[width=5cm,angle=-90]{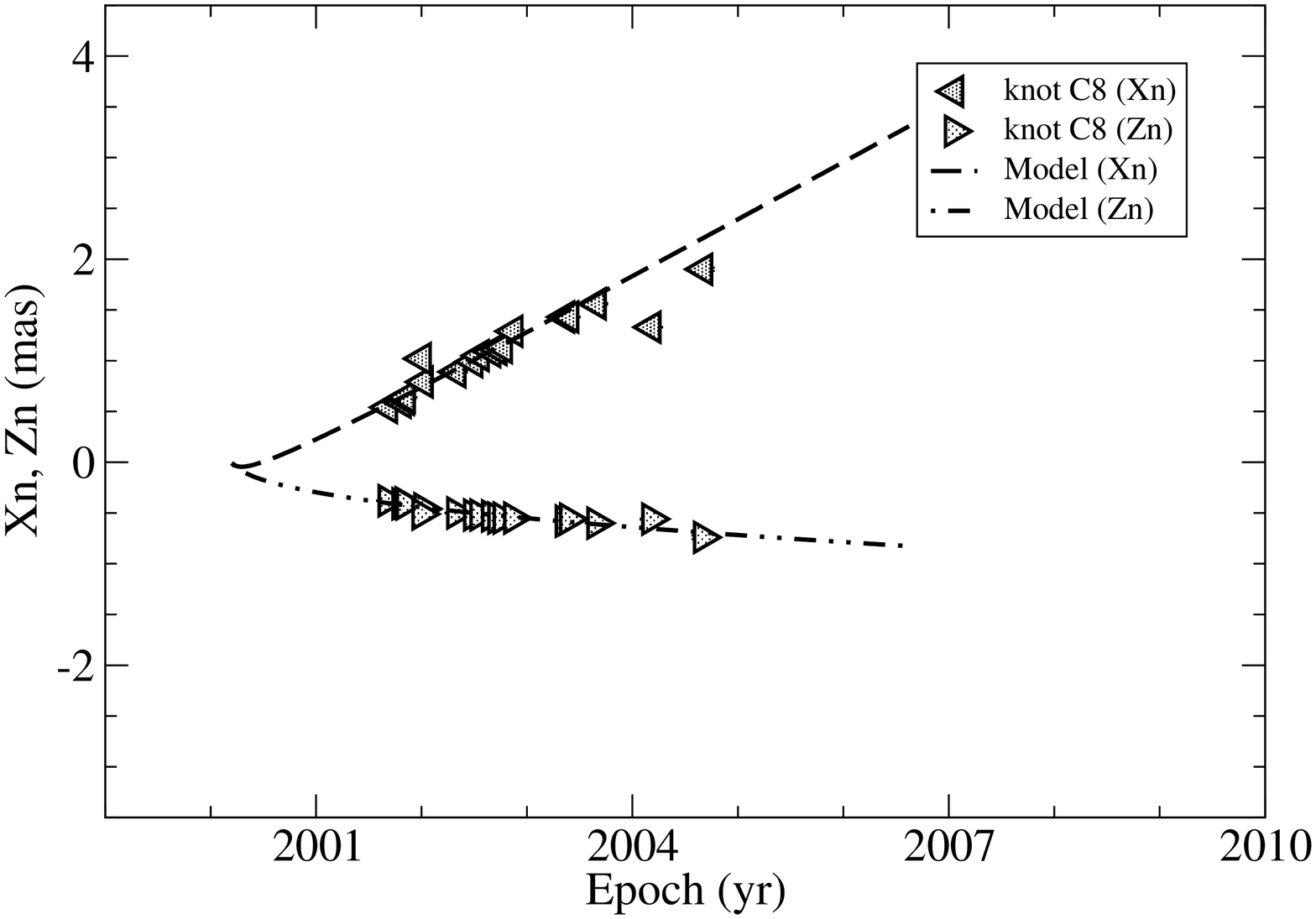}
     \includegraphics[width=5cm,angle=-90]{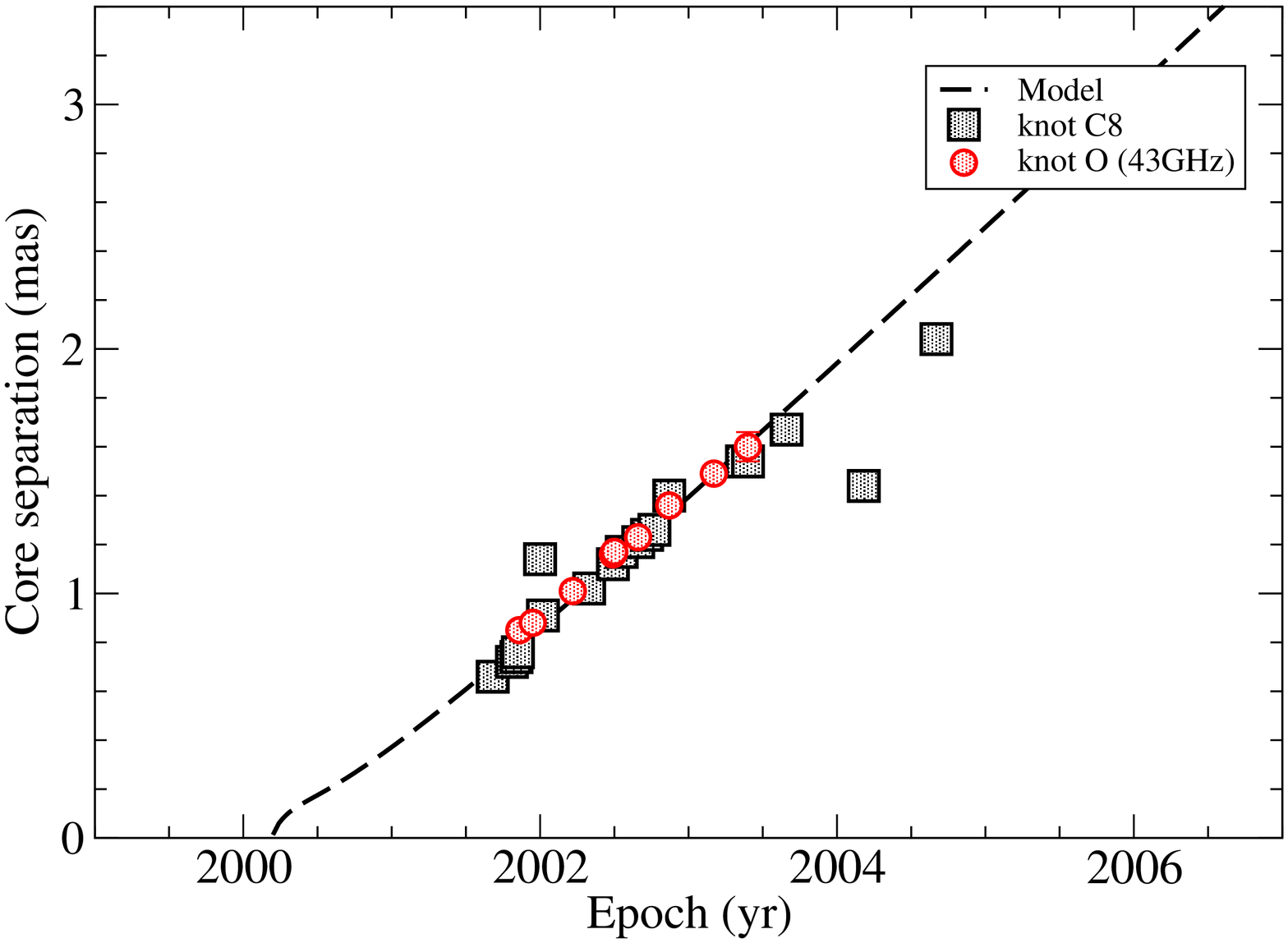}
     \includegraphics[width=5cm,angle=-90]{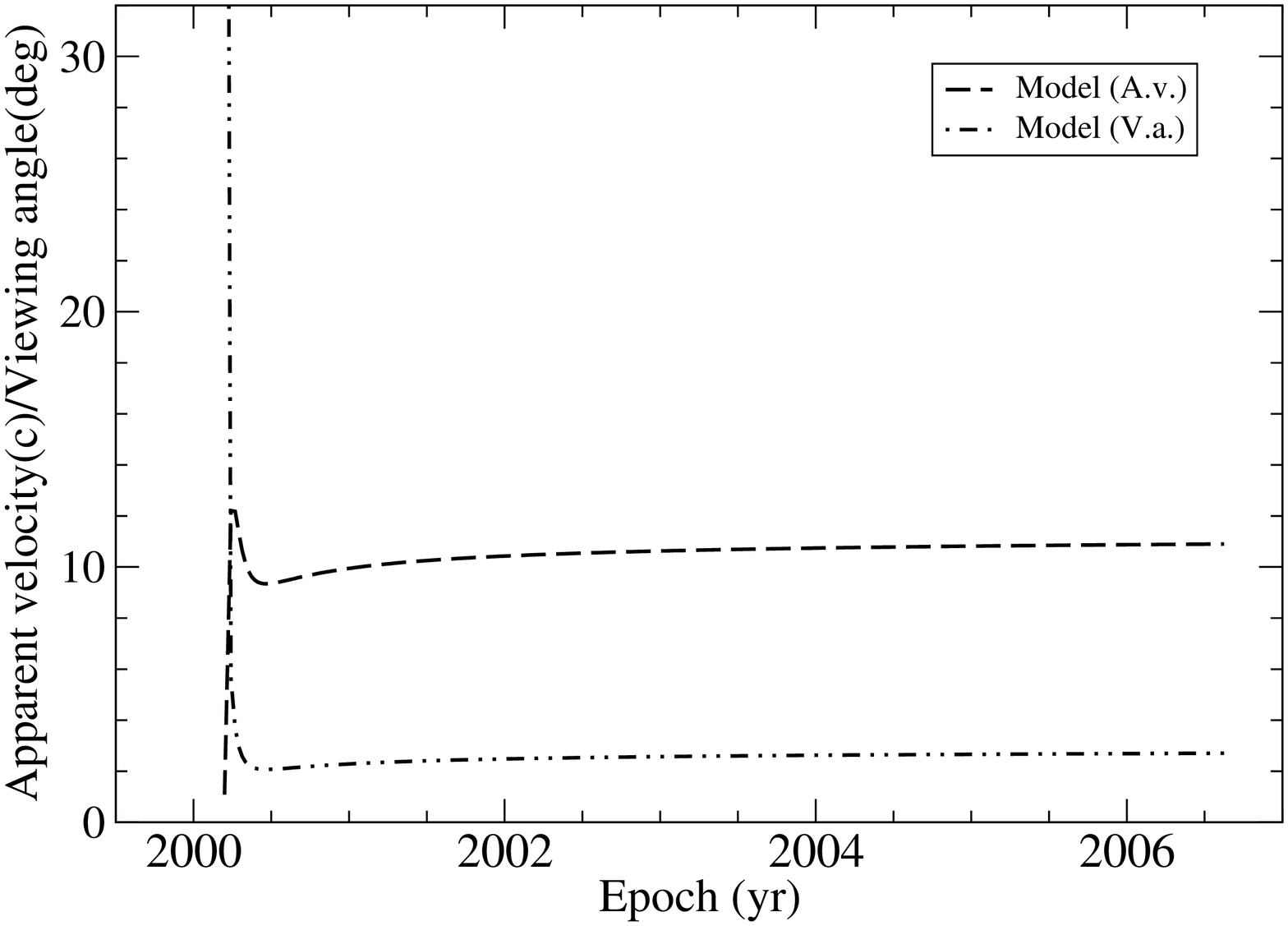}
     \includegraphics[width=5cm,angle=-90]{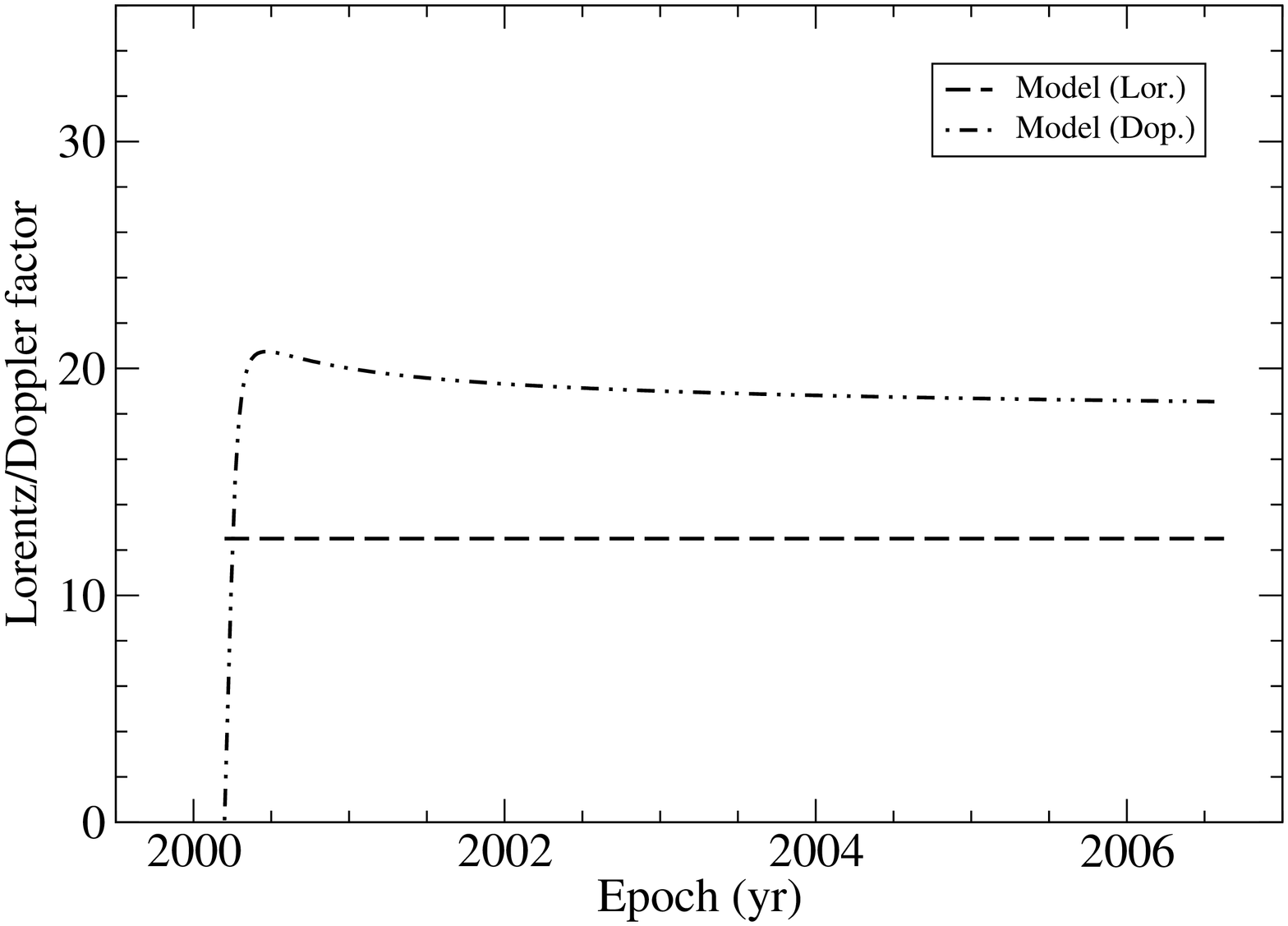}
     \includegraphics[width=5cm,angle=-90]{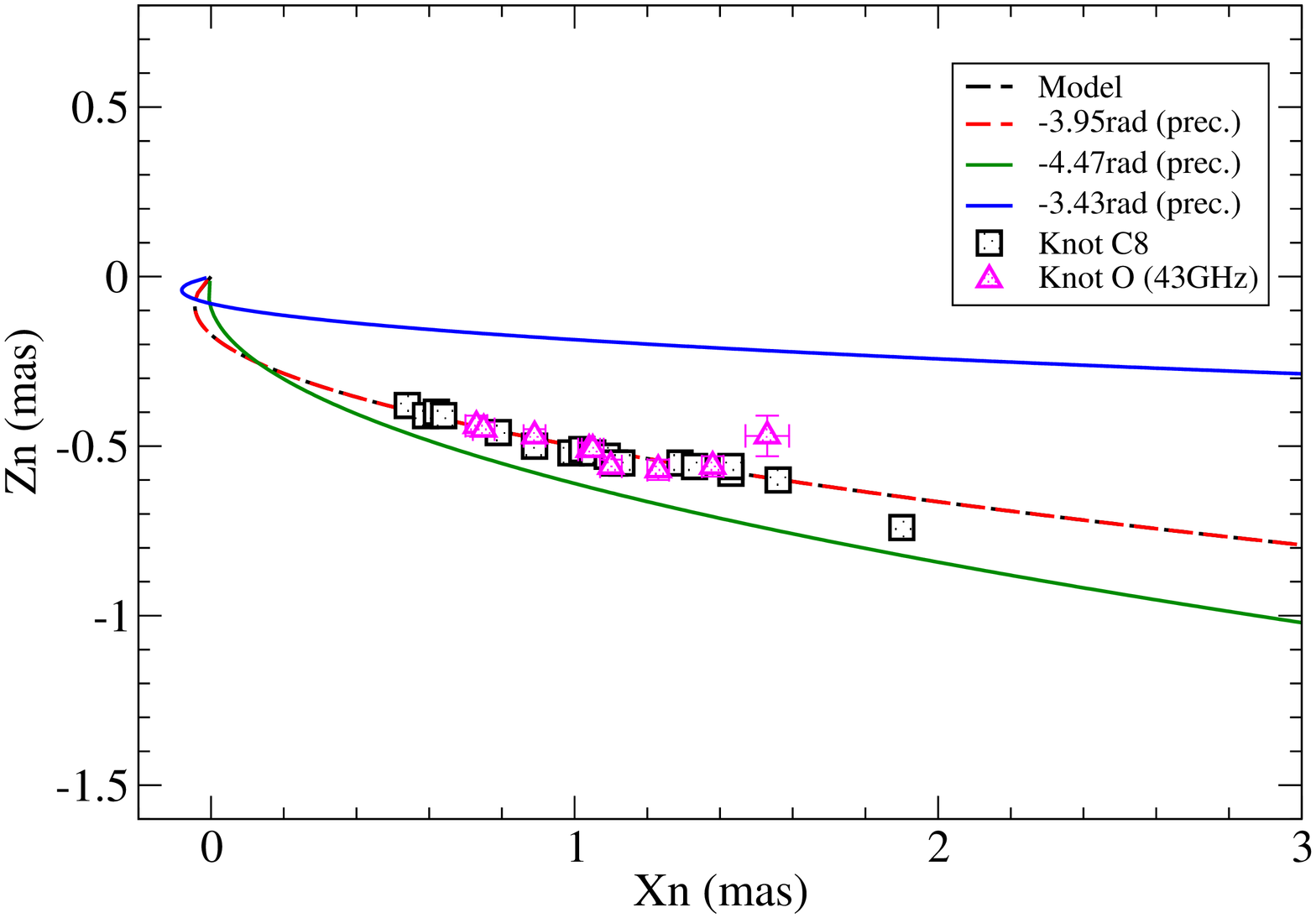}
     \caption{Model-fitting results for knot C8. As shown
    in top left panel, the trajectory observed by Agudo et al. at 43\,GHz
    (\cite{Ag12}, for knot-O, red points) is extremely well coincided with 
    the trajectory observed at 15\,GHz, and both are very well fitted by our
    precessing jet nozzle model with a parabolic trajectory. The ejection epoch measured
    by Agudo et al. was 2000.16$\pm$0.03, almost exactly equal to our modeled
    epoch 2000.20. The consistency of the observations at different frequencies
    for both knot C8 and knot C4 (Fig.11) provides convincing confirmation 
    for our parabolic precessing jet nozzle model with a 12\,yr precession 
    period for the northern jet
     and the double jet scenario for OJ287.}
    \end{figure*}
     \begin{figure*}
     \centering
     \includegraphics[width=5cm,angle=-90]{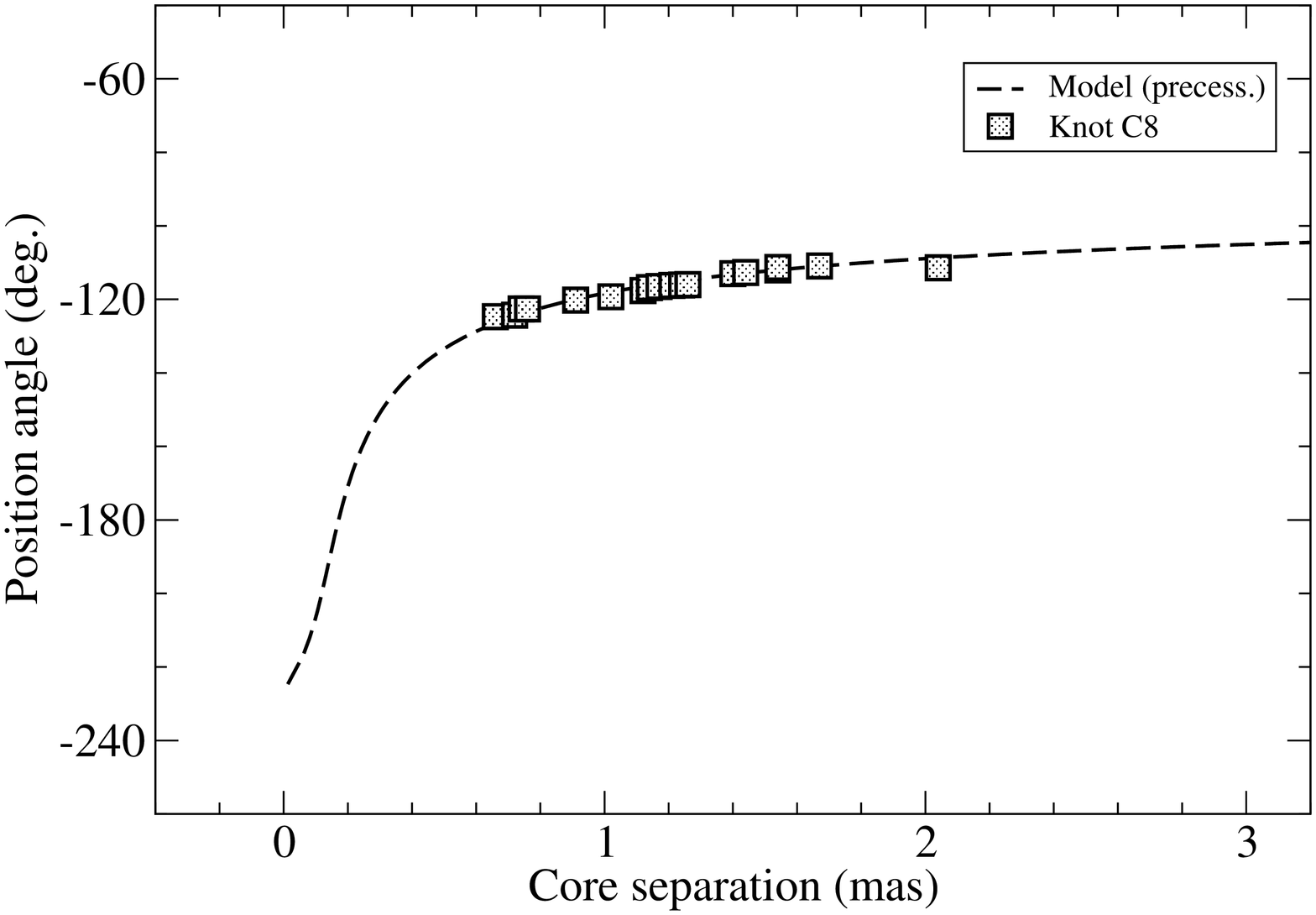}
     \includegraphics[width=5cm,angle=-90]{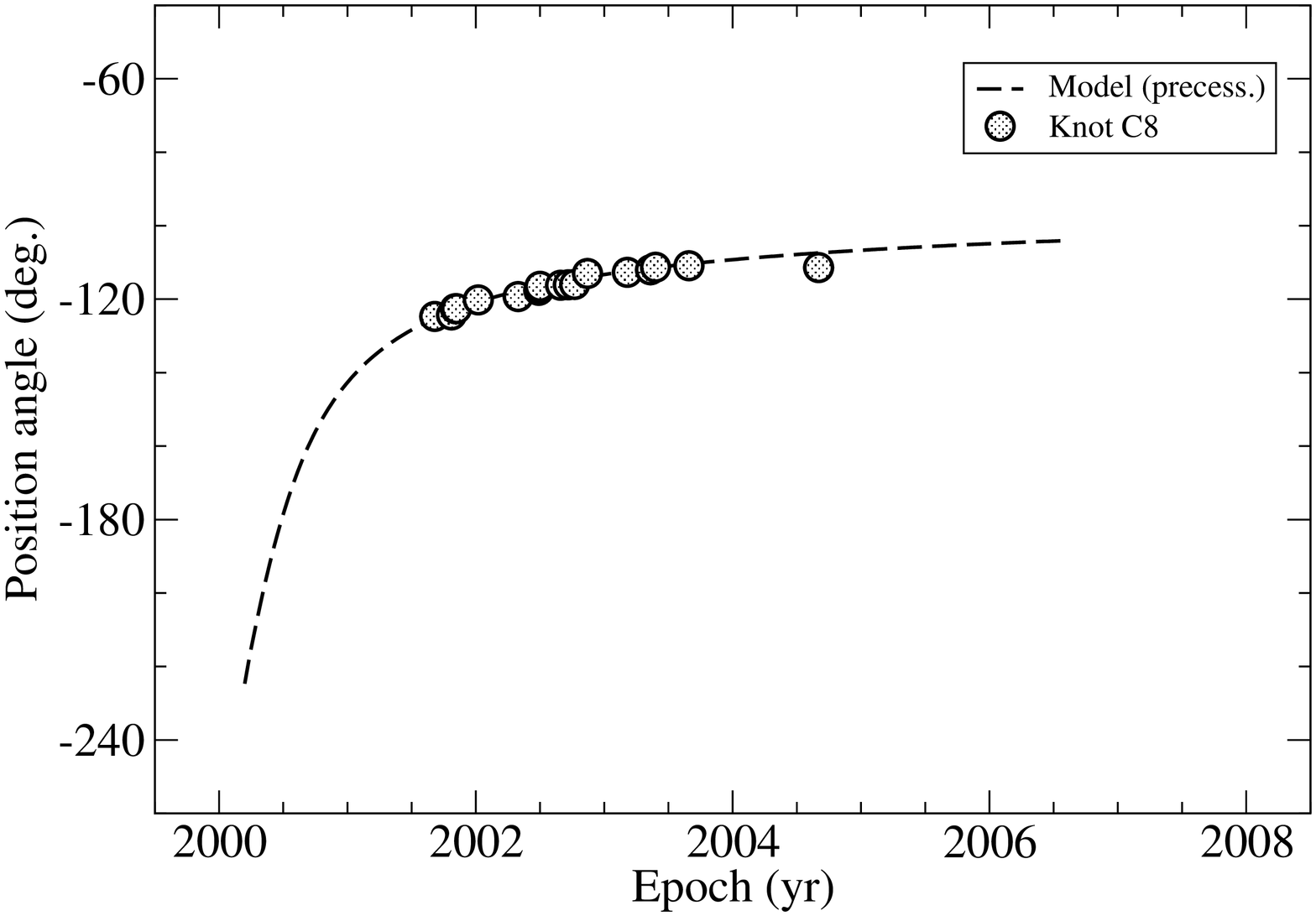}
     \caption{Modeled relations for knot C8: PA--$r_n$ and PA--t.
     The position angle changes along the precessing common parabolic
     trajectory from $\sim{-220^{\circ}}$ (at core separation $\sim$0.05\,mas)
     to $\sim{-120^{\circ}}$ (at core separation $\sim$0.8\,mas), demonstrating
     a large PA change of $\sim{100^\circ}$ from the innermost region to the
     outer region. During the observation period 2001.5--2004.0, its position 
    angle monotonically changed $\sim$${14^{\circ}}$ ($\pm$$0.4^{\circ}$): from
     $\sim{-125^{\circ}}$$\pm{0.3^{\circ}}$ to 
    $\sim{-111^{\circ}}$$\pm{0.3^{\circ}}$ (Britzen et al.
     \cite{Br18}). This clearly indicates  that its motion is non-ballistic and       ballistic precessing jet models seem inapplicable to OJ287.}
      \end{figure*}
     \subsection{Knot C8}
     The model-fitting results of the kinematic features for
     knot C8 is particularly important for our double jet scenario.

     Firstly we note that knot C8 was observed to be ejected at 2000.1, and
     knots C9 and C10 were ejected at 2001.80  and 2002.33 .
     The three knots were observed to have similar position angles within
     $r_n{\sim}$1.2\,mas at 15\,GHz: $-117.2^{\circ}$
     (knot C8), $-121.9^{\circ}$ (knot C9) and $-120.2^{\circ}$ (knot C10).
     This seems to indicate that they are ejected from the same
     jet consecutively. But through detailed analyzes we found that knot C8 
   has kinematic features distinct from those of knots C9 and C10: 
   (i) It moves apparently
     across the paths of knots C9 and C10; (ii) At core separations
     between 1.0\,mas and 1.6\,mas its trajectory  is curved becoming
     horizontal, typical for the components of the northern jet. This is
     different from the trajectories of the components C9 and C10
     which move ballistically \footnote{Knots C9 and C10 move along parabolic
     trajectories in space and their apparent ballistic motions are caused 
     by projection effects onto the plane of the sky.} within 
      core separation $r_n{<}$2\,mas with position angles $<{-120^{\circ}}$;
     (iii) its kinematics can be consistently
     interpreted with the knots (C4, C5, C6, C13U and C14) of the northern jet
     in terms of  a precessing jet nozzle model as shown in Figure 12.
     \footnote{Note that the values of parameters $a$ and $\psi$ for
     the northern jet are different from those for the southern jet.};
     (iv) In particular, knot C8 moved
     non-ballistically and its path (in the observed separation range 0.65\,mas
     to 2\,mas) can be extremely well  fitted  by the
     precessing common parabolic trajectory assumed for the northern jet
    (top left panel in Fig.12); (v) Interestingly, its trajectory observed 
    at 43\,GHz by Agudo et al. (\cite{Ag12}) is extremely well coincided 
    with the trajectory observed at 15\,GHz.
     The consistency of the trajectories observed at 15\,GHz and 43\,GHz
    may imply that there is no opacity effects between the two frequencies in 
    the outer jet regions (core separations from $\sim$0.5\,mas to 
    $\sim$2\,mas).
   
     The model-fitting results for knot C8 are shown in Figure 12.
    Its ejection epoch is model-fitted to be $t_0$=2000.20 
    and the  corresponding precession phase $\omega$=$-$3.95\,rad. 
    It can be seen that
     the trajectory, coordinates and core separation in the observed
     separation range 0.65\,mas to 2.0\,mas (Z=17.6\,mas to 49.1\,mas,
     or 79.0\,pc to 220.5\,pc) are all well fitted by the precessing nozzle 
     model.

     This is the best case verifying the validity
     of the parabolic trajectory pattern applicable to describe the knot's
     trajectory within radial distance $\simeq$200\,pc.

     In top left and bottom right panels of Fig.12, the 43\,GHz
      observation data points measured by Agudo et al.(\cite{Ag12}, for knot-O)
      are extremely well modeled. Combined with the 15\,GHz data, they
     convincingly show its non-ballistic motion and entire trajectory fitted
     perfectly by the parabolic precessing nozzle model (red dashed
     line in the top left panel of Fig.12).

     Moreover, the ejection epoch measured by Agudo et al. (\cite{Ag12})
      was 2000.16$\pm$0.03,
     which is almost exactly equal to our modeled epoch for knot C8 (2000.20),
     providing a strong support to our precessing nozzle model with a
     precession period of 12\,yr assumed for the northern jet. 

     In top left panel of Figure 12, the green and blue lines
    represent the modeled trajectories for precession phases
     $\omega{\pm}$0.63\,rad,
    indicating all the data-points within the position angle range defined by
    the two lines and 
   the precession period having been determined within
    an uncertainty of $\pm$1.2\,yr. In bottom right panel of Fig.12,
   the green and blue lines 
  represent the precessing common trajectories calculated for
   $\omega{\pm}$0.52\,rad, showing all the data-points within the position
   angle range defined by the two lines and 
   indicate its innermost
     precessing common trajectory having been observed. Thus knot C8
    is designated by symbol ``+'' in Table 6.

   In  Figure 13 the relations of the position angle vs core separation and 
   the position angle vs time are shown, indicating that the position angle 
   of knot C8 monotonically changed $\sim{14^{\circ}}$($\pm$$0.4^{\circ}$)
   along its trajectory during the observation period  2001.5--2004.0 
   (from --$125^{\circ}$($\pm$$0.3^{\circ}$) to
   --$110^{\circ}$ ($\pm$$0.3^{\circ}$); Britzen et al. \cite{Br18}). Thus
   the motion of knot C8 is definitely non-ballistic. Ballistic precessing 
   jet models seem  inapplicable to OJ287.
   \subsection{Knots C13U and C14}
     Knots C13U and C14 are very strong superluminal components. Their initial
     flux densities reached to $\sim$2--3\,Jy (at 15\,GHz), but with
     very slow apparent speeds. They might be associated with the optical
     double peaked outburst during the period of 2005--2007. The ejection 
     of components C13U and C14  may represent the re-starting of the activity
     of the northern jet after a $\sim$5\,yr quiescent period since the last 
     ejection of knot C8.

    The model-fitting results of the kinematics for knots C13U and C14
    are shown in Figures A.7 and A.8 of the Appendix, respectively. 
    For knot C13U (Fig.A.7),  its motion is modeled to 
    follow the precessing common parabolic trajectory within  core
    separation $r_n{\leq}$0.25\,mas \footnote{As shown in Fig.A.7
    (top left panel) a change
   of $\sim{50^{\circ}}$ in the position angle of knot C13U was observed during
    the period 2010.39--2014.67 (Britzen et al. \cite{Br18}.) Thus its motion
     was non-ballistic.}
    and its innermost parabolic trajectory 
    has been observed. For knot C14 (Fig.A.8),
    its motion is modeled to follow the precessing common parabolic trajectory
    within core separation $r_n{\leq}$0.27\,mas and its innermost parabolic
    trajectory has been observed. See captions of Fig.A.8 of the Appendix.
    \begin{table}
    \centering
    \caption{Northern jet: core separation ($r_n$) and the corresponding
    axial distance Z within  which the knots move along the precessing common
     parabolic trajectory. Symbol ``+'' denotes that the knots
    have been observed
    to follow the precessing common parabolic trajectory and symbol ``--''
    denotes the knots' innermost trajectory following  the precessing common
    parabolic trajectory  having not been observed (no observation data
    available).}
    \begin{flushleft}
    \centering
    \begin{tabular}{lrrrr}
    \hline
    Knot & $r_n$(mas) & Z(mas) & Z(pc) & status \\
    \hline
    C1 & 0.23 & 1.60 & 7.18  & -- \\
    C2 & 0.25  & 1.60 & 7.18 & -- \\
    C3 & 0.25 & 1.60 & 7.18 & + \\
    C4 & 0.16 & 0.8 & 3.59 & + \\
    C5 & 1.55 & 20 & 89.8 & + \\
    C6 & 1.19 & 15 & 67.4 & + \\
    C8 & 2.0 & 49.1 & 220 & + \\
    C13U & 0.25 & 2.0 & 9.0 & + \\
    C14 & 0.27 & 2.0 & 9.0 & + \\
    \hline
    \end{tabular}
    \end{flushleft}
    \end{table}
    \subsection{A brief summary for northern jet.}
     The main results  for the northern jet can  be  summarized as follows.

     \begin{figure*}
     \centering
     \includegraphics[width=5cm,angle=-90]{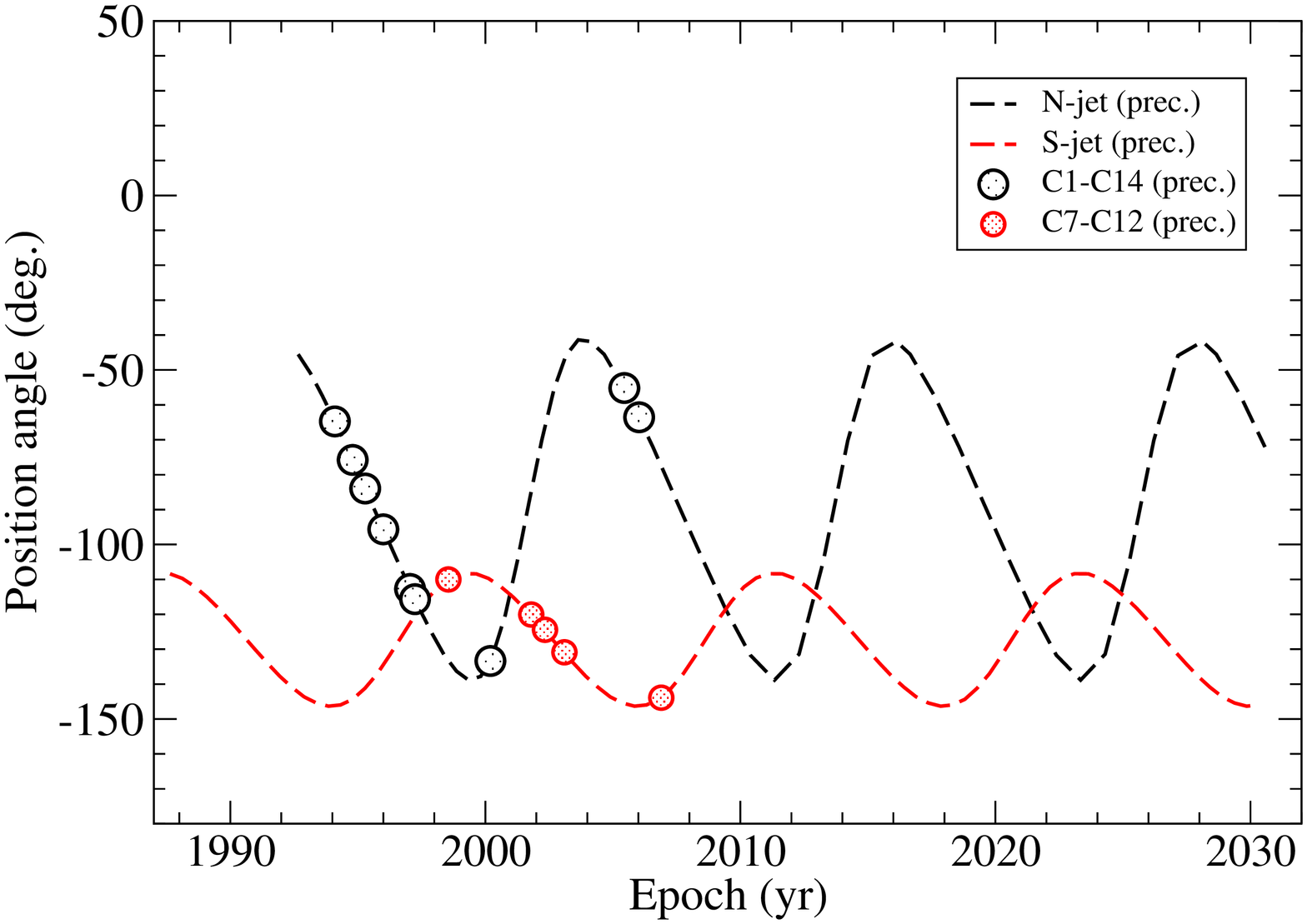}
     \includegraphics[width=5cm,angle=-90]{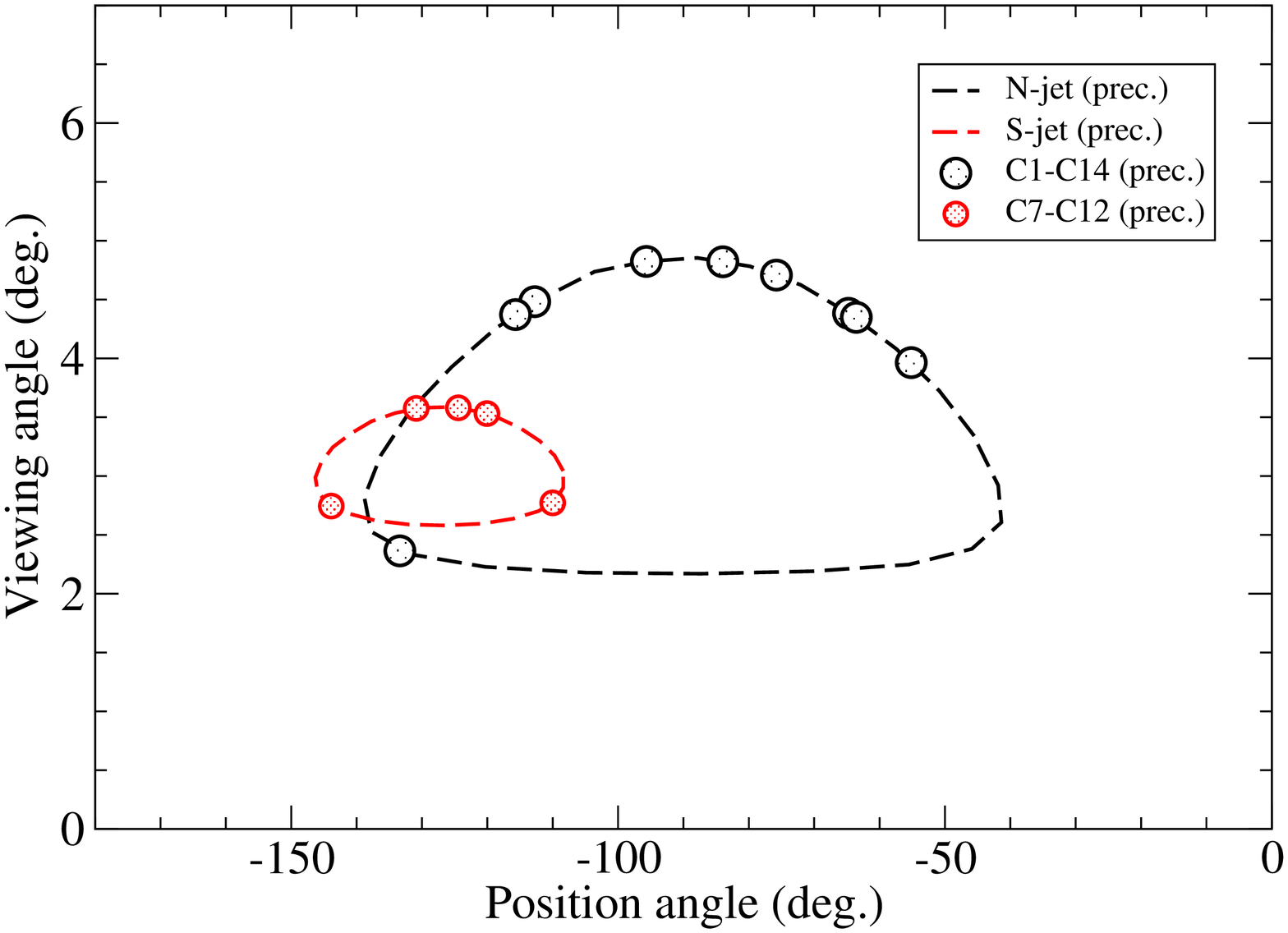}
     \includegraphics[width=5cm,angle=-90]{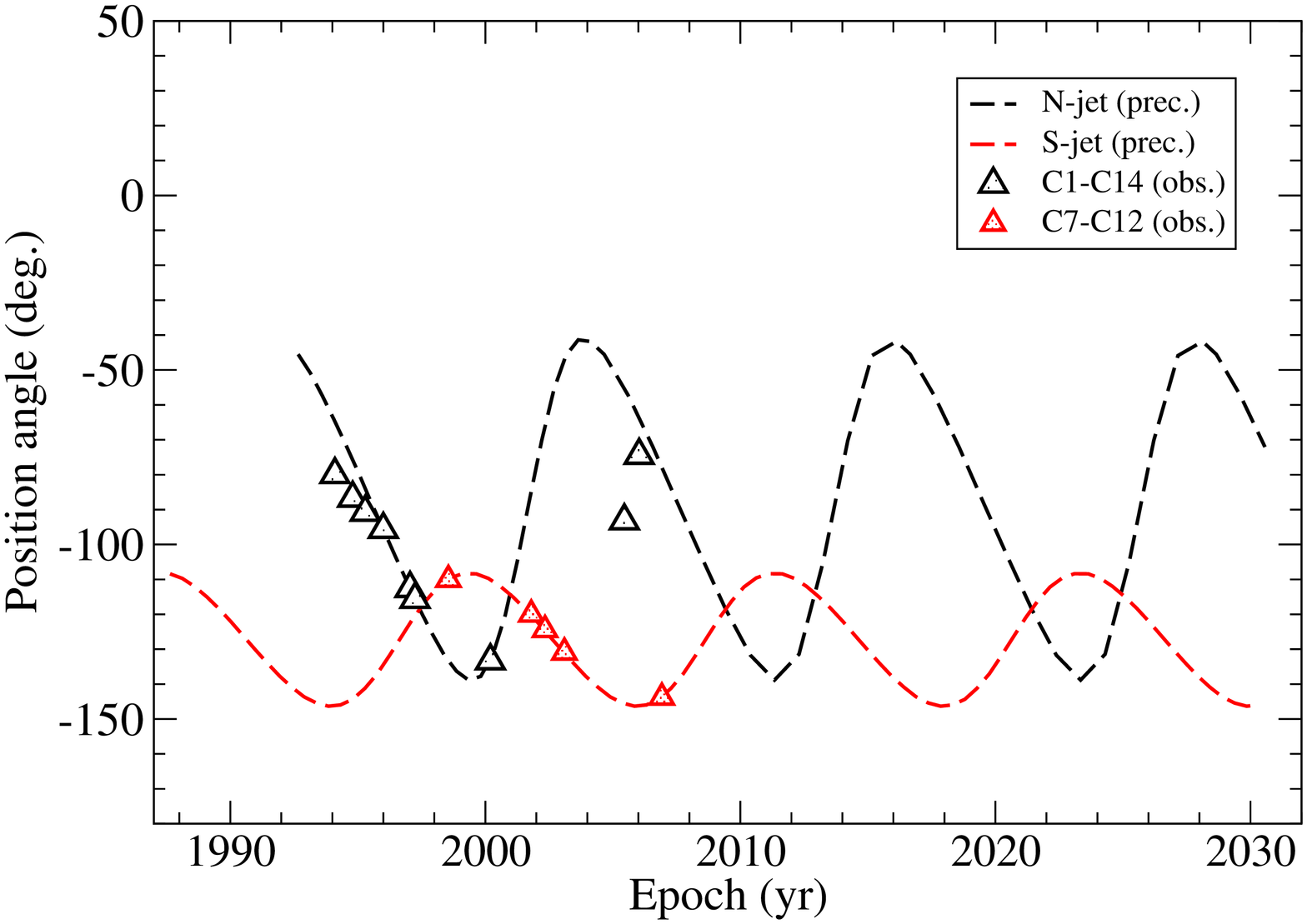}
     \includegraphics[width=5cm,angle=-90]{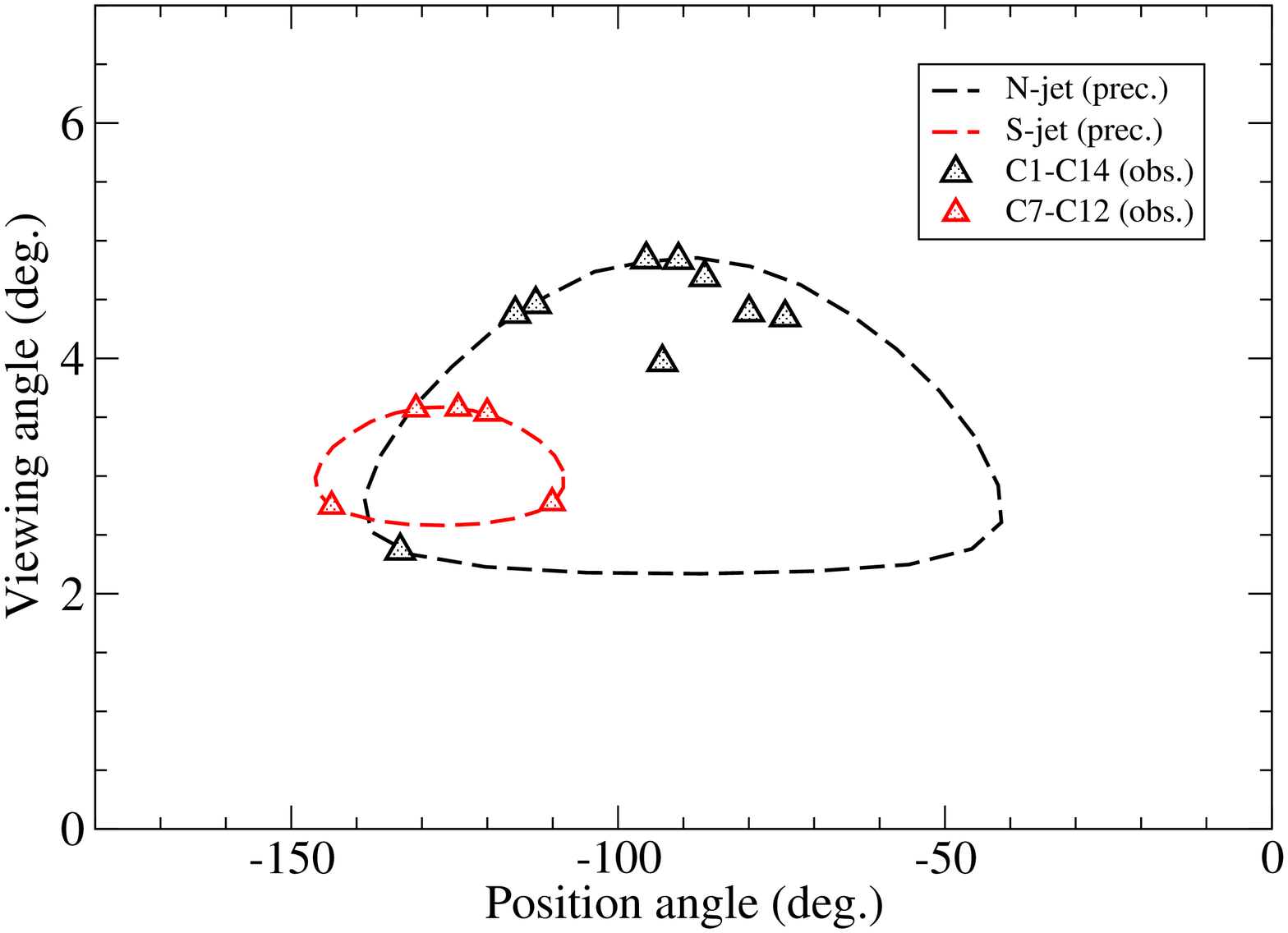}
     \caption{Top panels: modeled relations of position angle vs time 
    (left panel) and position angle vs viewing angle (right panel) for the
     superluminal knots (circles). Dashed lines denote the model simulations.
     Bottom  panels: observed relations of position angle vs time
    (left panel) and position angle vs viewing angle (right panel) for the
    knots (triangles). All relations are made
   at core separation $r_n$=0.5\,mas. The observation-data  clearly 
    reveal the trajectory curvatures of some knots.}
     \end{figure*}
      \begin{itemize}
     \item The kinematics of the nine superluminal components  can be
     consistently model-fitted in terms of a parabolic
     processing jet-nozzle model with a precession period of 12\,yr,
     same as that for the southern jet and the optical period
     (Sillanp\"a\"a \cite{Si88}, Valtonen et al. \cite{Va16}). The jet
     precesses clockwise, also similar to the southern jet.
    \item The modeled  jet cone spans  from PA=$-40^{\circ}$
    to PA=$-135^{\circ}$ (at core separation $r_n{\leq}$0.5\,mas) with the 
    precession axis at PA=${\sim}$--$80.5^{\circ}$, which is different from
     that of the southern jet precession axis at $\sim$$-130^{\circ}$ (see
    Figure 2). 
        \item Six out of the nine superluminal components (knots C1 to C6)
    have been observed downward curvatures in their outer  trajectories
     ($r_n{>}$2\,mas), opposite to the upward curvatures for the southern 
    jet components.
     \item For seven out of the nine knots (C3 to C14), their innermost 
    motion have been observed to follow the  precessing
    common parabolic trajectory (Table 6).
    \item The modeled ejection times are well consistent with 
   VLBI-measurements (Britzen et al. \cite{Br18}). In particular, 
   the ejection times modeled for
    knots C3, C4 and C8 are extremely well confirmed by the VLBI-measurements 
    at 43GHz, 22\,GHz and 8\,GHz (Agudo et al. \cite{Ag12}, Homan et al.
    \cite{Hom01}, Lister et al. \cite{Li98}, Tateyama et al. \cite{Ta99}),
    providing convincing evidence of the 12\,yr precession period. These
   are really posterior verifications, because we didn't use these data in the 
   construction of the precessing nozzle model.
    \item The kinematics of knot C8 can be consistently model-fitted with 
    the other eight components in terms of a  precessing jet nozzle model with
    model parameters, different from those for the southern jet. Its outer
    trajectory (observed at 15\,GHz and 43\,GHz) was extremely well fitted 
    by the modeled parabolic trajectory.  Moving along the parabolic trajectory
    knot C8 should have changed its position angle from $\sim$--$220^{\circ}$
    to $\sim$--$120^{\circ}$ after its emergence from the core (see Fig.13).
    Similar behavior has been observed in the motion
    of the superluminal knots in blazar 3C279 (e.g., knots C30--C32; Qian
    et al. \cite{Qi18}, Jorstad et al. \cite{Jor17}, Lu et al. \cite{Lu13}).
   \item We find that during the period of $\sim$1997--2007, the northern
   jet and southern jet launched superluminal components alternatively. Our
   model-fitting procedure disentangled the kinematic features of these 
   components, thus demonstrating the 12\,yr precession period 
   for both the jets. 
    \end{itemize}
   \section{Correlation between optical outbursts and radio knot ejections}
     As Britzen et al. (\cite{Br18}) and Tateyama et al. (\cite{Ta99}) point
     out that there is a notable similarity between the optical and radio
    light curves, even small variations on the radio light curves coinciding
    with the optical flares. And the optical double-peaked outbursts are
    associated with the emergence of superluminal components from the
    core (Valtaoja et al. \cite{Val20}). We collected relevant data from 
    the literature (Gabuzda et al. \cite{Ga89}, Tateyama et al. \cite{Ta99},
     Valtaoja et al., \cite{Val20}, Valtonen et al. \cite{Va08}, Britzen et al.
     \cite{Br18}) and made the Table 7 to investigate the association of 
    the double-peaked optical outbursts and radio component ejections.
    It is found that all the double-peaked optical outbursts are associated 
    with the emergence of superluminal components launched from the northern 
   jet, which has been working for more than 40 years. In contrast, the
    emergence of knots C7--C12 launched from the southern jet had nothing to
    do with the appearance of the double-peaked optical outbursts.

     \begin{table*}
     \centering
     \caption{Association of the ejection of superluminal components and the
      periodic  double-peaked optical outbursts: starting time of
     the double-peaked optical outbursts, radio knot ID, the ejection time of
     the superluminal knots, ejection position angle, jet ID. Symbols
     G, T and B ahead of the knot identification represent the author of
     references: G -- Gabuzda et al. (\cite{Ga89}), T -- Tatayama et al.
     (\cite{Ta99}), B -- Britzen et al. (\cite{Br18}). For the 2015.87
     optical outburst, we assume that the ejection of its associated radio knot
     (designated as knot-X here) occurred in 2014.67 when the flux density
     of the core (component-k, Britzen et al. \cite{Br18}) increased to 
     $\sim$4\,Jy. Its ejection position angle
     is assumed to be --$58^{\circ}$ which was the position angle of
     the double structure observed in 2017.19. For comparison, 
     the non-periodic outburst observed in 2001--2003 is also listed.}
     \begin{flushleft}
     \centering
     \begin{tabular}{lrrrr}
     \hline
     Starting times & radio knot & ejection epoch & ejection PA & jet ID\\
    \hline
     1971.08/1972.94 &  G--K1   & 1969.7 & $-98^{\circ}$ & N-jet \\
     1982.96/1984.10 &  T--K3   & 1982.3 & $-100^{\circ}$ & N-jet \\
     1994.69/1995.84 &  B--C1   & 1994.10 & $-90^{\circ}$ & N-jet\\
                     &  B--C2   & 1994.80 & $-90^{\circ}$ & N-jet\\
                     &  B--C3   & 1995.29 & $-90^{\circ}$ & N-jet\\
                     &  B--C4   & 1995.50 & $-90^{\circ}$ & N-jet \\
     2005.75/2007.69 &  B--C13U  & 2005.45 & $-75^{\circ}$ & N-jet\\
                     &  B--C14   & 2006.03 & $-64^{\circ}$ & N-jet \\
     2015.87         &  B--X   & 2014.67  & $-58^{\circ}$ & N-jet \\
     \hline
     2001--2003 & B--C8 & 2000.1 & $-115^{\circ}$ & N-jet\\
                & B--C9 & 2000.8 & $-120^{\circ}$ & S-jet\\
                & B--C10 & 2002.5 & $-120^{\circ}$ & S-jet\\
                & B--C11 & 2002.9 & $-130^{\circ}$ & S-jet\\
     \hline
     \end{tabular}
     \end{flushleft}
     \end{table*}
      Here we would pose the interesting and important question: why the 
    northern jet is much active than the southern jet?  This phenomenon is
    very similar to that observed in blazar 3C279 (Qian et al. \cite{Qi18})
   and may be explained similarly by HD/MHD theories for cavity-accretion
    processes in near equal-mass binary black hole systems. According to 
    hydrodynamic and magnetohydrodynamical simulations of the dynamics 
    for these supermassive binary systems (Tanaka \cite{Tan13}, Shi et al. 
    \cite{Sh12}, Artymovicz \cite{Art98}),
    gas streams preferentially accrete onto the secondary black hole
   and the jet produced by the secondary hole will be much more active than
   the jet produced by the primary hole. Thus in our double-jet scenario for
   OJ287,  the northern jet may be produced by the secondary hole.
   This attribution seems verified by the aperture width of the northern jet 
   which is much larger than the aperture of the southern jet 
   (ascribed to the primary hole).
    \section{Summary}
     Our precessing double-jet nozzle model involves two assumptions. (1)
    OJ287 has a double-jet structure and superluminal components are ejected 
    form each of the precessing jet nozzle; (2) the superluminal knots move
    along respective common parabolic trajectories.
   \footnote{Generally, precessing common trajectories can also be 
   rectilinear, conical or helical.} These assumptions are based on some 
   available observations  and can 
   be understood  in terms of magnetohydrodynamic theory for the formation,
   collimation and acceleration of relativistic jets in blazars 
   (Qian et al. \cite{Qi17}, Blandford \& Znajek
   \cite{Bl77}, Blandford \& Payne \cite{Bl82}, MacDonald \& Thorne
    \cite{Mac82}, McKinney et al. \cite{Mc12}, Beskin \& Zheltoukhov
   \cite{Be13}, Asada \& Nakamura \cite{As12}, Nakamura \& Asada \cite{Na13}, 
   Mizuno et al. \cite{Miz15},  Marti et al. \cite{Mart16}, Mutel \& Denn 
  \cite{Mu05}, Stirling et al. \cite{St03}, Cohen et al. \cite{Co16},
   Polko et al. \cite{Po13},
    \cite{Po14}, Vlahakis \& K\"onigl \cite{Vl04}).
    
     We suggest that blazar OJ287 may have a double-jet structure
    produced by a supermassive black hole binary. Its northern jet and southern
    jet have different orientations and cone sizes in 3-dimensional space. Both
     relativistic jets precess clockwise with the same period 
    of 12\,yr, closely equal to the period of optical double-peaked 
    outbursts. In Figure 14 the modeled and observed relations of jet position
    angle vs time and vs viewing angle are shown to demonstrate the relative
    distributions of the ejection of the superluminal knots from the two jets.

    It is particularly important to note that the northern jet is much more
    active than the southern jet, producing all the five optical double-peaked
    outbursts. This preferential behavior of the northern jet  is very similar
    to the phenomenon observed in blazar 3C279 (Qian et al. \cite{Qi18}, Cheung 
    \cite{Che02}, de Pater \& Perley \cite{deP83}). In the double-jet
    structure of 3C279, one jet has been very active and extends to kpc-scales,
    but the other jet has no trails on kpc-scales.

   The differential activities of the two jets derived for both 3C279 and OJ287
   could be well understood in terms of HD/MHD theory for the accretion 
   processes occurring in supermassive black hole binary systems. According to
   the HD/MHD simulation results for cavity-accretion processes in near 
   equal-mass binary systems, gas streams preferentially accrete onto
   the secondary hole (Artymowicz \& Lubow \cite{Ar96}, Cuadra et al.
   \cite{Cu09}, Shi et al. \cite{Sh12}, Tanaka \cite{Ta13}, D'Orazio et al.
   \cite{Do13}). This naturally explains why the northern jet of OJ287 
   has been highly active in producing all the major
   optical double-peaked outbursts during the past $\sim$45 \,years (see Table
   7), if the northern jet is identified as the jet produced by the secondary
   hole.

    \begin{table}
    \centering
    \caption{Estimations of the masses of the supermassive black hole 
   binary as a function of parameter $\epsilon$ (viewing angle of the jet 
   axis of the secondary hole).}
    \begin{flushleft}
    \centering
    \begin{tabular}{lrrrr}
    \hline
    $\epsilon$ & $2^{\circ}$ & $3^{\circ}$ & $4^{\circ}$ & $5^{\circ}$ \\
    $M_8$ & 0.88 & 2.96 & 4.32 & 13.7 \\
    $m_8$ & 0.26 & 0.89 & 1.30 & 4.11 \\
    \hline
    \end{tabular}
    \end{flushleft}
    \end{table}
   Since both jets have the same precession period of 12\,yr, the precession
   of the jets may be originated from the binary motion: the direction of the 
  jets is modulated by the change of their orbital velocity directions relative
 to the observer (e.g., Roos et al. \cite{Ro93}, Kaastra \& Roos \cite{Kaa92},
   Qian et al. \cite{Qi17}, \cite{Qi18}).
  In this case the total mass and mass-ratio of the binary
  could be estimated from the observed jet apertures (see Fig.14), which
  depend on the parameter $\epsilon$ and the orbital velocities of the holes.
  If a jet produced in a binary system has a precession (or swing) 
   caused by the orbital motion of its associated hole, then its jet-aperture
   should be approximately equal to  $V_{orb}$/$V_j$ ($V_{orb}$--the orbital
   velocity of the hole and $V_j$$\approx$c--the velocity of the jet;
   e.g., Artymovicz \& Lubow \cite{Ar96}). Thus
  the mass ratio q=m/M of the binary could be approximately estimated as the
   ratio of the observed jet apertures. For OJ287 this ratio is on order 
   of $\sim$0.3 
   (see Fig.14). The total mass m+M of the binary could be estimated from the
   following formula (see Qian et al. \cite{Qi17}, \cite{Qi18}):
   \begin{equation}
     {{M_8}+{m_8}}{\simeq}1.02{\times}{10^4}{(\frac{T_{orb}}{\rm{yr}})}
                         [(1+q){{\beta}_{orb,m}}]^{3}
   \end{equation}
   ${\beta}_{orb,m}$ is the orbital velocity of the secondary hole, 
   approximately equal to $\tan{\eta}$, $\eta$ is half the northern jet cone 
  aperture (de-projected). $T_{orb}$ is the orbital period in the source frame.
   In the present model for OJ287 $T_{orb}$$\simeq$9.2\,yr,
    $\eta$$\simeq$$2^{\circ}$. 
   In Table 8 the estimations of the masses of the 
   binary holes are listed as function of parameter $\epsilon$. It can be seen
   that the total mass is $\stackrel{<}{_\sim}$1.5${\times}$${10^9}{M_{\odot}}$,
   if $\epsilon{<}{5^{\circ}}$. Obviously, we have found a new and independent
   method to determine the masses of a supermassive black hole binary through
   the measurement of its double jet precession.
  
    As a supermassive binary system should emit gravitational waves
    (Einstein \cite{Ei16}, \cite{Ei18}) during its in-spiral process and
    coalescence. Since the binary system in OJ287 may have an orbital motion
    on a sub-parsec scale, there might have the  possibility
     to follow its orbital motion in the sky, e.g., by using GRAVITY 
    instrument (Eisenhauer et al. \cite{Eis11}).
   \section{Discussion}
    The phenomena observed in OJ287 at optical and radio wavelengths
   have been extensively investigated and interpreted by many authors from
   various aspects (e.g., Sillanp\"a\"a et al. \cite{Si88};
   Lehto \& Valtonen \cite{Le96}; Sundelius et al. \cite{Su97};
   Valtaoja et al. \cite{Val20}; Villforth et al. \cite{Vil10};
   Villata et al. \cite{Villa98}; Tanaka \cite{Ta13}, Qian \cite{Qi15},
    Britzen et al. \cite{Br18}). But several basic issues still need to be
   clarified: e.g., the mass and mass-ratio of the binary system,  jet
   precession period, nature of optical outbursts, interaction between the
   binary black holes and the accreting material and magnetic field,
   accreting flow patterns, correlation between the
   optical and radio variability, association of the optical outbursts with
   the formation and emergence of superluminal knots and so on
   (e.g., referring to the comments made by Villforth et al. \cite{Vil10}
   on most of the existing binary black hole models).

   For demonstrating the main physical processes involved in the OJ287
   phenomena, we will concentrate on the comparison of three existing models.
   \subsection{Precessing binary model}
    In the precessing binary model Lehto \& Valtonen (\cite{Le96})
   concentrated on the interpretation of the four distinct features of the 
   OJ287 phenomena: (1)
   periodic double-peaked outbursts with a period of $\sim$11--12\,yr; (2)
   the time-intervals of $\sim$1--2\,yr between the two peaked outbursts; 
   (3) the first flares of the double-outbursts having sharp
   rising phases with time-scales of $\sim$10 days and zero polarization;
    (4) non-periodic optical flares with high polarization degrees.

    This is a binary black hole model with an extremely small mass ratio
    (m:M\,=\,0.007:1) and a high orbital inclination.
   According to this model, the regular orbital motion of the
   secondary black hole around the primary  provides the periodicity 
   of $\sim$12\,yr. The two disk-crossings per pericenter passage of the 
   secondary hole penetrating the accretion-disk of the primary hole cause
   the double-peaked outbursts with a time interval of 1--2\,yr. The first 
   optical flares with steep rising phases are assumed to be produced by 
   the evolving gas-bubbles torn out from the primary disk. The emission of the
   gas-bubbles are originated from bremsstrahlung mechanism and is
    non-polarized (zero polarization degree). For interpreting the 
   non-periodic optical outbursts, the model assumes that the tidal 
   disturbances in the primary disk induced by the secondary hole impactings 
   during its pericenter passages provide enhanced mass-accretion onto the
   primary hole, resulting in the ejection of superluminal components from the
   primary jet and producing synchrotron flares. Thus these non-periodic 
  outbursts are highly polarized. However, this 
   "double-mechanism" (bremsstrahlung-synchrotron) model can
   not explain the extreme stability of the V--R color index observed 
   during the the OJ-94 project (Takalo \cite{Tak96}), as pointed out by 
   Sillanp\"a\"a et al. (\cite{Si96a}, \cite{Si96b}) and Pietil\"a
    (\cite{Pie98}). This model has been claimed to be able to test the general
   relativity effects possibly occurred in the binary system of OJ287, because
   the motion of the secondary and the timing of the impactings can be
    precisely determined by the post-Newtonian orbital solution (Valtonen
   et al. \cite{Va11}).
   \subsection{Cavity-flare model}
    Tanaka (\cite{Tan13}) proposed a cavity-flare model to
   explain the periodic double-peaked optical outbursts. This is a binary
   black hole model with a moderate mass ratio (m:M\,=\,0.25:1) and a coplanar
   orbital  motion. This model is based on HD/MHD simulations
   for near equal-mass binary black hole systems (e.g., Artymovicz \cite{Art98},
   Hayasaki et al. \cite{Ha08}, Cuadra et al. \cite{Cu09}, Shi et al.
    \cite{Sh12}, Sesana et al. \cite{Se12}, Roedig et al. \cite{Ro12}, D'Orazio
   et al. \cite{Do13}, Tanaka \cite{Tan13}). Numerical simulations of the
    interaction between a supermassive black hole binary and its circumbinary
   disk show that (1) triple disks could exist in the supermassive binary
   system, because  the mass transfer from
   the circumbinary disk would form accretion
   disks around both black holes and  relativistic jets launched from  both the
   holes; (2) The tidal torques of the binary during the orbital motion would 
   suppress  the accretion rates onto the binary black holes surrounded by a 
   low-density cavity. During the pericenter passage of the secondary hole 
   two elongated accretion gas streams would leak into the cavity,
    accreting onto one or both SMBHs and  producing a double-peaked thermal
   optical outburst (per orbital period of 12\,yr). However, Tanaka's 
   model didn't
   discuss the explanation of the non-periodic optical flares and timing of
   the outbursts. For understanding the entire
   phenomena observed in OJ287 in terms of  cavity-flare model, more
   HD/MHD simulations and theoretical research are imperatively needed.
   \subsection{Relativistic jet models}
   In contrast to the precessing binary model and the cavity-flare
   model, both of which invoke a combination of bremsstrahlung and synchrotron
   mechanisms to explain the optical outbursts (periodic and non-periodic),
   relativistic jet models only invoke synchrotron mechanism to explain
   the optical outbursts. That is, all the optical outbursts (periodic
   doubled-peaked flares and non-periodic flares) are assumed to be 
   originated from the
   relativistic jets produced by the primary and secondary black holes.
   Villata et al. (\cite{Villa98}) proposed a double jet model to explain the
   periodic double-peaked optical outbursts, regarding both outbursts being
   synchrotron flares originated from the double jets. Qian (\cite{Qi15})
   investigated the possibility that the double-peaked optical outbursts could
   be produced by light-house effects in the jet or superluminal knots moving
   across two re-collimation shocks. Villforth et al. (\cite{Vil10}) suggested
   a ``disk magnetic-breathing'' model and assumed that the massive accretion
   of poloidal field causes the optical double-peaked outbursts.

   Recently, Britzen et al. (\cite{Br18}) proposed an elaborated jet model to
   explain the phenomena in OJ287.
   They suggested that the radio jet produced by the primary black hole is
   precessing and rotating, and discussed  a precession/nutation mechanism,
   showing that the jet kinematics as  well as the optical and radio light 
   curves can be interpreted in terms of
   geometric effects and Doppler beaming. Newtonian-driven precession and/or
   Lense-Thirring effect are suggested to explain the time scale of the jet
   precession.

   Relativistic jet models assume that the periodic double-peaked optical
   outbursts are originated from jet synchrotron emission. Except this 
   assumption distinct from the precessing binary model (Lehto-Valtonen model)
   and the cavity-flare model (Tanaka-model),
    relativistic jet models are similar to these binary models in understanding
   the other aspects of the phenomena in OJ287. Thus we will concentrate on the
  interpretation of the periodic double-peaked optical outbursts
   in terms of relativistic jet models and propose a tentative framework to 
   understand the entire phenomena in OJ287.
    \subsection{implications of present work}
   The present work may have provided some new aspects viewing the OJ287
   phenomena, helping to investigate the interpretation of the origin of the
   periodic double-peaked optical outbursts.

   \subsubsection{Periodicity and double structure}
    As shown in Sec.~9, in our double-jet scenario proposed for OJ287,
   the mass ratio m:M of the binary is estimated to be in the 
   order of $\sim$0.3:1.
   Therefore the phenomena observed in OJ287 should be investigated by applying
   HD/MHD theories of the physical processes (e.g., 
   cavity-accretion/jet-formation) in near equal-mass 
   supermassive black hole systems. We would use the results obtained in
   Tanaka (\cite{Ta13}) for reference.\footnote{Unfortunately, theoretical 
  research and numerical simulations for physical processes in near equal-mass 
  binary systems are far from sufficient for studying the phenomena in 
  OJ287 and other blazars.}
  
    We suggest speculatively that the periodic double-peaked optical
    outbursts are produced through the cavity-accretion processes, in which
   two gas-streams leaked from the circumbinary disk (per periastron passage
   of the secondary hole) accrete onto the binary holes consecutively,
   causing double-peaked optical outbursts with time interval $\sim$1--2\,yr
   (e.g., Hayasaki et al. \cite{Ha08}, Artymovicz \cite{Art98}).
   Different from the Tanaka's cavity-flare model (\cite{Tan13}), here we 
   might assume that the enhanced mass-accretion onto the binary holes 
   could be converted to the ejection  of superluminal components from 
   both jets through jet-formation mechanisms (e.g., mechanisms
   suggested by Blandford-Znajek \cite{Bl77} and Blanford-Payne 
   (\cite{Bl82}), causing a pair of synchrotron optical outbursts. 
   Because the  precession of the jets are caused by the orbital motion of 
   the binary, the orbital motion of the binary holes naturally provides
   the 12\,yr precession period of the jets. Occasionally, gas 
   streams accreted onto the binary might produce the non-periodic flares.

   This ``single-mechanism'' scenario seems to be supported by the 
   radio/optical variability
    studies. It has been found that the optical
   variability is highly correlated with the radio variability and the
   emergence of the superluminal components (Tateyama et al. \cite{Ta99},
   Britzen et al. \cite{Br18}). Upon detailed inspection it has been
   found that the optical double-peak outburst structures  had  correspondent
   radio double-bump  burst structures, as summarized in Table 9. The optical
   and radio  light curves given in Britzen et al. (\cite{Br18}) reveal that
   the non-periodic optical double-peaked outburst during 2001.8--2003.1 
   also had a corresponding radio double-bump burst structure. This seems
   to imply that the nature of the periodic optical double-peaked outbursts
  could be essentially similar to that of the non-periodic double-peak
  outbursts. Both periodic and non-periodic optical outbursts could be produced
   by the same mechanism: synchrotron radiation. The most persuasive argument
    for this ``single mechanism scenario'' had already
   been suggested by Sillanp\"a\"a et al. (\cite{Si96a}): the extreme
   stability of the V-R color index measured by the OJ-94 monitoring project
   (Takalo \cite{Tak96}) supporting the same energy production mechanism during
   a period of $\sim$2.3\,yr (1993.8--1996.1). We note that during the OJ-94
    project the major double-peaked optical outbursts (starting at 1994.65 
   and 1995.75 and claimed as thermal flares) were contemporarily
    observed with many synchrotron flares. 
   The stability of the V-R color index seems to
   clearly imply that all the optical flares observed during the OJ-94
   project should be originated from the same energy production mechanism.
   The solely possible mechanism is synchrotron with a power-law electron
   energy spectral index of $\sim$2.90 at V-R band. Any suggestion of dual
   mechanism (e.g. a combination of bubble-producing thermal flares and
   synchrotron flares) would be difficult to explain this color stability,
   as Sillanp\"a\"a et al. (\cite{Si96a}) and Pietil\"a (\cite{Pie98})
   commented.
   \subsubsection{Interpretation of optical light-curves}
   In this work  we have found that the superluminal knots ejected from
   both the relativistic jets are moving along trajectory of parabola-like
   shape. This would result in large changes of their Doppler factor near 
   the core.
    \begin{table*}
    \centering
    \caption{Correspondence between  the periodic double-peaked optical
    outburst structure and  the double-bumped radio burst structure. In some
    cases simultaneous optical and radio outbursts are observed and usually
    the double-bump radio bursts were observed to be delayed to the optical
    outbursts. Synchrotron self-absorption effects could cause the radio bursts
    at low frequencies (e.g., $<$15\,GHz) to be not observed, but at higher
    frequencies ($>$37\,GHz--90\,GHz) the optical-radio connection could still
    be observed (Valtaoja et al. \cite{Val20}). The 1994.7 radio burst may be
    an example: at 90\,GHz it shows a double bump, but at 15/22\,GHz the first
    bump was not observed. For comparison, we also list the non-periodic
    outburst observed in 2001--2003 that could be produced in the southern
    jet.}
    \begin{flushleft}
    \centering
    \begin{tabular}{lrr}
    \hline
    optical  &  radio  & jet ID\\
    \hline
    1971.2--1974.0 (three peaks)   &  1971.5--1976.6 (three bumps, 8\,GHz) &
                                                            N-jet \\
    1971.2--1972.7 (double peak)   &  1971.5--1974.3 (double bump, 8\,GHz) &
                                                            N-jet \\
    1983.0--1984.5 (double peak)   &  1983.0--1984.5 (double bump, 15\,GHz) &
                                                             N-jet \\
    1994.6--1996.3 (double peak)   &  1994.6--1996.5 (double bump, 15\,GHz) &
                                                              N-jet \\
    2005.8--2008.5 (double peak)   &  2005.8--2009.5 (double bump, 15\,GHz) &
                                                              N-jet \\
    2015.9--2016.5 (first peak)    &  2016.0--2017.1 (first bump, 15\,GHz) &
                                                               N-jet  \\
    \hline
    2001.8--2003.1 (double peak)   &  2002.0--2005.1 (double bump, 15\,GHz) &
                                                               S-jet \\
    \hline
    \end{tabular}
    \end{flushleft}
    \end{table*}
    \begin{table*}
    \centering
    \caption{Comparison of the predictions (timings) of outbursts by different
     models. Data on models (orbit-1, orbit-2 and non-GR) are taken from
    Valtonen (\cite{Va07}), data on "new model" is taken from Valtonen et al.
    (\cite{Va16}). Adjustment of time delays has been used to make the timing
    of the 2015.87 optical outburst. The non-GR model seems to have a more
    accurate prediction than the orbit-1 model, but similar to orbit-2 model.}
    \begin{flushleft}
    \centering
    \begin{tabular}{lrrrr}
    \hline
    Model type  & orbit-1  & orbit-2 & non-GR & new model\\
    \hline
    Impacting time & 2013.53 & 2014.58 & -- & 2013.45\\
    Time-delay (yr) & 2.82 & 1.35 & -- & 2.42\\
    Predicted time & 2016.35 & 2015.93 & 2015.82 & 2015.87\\
    Observed & 2015.87 & 2015.87 & 2015.87 & 2015.87 \\
    \hline
    \end{tabular}
    \end{flushleft}
    \end{table*}
    \begin{figure*}
    \centering
    \includegraphics[width=5cm,angle=-90]{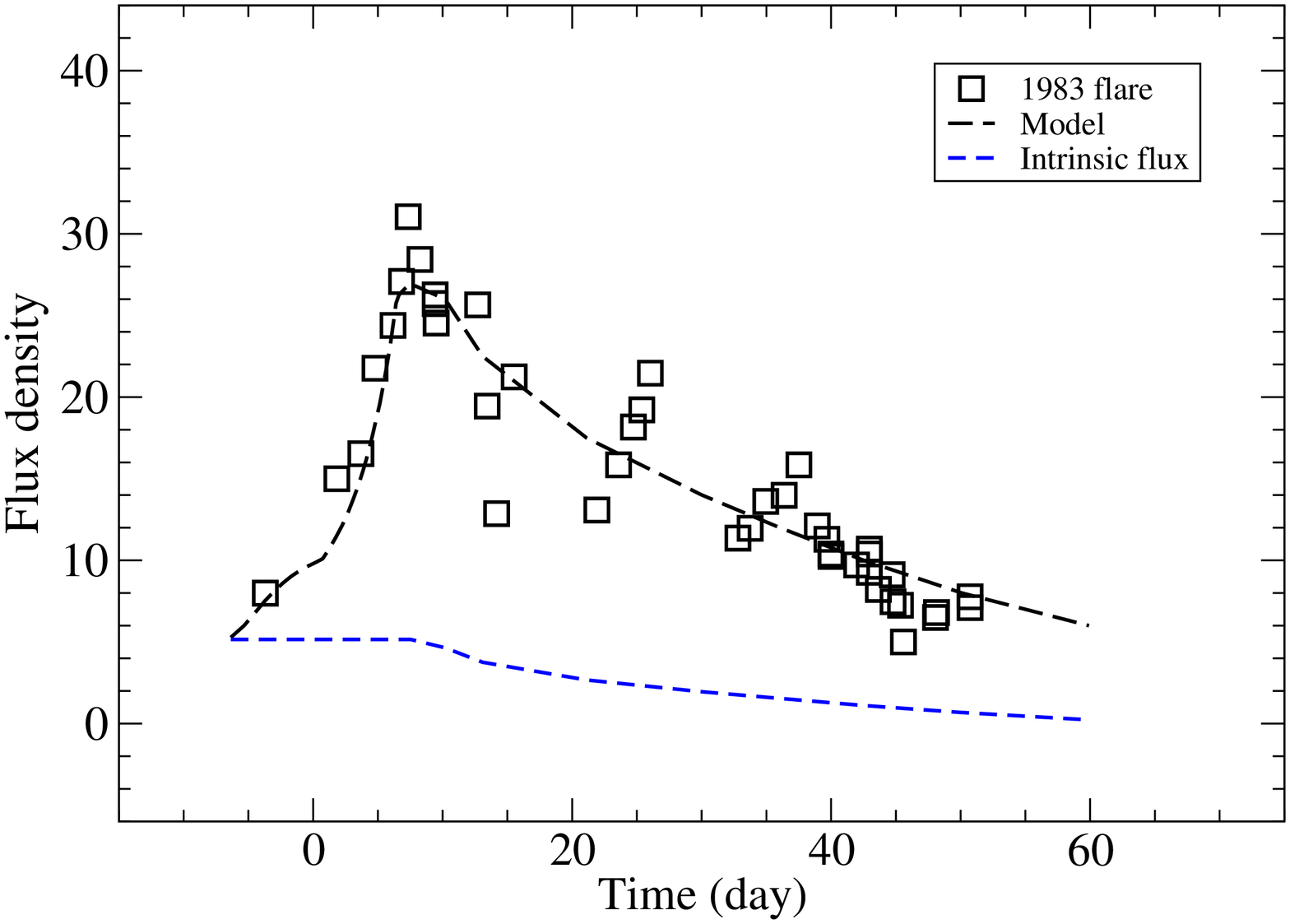}
    \includegraphics[width=5cm,angle=-90]{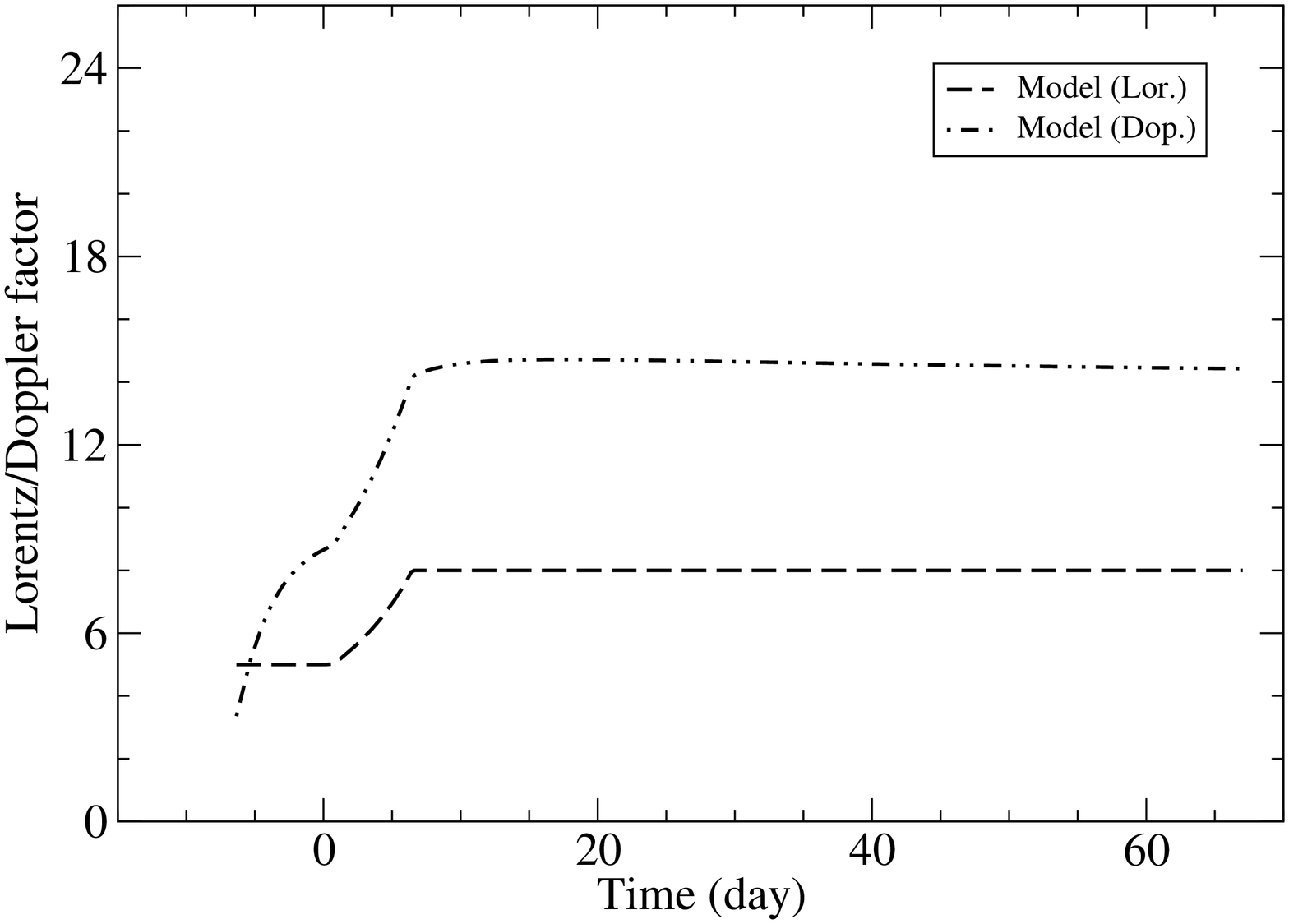}
    \includegraphics[width=5cm,angle=-90]{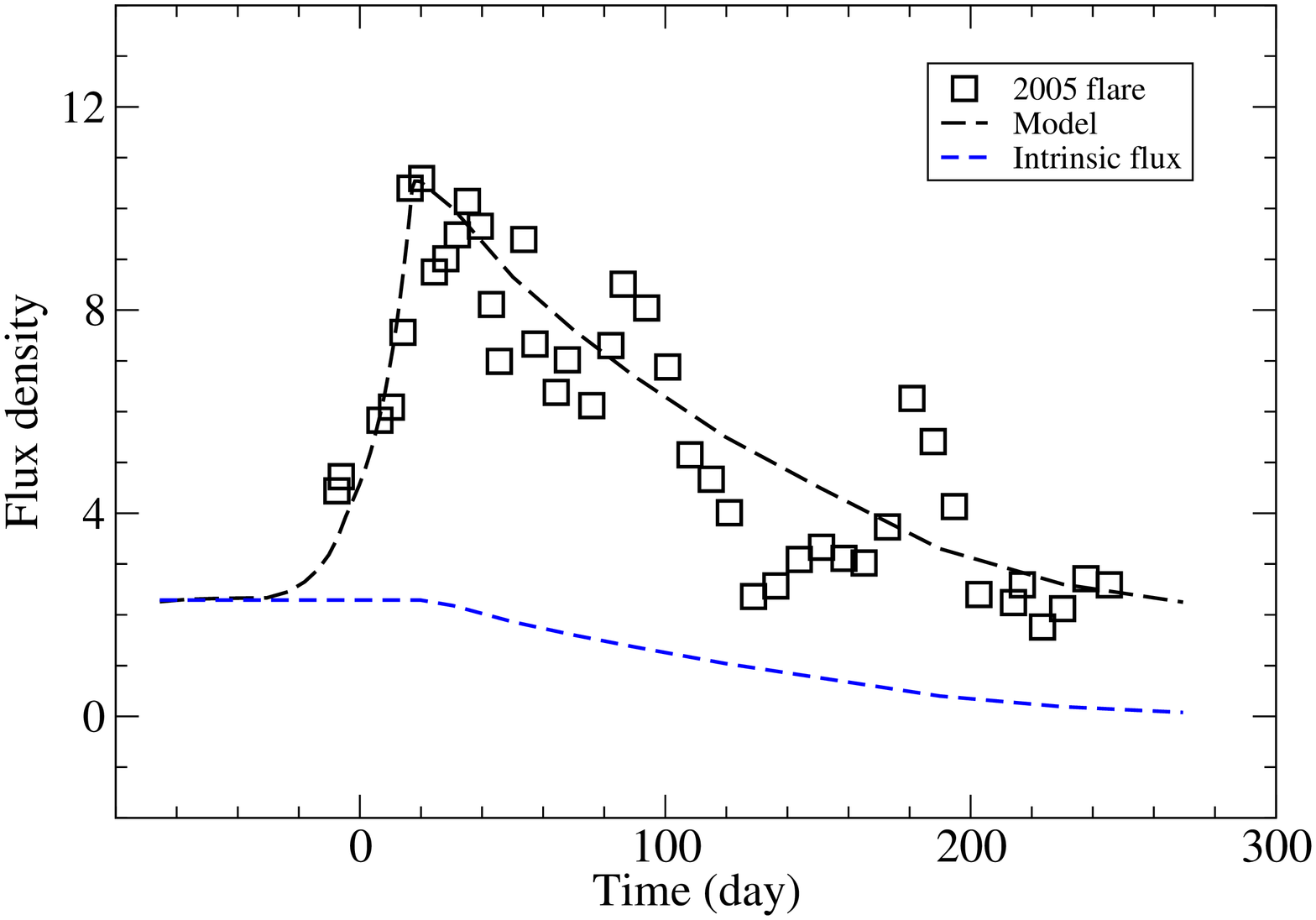}
    \includegraphics[width=5cm,angle=-90]{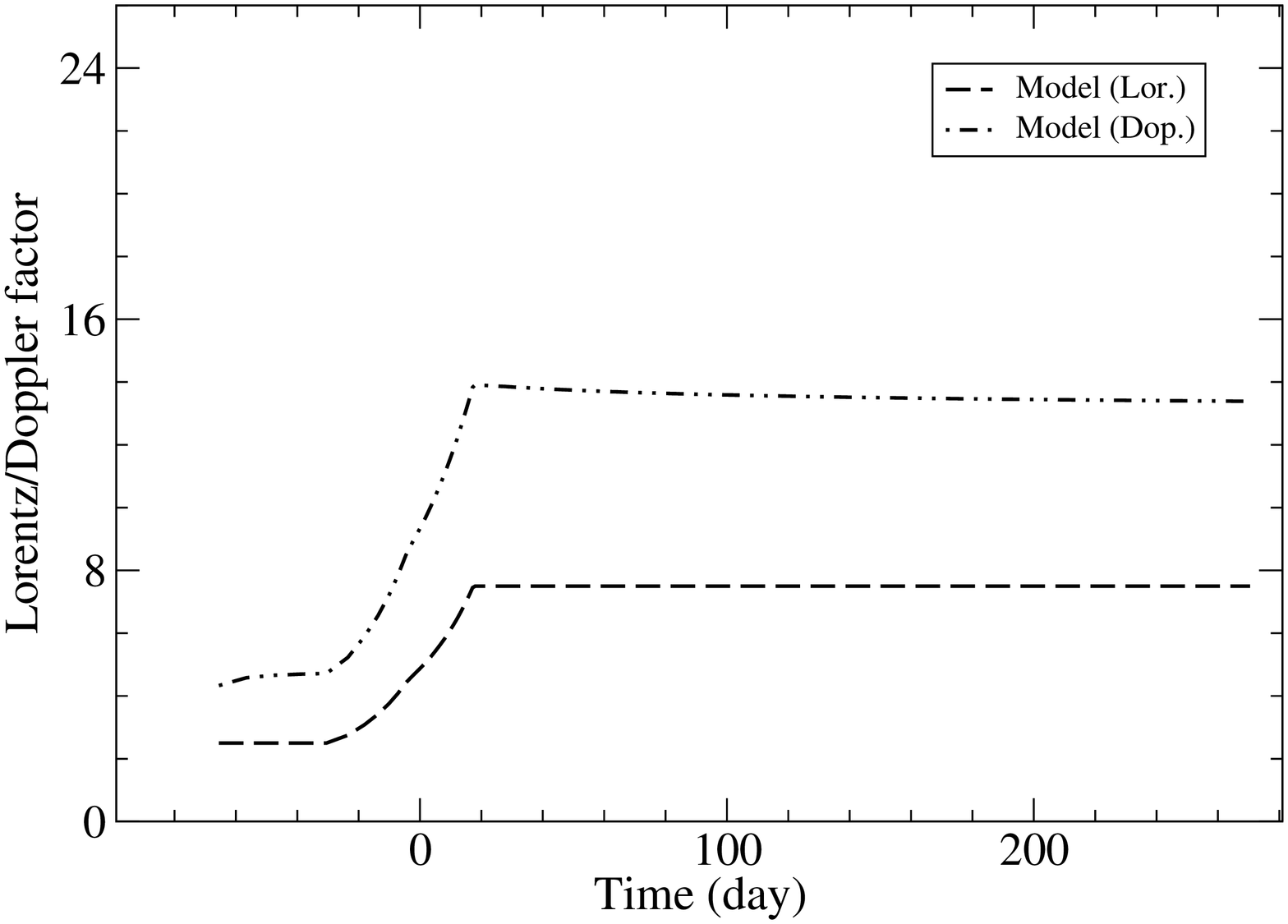}
    \includegraphics[width=5cm,angle=-90]{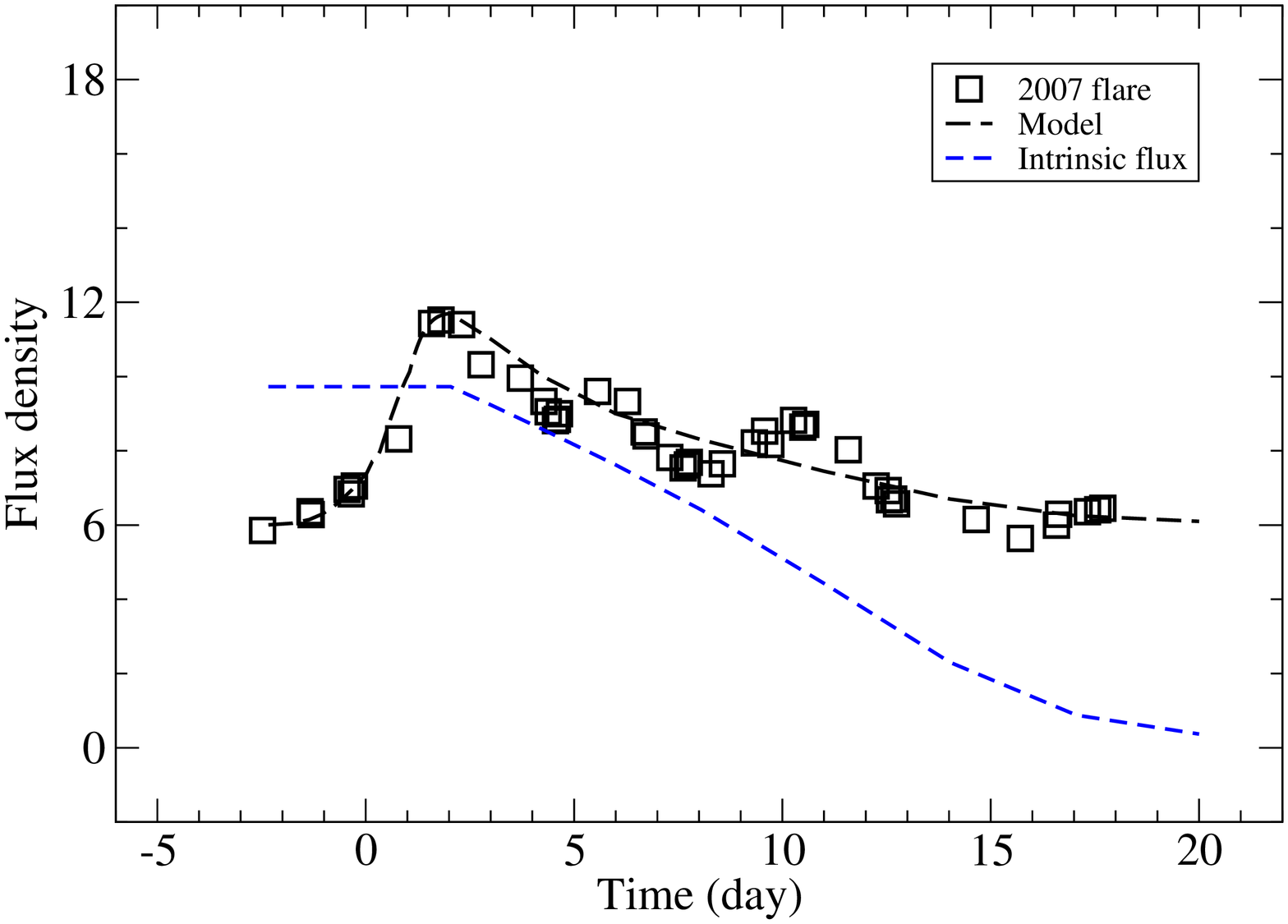}
    \includegraphics[width=5cm,angle=-90]{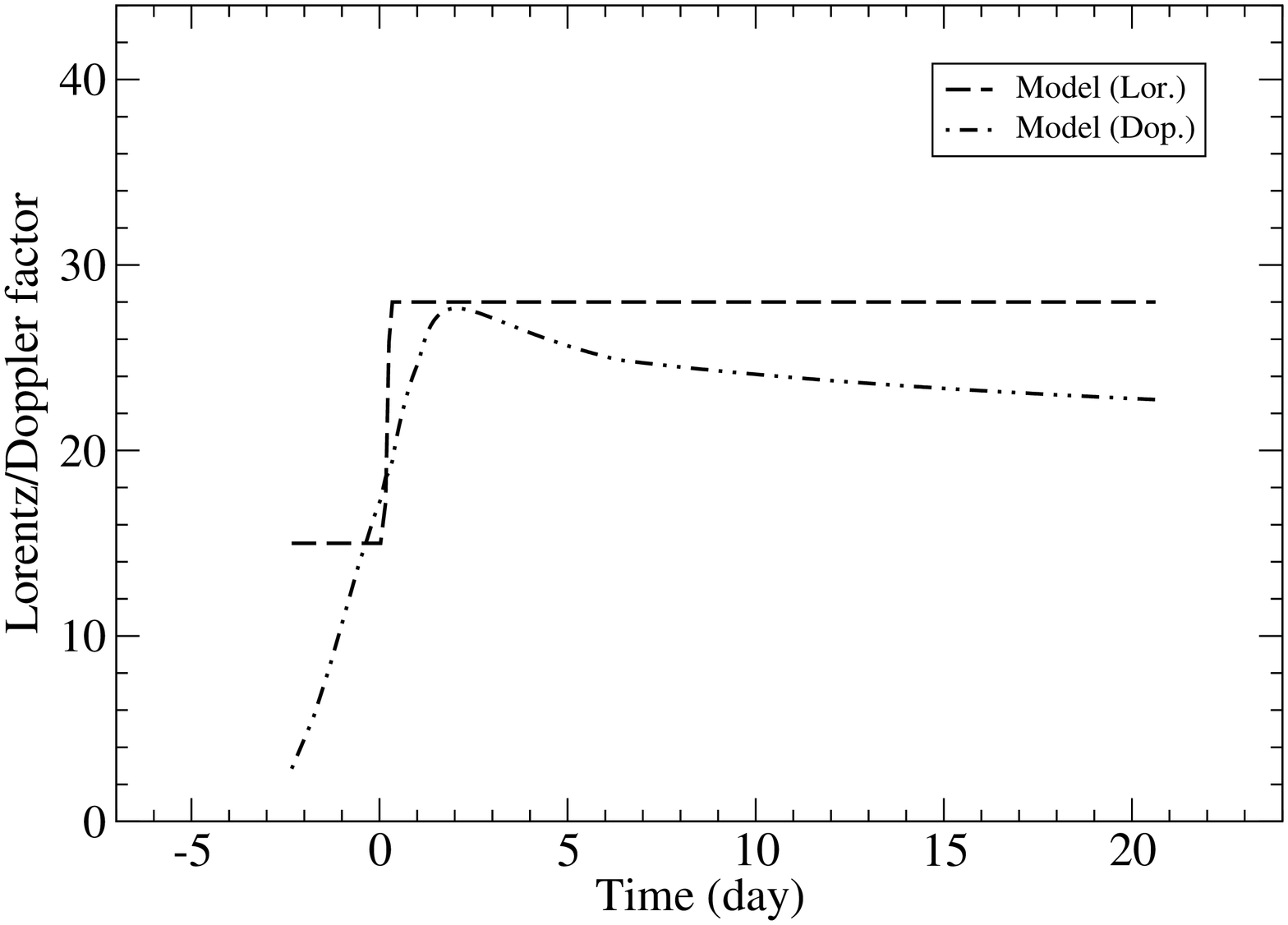}
    \caption{Model simulation of the light curves for the three major optical
    outbursts occured at 1983.0, 2005.76, 2007.70 (claimed as thermal 
    optical flares). The initial sharp rises are
    originated from the Doppler boosting effects (left panels) when the optical
    knots moving along the parabolic trajectories. The right panels show the
    modeled Lorentz/Doppler factors. The variations of the intrinsic
    flux density of the optical knots are shown in the  left panels
    (blue lines) and the flux units are $10^{-4}$\,mJy, $10^{-4}$\,mJy and
    ${10^{-6}}$\,mJy, respectively. The source base-levels are 5\,mJy,
     2\,mJy and 6\,mJy, respectively. The flux units for the 
     observed and modeled light curves are 1\,mJy.} 
    \end{figure*}
    \begin{figure*}
    \centering
    \includegraphics[width=5cm,angle=-90]{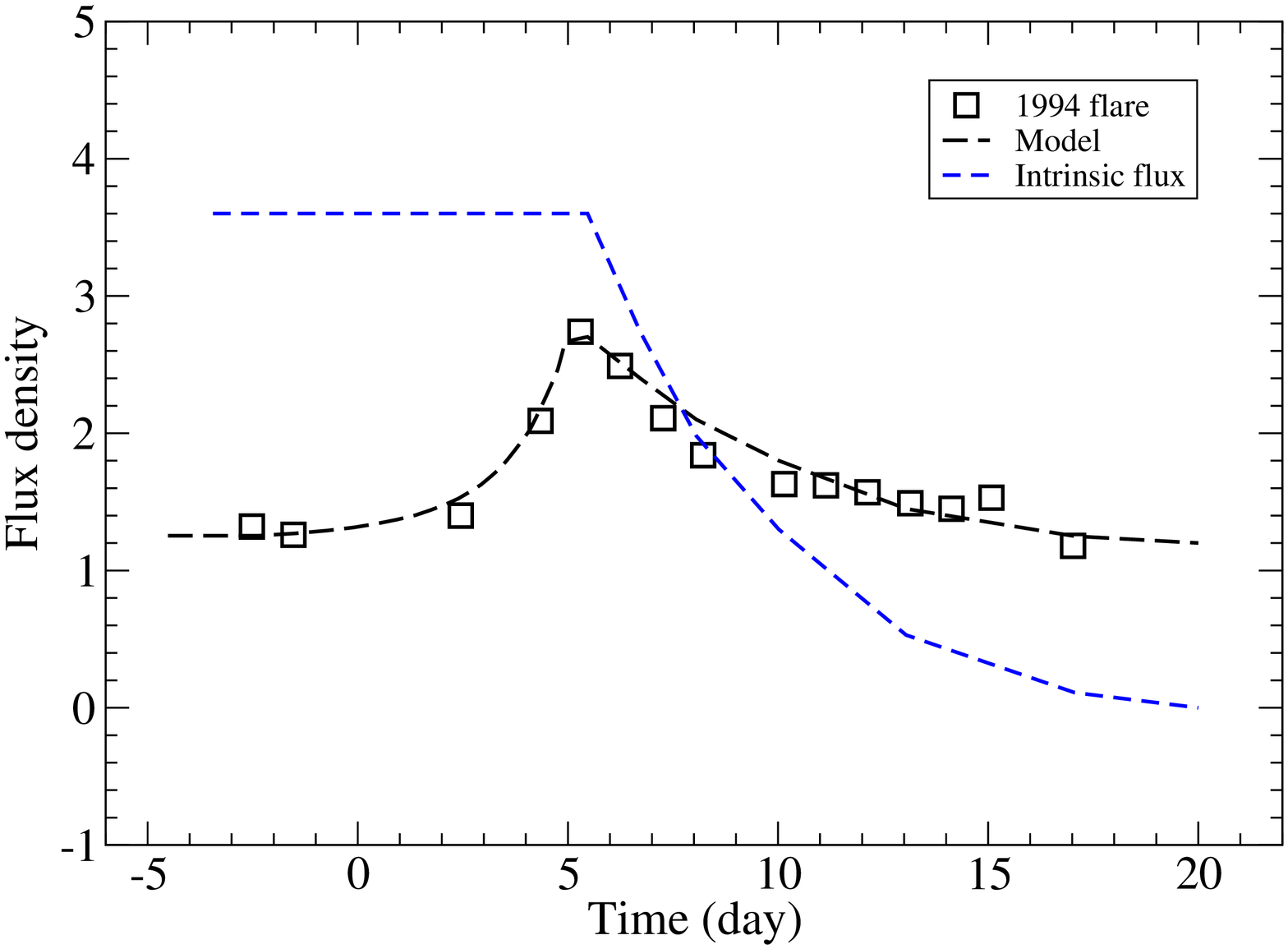}
    \includegraphics[width=5cm,angle=-90]{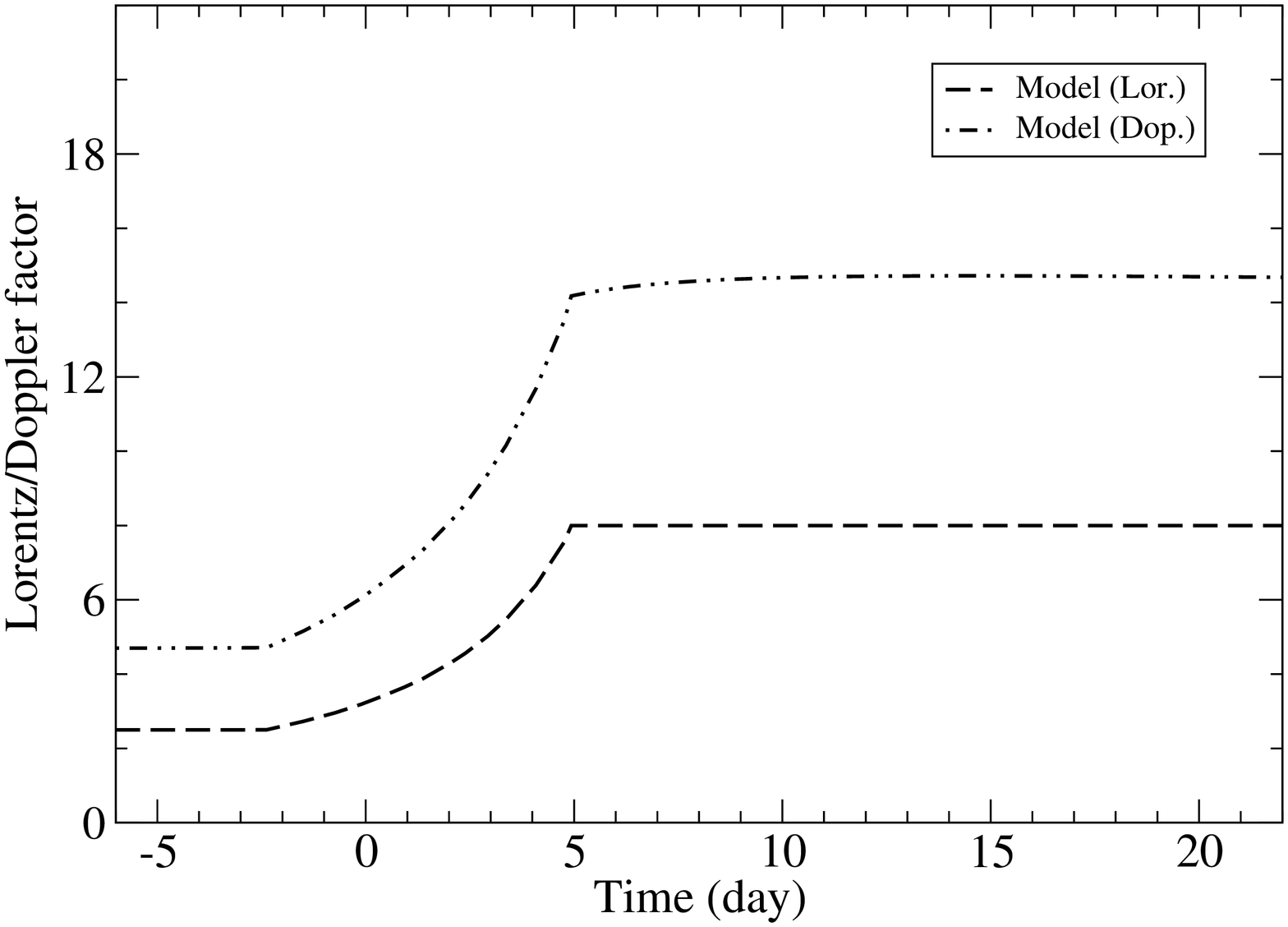}
    \caption{Model simulation of the light curve for a typical synchrotron
    flare occurred at 1994.02. The initial sharp rise of the light curve is
    modeled as due to the Doppler boosting when the optical knot moving along
    the parabolic trajectory. The right panel shows the modeled Lorentz/Doppler
    factor. The variation of the intrinsic flux density is shown
     in the left panel (blue line) and the flux unit is $10^{-5}$\,mJy.
    The source base-level is 1.2\,mJy. The flux units for the observed and
    modeled light curves are 1\,mJy.}
    \end{figure*}
    Thus the sharp rise phase of the periodic double-peaked optical outbursts
   could be interpreted in terms of Doppler boosting. In Figure 15 we show
   the results of our simulations by relativistic jet models for the light
   curves of the three optical outbursts occurred at 1983.0, 2005.76, 2007.70.
   \footnote{The data are collected from Valtonen et al. (\cite{Va08}).}
   
    In fact,
   using relativistic jet models, the evolution of the optical flux density
   of a superluminal knot can be write as:\\
   \begin{equation}
      {S(t)}={S_{in}}(t){\times}[\delta(t)]^{p+\alpha} ,
    \end{equation}
    $S_{in}$(t) is intrinsic flux density, p=3 for individual knots 
   (Lind \& and Blandford \cite{Lin85}) and 
    $\alpha$$\simeq$1.0 is optical spectral index :
    $S_{\nu}$$\propto$${\nu}^{-\alpha}$. We assume that the optical knots are
    ejected along parabolic trajectories as the radio knots, but having smaller
    scale sizes. We choose a model trajectory for the optical knot having
    parameters similar to those for radio knot C8 of the
    northern jet ($\epsilon$=$3^{\circ}$, $\psi$=0.0\,rad, $\omega$=-3.95\,rad,
    $x$=0.5), but a smaller scale size $a$=0.0402$[\rm{mas}]^{1/2}$.
    We also assume
    that during the rising phases the intrinsic optical flux densities are
    constant and thus the sharp rises of the flux density are fully determined
    by the Doppler boosting. But after reaching the maxima of the light curves
     the optical flux variations could be explained in terms of the intrinsic
     evolution $S_{in}$(t) of the optical knots (due to
    $\delta(t)$$\simeq$constant.).\footnote{Intrinsic optical variations are
    usually caused by particle acceleration, energy dissipation due to
    expansion and synchrotron/inverse-Compton losses  within the optical 
    knots and other processes, referring to Qian \cite{Qi96a}, \cite{Qi96b},
    \cite{Qi97}. Generally, in order to decompose the contributions from
    Doppler boosting and intrinsic variations, multi-wavelength
    light curves are required.} The model-fitting results of the light curves 
    for the three optical outbursts observed in 1983.0, 2005.76 and 2007.70
    are shown in Figure 15. It can be seen that the entire optical light curves
    of the three flares (all claimed to be thermal outbursts)  can be well 
    interpreted in terms of relativistic jet models. This is the first time 
   that the double-peaked optical outbursts are interpreted in terms of
    the superluminal motion of optical knots ejected from the optical core.
    In Figure 16 we
    also show the model-fitting
    results of the optical light curve for the typical synchrotron flare
   observed in 1994.02 (Fig. 4 in Sillanp\"a\"a et al. \cite{Si96b}; OJ-94
    project data). Obviously, the four flaring events have very similar
    behaviors and can be interpreted in terms of the evolution of superluminal
    knots ejected from relativistic jets.

    We point out that the results shown in Figures 15 and 16  may be 
    instructive and significant. In the precessing binary model (Lehto 
    \& Valtonen \cite{Le96}, Valtonen et al. \cite{Va17}), 
    the 2005.76 and 2007.70 outbursts (as a double-peaked outburst) and 
   the 1983.0 outburst were all
   assumed to be originated from the bremsstrahlung (thermal emission) of
   the gas-bubbles torn out from the disk of the primary hole when the
   secondary impacting onto the primary disk. Very complicated calculations
   of the disk-crossing processes and the formation and evolution of the
   gas-bubbles were made for model-fitting their light curves (e.g.,
   Lehto \& Valtonen \cite{Le96}, Pihajoki \cite{Pi16}, Pihajoki et al.
   \cite{Pi13a}).   But in our relativistic jet models their optical light
   curves can be relatively simply explained  in terms of the flux evolution
   of the optical knots
   ejected from the central supermassive binary black hole system
   along parabolic trajectories with Lorentz factors consistent with those
   obtained by radio VLBI-measurements. Moreover, the rising phase of the 2007
   outburst had a very short time-scale of $\sim$5 days and the 2005 outburst
   had a rising time-scale of $\sim$50\,days. This difference in rising
    time-scales could be simply explained in terms of the optical knots having
   different Lorentz factors.
    \subsubsection{Low polarization degrees}
    The interpretation of the light-curves for the four outbursts in terms of 
   relativistic jet models given above is very encouraging. We then can 
   consider the possibility of low polarization degrees for the double-peaked 
   optical outbursts. This might occur during the pericenter passages of the 
   secondary hole, when the gravitational (tidal) torques induce strong 
   turbulent magnetized mass flows accreting onto both black holes and, through
    jet-formation processes, result in ejection of
    superluminal knots (plasmons or shocks, Marscher \& Gear \cite{Ma85},
   Pacholczyk \cite{Pa70}, Qian \cite{Qi97}) with almost random
     magnetic fields (turbulent energy
    $>>$ magnetic energy). The optical
    flares produced by these knots will be synchrotron flares with very low
    polarization (e.g., a few percent, but not zero; Marscher et al.
   \cite{Ma08}, Marscher \cite{Ma14}, Burn \cite{Bu66}): in this case
    polarization degree
    p=$p_0$${B_u}^2$/(${B_u}^2$+${{B_r}^2}$),
     $B_u$ and $B_r$ are ordered and random field strength,
    respectively, $p_0$$\simeq$0.75 -- typical polarization degree of a
    synchrotron source with a uniform field (Burn \cite{Bu66},
    Pacholczyk \cite{Pa70}).
    \footnote{This is just
   a tentative suggestion and should be tested by  further polarization
   observations.}
   
    We have noticed that there are some clues supporting this
   consideration. For example, the radio bursts associated with the major
   optical outbursts occurred in 1983.0,
   1994.7 and 1995.9 were all low-polarized flares, showing distinct
   minimal polarization degrees of $\sim$2\% (Figs.\,5 and 7 in Valtaoja et al.
    \cite{Val20}), possibly indicating the extremely turbulent conditions 
   of the radio knots (Pacholczyk \cite{Pa70}). Moreover,
   the 2015.87 major optical outburst was observed to have an almost constant
   polarization degree of $\sim$6\% that is quite different from the
   extremely low polarization degrees of $\leq$2\% observed for the previous
   optical outbursts. A superposition of three
    constituents are assumed to explain this polarization degree:
   the major bremsstrahlung (thermal) outburst with zero polarization
   (as predicted  by the disk-impact model), a synchrotron flare with a
   polarization of 40\% and a base-level component with a polarization of 10\%
   (Valtonen et al. \cite{Va16}, \cite{Va17}). This assumption of
   multi-component polarization structure might pose questions: is it this
   decomposition unique? Is it possible that the optical outburst itself has
   a low (but not purely zero) polarization degree?  This problem
  seems not easily  solved and more polarization observations in radio and 
   optical bands are needed to investigate this possibility.
   \subsubsection{Timing of outbursts}
   As shown by Tanaka (\cite{Ta13}), cavity-accretion models could provide
   interpretation for the double-peaked optical flares per binary orbit: that 
   is, the 12\,yr periodicity and the 1--2\,yr time--interval. However, 
   due to the temporal stochasticity in the  accretion dynamics they 
   cannot  precisely constrain the flare timing. This is quite different from
   the precessing binary model of Lehto \& Valtonen (\cite{Le96}), which 
   strongly constrains the flare timing based on the binary orbit solution.
   In order to interpret the phenomena in OJ287 in terms of cavity-accretion
   (or cavity-flare) models, more  
   MHD-simulations and investigations of the physical processes in 
   near equal-mass binary systems would be required to search for appropriate
   ``multi-parameter solutions'', including those parameters
   describing the binary orbit, black hole masses, circumbinary-disk structure,
   interaction between the binary and circumbinary disk, formation of black 
   hole disks, leakage of a pair-stream per periastron passage and periodically
   enhanced mass accretion onto the binary holes (referring to Hayasaki et al. 
   \cite{Ha08} and references therein).  In fact, in the precessing binary
   model (Valtonen et al. \cite{Va07}, Valtonen et al. \cite{Va16}),
   the timing of optical outbursts
   not only depends on the binary orbit model, but also  on the choice
   of the disk model for estimating the delay times, especially for the disk
   crossings faraway from the primary hole (Pietil\"a \cite{Pie98}, Pihajoki
    \cite{Pi16}). This can be seen in Table 9, where the predictions of the
  2015.87 flare by four different models are compared.  It clearly shows
   that the flare-timing prediction by the
    non-gravitational model being better than that by both the models
    'orbit-1" and "orbit-2", and also indicates that the "new model" adopted
   a delay time of 2.42\,yr (instead of 2.82\,yr of model "orbit-1"), making
    it to get a more accurate timing of the 2015.87 optical outburst.
    Thus more HD/MHD studies and numerical simulations on the physical
    processes in supermassive binary systems are imperatively required to
    search multi-parameter solutions, specific for the phenomena in OJ287.
   Actually, some studies performed (e.g.) by Hayasaki et al. (\cite{Ha08}) and
   D'Orazio et al. (\cite{Do13}, Farris et a. \cite{Fa14}) are promising 
   to solve the problems involved in OJ287. It seems that spin of the binary 
   holes should be included in MHD simulations, because holes' spins may play
   important roles in sustaining hole-disks and forming holes' magneto-spheres,
   and thus forming relativistic jets (or AGN).      
     \section{Conclusion}
   In a brief summary, this work may have provided a new and alternative
   insight into the phenomena in OJ287 and  significant information about its
  physical processes.

    Based on the results obtained in this paper, it seems that
    relativistic jet models, as described
  above, may be potentially able to provide a fully and reasonable
  explanation of the entire emission properties and kinematic features observed
   in blazar OJ287 in a compatible framework. Actually, we have proposed a
   comprehensive framework of relativistic jet models to understand the
   phenomena observed in blazar
   OJ287. It contains the following ingredients.
  \begin{itemize}
   \item OJ287 may have a supermassive black hole binary in its center with a
   moderate mass ratio and two relativistic jets are produced by the primary
   and secondary holes;
   \item Through the cavity-accretion mechanism, the orbital motion of the
    binary in an eccentric orbit creates periodic (or regular)
    enhanced-accretion events (of mass
    and magnetic field) per periastron passage (due to disturbances
    induced by gravitational torques) with two gas-streams
    accreting  onto the binary. Occasional and non-periodic accretion events
    are also occurring during the orbital motion.
   \item The periodic accretion events, through jet formation mechanisms,
   will be converted into the consecutive ejections of two superluminal optical
    knots, resulting in the periodic double-peaked optical outbursts
   of a period $\sim$12\,yr with an double-peak time-interval of
   $\sim$1--2\,yr. 
   Similarly, the non-periodic accretion events will also produce usually
   observed  superluminal knots and their synchrotron emission;
   \item All the superluminal optical knots produced by these enhanced
  accretion events are ejected from the optical core along precessing parabolic
  trajectories, thus their light curves, especially the fast rising phases,
  can be explained by Doppler boosting, taking their intrinsic flux evolution
  into account (Figures 15 and 16);
   \item Both the optical and radio outbursts are 
   produced in the relativistic
   jets and originated from synchrotron emission. The optical variations are
   tightly correlated with the radio variations and the the emergence of
   superluminal radio knots.
  \item In addition to the optical outbursts, there is a base-level emission
  varying with longer time-scales, which could be related to the
  ``quiescent jet'' (or ``optical polarization core'') suggested by Villforth
   et al. (\cite{Vil10}). Thus the observed optical (and radio) light curves
   are constructed from the summation of the light curves of the individual
   superluminal knots and the base-level emission.
  \end{itemize}
   Although this framework is mostly qualitative, it satisfies the
   requirement of ``single mechanism'' and the entire phenomena observed in
   blazar OJ287 are ascribed to the relativistic jets and
    unified with the
   physics in other blazars. However, detailed investigations of
   HD/MHD processes of cavity-accretion, periodic gas-streaming onto
   the binary holes and interaction between the accretion and jet
   formation in the binary black hole systems for near equal-mass
   cases are required, especially for understanding the mechanism of the
   double-peaked optical outbursts. In addition, seeking suitable
   multi-parameter solutions for orbital motion and  flare-timing, and
   studying outburst polarization, accretion and knot ejection, etc. are also
   needed. Certainly, for interpreting the OJ287 phenomena, the main subjects
    would be: (1) how to take the effects of spins of the binary into 
   consideration in the cavity-accretion scenario. Spin of the binary
   holes may play important role in sustaining the disks around the binary 
   holes and re-collimating the material flung out from the cavity along the 
   direction perpendicular to the circumbinary disk; (2) how to deal with the 
   accretion of magnetic field onto the holes and the structure of the
    magneto-spheres around the binary holes; (3) how to deal with the
   formation of relativistic jets from both black holes; (3) how to deal with
   the interaction between the cavity-accretion and the relativistic jets.

    We would like to emphasize that the model-fitting results of the
   parsec-scale kinematics for OJ287 given  in
   this paper were obtained only from the detailed analysis  of the
   kinematics of the radio superluminal components and thus are independent
   of any existing proposed theoretical models for interpreting the optical
   phenomena observed in OJ287 (light curves and variability in polarization).
    These results, if correct, may be worth
   being taken into consideration in constructing better theoretical models.
   \begin{acknowledgement}
    I gratefully thank Dr.~S.~Britzen (Max-Planck Institute f\"ur 
   Radioastronomie) for providing the 15\,GHz VLBI-data on OJ287. 

   \end{acknowledgement}

   \begin{appendix}
    \section{Model fitting results of southern jet (for knots C11 and  C12)
    and northern jet (for knots C2, C3, C5, C6, C13U and C14.}
    
    \begin{figure*}
    \centering
    \includegraphics[width=5cm,angle=-90]{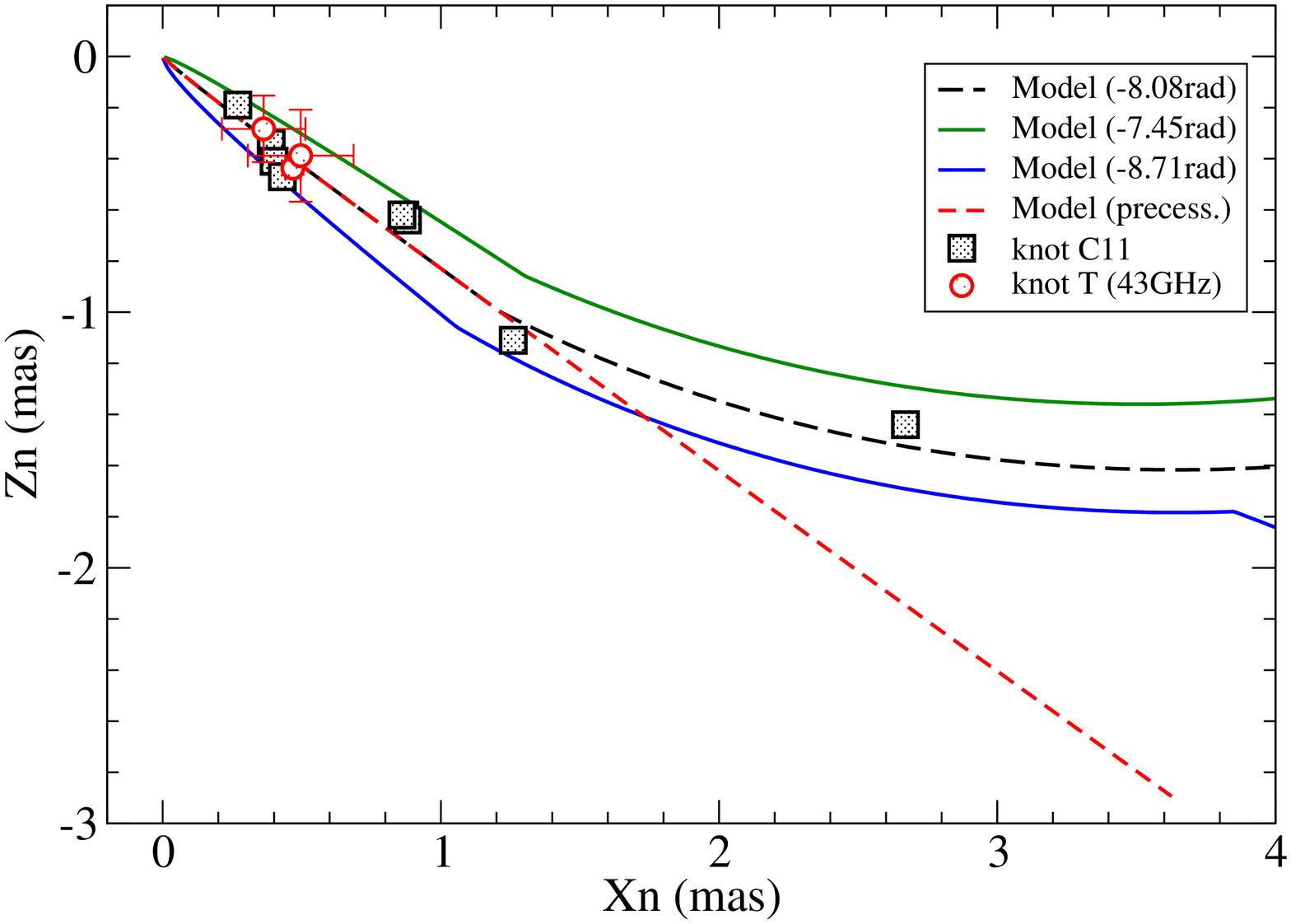}
    \includegraphics[width=5cm,angle=-90]{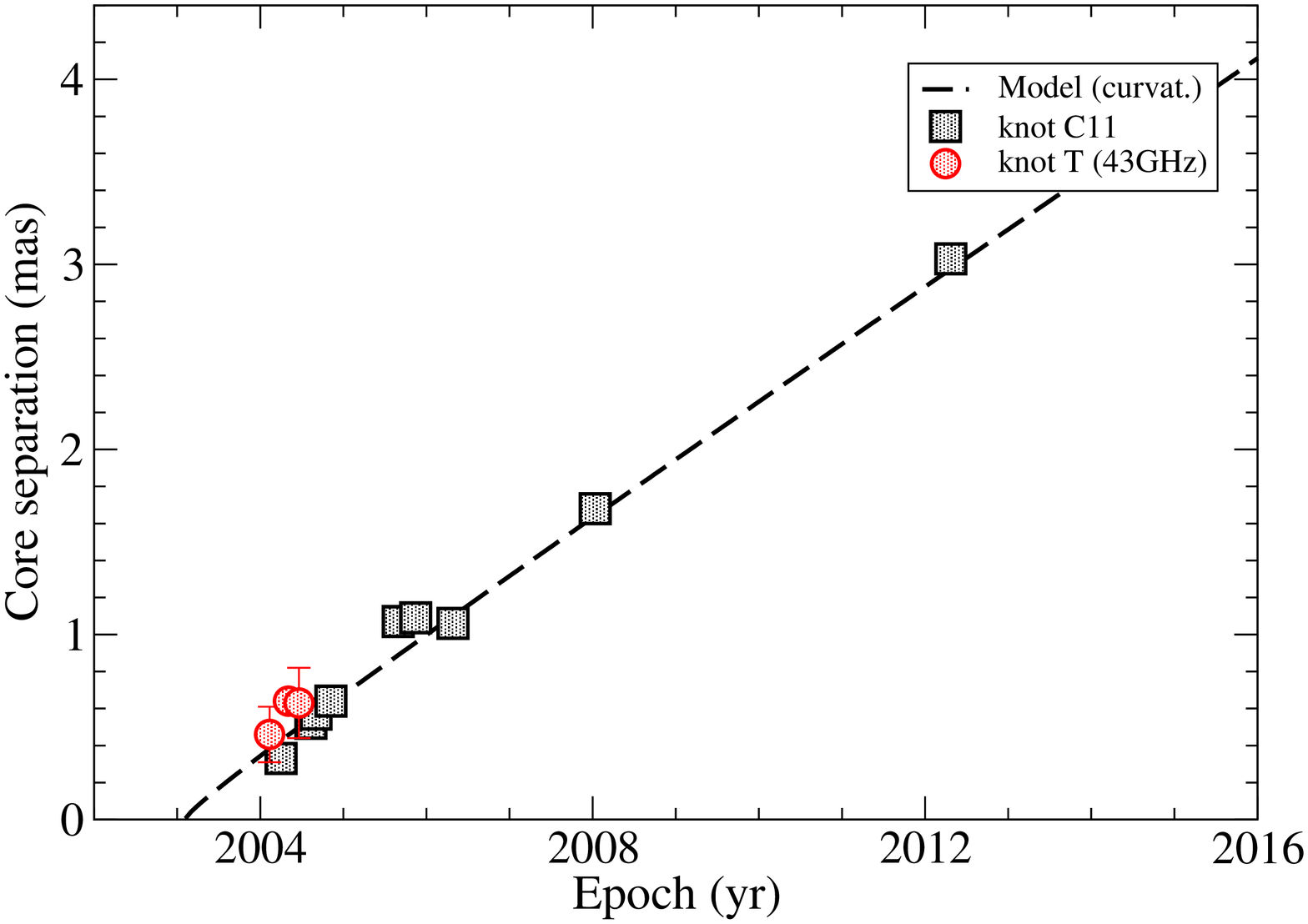}
    \includegraphics[width=5cm,angle=-90]{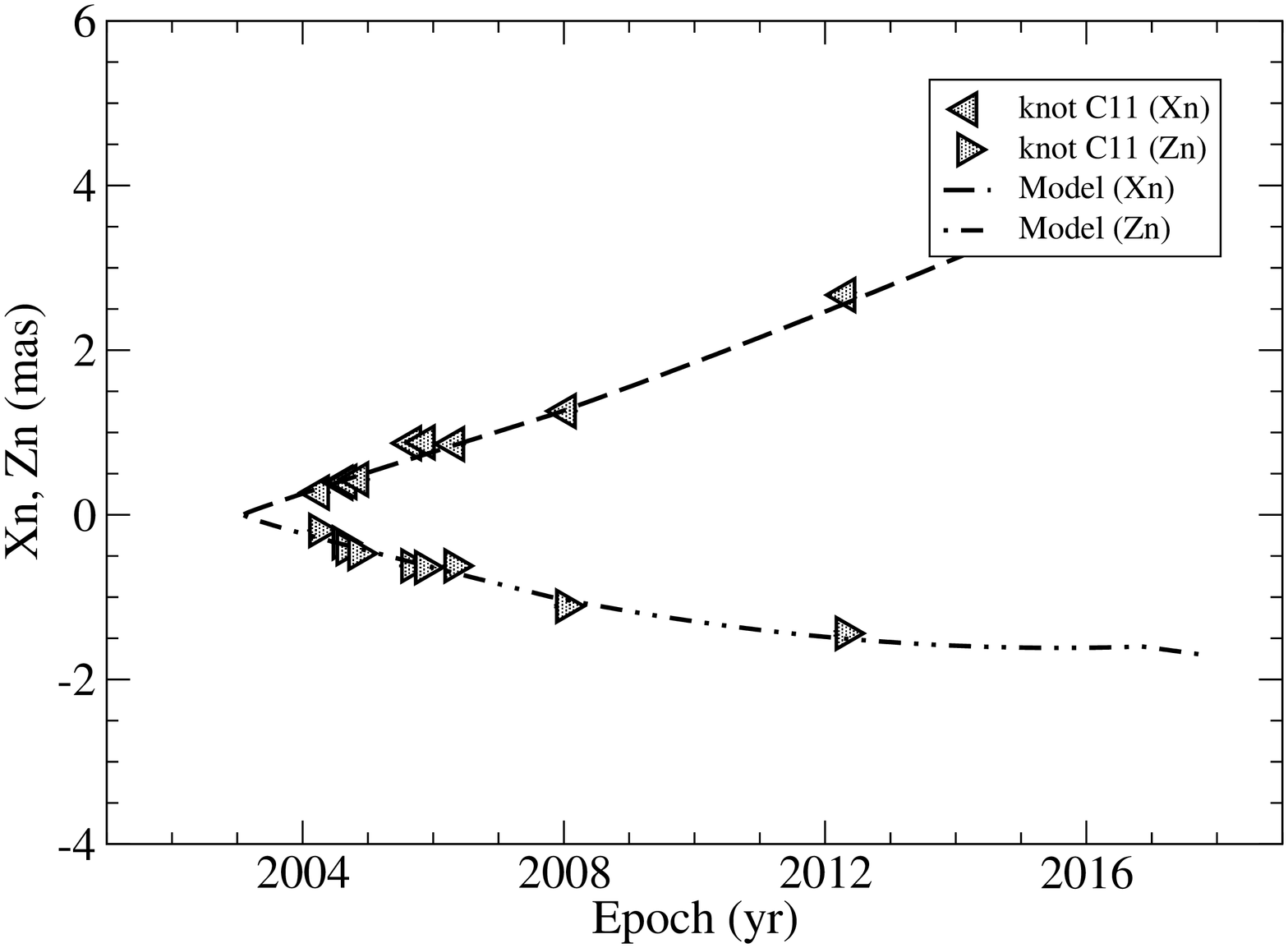}
    \includegraphics[width=5cm,angle=-90]{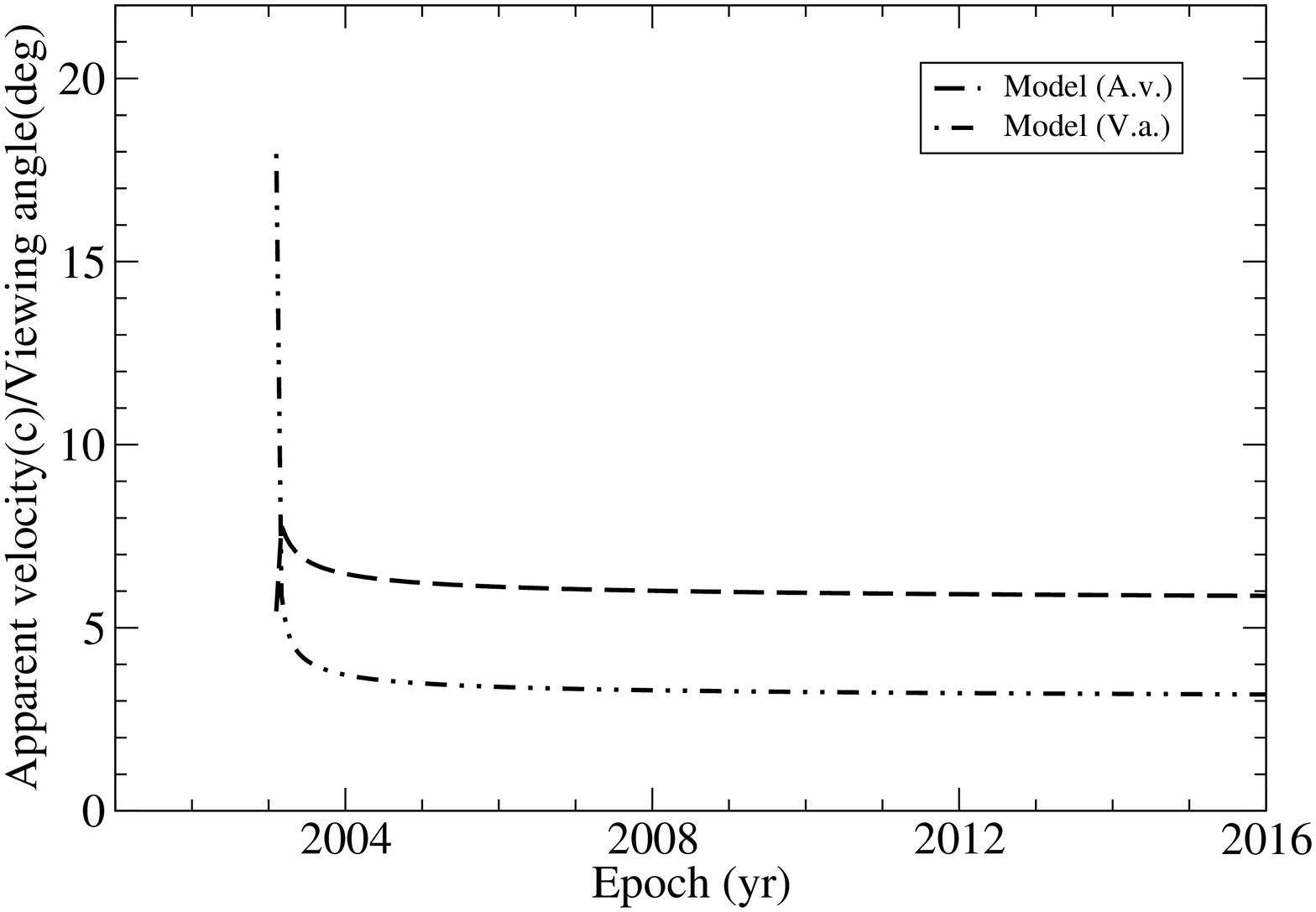}
    \includegraphics[width=5cm,angle=-90]{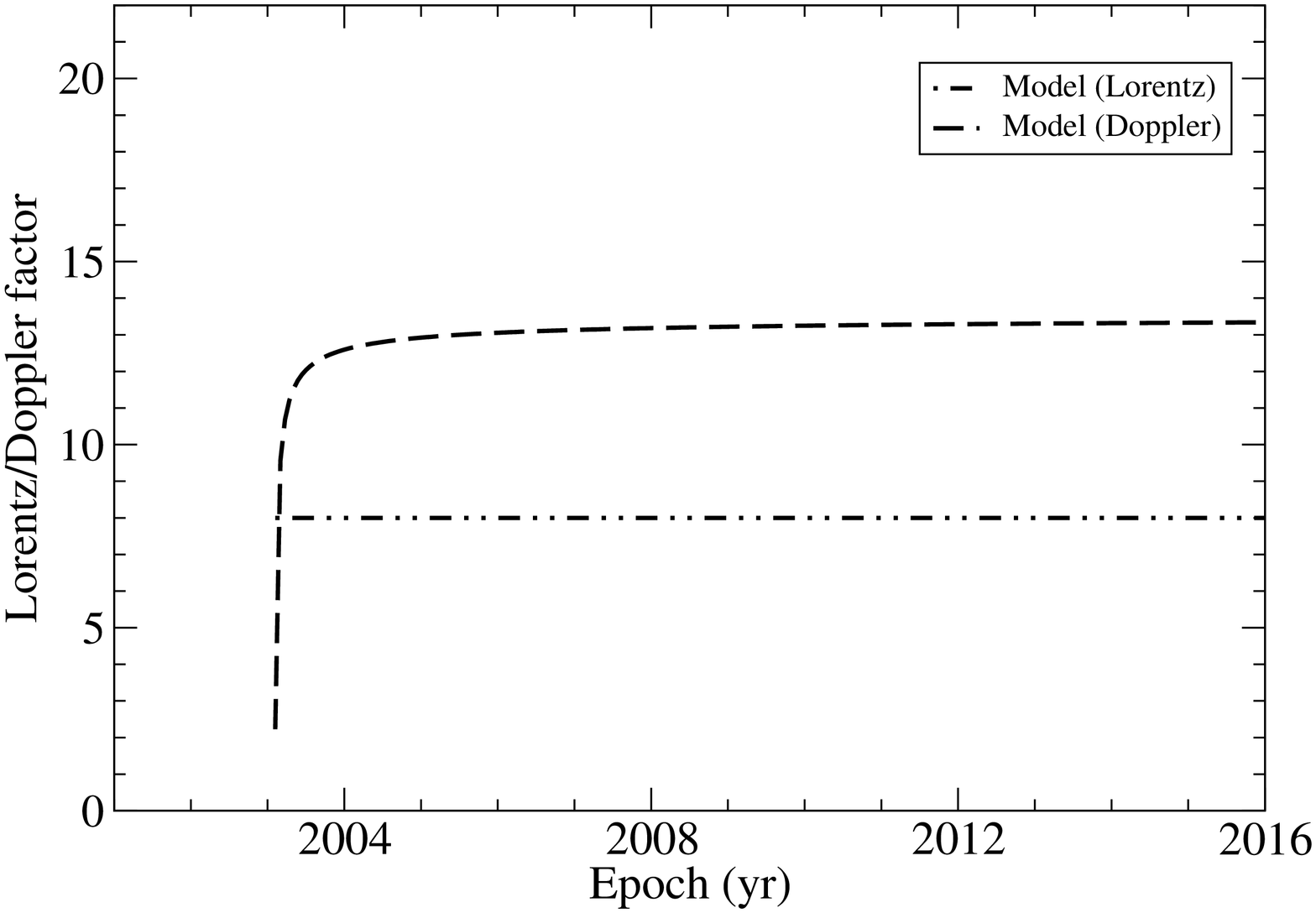}
    \includegraphics[width=5cm,angle=-90]{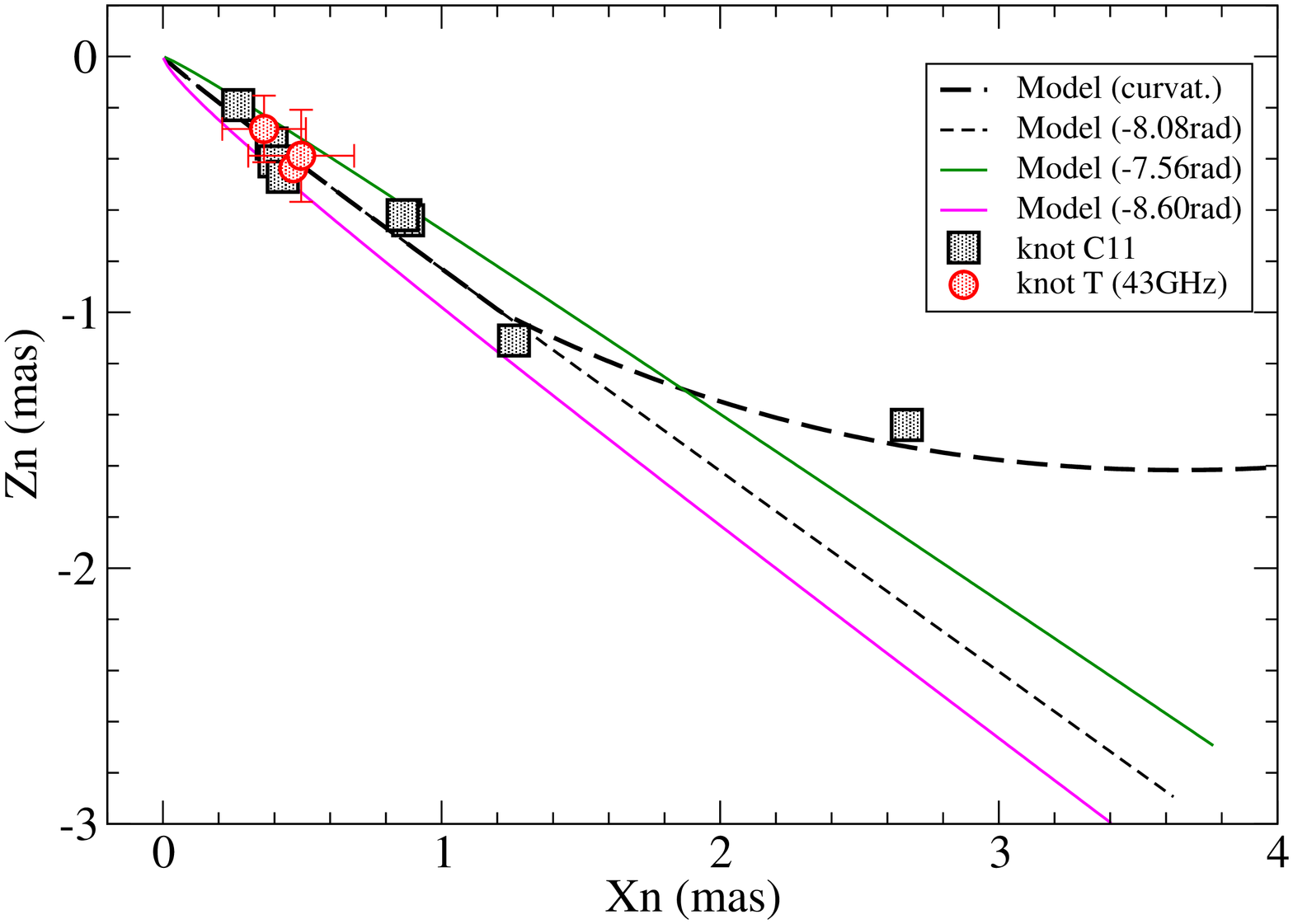}
    \caption{Model-fitting results of the kinematic features for knot C11. The
    entire modeled trajectory is denoted by the black dashed line in top left
    panel. The green and blue lines represent the modeled trajectories 
    calculated for precession phases $\omega$=--8.08$\pm$0.63\,rad, showing all
    the data points being within the position angle range defined by the two 
    lines and indicating the precession period having been determined within
    an uncertainty of $\sim{\pm}$1.2\,yr. The 43\,GHz data given in Agudo 
    et al. (\cite{Ag12} for knot-T) are also well fitted by the model. The 
    green and blue lines in bottom right panel represent the precessing common
    trajectories calculated for precessing phases $\omega{\pm}$0.52\,rad, 
    showing most of the data points being within the position angle range
    defined by the two lines and indicating its innermost common parabolic 
    trajectory having been observed. Thus knot C11 is designated by symbol
    ``+'' in Table 3.}
    \end{figure*}

    \begin{figure*}
    \centering
    \includegraphics[width=5cm,angle=-90]{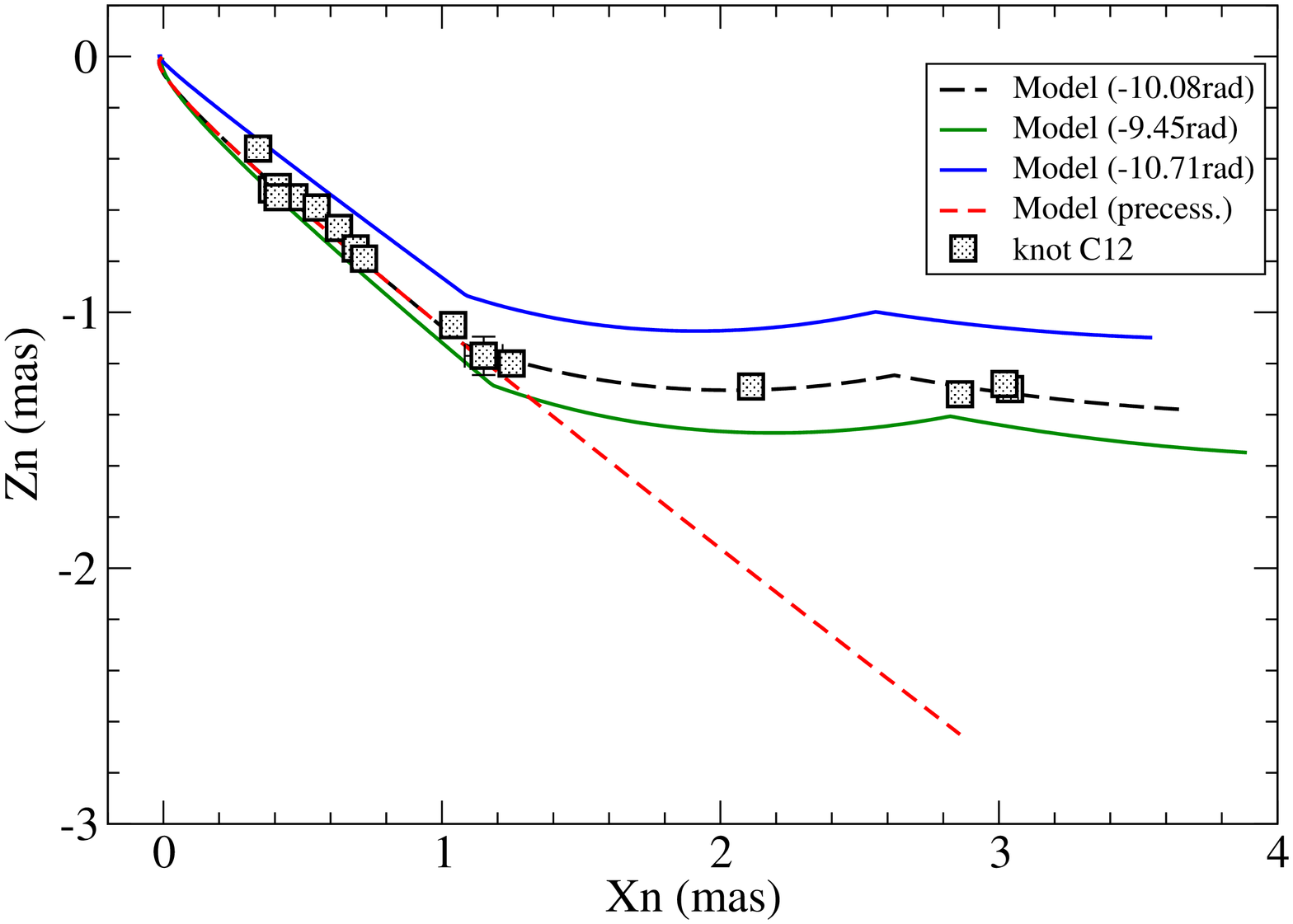}
    \includegraphics[width=5cm,angle=-90]{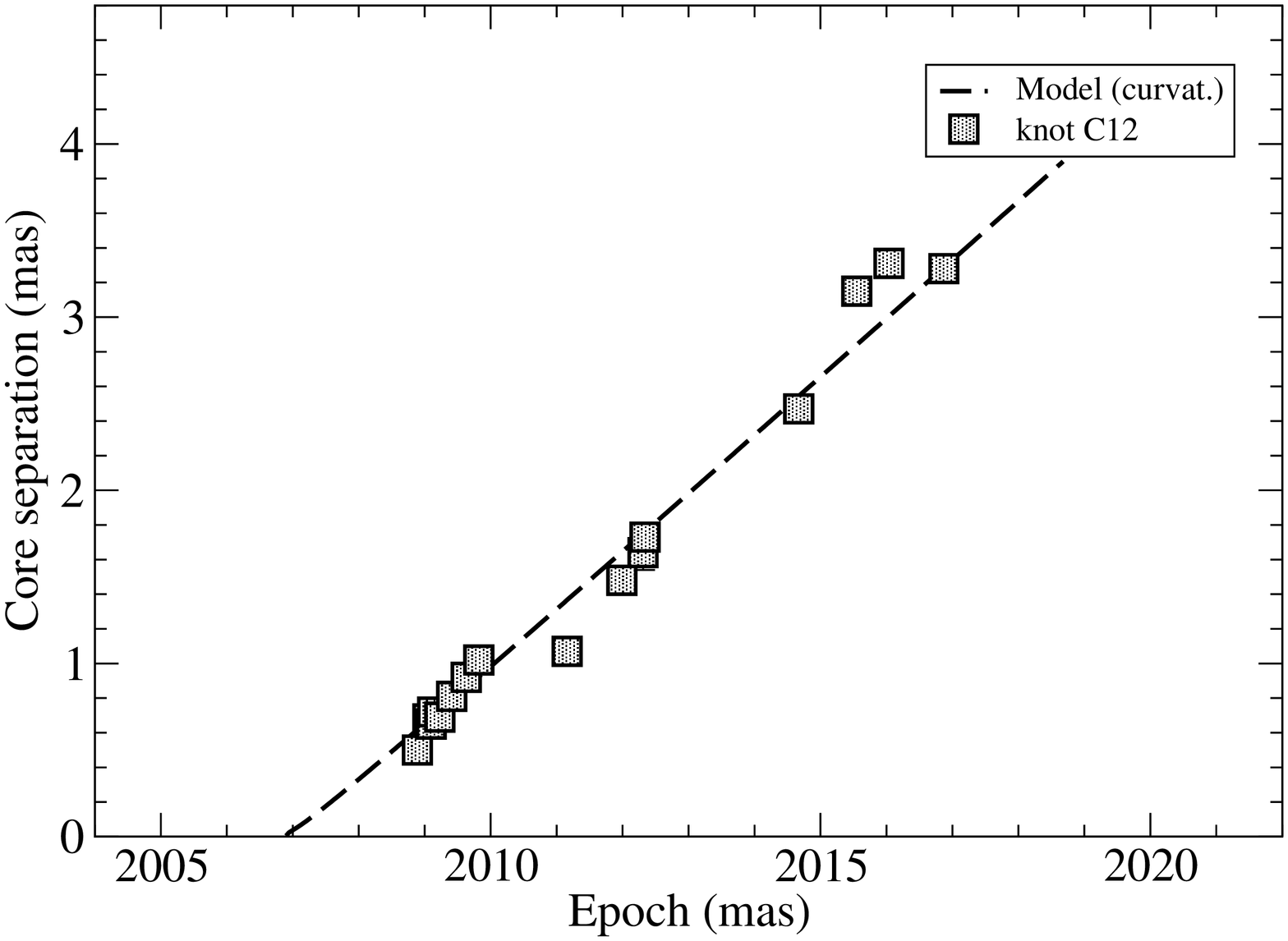}
    \includegraphics[width=5cm,angle=-90]{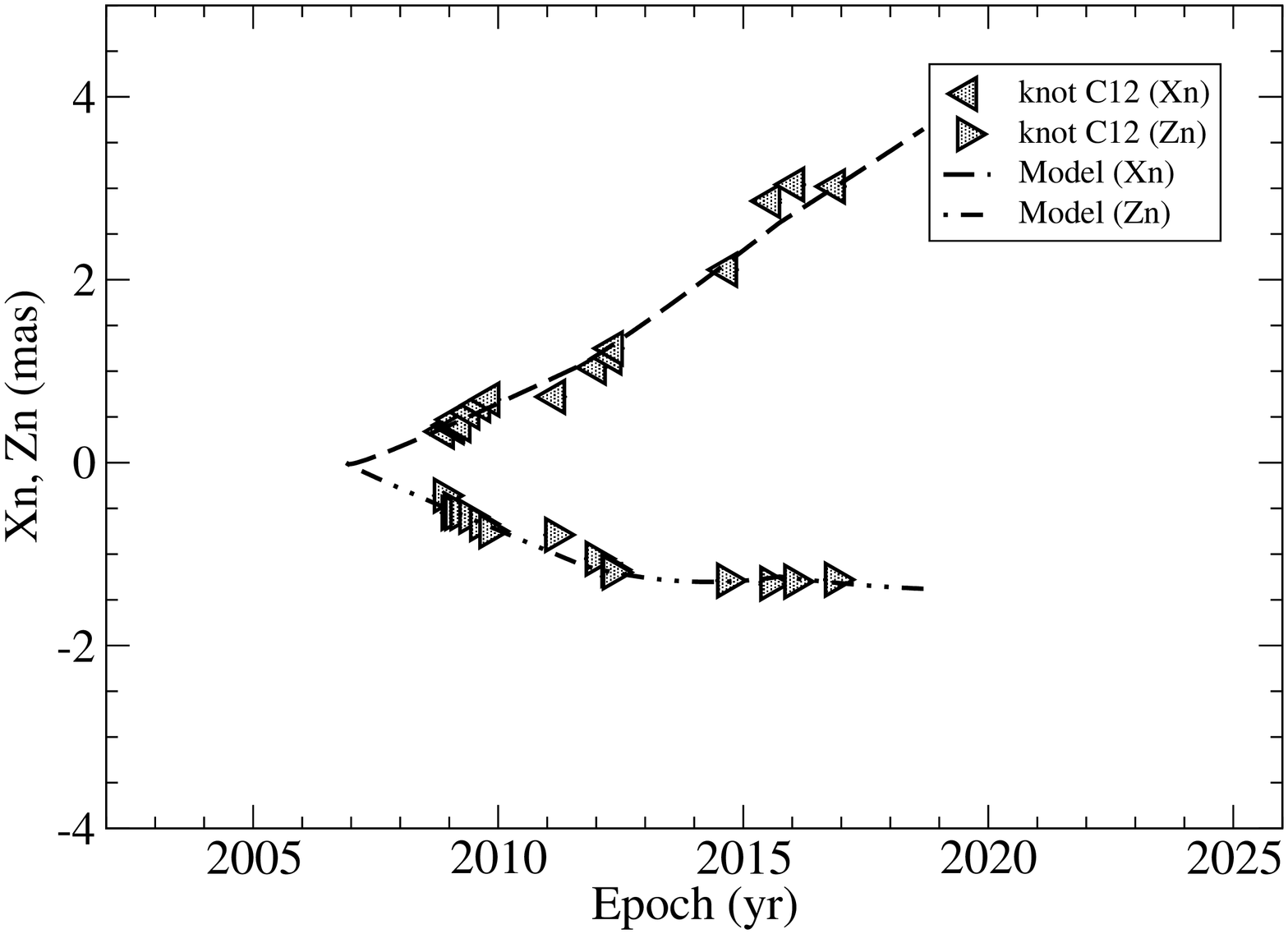}
    \includegraphics[width=5cm,angle=-90]{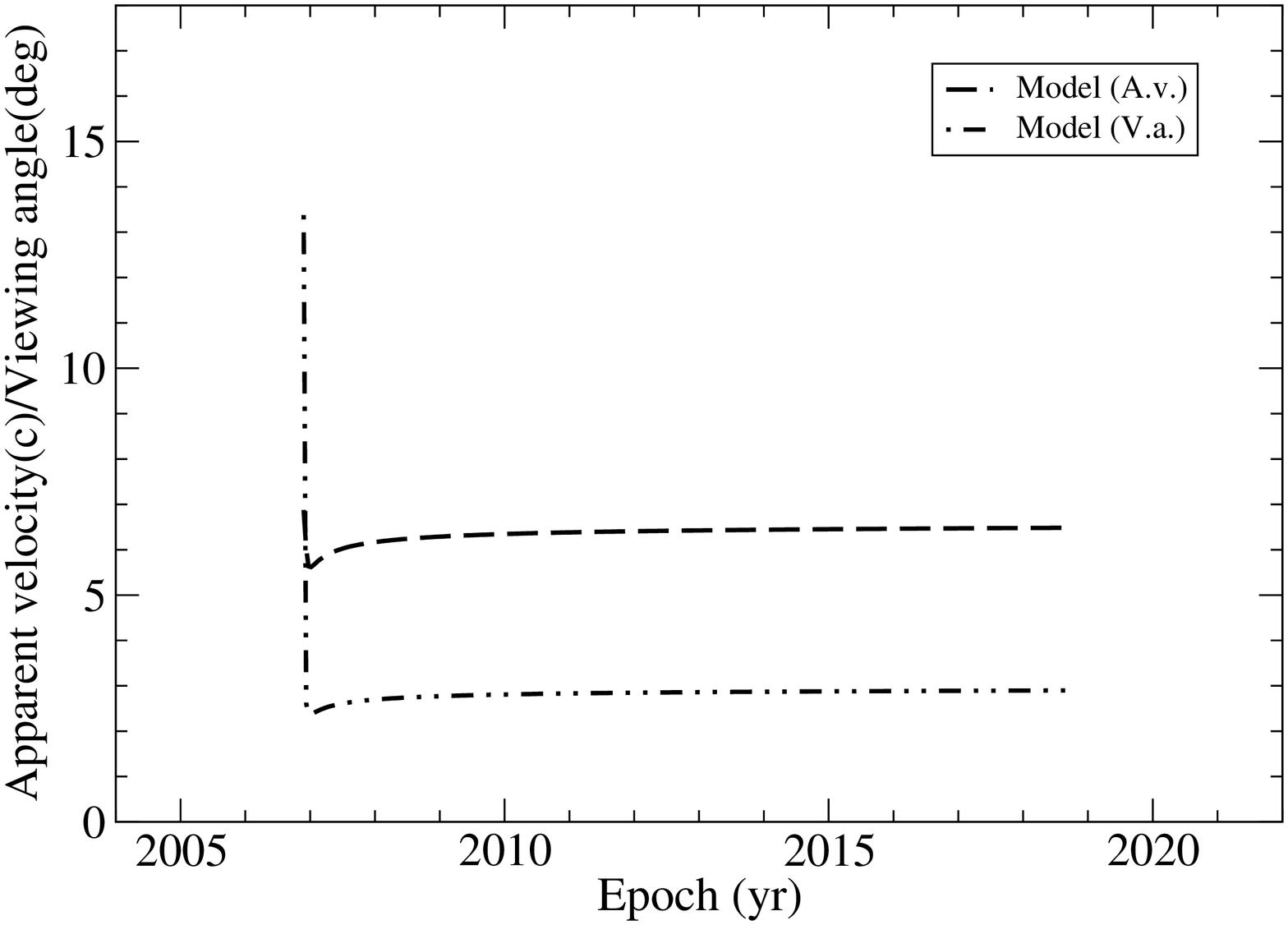}
    \includegraphics[width=5cm,angle=-90]{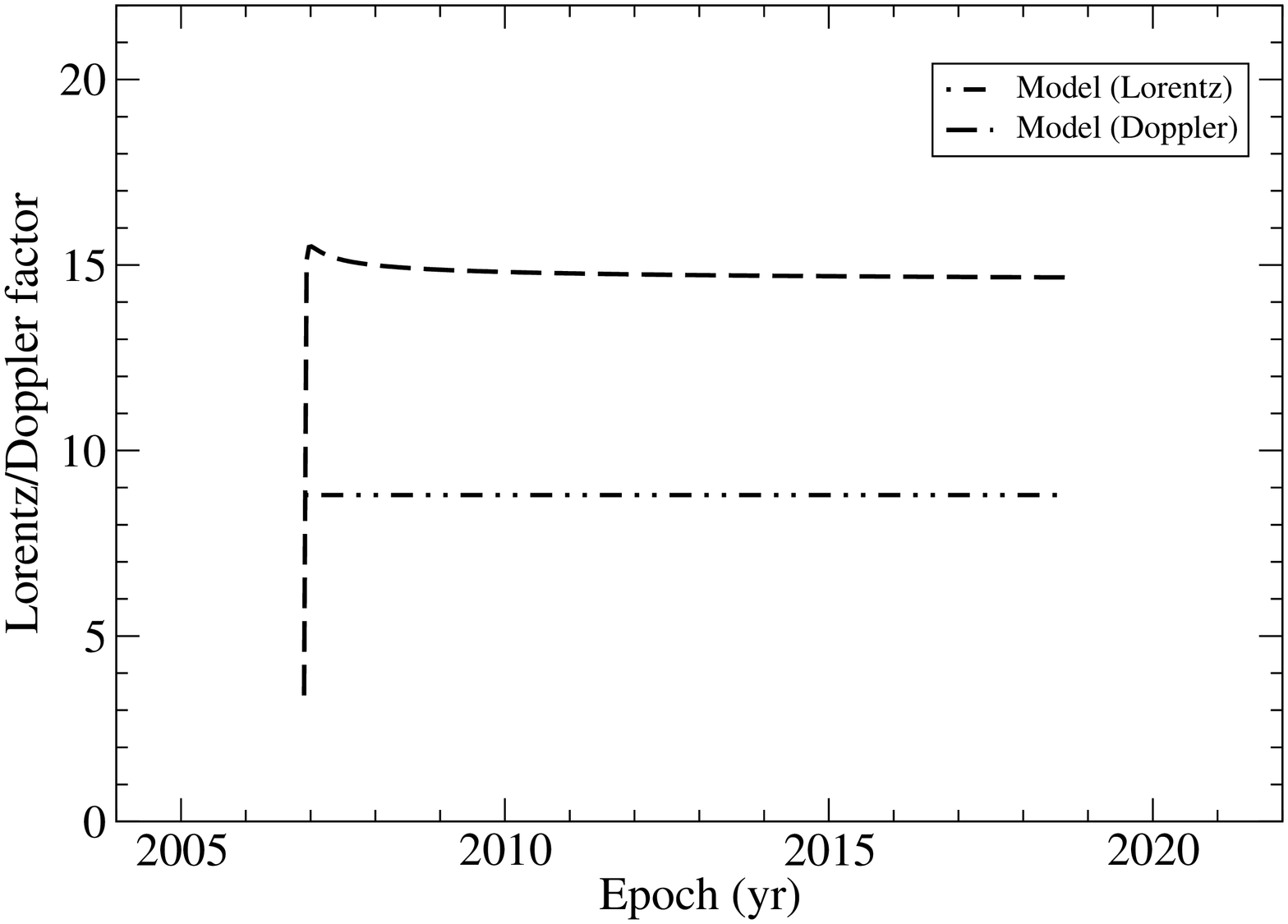}
    \includegraphics[width=5cm,angle=-90]{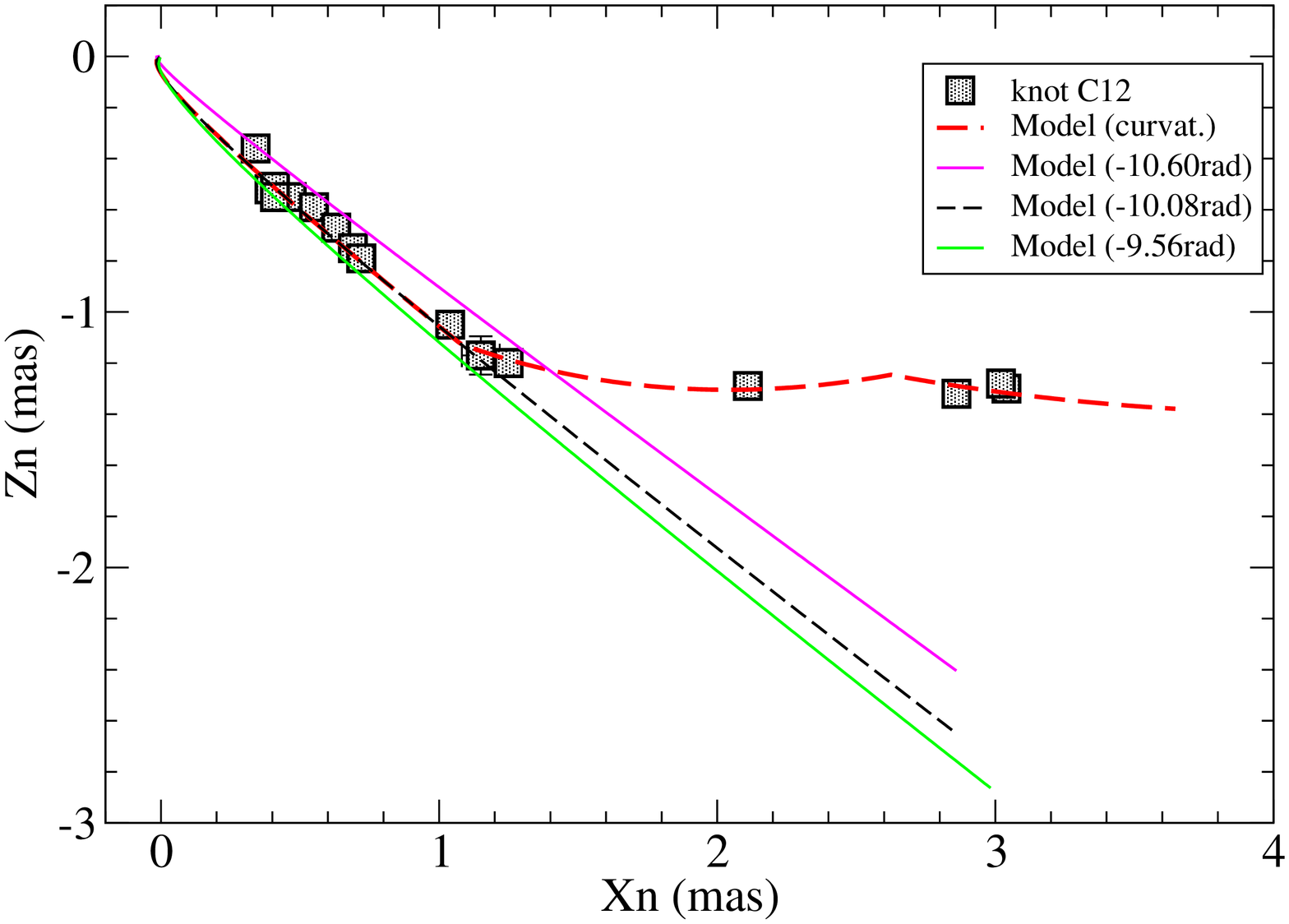}
    \caption{Model-fitting results of the kinematic features for knot C12.
   The entire modeled trajectory is denoted by the black dashed line in top 
   left panel. The green and blue lines represent the modeled  trajectories
   calculated for precession phases $\omega$=--10.08$\pm$0.63\,rad, showing all
   the data points being within the position angle range defined by the
   two lines and indicating the precession period having been determined within
   an uncertainty of $\sim{\pm}$1.2\,yr. The green and 
   blue lines  in bottom right panel represent the precessing common 
  trajectories calculated for precession phases $\omega{\pm}$0.52\,rad, showing
   most of the data points being within the position angle range defined by 
  the two lines and indicating its innermost precessing common parabolic 
  trajectory having been observed. Thus knot C12 is designated by symbol ``+''
   in Table 3.}
    \end{figure*}

    \begin{figure*}
   \centering
    \includegraphics[width=5cm,angle=-90]{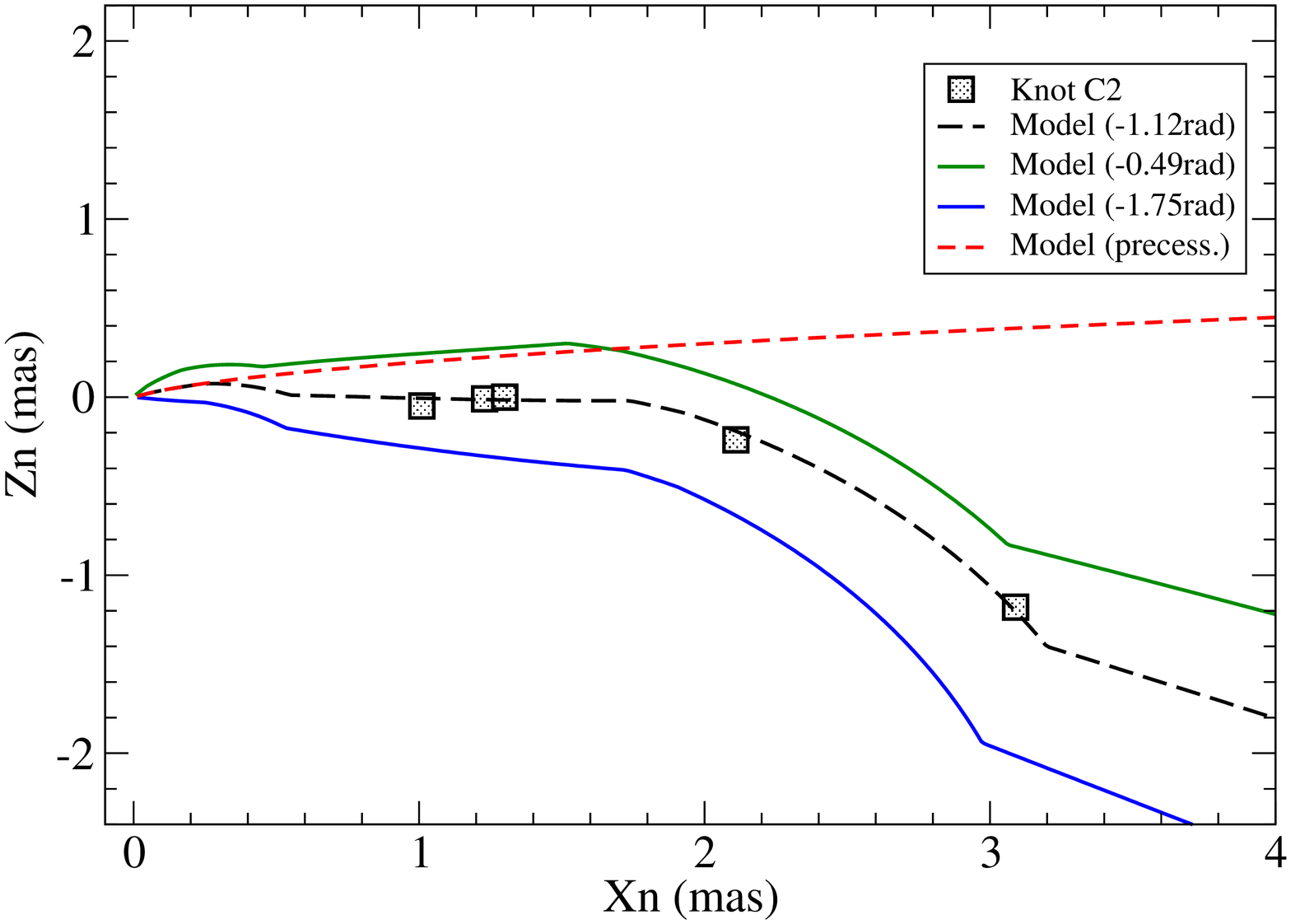}
    \includegraphics[width=5cm,angle=-90]{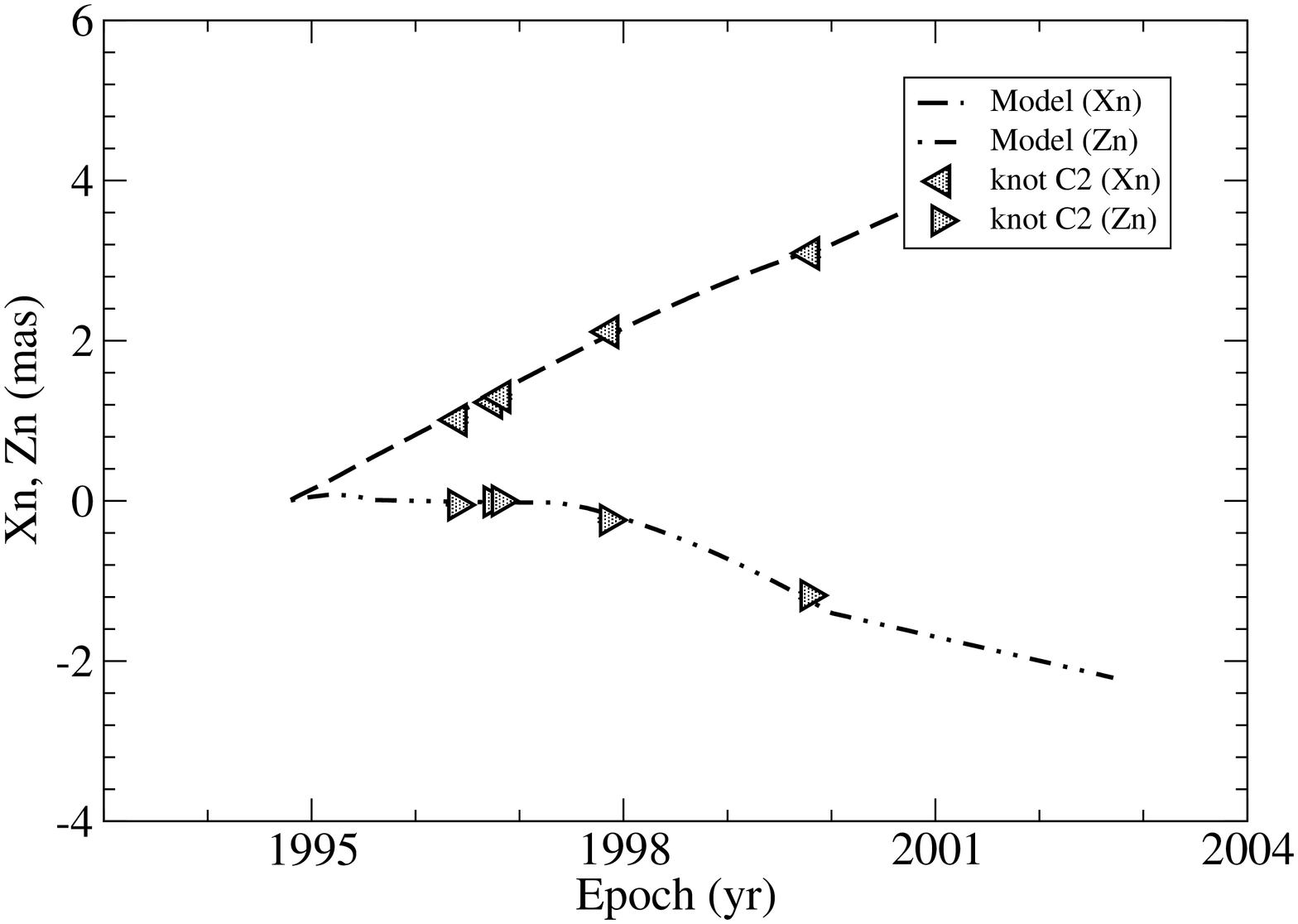}
    \includegraphics[width=5cm,angle=-90]{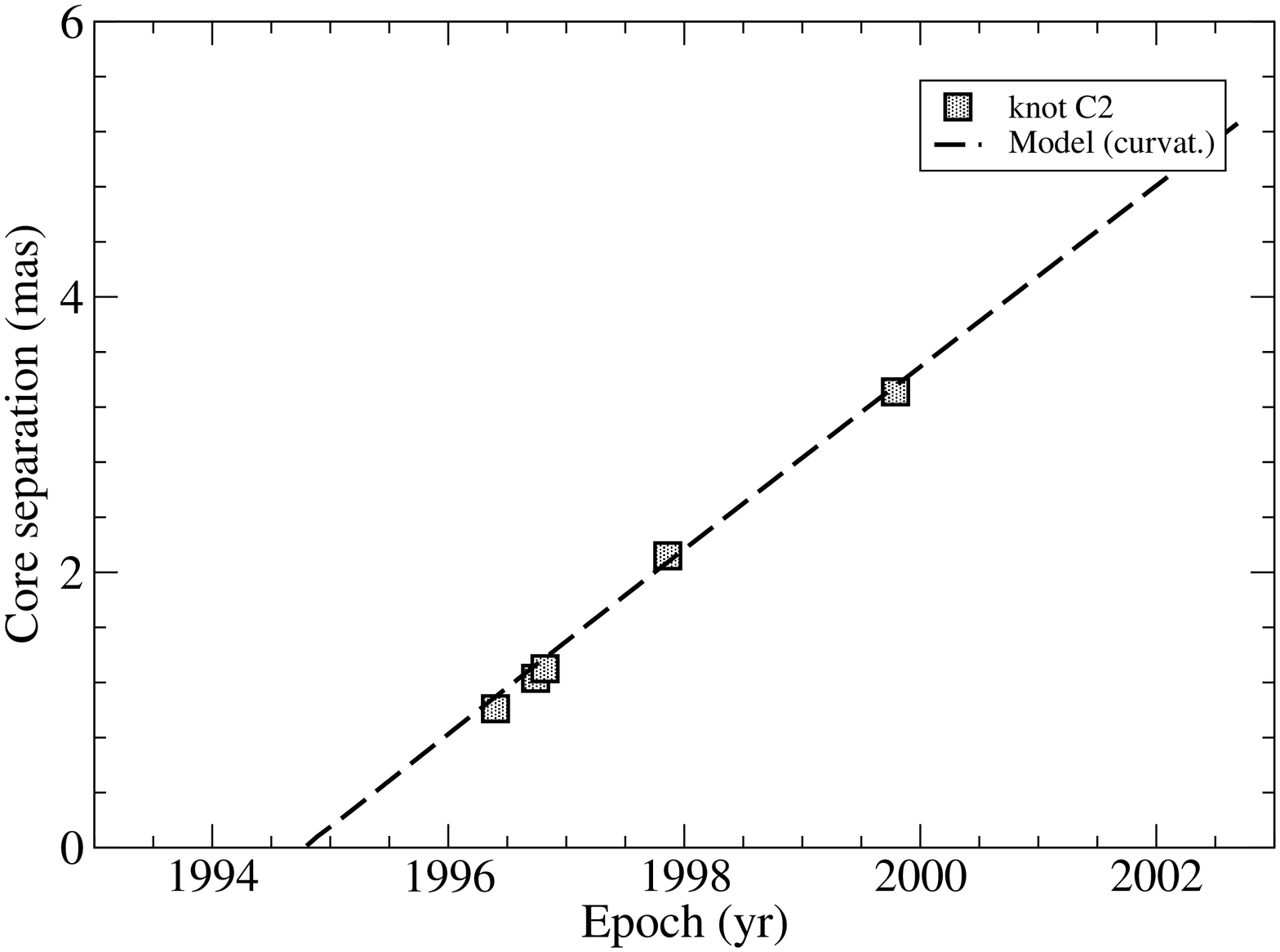}
    \includegraphics[width=5cm,angle=-90]{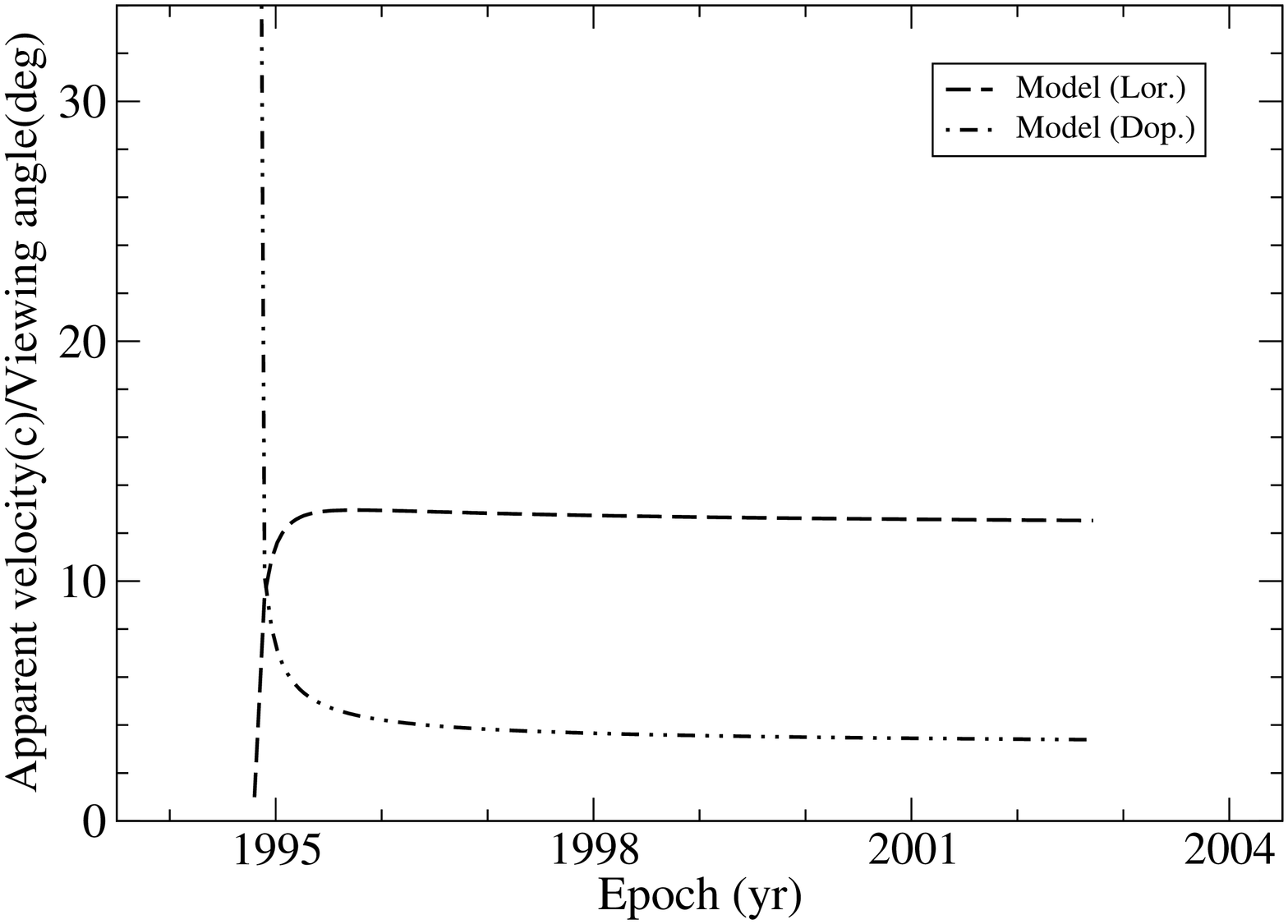}
    \includegraphics[width=5cm,angle=-90]{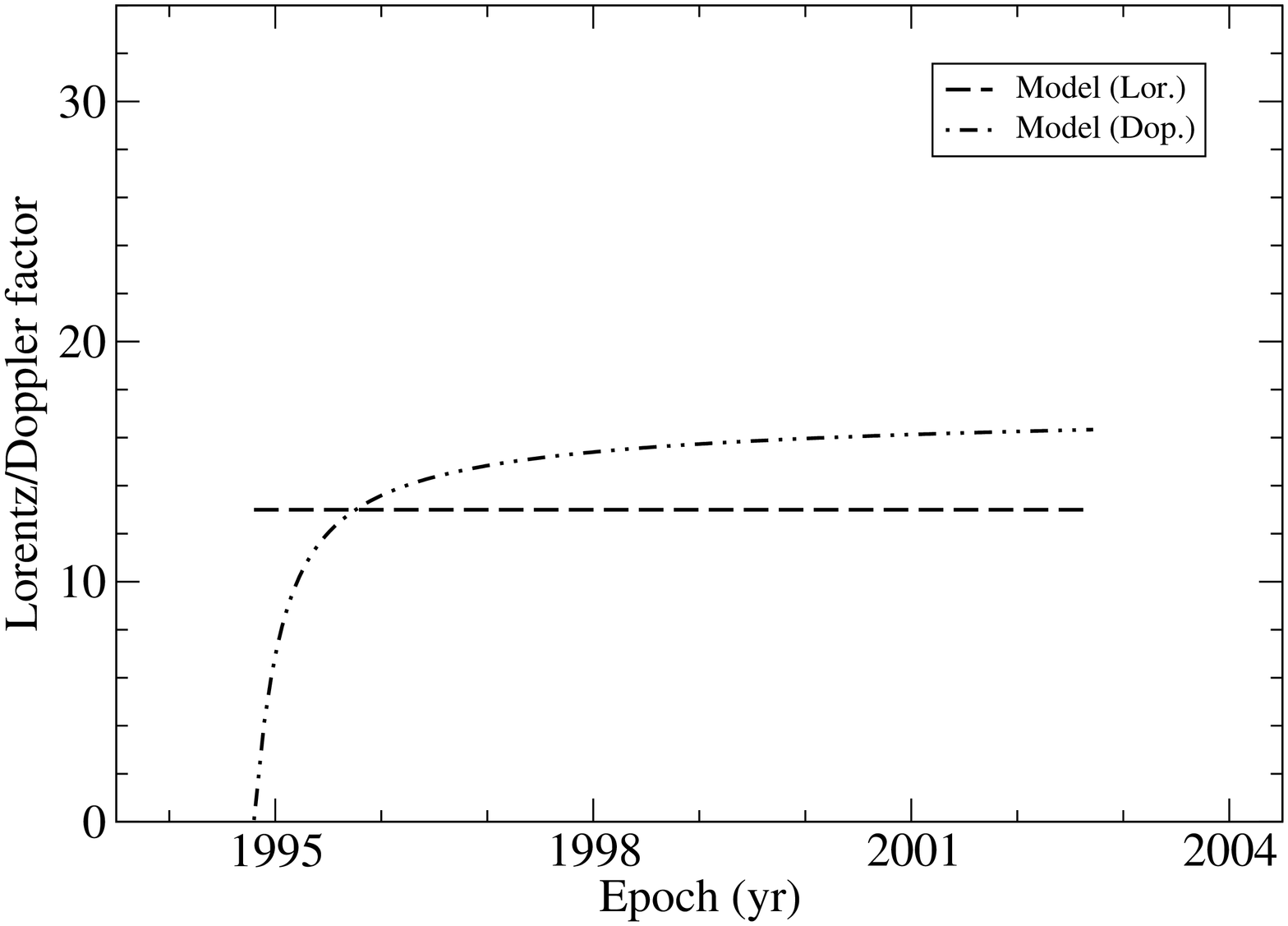}
    \includegraphics[width=5cm,angle=-90]{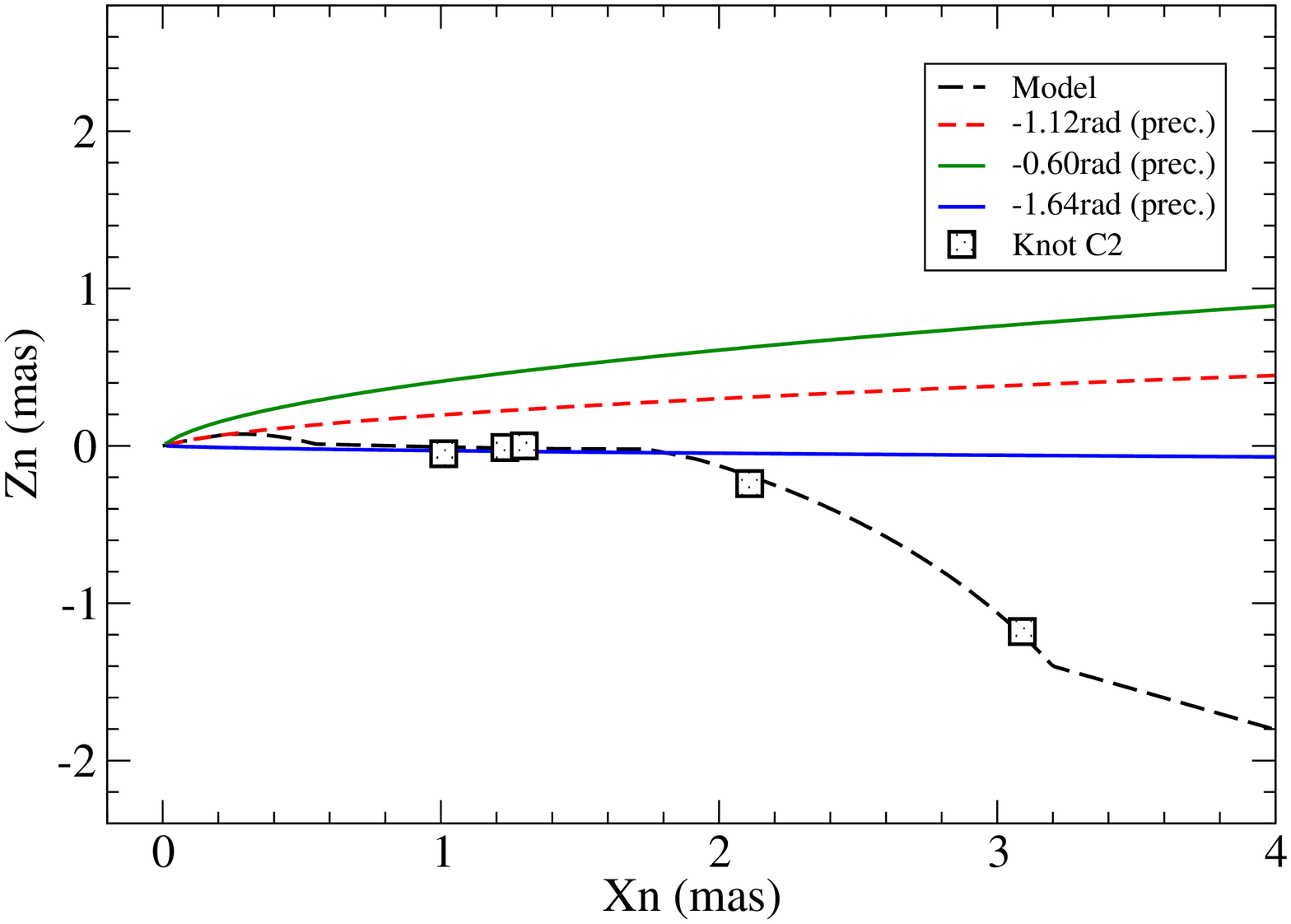}
    \caption{Model-fitting results of the kinematic features for knot C2.
    Within core separation $r_n$=0.3\,mas, its motion is modeled to follow the
    precessing  common parabolic trajectory (red dashed line in top left panel.
    The entire modeled trajectory is shown by the black dashed line. The green 
    and blue lines represent the modeled trajectories calculated for precession
    phases $\omega{\pm}$0.63\,rad, showing  most of the data points within
    the position angle range defined by the two lines and  the 
    precession period having been determined within an uncertainty of 
    $\sim{\pm}$1.2\,yr. In bottom right panel, the green and blue lines 
    represent the precessing common trajectories calculated for 
    $\omega{\pm}$0.52\,rad, showing no data points within the position angle 
    range defined by the two lines and its innermost precessing common 
    trajectory having not benn observed (no observation data available). Thus
    knot C2 is designated  by symbol ``--'' in Table 6.}
    \end{figure*}

    \begin{figure*}
    \centering
    \includegraphics[width=5cm,angle=-90]{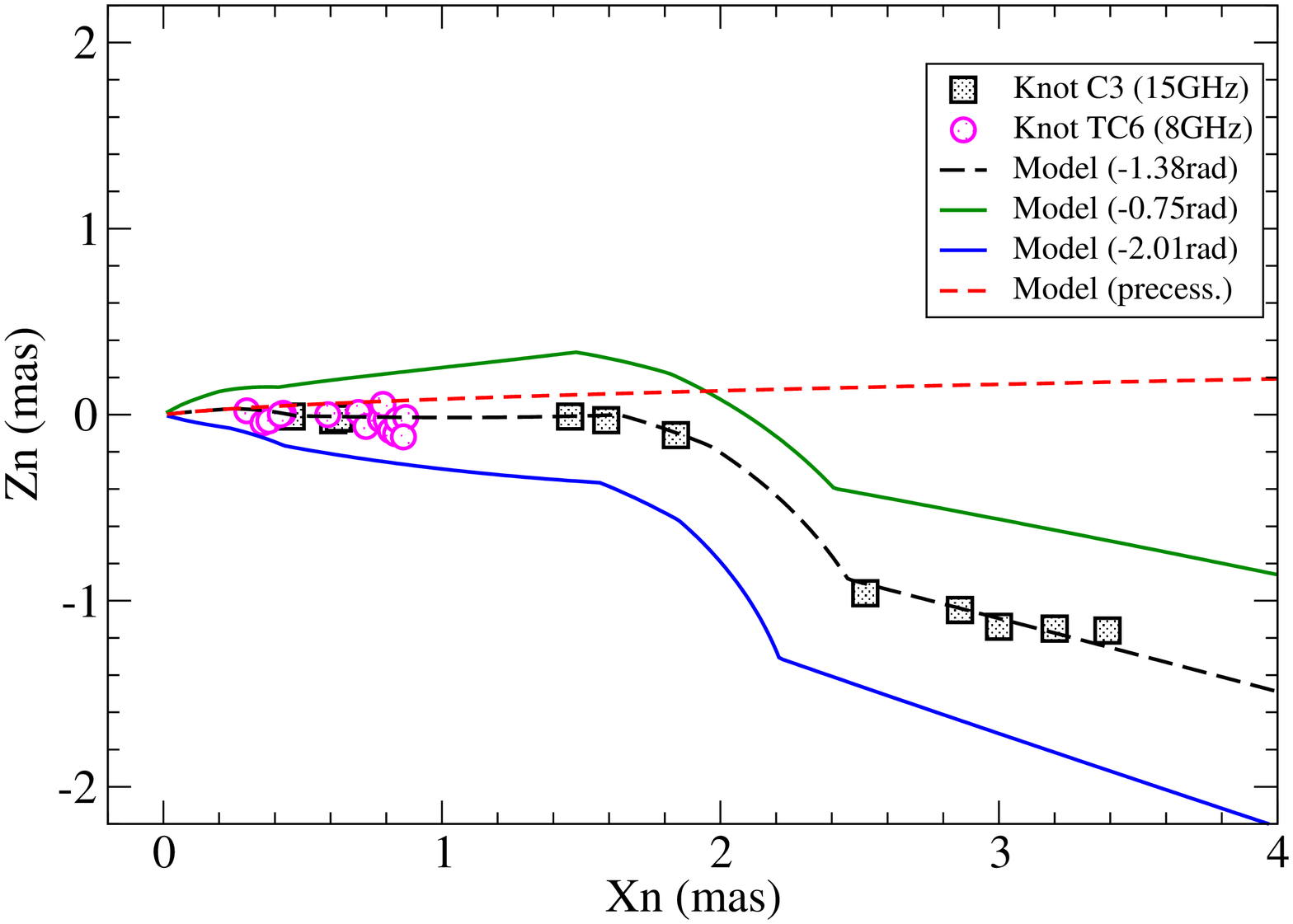}
    \includegraphics[width=5cm,angle=-90]{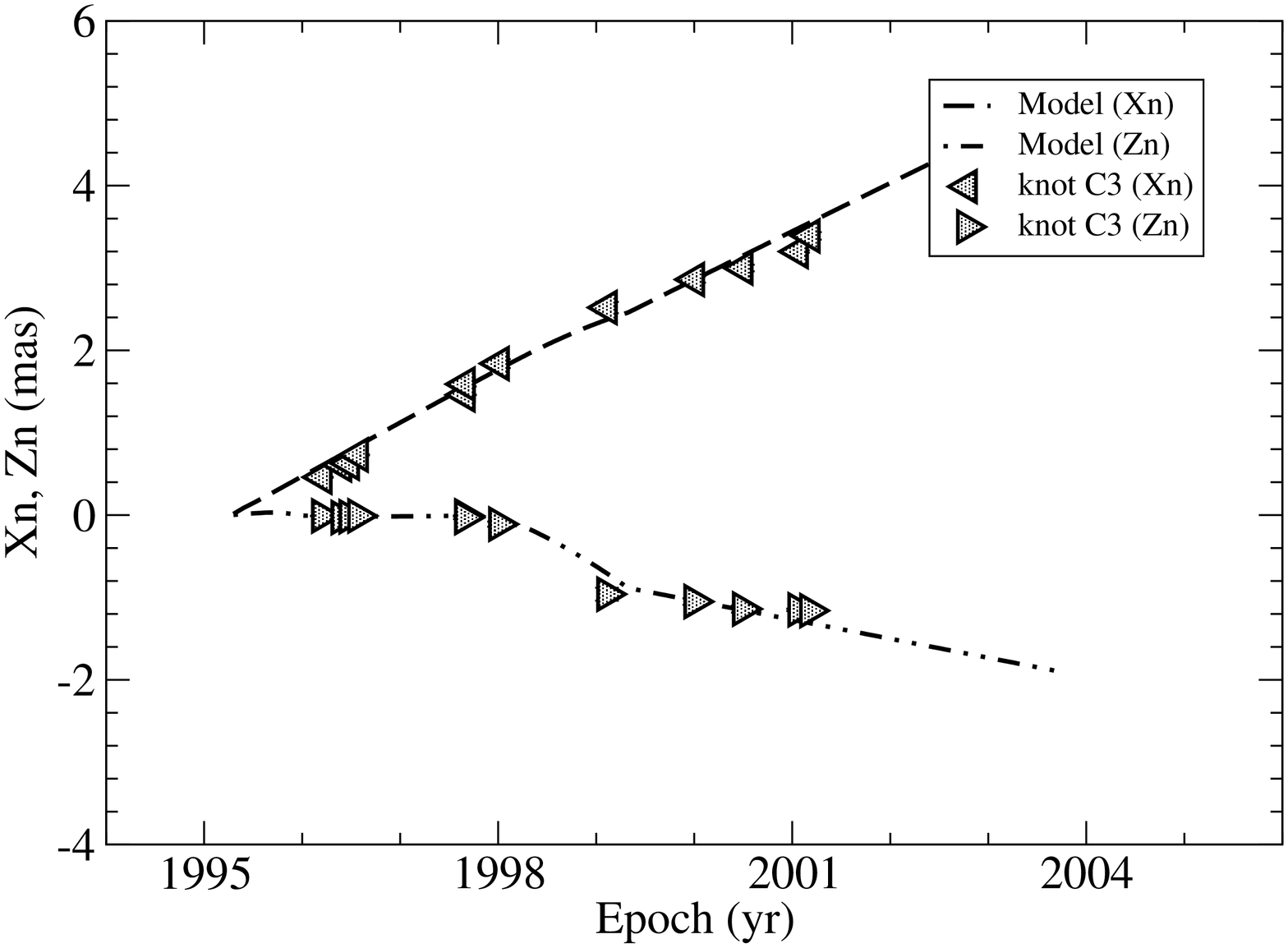}
    \includegraphics[width=5cm,angle=-90]{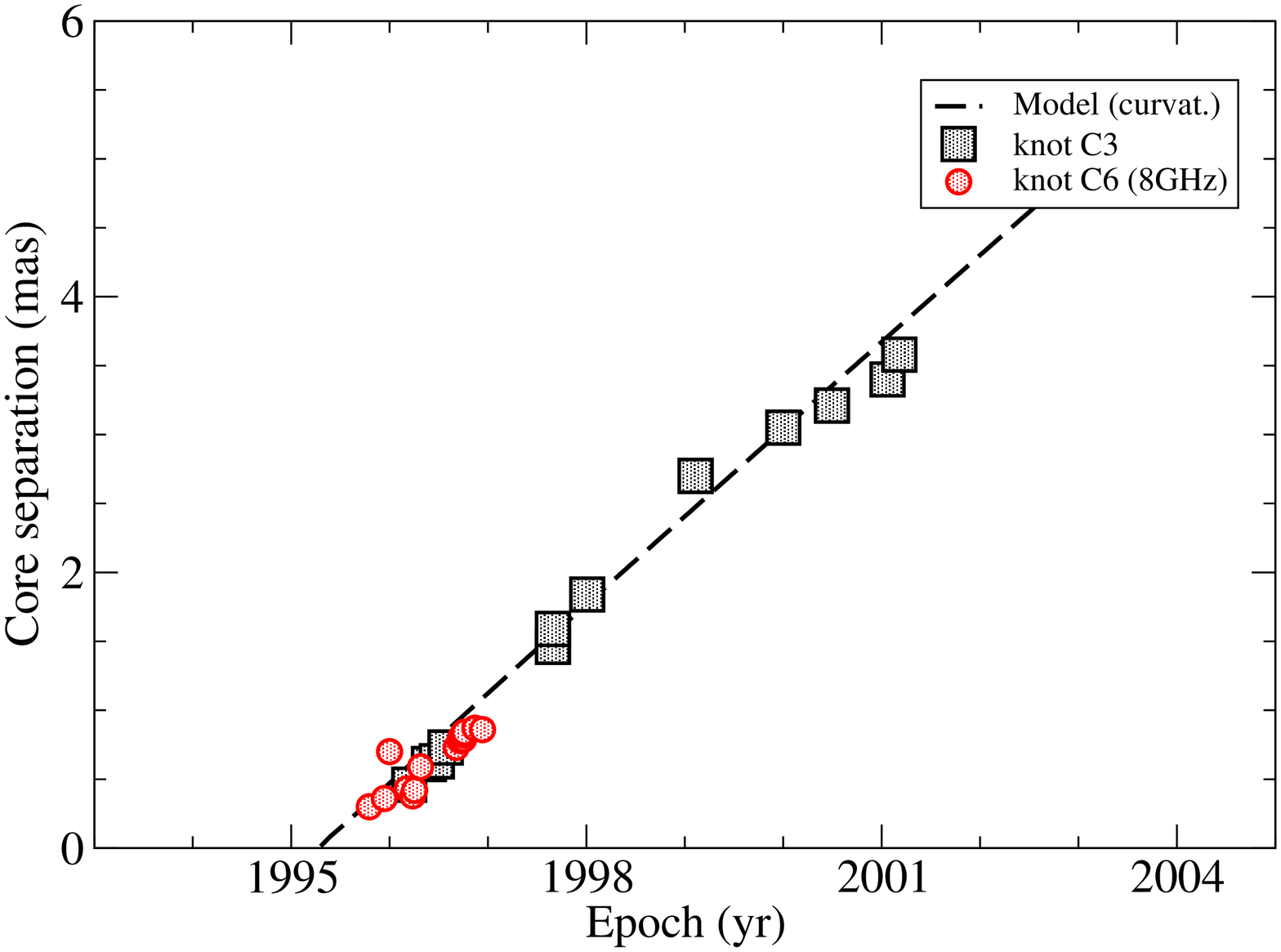}
    \includegraphics[width=5cm,angle=-90]{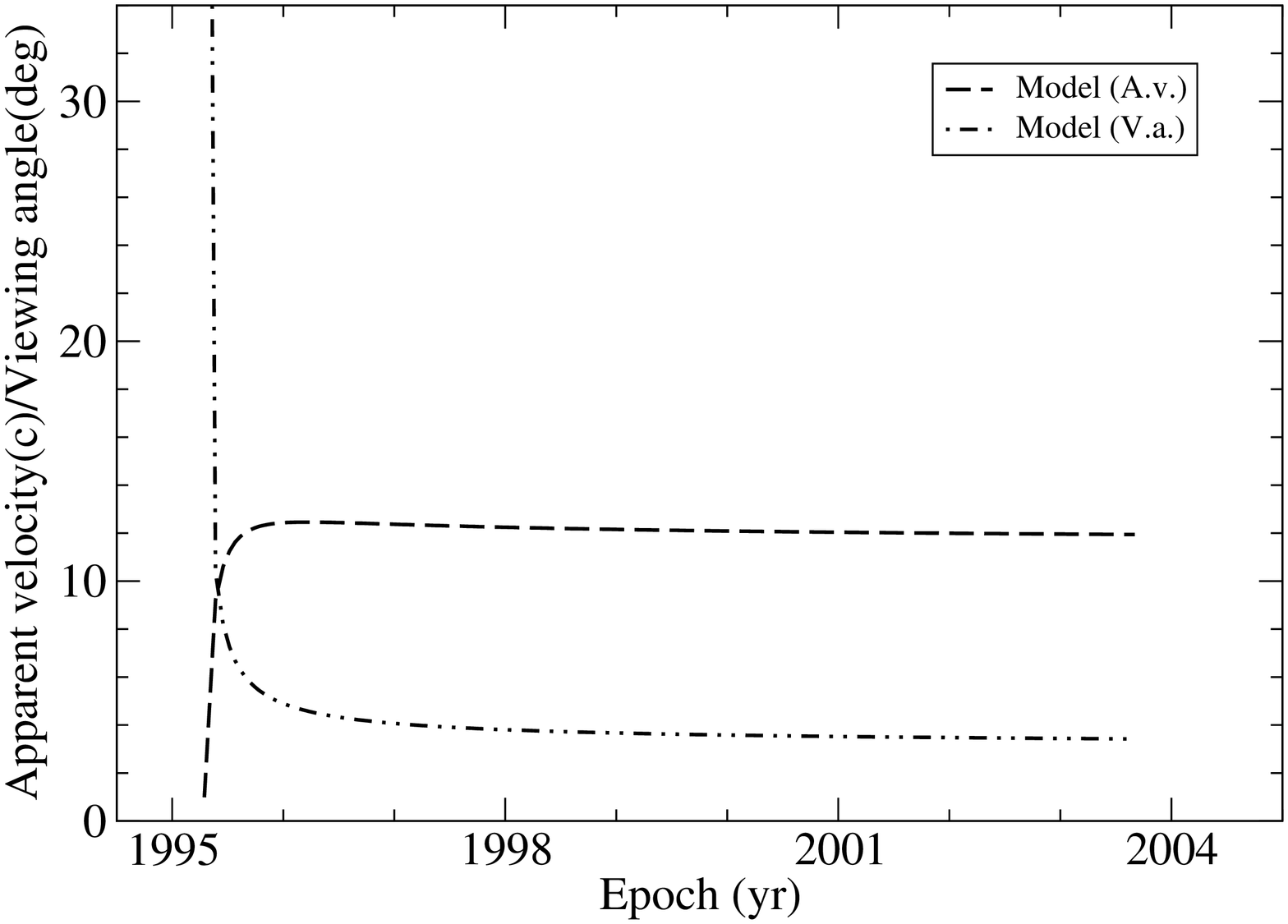}
    \includegraphics[width=5cm,angle=-90]{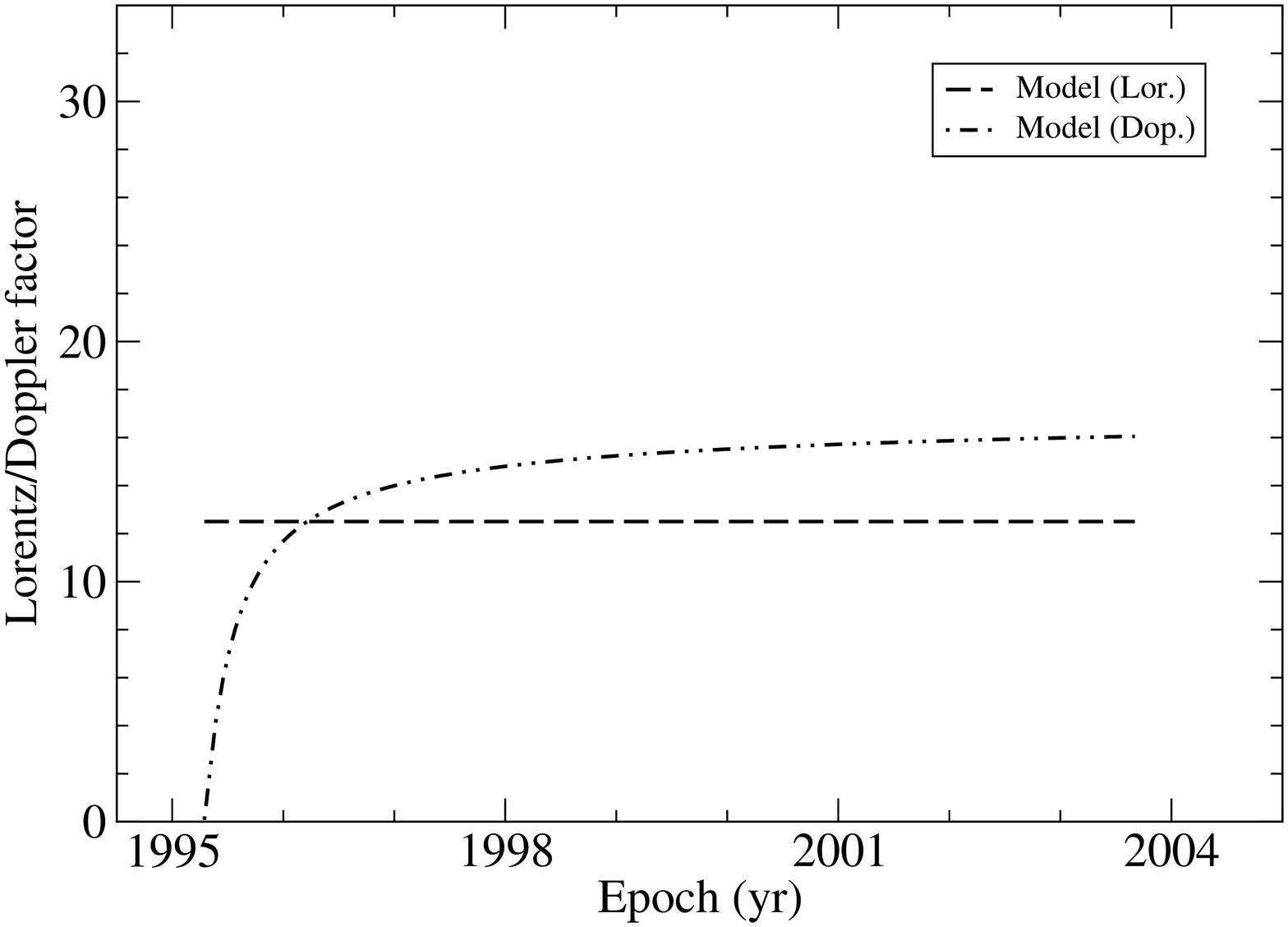}
    \includegraphics[width=5cm,angle=-90]{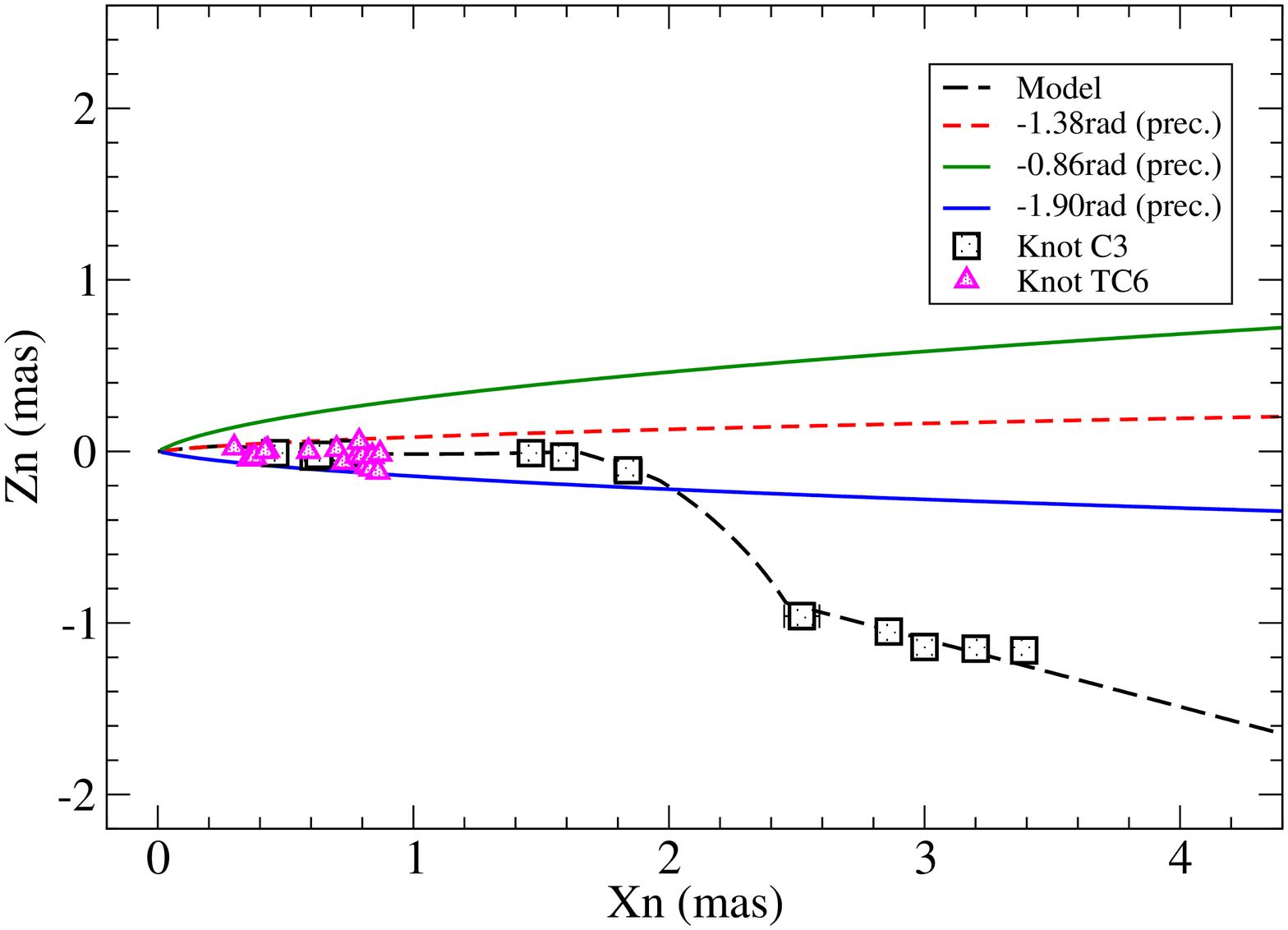}
    \caption{Model-fitting results of the kinematic features for knot C3.
   Its motion was modeled  to follow the precessing common parabolic trajectory
   within core separation $r_n$=0.25\,mas.  The entire
   trajectory is shown by the black dashed line in top left panel. The 
   green and blue lines in bottom right panel represent  the precessing 
   common trajectories calculated for ${\omega}\pm$0.52\,rad, showing  a number 
   of data points within the position angle range defined by te two lines and 
   indicating its innermost precessing parabolic trajectory having been 
   observed. Thus knot C3 is designated by symbol ``+'' in Table 6. The 8\,GHz 
   data given in Tateyama et al. (\cite{Ta99} for knot-C6, designated as 
   knot-TC6 here) are well fitted by the model. The ejection epoch 1995.4
   derived by Tateyama et al. is well consistent with our modeled epoch of
   1995.29, providing support for the 12\,yr precession period for the northern
   jet.}
    \end{figure*}

    \begin{figure*}
    \centering
    \includegraphics[width=5cm,angle=-90]{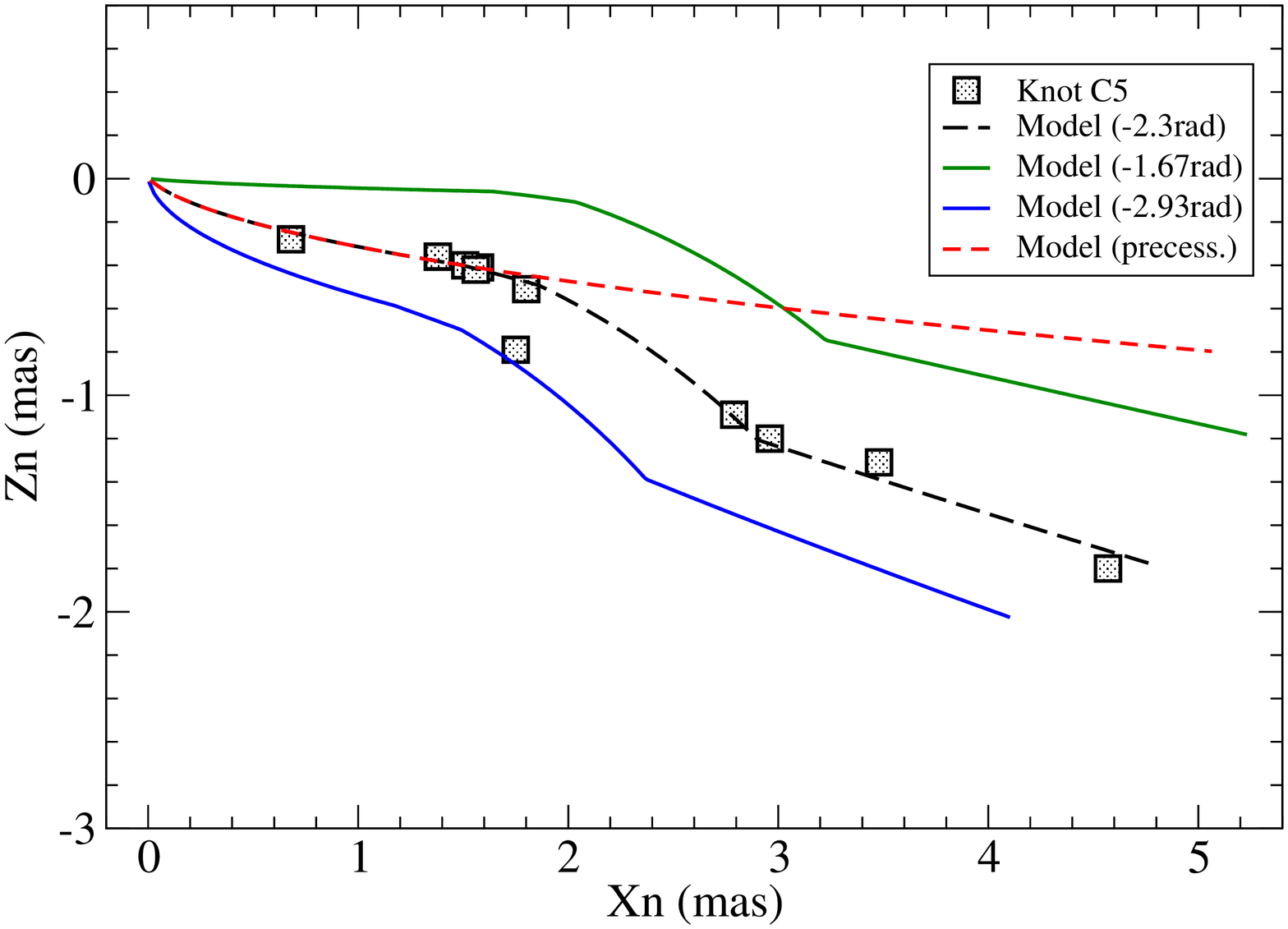}
    \includegraphics[width=5cm,angle=-90]{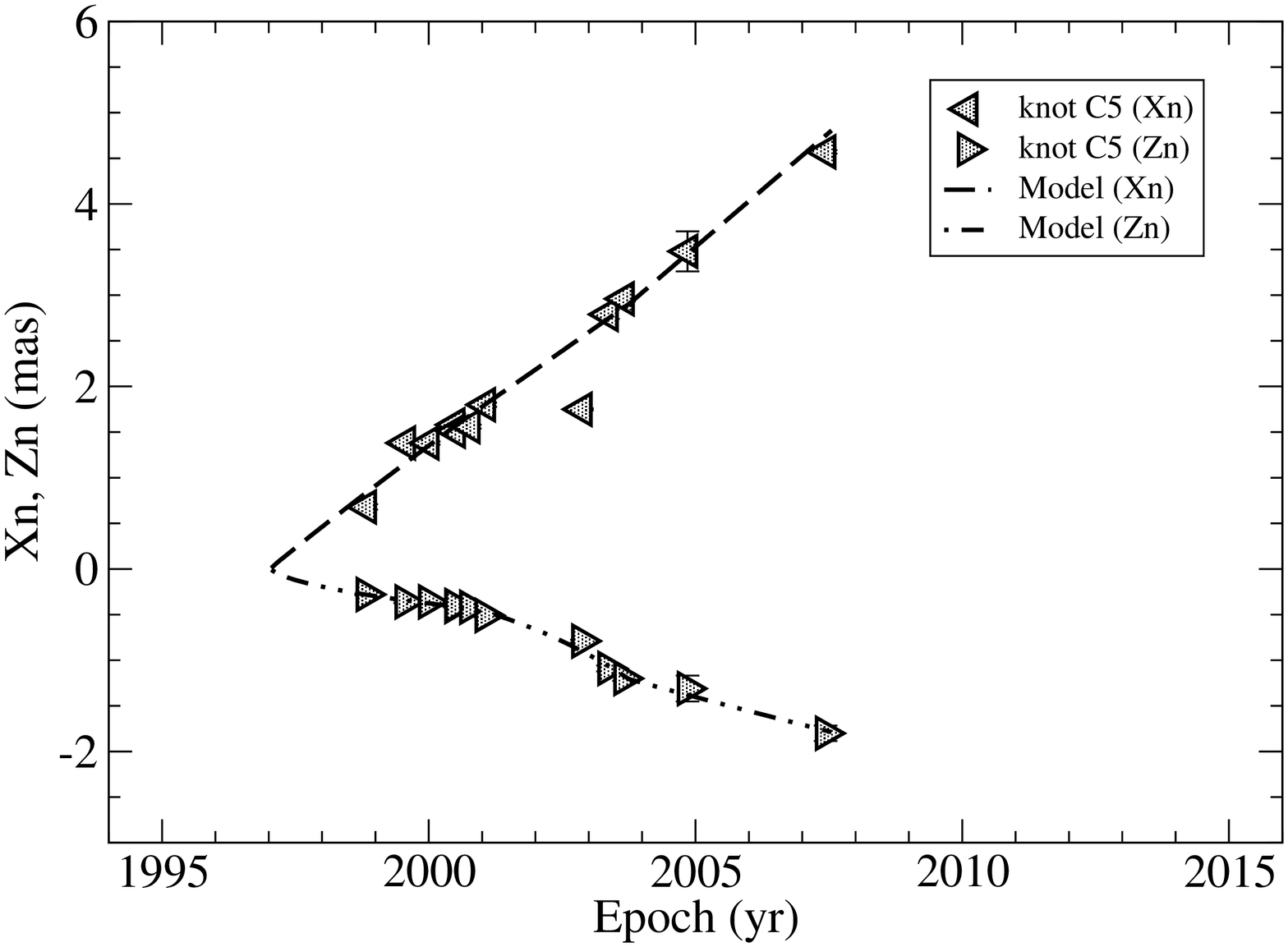}
    \includegraphics[width=5cm,angle=-90]{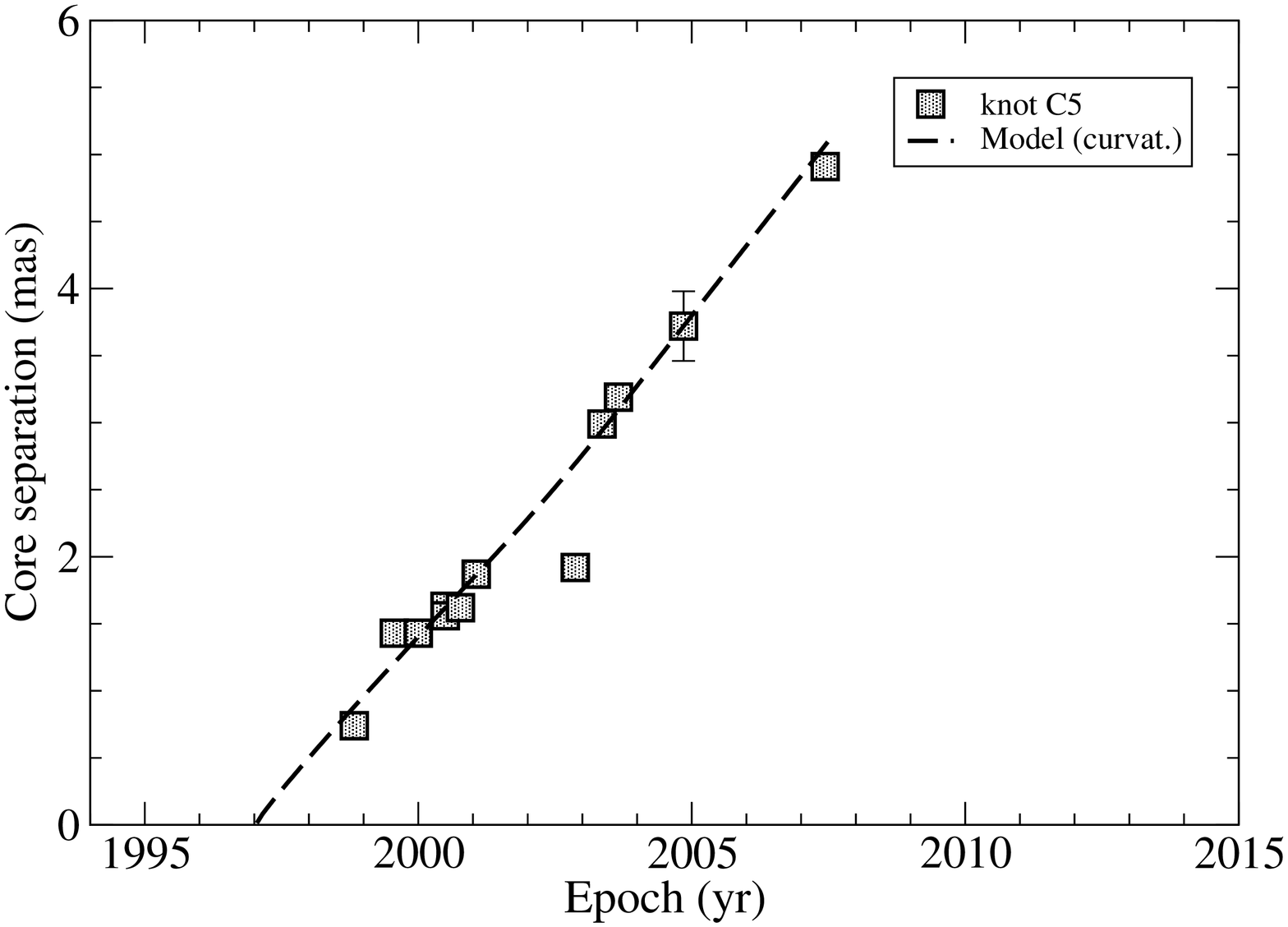}
    \includegraphics[width=5cm,angle=-90]{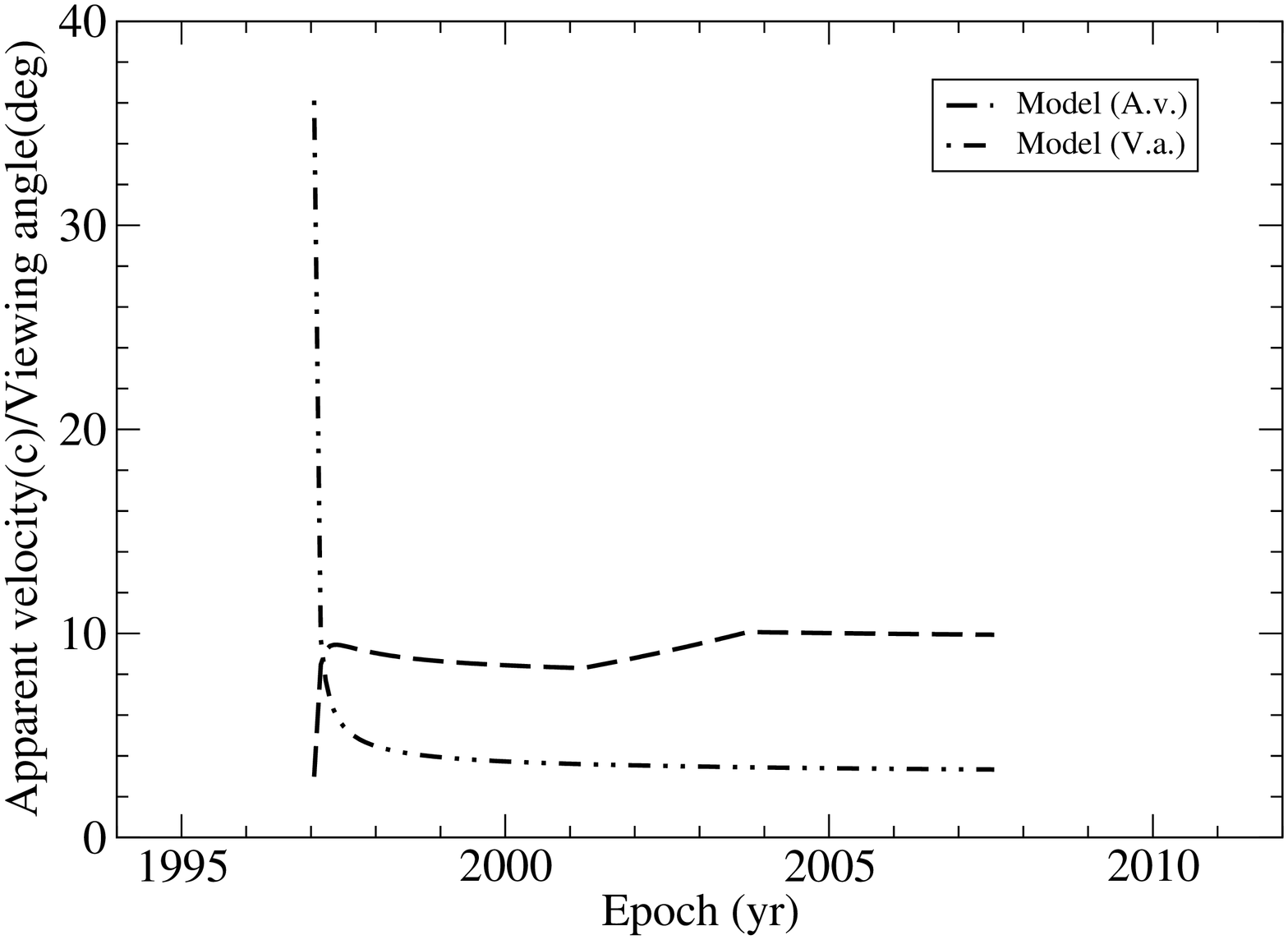}
    \includegraphics[width=5cm,angle=-90]{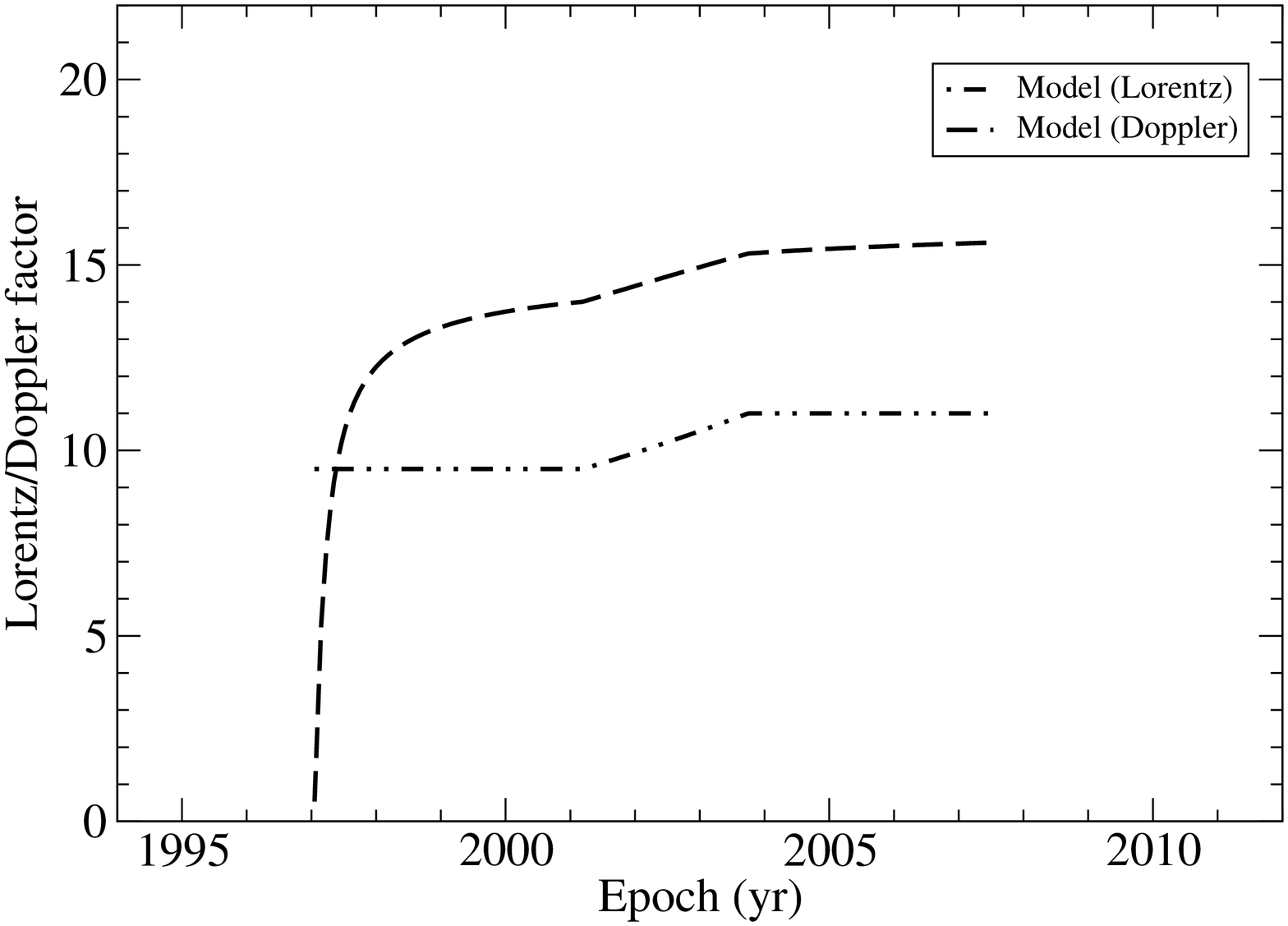}
    \includegraphics[width=5cm,angle=-90]{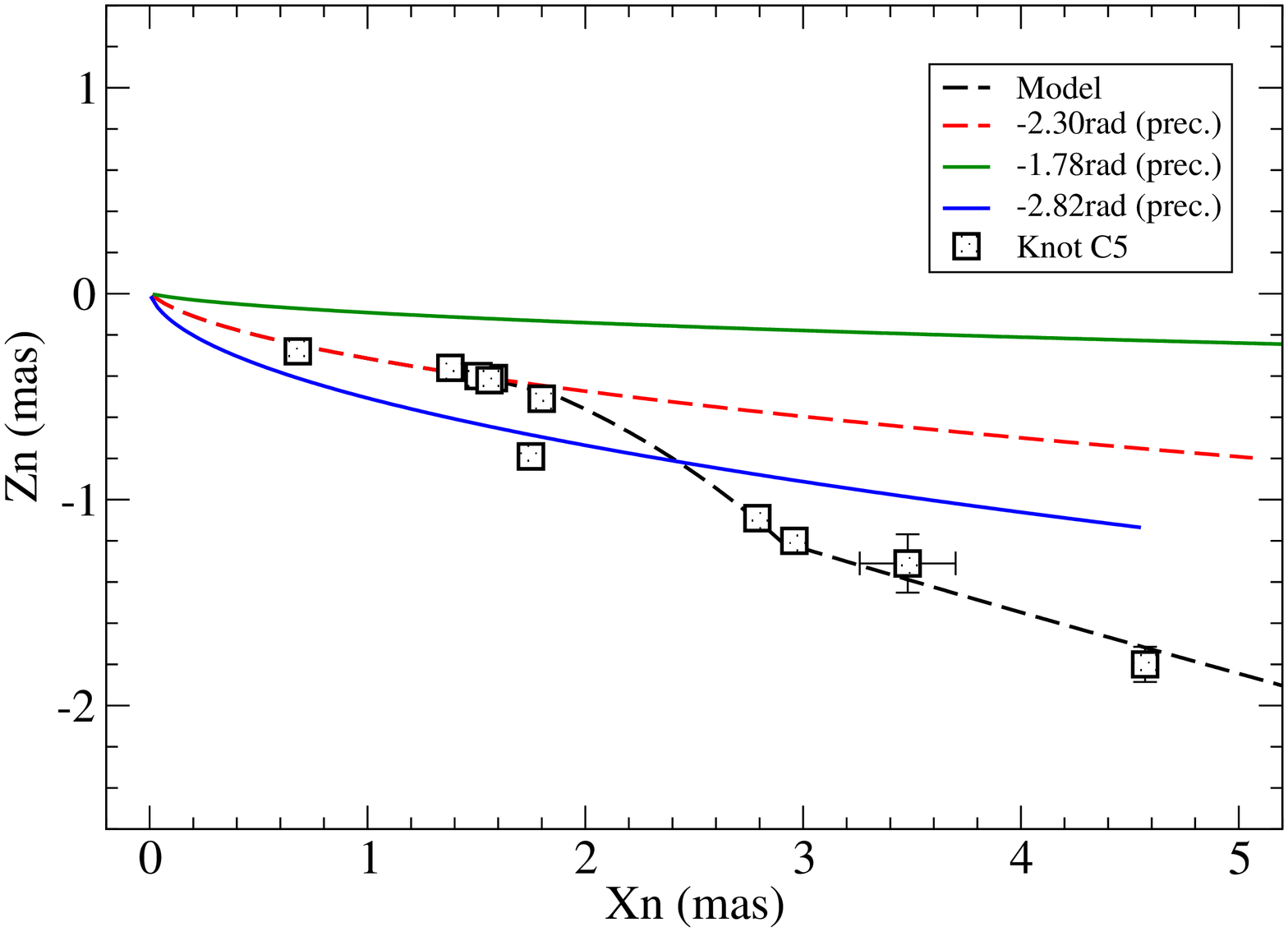}
    \caption{Model-fitting results of the kinematic features for knot C5. Its
  modeled ejection epoch $t_0$=1997.05 and the corresponding precession
  phase $\omega$=--2.3\,rad. Within core separation $r_n$=1.55\,mas knot C5
  is modeled to move along the precessing common parabolic trajectory (red
 dashed line in top left panel). The entire modeled trajectory is shown by the
  black dashed line. The green and blue lines represent the modeled
 trajectories calculated for precession phases $\omega{\pm}$0.63\,rad, showing
 all the data points being within the position angle range defined by the two
 lines and indicating the precession period having been determined within an 
 uncertainty of $\sim{\pm}$1.2\,yr. In bottom right panel the green and blue 
 lines represent the precessing common trajectories calculated for
 $\omega{\pm}$0.52\,rad, showing a number of data points being within the 
 position angle range defined by the two lines and indicating its innermost
  precessing parabolic trajectory having been observed. Knot C5 is thus 
 designated by symbol ``+'' in Table 6.}
    \end{figure*}

    \begin{figure*}
    \centering
    \includegraphics[width=5cm,angle=-90]{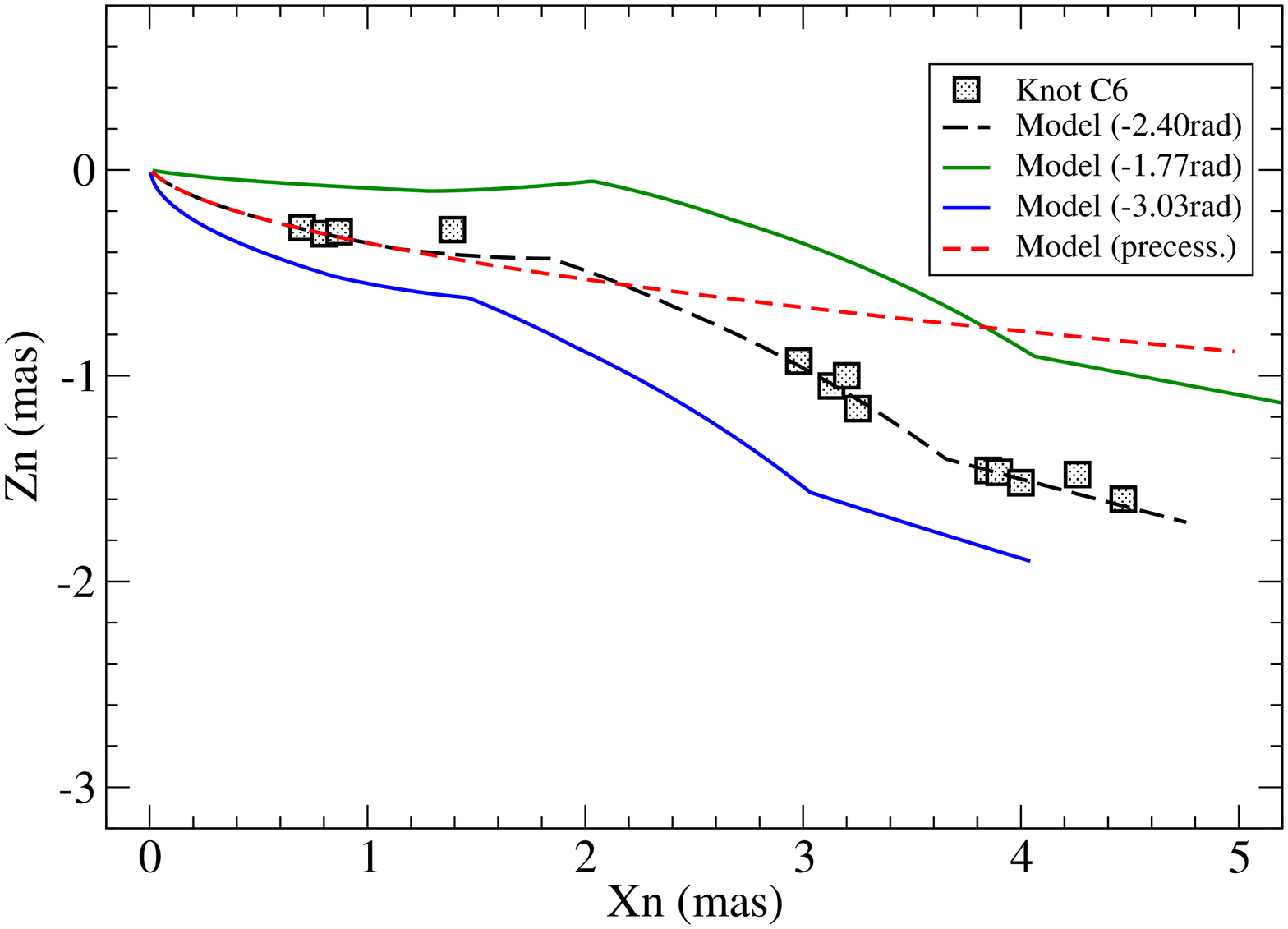}
    \includegraphics[width=5cm,angle=-90]{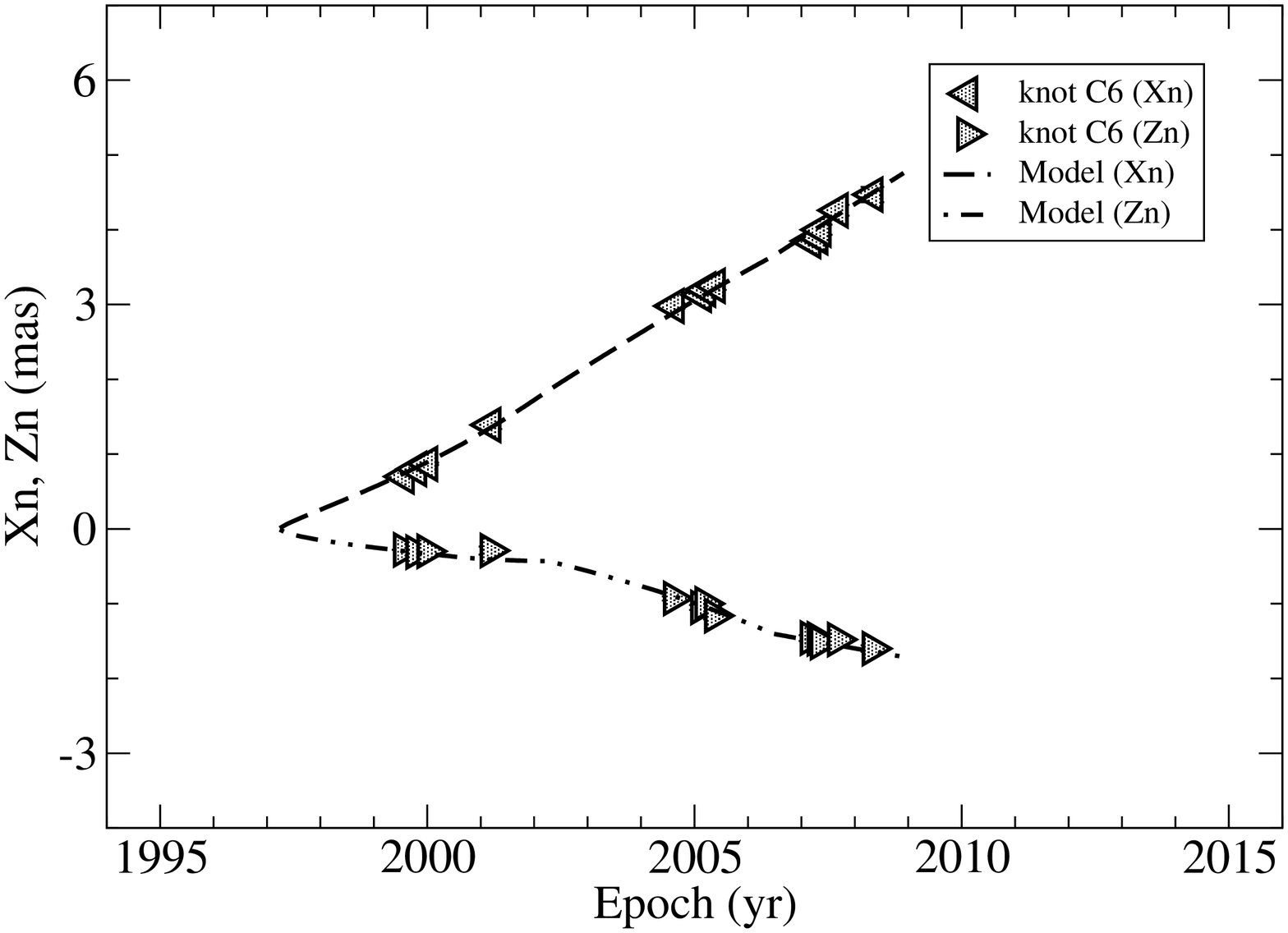}
    \includegraphics[width=5cm,angle=-90]{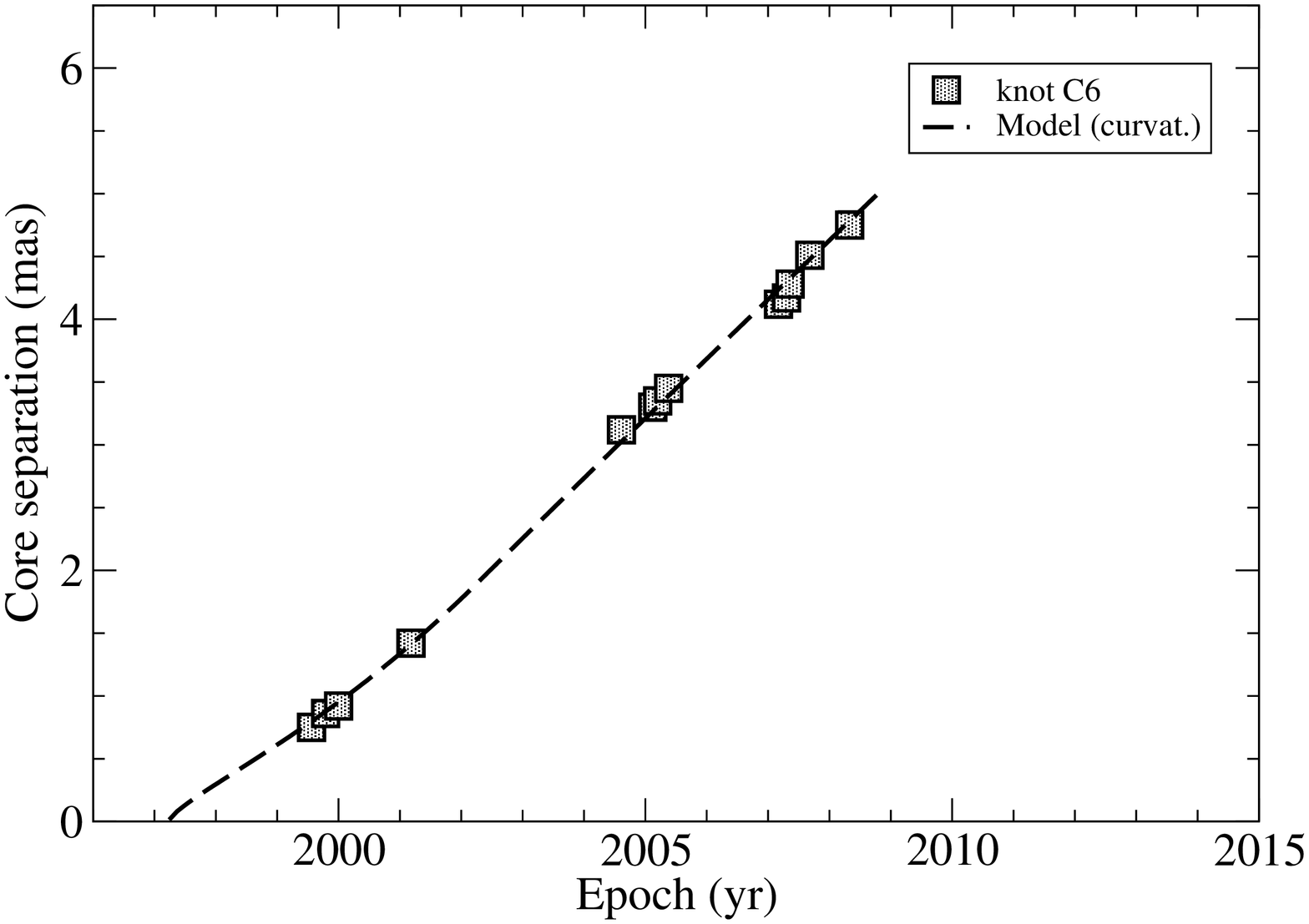}
    \includegraphics[width=5cm,angle=-90]{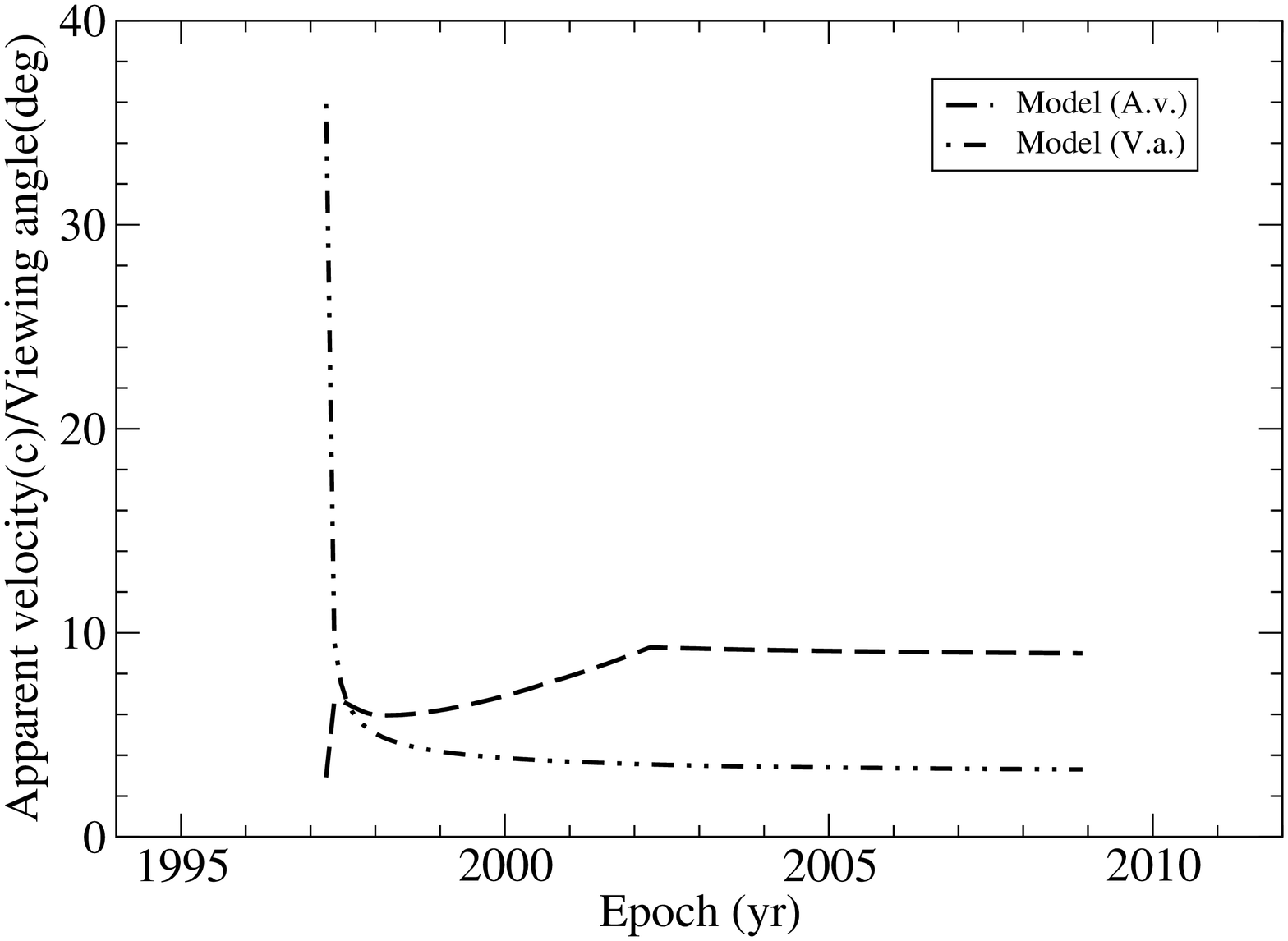}
    \includegraphics[width=5cm,angle=-90]{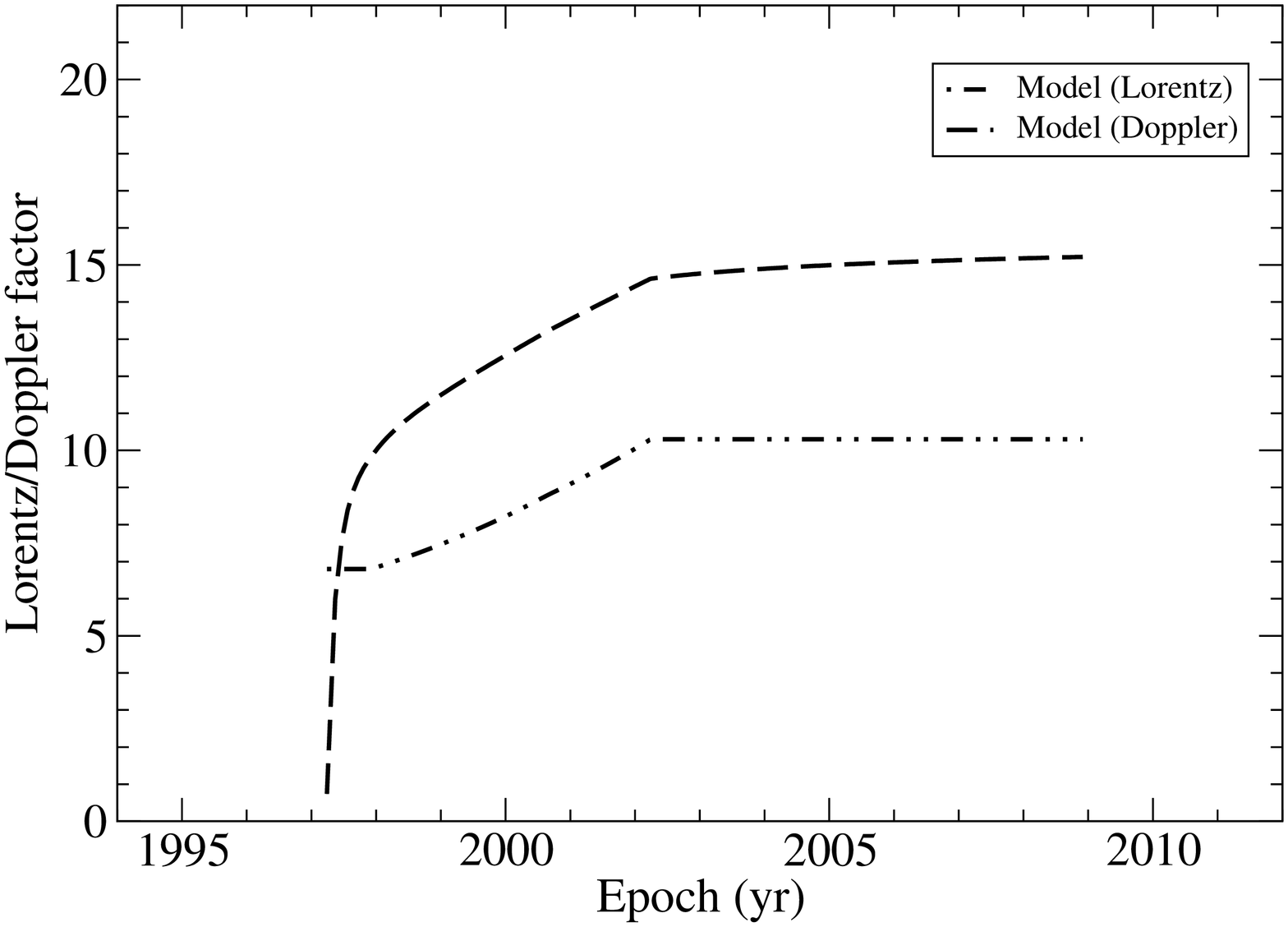}
    \includegraphics[width=5cm,angle=-90]{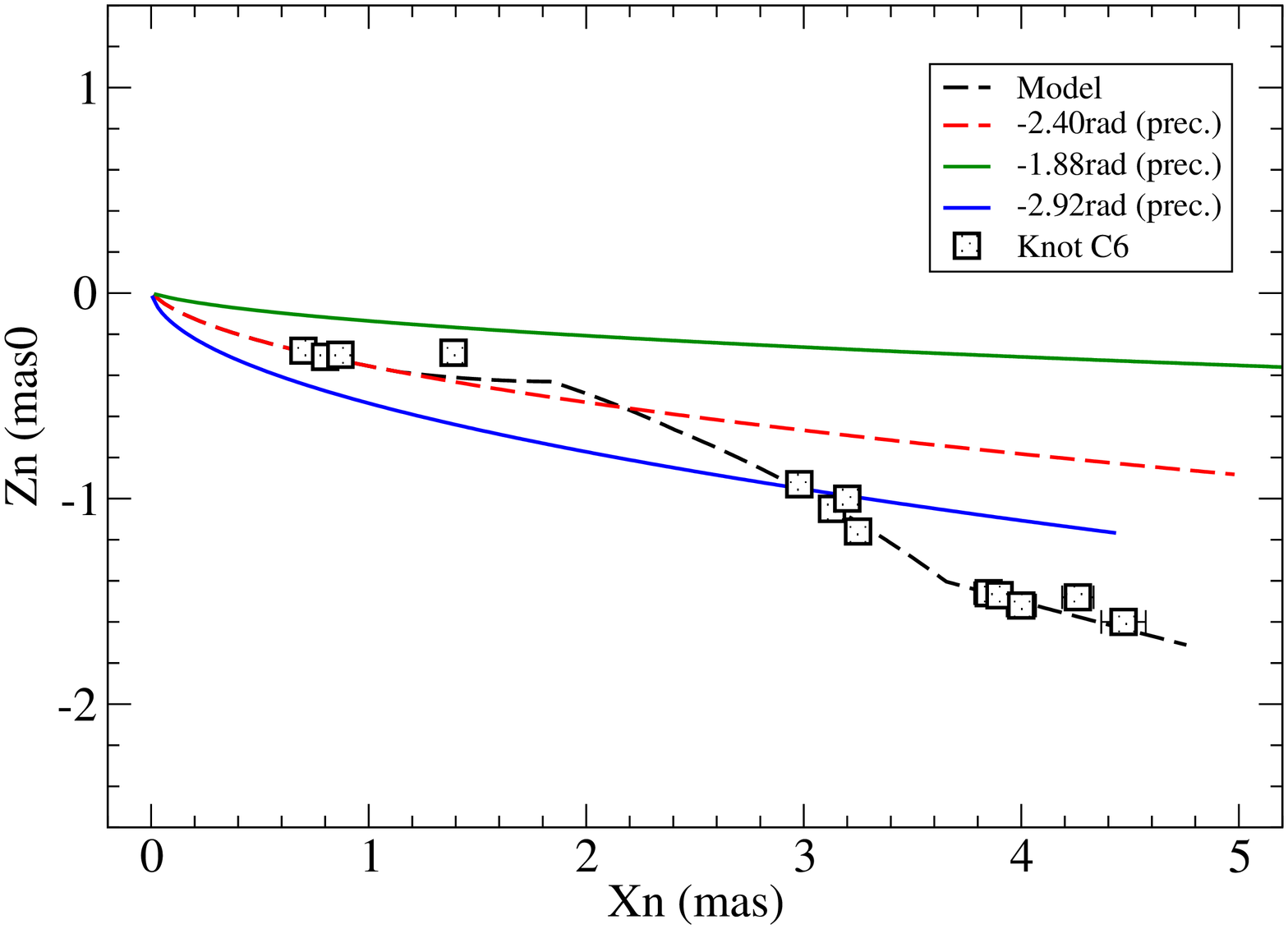}
    \caption{Model-fitting results of the kinematic features for knot C6. its 
    modeled ejection epoch $t_0$=1997.24 and the corresponding precession phase
   $\omega$=--2.40\,rad. Within core separation $r_n$=1.19\,mas knot C6 is 
   modeled to move along the precessing common parabolic trajectory (red dashed
   line in top left panel). The entire modeled trajectory is shown 
   by the black dashed line.   The green and blue lines represent the modeled
  trajectories calculated for precession phases $\omega{\pm}$0.63\,rad, showing
  all the data points being within the position angle range defined by the two
   lines and indicating the precession period having been determined within
  an uncertainty of $\sim{\pm}$1.2\,yr. In bottom right panel the green
   and blue lines represent the precessing common trajectories calculated for
   $\omega{\pm}$0.52\,rad, showing a number of data points being within the
    position angle range defined by the two lines and indicating its innermost
   precessing parabolic trajectory having been observed. Knot C6 is thus 
   designated by symbol ``+'' in Table 6.}
     \end{figure*}

  \begin{figure*}
  \centering
  \includegraphics[width=5cm,angle=-90]{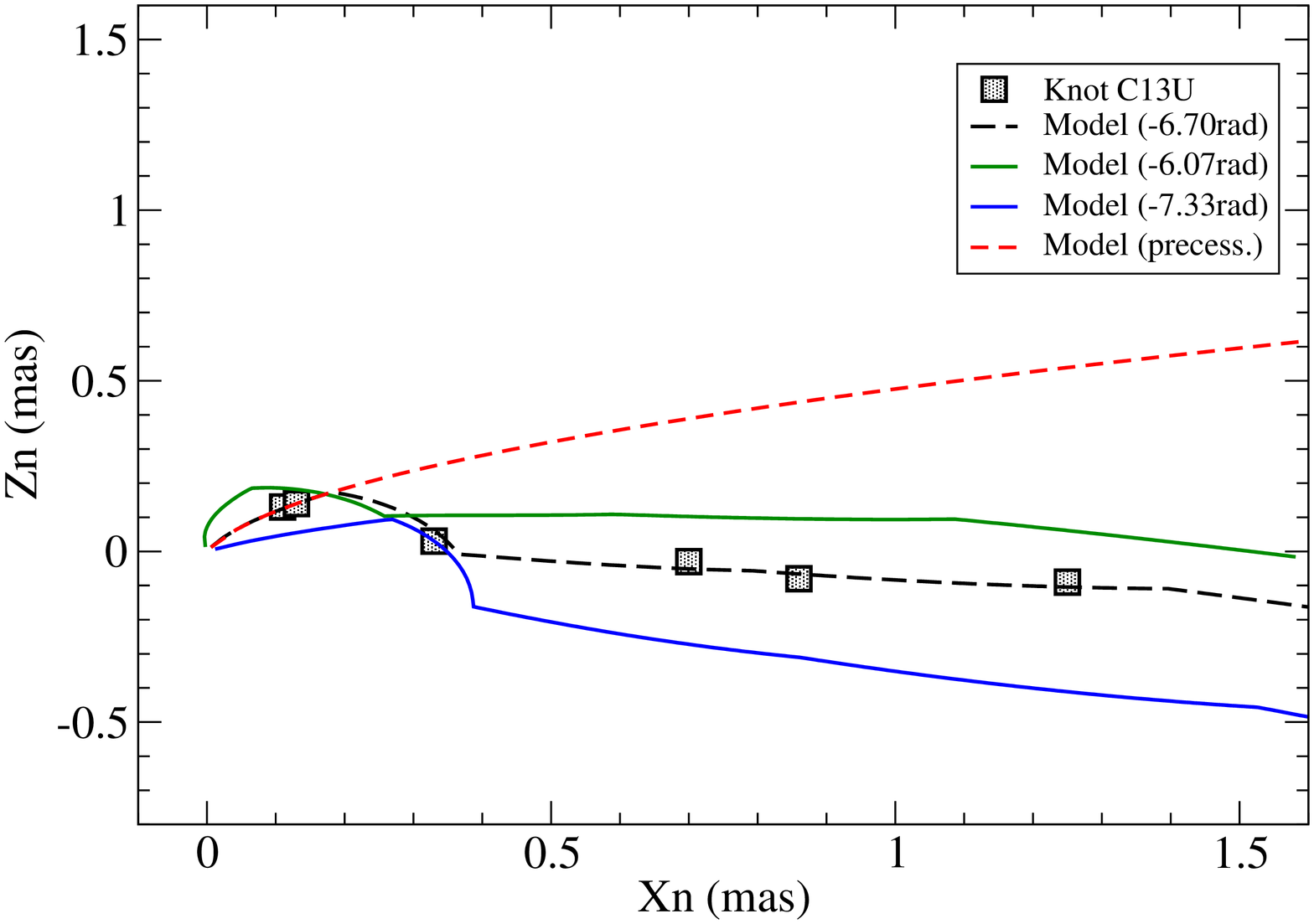}
  \includegraphics[width=5cm,angle=-90]{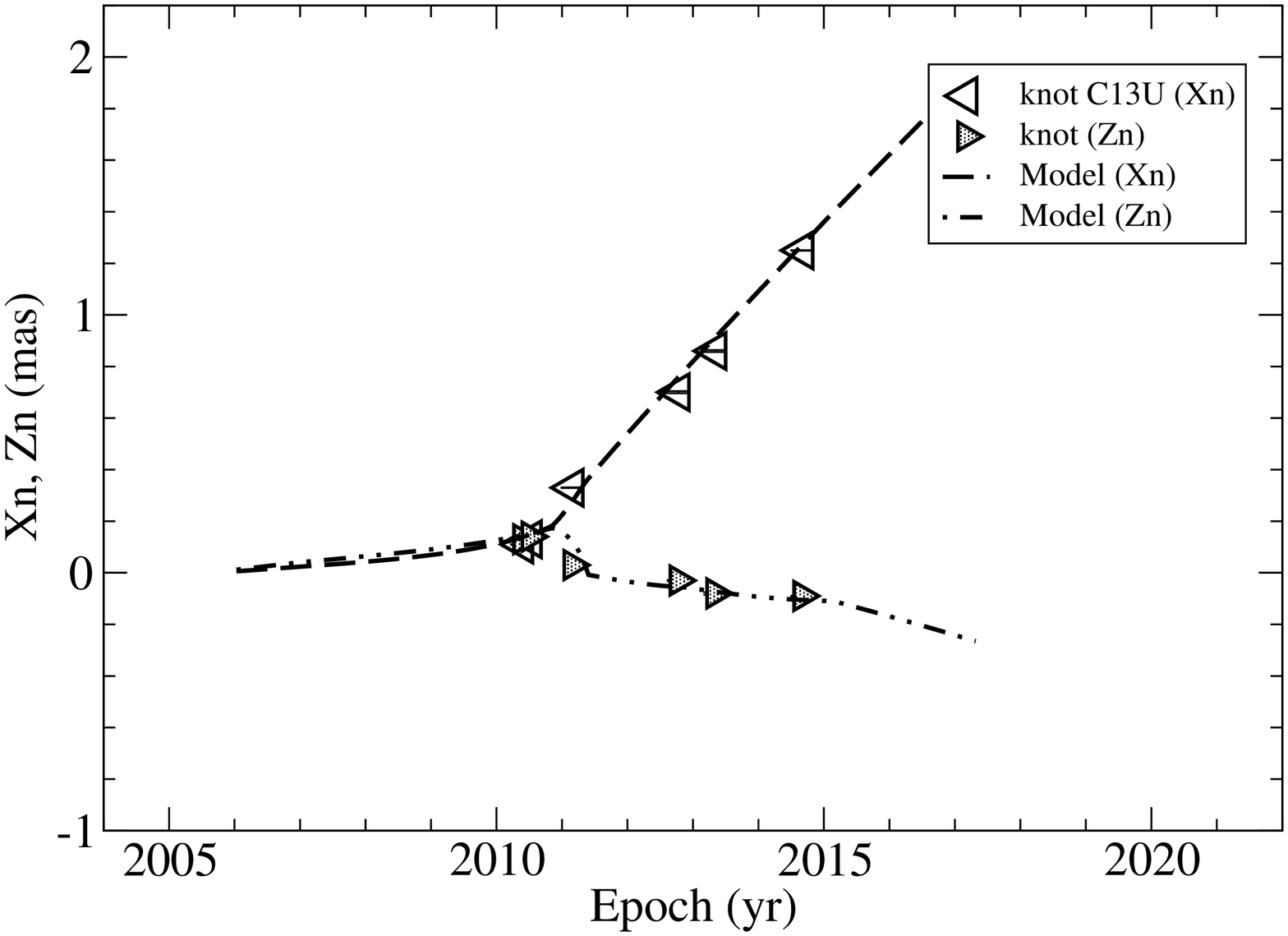}
  \includegraphics[width=5cm,angle=-90]{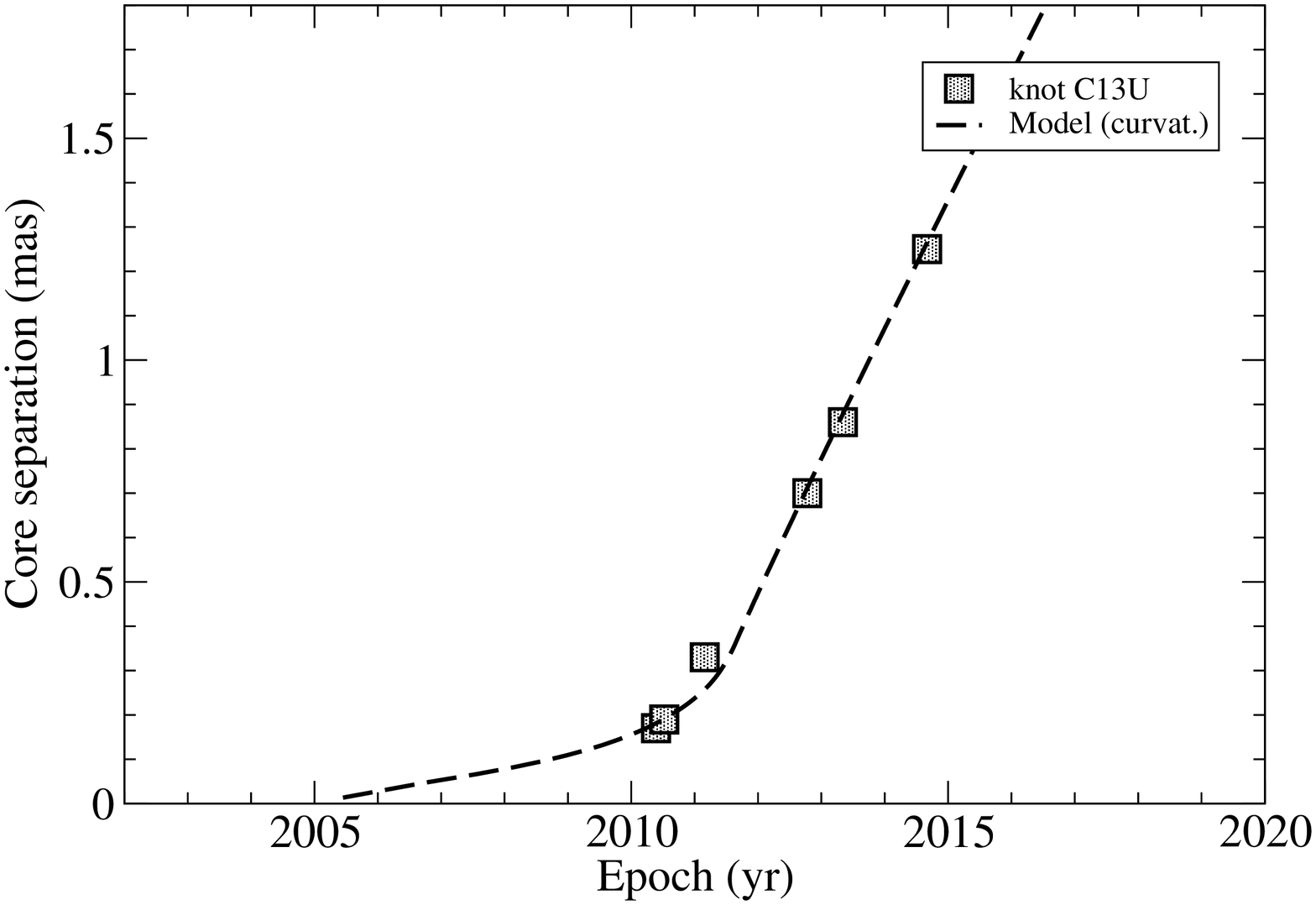}
  \includegraphics[width=5cm,angle=-90]{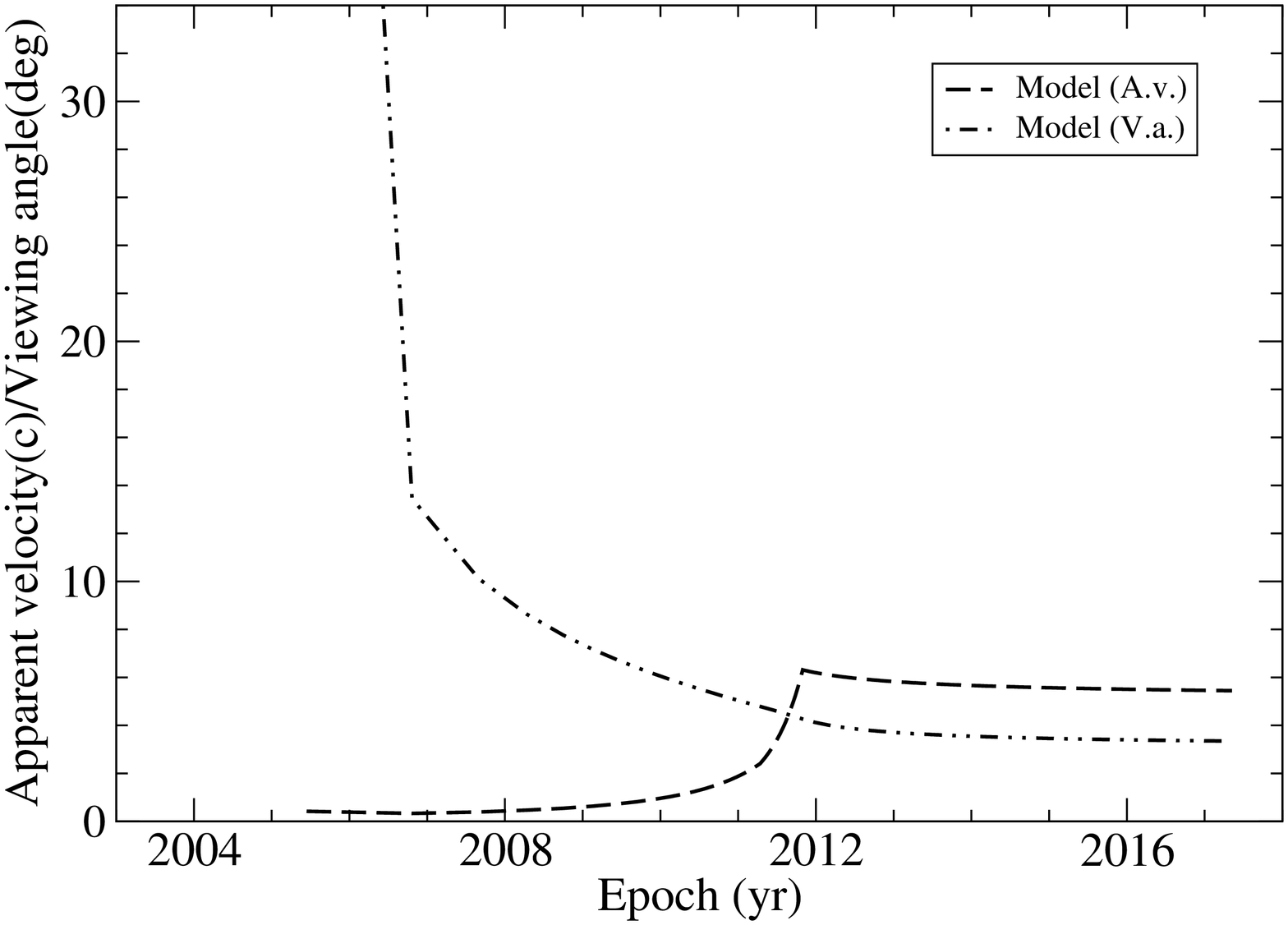}
  \includegraphics[width=5cm,angle=-90]{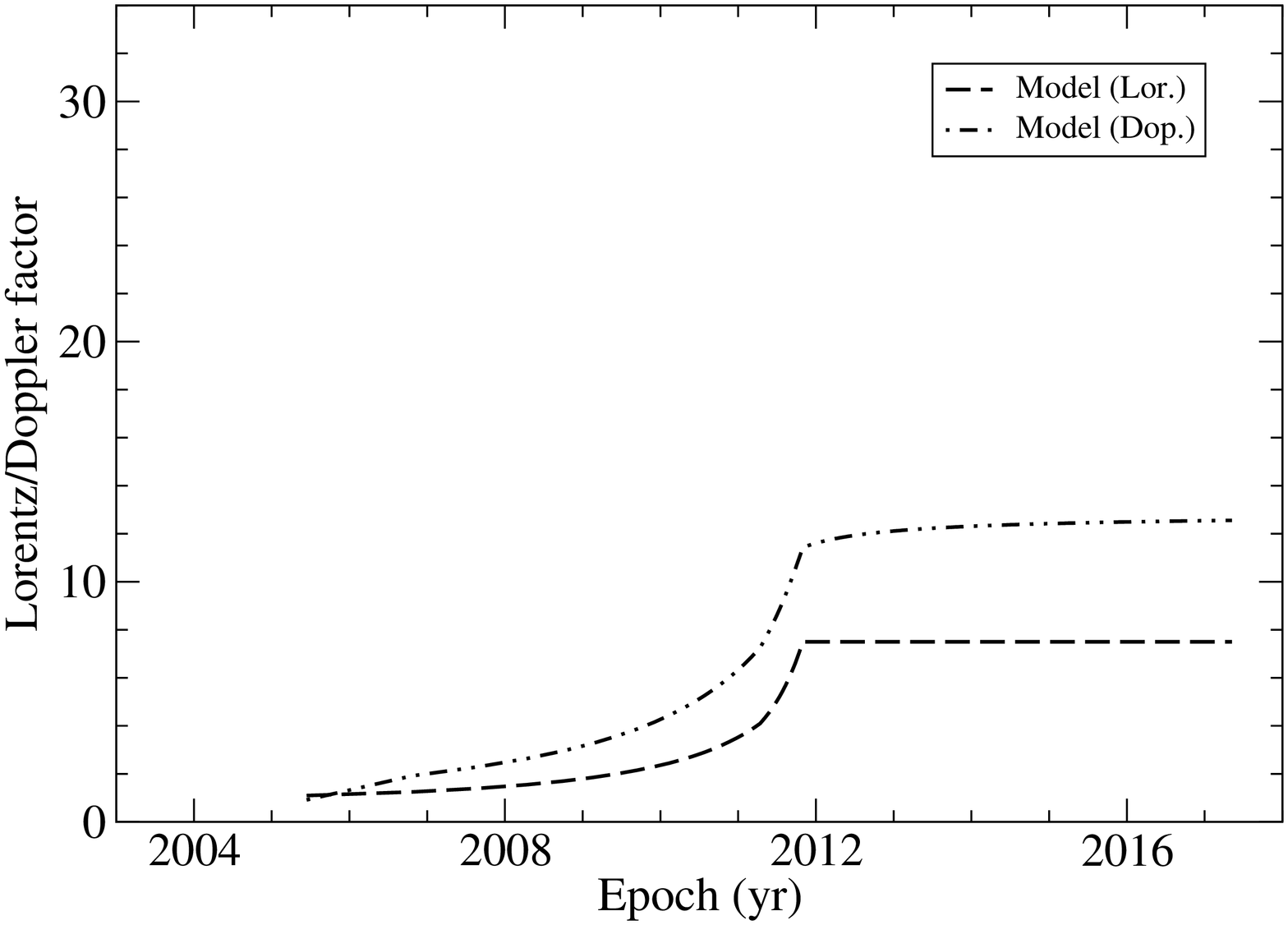}
  \includegraphics[width=5cm,angle=-90]{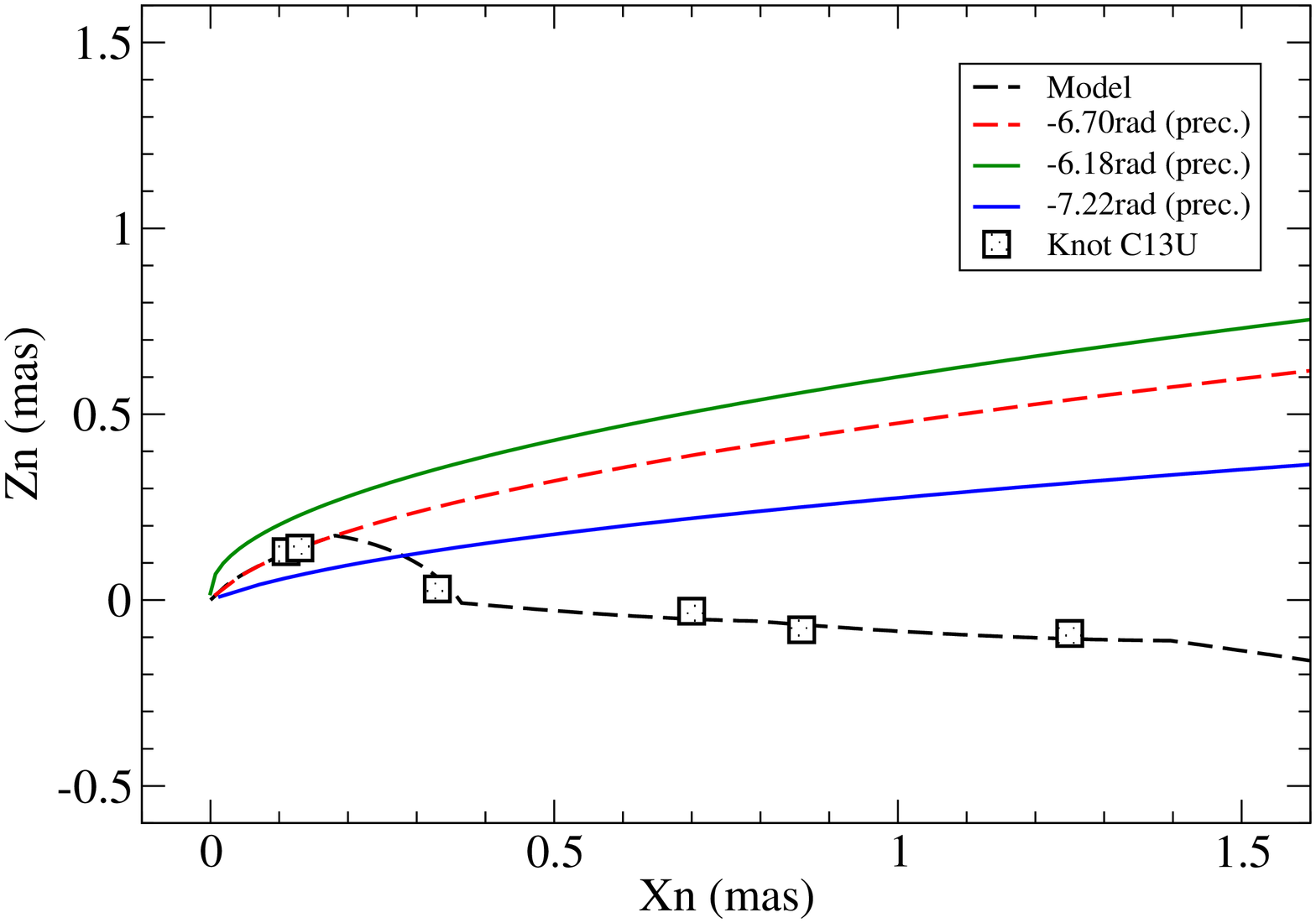}
  \caption{Model-fitting results of the kinematic features for C13U. Its 
   ejection time is modeled as $t_0$=2005.45 and the corresponding precession
   phase $\omega$=--6.70\,rad. Within core separation $r_n$=0.25\,mas knot C13U
   is modeled to move along the precessing common parabolic trajectory (red 
   dashed line in top left panel). Beyond this separation changes in parameter
    $\psi$ (or trajectory curvatures) are introduced to explain its outer
   trajectory. The entire modeled trajectory is shown by the black dashed line.
   During the period 2010.39--2014.67 the position angle of knot C13U
   changed from $\sim{-42^{\circ}}$ to $\sim{-95^{\circ}}$, showing its 
   motion along a non-ballistic trajectory (Britzen et al. \cite{Br18}).
   The green and blue lines represent the modeled trajectories calculated for
   precession phases $\omega{\pm}$0.63\,rad, showing all the data points being
   within the position angle range defined by the two lines and indicating
   the precession period having been determined within an uncertainty of 
   $\sim{\pm}$1.2\,yr. In bottom right panel, the green and blue lines
   represent the precessing common trajectories calculated for 
   $\omega{\pm}$0.52\,rad, showing a few data points being within the position
   angle range defined by the two lines and its innermost precessing trajectory
   having been observed. Thus knot C13U is designated by ``+'' in Table 6.}
  \end{figure*}

  \begin{figure*}
  \centering
  \includegraphics[width=5cm,angle=-90]{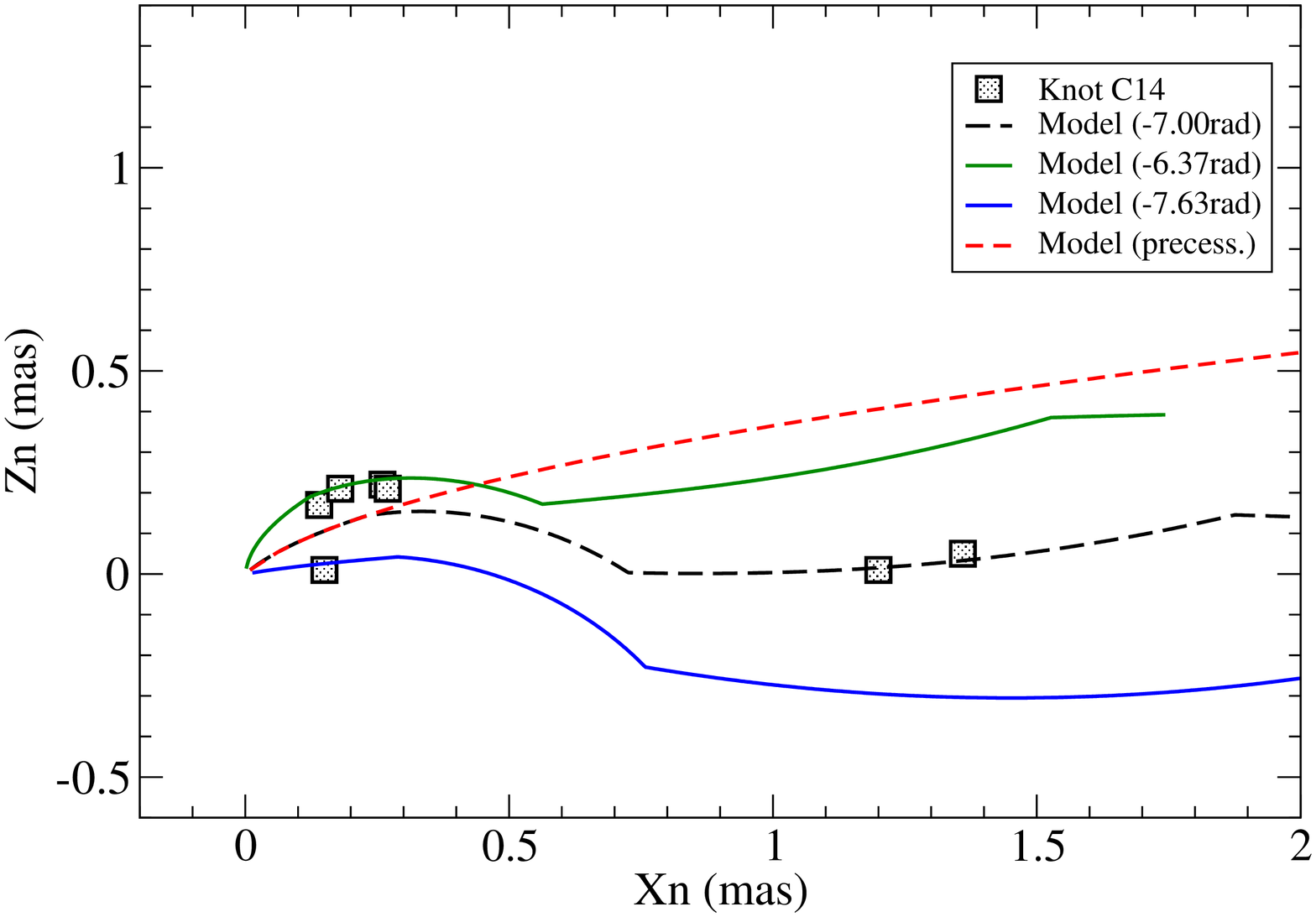}
  \includegraphics[width=5cm,angle=-90]{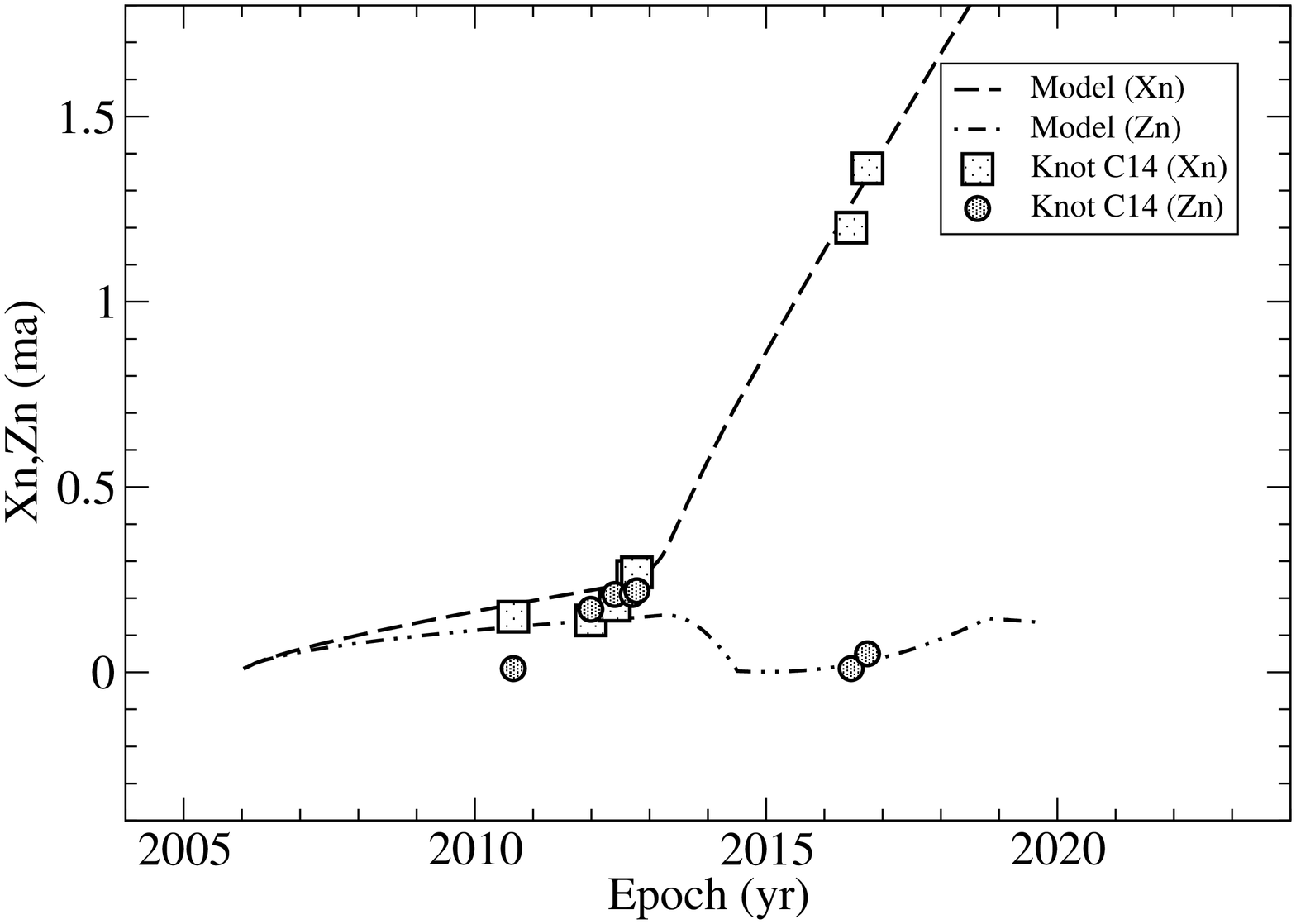}
  \includegraphics[width=5cm,angle=-90]{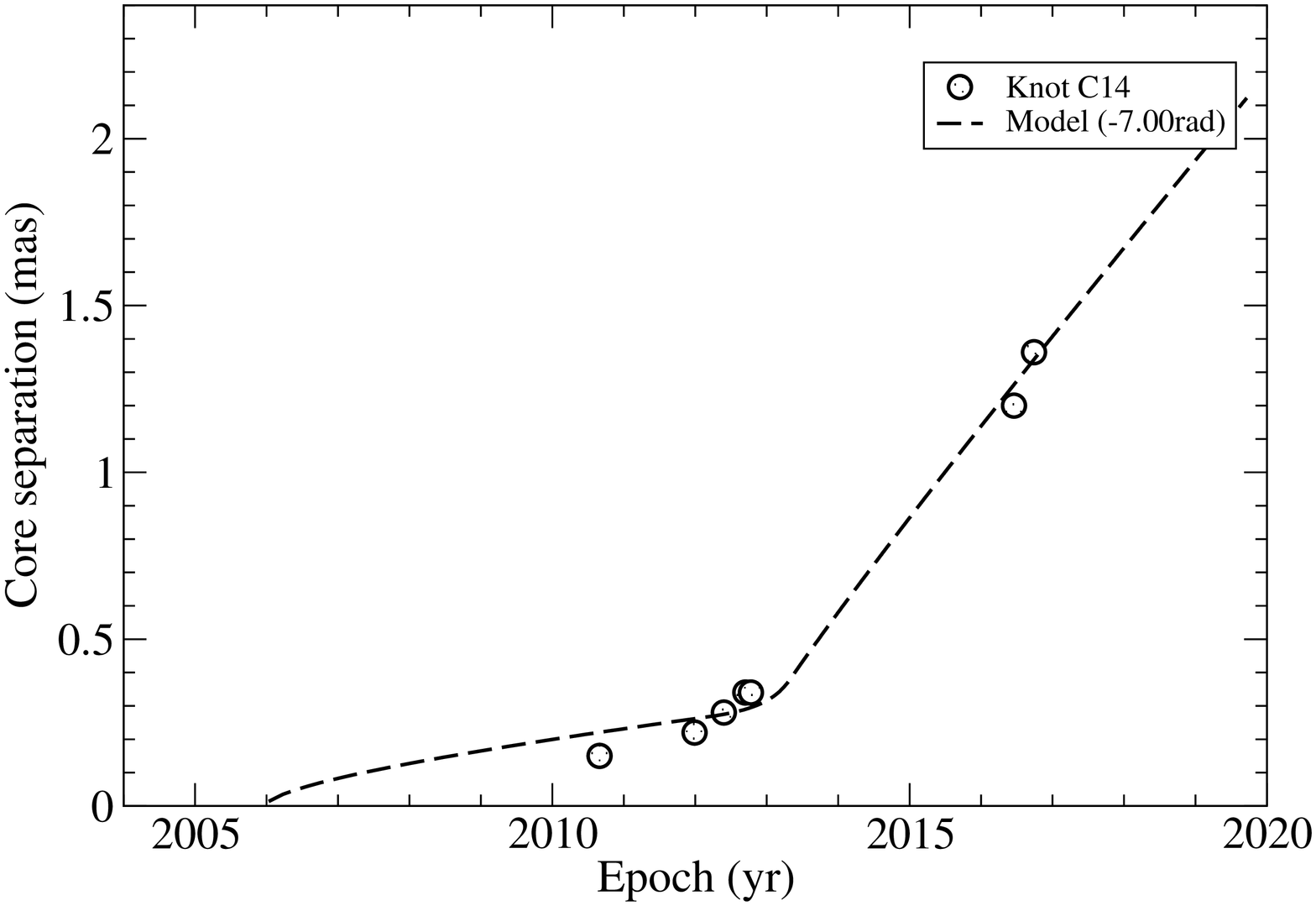}
  \includegraphics[width=5cm,angle=-90]{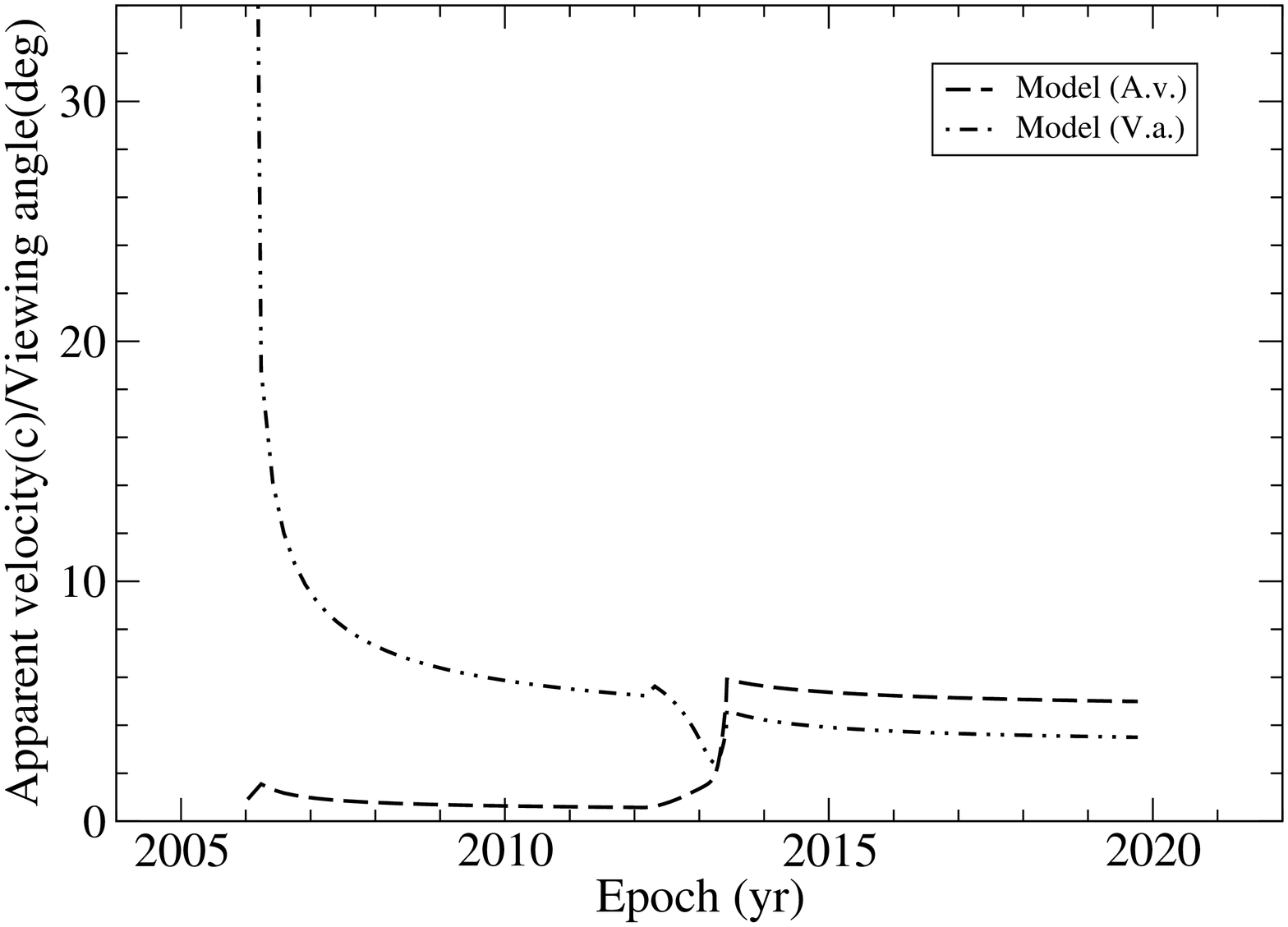}
  \includegraphics[width=5cm,angle=-90]{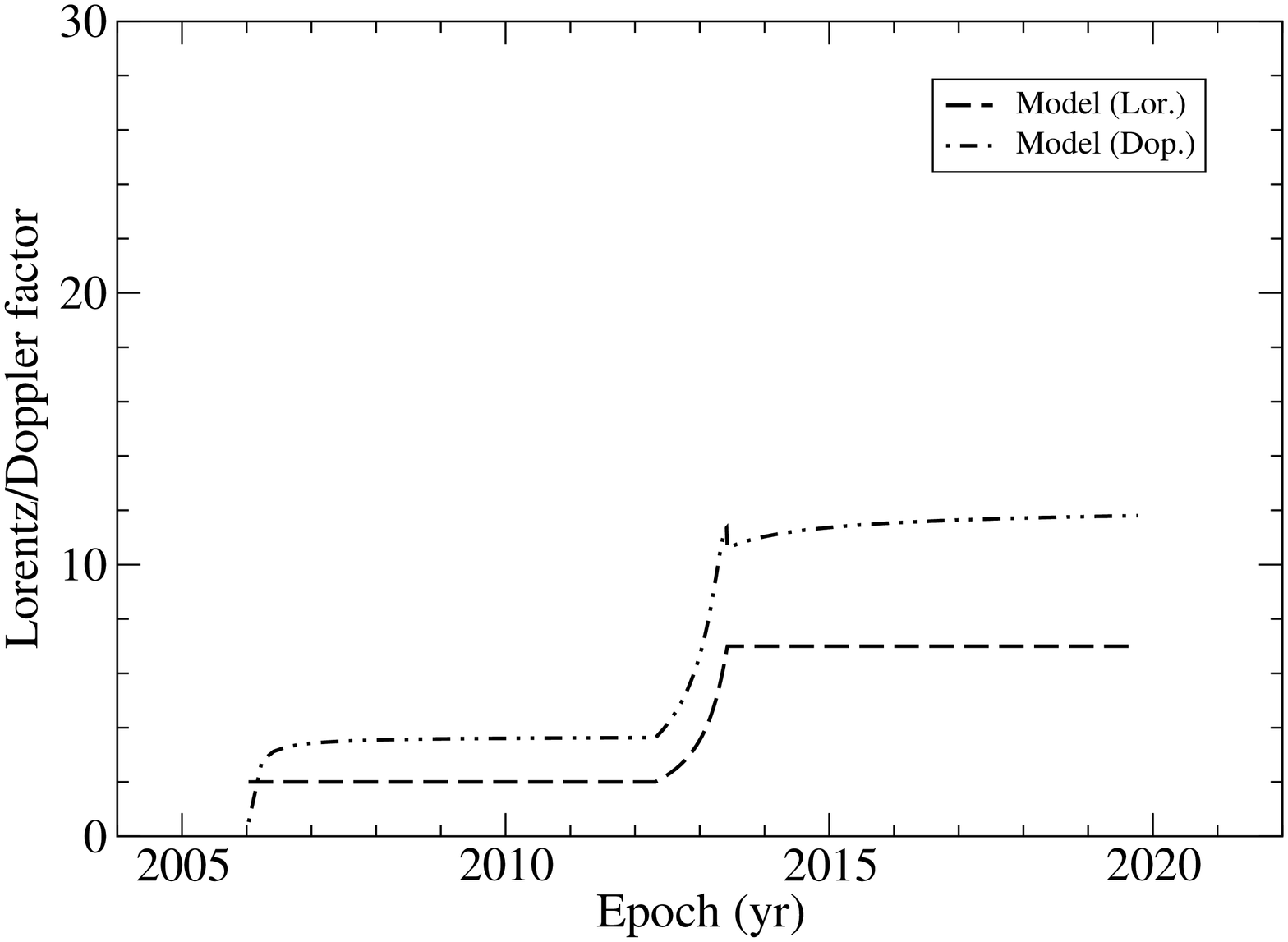}
  \includegraphics[width=5cm,angle=-90]{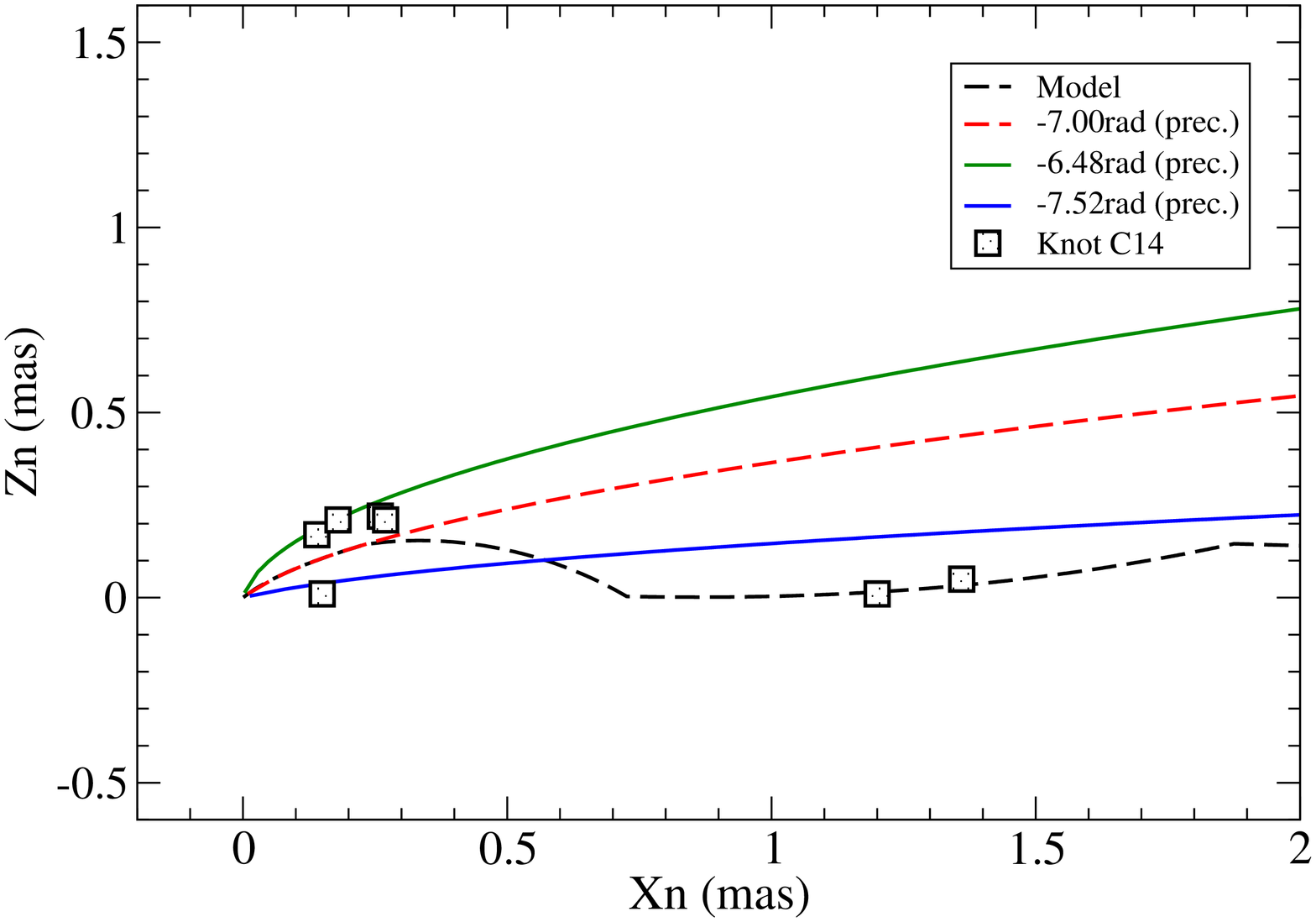}
   \caption{Model-fitting results of the kinematic features for knot C14.
     its ejection time is modeled as $t_0$=2006.03 and the corresponding 
    precession phase $\omega$=--7.00\,rad.
    Within core separation $r_n$=0.27\,mas knot C14 is modeled to  move along
    the precessing common parabolic trajectory (red dashed line in 
      top left panel). Beyond this separation changes in parameter $\psi$
   (or trajectory curvatures) are  introduced
   to explain its outer trajectory. The entire modeled trajectory is shown by
   the black dashed line. The green and blue lines represent the modeled
   trajectories calculated for precession phases $\omega{\pm}$0.63\,rad,
   showing all the data-points being within the position angle range defined by
   the two lines and indicating the precession period having been determined
    within an uncertainty of $\sim{\pm}$1.2\,yr. In bottom right panel, 
   the green and blue
  lines represent the precessing common trajectories calculated for
    $\omega{\pm}$0.52\,rad, showing a few data-points being within the
   position angle range defined by the two lines and its innermost precessing
   trajectory having been observed. Thus knot C14 is designated by 
   symbol ``+'' in Table 6.}
  \end{figure*}
  \end{appendix}

  \end{document}